\documentclass[oneside,english]{scrbook}
\usepackage[T1]{fontenc}
\usepackage[latin9]{inputenc}
\usepackage{color}
\usepackage{babel}

\usepackage{verbatim}
\usepackage{amsmath}
\usepackage{graphicx}
\usepackage{amssymb}
\usepackage{esint}
\usepackage[unicode=true, 
 bookmarks=true,bookmarksnumbered=false,bookmarksopen=false,
 breaklinks=false,pdfborder={0 0 1},backref=false,colorlinks=true]
 {hyperref}
\hypersetup{pdftitle={Quantum Fluctuations of Vector Fields and the Primordial Curvature Perturbation in the Universe},
 pdfauthor={Mindaugas Karciauskas},
 pdfsubject={PhD thesis},
 pdfkeywords={Cosmology, inflation, gravity, cosmological perturbations, vector fields}}
 
\makeatletter

\providecommand{\tabularnewline}{\\}

\usepackage{setspace} 
\hypersetup{linktocpage=true,linkcolor=blue,citecolor=blue,hyperfootnotes=false}
\title{Quantum Fluctuations of Vector Fields and the Primordial Curvature Perturbation in the Universe}
\author{\LARGE{Mindaugas Kar\v{c}iauskas}\\ \small{MSc, BSc}}
\date{\vspace{1cm} \includegraphics[width=80mm]{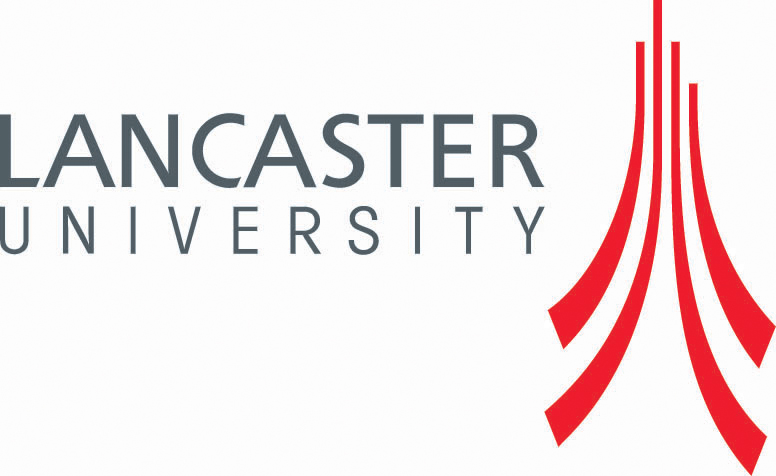}}
\publishers{ \textbf{December 2009} \\ \vspace{10pt} \normalsize{This thesis is submitted in partial fulfillment of the requirements for the degree of Doctor of Philosophy. No part of this thesis has been previously submitted for the award of a higher degree.}}
\dedication{\hspace{10pt}}

\makeatother

\begin{document}
\global\long\def\d#1#2{\mathrm{d}^{#2}#1}
 \global\long\def\mpl{m_{\mathrm{Pl}}}
 \global\long\def\pl{\parallel}

\global\long\def\fnle{f_{\mathrm{NL}}^{\mathrm{equil}}}
 \global\long\def\fnll{f_{\mathrm{NL}}^{\mathrm{local}}}
 \global\long\def\fnl{f_{\mathrm{NL}}}
 \global\long\def\fnlei{f_{\mathrm{NL,\, iso}}^{\mathrm{equil}}}
 \global\long\def\fnlli{f_{\mathrm{NL},\,\mathrm{iso}}^{\mathrm{local}}}
 \global\long\def\fnli{f_{\mathrm{NL}}^{\mathrm{iso}}}

\global\long\def\Ge{\mathcal{G}^{\mathrm{equil}}}
 \global\long\def\Gl{\mathcal{G}^{\mathrm{local}}}

\global\long\def\ket#1{\left|#1\right\rangle }
 \global\long\def\bra#1{\left\langle #1\right|}
 \global\long\def\tx{\left(t,\mathbf{x}\right)}
 \global\long\def\dt#1{,#1}
 \global\long\def\cp{\mathrm{c.p.}}

\global\long\def\el#1#2{e_{#1}^{\mathrm{L}#2}}
 \global\long\def\er#1#2{e_{#1}^{\mathrm{R}#2}}
 \global\long\def\eln#1{e_{#1}^{||}}
 \global\long\def\Te{T_{ij}^{\mathrm{even}}}
 \global\long\def\To{T_{ij}^{\mathrm{odd}}}
 \global\long\def\Tl{T_{ij}^{\mathrm{long}}}

\global\long\def\kh{\hat{\mathbf{k}}}
 \global\long\def\khp{\hat{\left(\mathbf{k}\right)}}
 \global\long\def\kp{\left(\mathbf{k}\right)}
 \global\long\def\kbi#1{\mathbf{k}_{#1}}
 \global\long\def\ki#1{k_{#1}}

\global\long\def\Pp{\mathcal{P}_{+}}
 \global\long\def\Pm{\mathcal{P}_{-}}
 \global\long\def\Pl{\mathcal{P}_{||}}
 \global\long\def\Pz#1{\mathcal{P}_{\zeta}^{\mathrm{\mathrm{#1}}}}

\global\long\def\oi{\Omega_{\mathrm{ini}}}
 \global\long\def\ohw{\hat{\Omega}_{W}}
 \global\long\def\ow{\Omega_{W}}
 \global\long\def\od{\Omega_{\mathrm{dec}}}
 \global\long\def\me#1{M^{#1}}

\global\long\def\zw{\zeta_{W}}
 \global\long\def\zg{\zeta_{\gamma}}
 \global\long\def\zend{\zeta_{\mathrm{end}}}
 \global\long\def\zinf{\zeta_{\mathrm{inf}}}
 \global\long\def\zt{\tilde{\zeta}}

\global\long\def\rg{\rho_{\gamma}}
 \global\long\def\rw{\rho_{W}}
 \global\long\def\nm{\alpha}

\global\long\def\lsim{\lesssim}
 \global\long\def\gsim{\gtrsim}
 \global\long\def\gw{\Gamma_{W}}
\global\long\def\dg{\mathcal{D}_{g}}

\pagenumbering{roman}\maketitle

\phantom{.} \vspace{207pt} \thispagestyle{empty}\singlespacing

\begin{center}
\parbox[b]{0.455\columnwidth}{%
{}``What he {[}a scientist{]} is really seeking is to learn something
new that has a certain fundamental kind of significance: a hitherto
unknown lawfulness in the order of nature, which exhibits \emph{unity}
in a \emph{broad range} of phenomena. Thus, he wishes to find in the
reality in which he lives a certain oneness and totality, or wholeness,
constituting a kind of \emph{harmony} that is felt to be beautiful.
In this respect, the scientist is perhaps not basically different
from the artist, the architect, the music composer, etc., who all
want to \emph{create} this sort of thing in their work.''

\begin{flushright}
David Bohm
\par\end{flushright}%
}
\par\end{center}

\pagebreak{}

\thispagestyle{empty}\onehalfspacing~

\pagebreak{}

\chapter*{Acknowledgments}

This thesis is a culmination of more than three years spend in England
as a PhD student, which signify not only my scientific achievements
but also people I met and moments experienced together. Although written
by me, these pages include contributions from all these people without
whom this would be just a collection of blank paper. Some of contributions
are too subtle to put into words, but each of them is essential and
invaluable. It is impossible to name everyone, so I sincerely apologize
those whose name is not mentioned here.

First and foremost I want to say thank you to Milda Bei\v{s}yt\.{e}.
Using letters and words it is impossible to express my infinite gratitude
to you. You filled my every moment with life. Without you everything
would be so much different. 

I will always be indebted to my family: my father Algirdas Kar\v{c}iauskas,
my mother Aldona Kar\v{c}iauskien\.{e}, my brother Algirdas and
his family as well as my grandmothers. Although they were far away,
I always felt their unconditional support very closely. \emph{A\v{c}i\={u}
jums labai}.

I feel extremely lucky to have Konstantinos Dimopoulos as my supervisor.
First of all, I am very grateful for a PhD project he assigned to
me. It allowed to experience a real thrill, despair and excitement
of research and discovery. I want to thank him for endless discussions
about cosmology, science and life in general. They enriched my knowledge
enormously and made me rethink many things which I took for granted.
His guidance through the maze of all aspects of becoming a doctor
was vital in many cases. Following the example of his previous student,
I want to say as well $\varepsilon\upsilon\chi\alpha\rho\iota\sigma\tau\omega$
$\pi o\lambda\upsilon$, $K\omega\sigma\tau\alpha$!

My PhD years, and especially the last one, would have been much poorer
without David Lyth. I learned so many things from him and not only
about cosmology. I am very grateful for his patience with my intelligent
and not at all intelligent questions and for his encouragements. 

I am very much obliged to each member of the Cosmology and Astroparticle
Physics Group for a particularly friendly and supportive atmosphere.
Thank you John, Anupam, Kaz, Narendra, Juan, Chia-Min, Rose, Jacques,
Francesca and Philip.

A very special thank you I would like to say to Rasa Bei\v{s}yt\.{e}.
I cannot imagine how I would have gone through my thesis writing if
not you.

I want to thank two dear friends I made in Lancaster. Thank you Ugn\.{e}
Grigait\.{e} for many unforgettable moments we shared together exploring
the English life and beauties of the Lake District. And thank you
Art\={u}ras Jasiukevi\v{c}ius, you were a close and much needed
companion in experiencing cultural shocks during my first year in
England.

It is difficult to express my gratefulness to Paul Taylor and Rev.
Wilfrid Powell for sharing their path. It was an island when drowning
seemed the only option and it is a constant source of hope.

In my last year I met a person with endless optimism and enthusiasm
- Annmarie Ryan. Thank you for your drumming classes and thanks for
all the group with whom we have been creating and sharing moments
of beauty with African rhythms.

It would be unfair if I didn't mention my friends and comrades back
in Lithuania. Knowing and feeling their presence was a great support
while in England and it was always a great fun to visit them. Unfortunately,
thickness limitations for this thesis does not allow to name each
of you separately. But I am sure you know who you are. Thank you very
much!

The final touches and improvements to this thesis were made due to
my examiners Dr. John McDonald and Prof. Anne-Christine Davis. Thank
you for this and for approving me to become a Doctor of Philosophy.

And finally I want to thank Physics Department of Lancaster University.
Without their financial support this thesis and these acknowledgments
would not even exist. I also want to thank the Faculty of Science
and Technology and the William Ritchie travel fund for supporting
my travels to conferences.

~

\noindent \emph{1 September 2010}

\pagebreak{}

\chapter*{Abstract}

The successes and fine-tuning problems of the Hot Big Bang theory
of the Universe are briefly reviewed. Cosmological inflation alleviates
those problems substantially and give rise to the primordial curvature
perturbation with the properties observed in the Cosmic Microwave
Background. It is shown how application of the quantum field theory
in the exponentially expanding Universe leads to the conversion of
quantum fluctuations into the classical field perturbation. The $\delta N$
formalism is reviewed and applied to calculate the primordial curvature
perturbation $\zeta$ for three examples: single field inflation,
the end-of-inflation and the curvaton scenarios.

The $\delta N$ formalism is extended to include the perturbation
of the vector field. The latter is quantized in de Sitter space-time
and it is found that in general the particle production process of
the vector field is anisotropic. This anisotropy is parametrized by
introducing two parameters $p$ and $q$, which are determined by
the conformal invariance breaking mechanism. If any of them are non-zero,
generated $\zeta$ is statistically anisotropic. Then the power spectrum
of $\zeta$ and the non-linearity parameter $\fnl$ have an angular
modulation.

This formalism is applied for two vector curvaton models and the end-of-inflation
scenario. It is found that for $p\ne0$, the magnitude of $\fnl$
and the direction of its angular modulation is correlated with the
anisotropy in the spectrum. If $\left|p\right|\gtrsim1$, the anisotropic
part of $\fnl$ is dominant over the isotropic one. These are distinct
observational signatures; their detection would be a smoking gun for
a vector field contribution to $\zeta$.

In the first curvaton model the vector field is non-minimally coupled
to gravity and in the second one it has a time varying kinetic function
and mass. In the former, only statistically anisotropic $\zeta$ can
be generated, while in the latter, isotropic $\zeta$ may be realized
too. Parameter spaces for these vector curvaton scenarios are large
enough for them to be realized in the particle physics models. In
the end-of-inflation scenario $\fnl$ have similar properties to the
vector curvaton scenario with additional anisotropic term.

\pagebreak{}

\thispagestyle{empty}~

\pagebreak{}

\tableofcontents{}

\pagebreak{}

\thispagestyle{empty}~

\pagebreak{}

\chapter{The Hot Big Bang and Inflationary Cosmology}

\pagenumbering{arabic}

In the first Chapter of this thesis we start by reviewing briefly
the Hot Big Bang (HBB) model of the Universe. It is one of the greatest
achievements of the last century in understanding the structure and
evolution of the Universe from the first second until today, $13.7\times10^{9}$
years later. Predictions of the HBB model are in very impressive agreement
with the observed distribution of the large scale structure and with
the abundance measurements of the light elements. However, in addition
to the dark matter and dark energy problems, the HBB model suffers
from the need to fine tune initial conditions. The latter motivates
us to look at the earlier stage of the evolution of the Universe.
Currently the most popular and most predictive paradigm for this epoch
is the inflationary scenario, which is introduced in section~\ref{sec:Inflation}.

\section{Kinematics of HBB\label{sec:Kinematics-of-HHB}}

The HBB model relies on a hypothesis called the Cosmological Principle
which states that the Universe is spatially homogeneous and isotropic
on sufficiently large scales. 

The physical model of the Universe, is divided into two parts. One
part describes the large scale behavior of the system and possesses
high degree of symmetries so that mathematical models become simple
and equations relatively easy to calculate. This is a background model.
The second part deals with the deviations from the simplistic description
of the background. These deviations are considered to be small compared
to the background values. They don't influence the large scale behavior
of the system: only the region much smaller than the scale on which
background is defined. 

The Cosmological Principle is a hypothesis about the properties of
the background distribution of matter in the Universe. The background
is defined as the smeared-out distribution of matter, with smearing
performed on large enough scales so that the distribution appears
smooth. However, a priory it is not clear that such scales do exist.
It might be that probing larger and larger cosmological scales, we
constantly discover new structures. This would happen if galaxies
are distributed hierarchically at all distances as in the fractal
Universe. In such a Universe probing larger and larger distances we
find galaxies, clusters of galaxies, clusters of clusters of galaxies
and so on.%
\begin{figure}
\begin{centering}
\includegraphics[width=7.5cm]{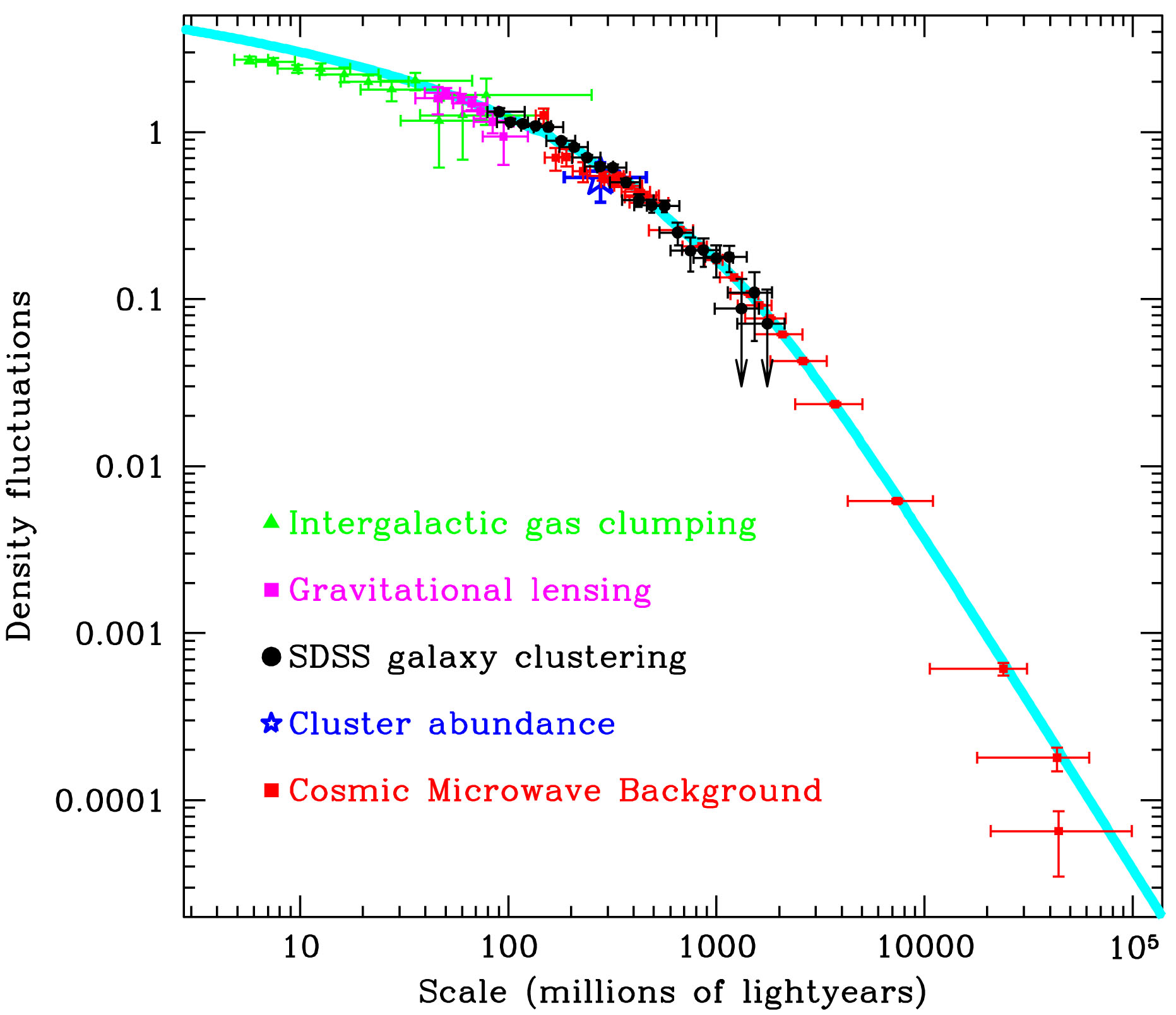}
\par\end{centering}

\caption{\label{fig:smoothing_and_perturbations}Density fluctuations as the
function of the size of the smoothing scale. The thick line represents
a model with a scale invariant power spectrum and cosmological parameters
$\Omega_{m}=0.28$, $H=72\,\mathrm{km}\,\mathrm{s}^{-1}\,\mathrm{Mpc}^{-1}$
and $\Omega_{b}/\Omega_{m}=0.16$, $\tau=0.17$, were $\Omega_{m}$
and $\Omega_{b}$ is the dark matter and baryon density parameters
respectively, $H$ is the Hubble parameter in physical units and $\tau$
is the optical depth \cite{MaxTegmark_Web}. }

\end{figure}
However, in the real Universe there is a scale at which the hierarchical
structure stops and the Universe may be considered smooth. From the
Figure~\ref{fig:smoothing_and_perturbations} one can see that the
Universe looks smoother and smoother if we probe it on larger scales.
At around few hundreds Mpcs, which correspond to the size of largest
superclusters, perturbations becomes smaller than the background value.
At these scales separation of the matter distribution into the smooth
background value and small perturbations is well justified.

The isotropy hypothesis of the Cosmological Principle is supported
by observations too. The strongest evidence comes from the measurements
of the temperature irregularities of the Cosmic Microwave Background
(CMB). The COBE satellite was the first to measure these irregularities
\cite{Smoot_etal(1992)COBE}. They showed that the anisotropy in the
temperature distribution is only of order $\Delta T/T\sim10^{-5}$.
In addition the evidence for the isotropy of the Universe is further
supported by the galaxy redshift surveys, measurements of peculiar
velocities of galaxies, distribution of radio galaxies, X-ray background
and the Lyman-$\alpha$ forest \cite{Lahav_(2000)CosmPrinc}. Another
assumption of the cosmological principle, the homogeneity of the Universe,
is an inevitable conclusion if we assume the validity of the Copernican
Principle. This principle states that our location in the Universe
is not central or somehow special. Combined with the evidence of the
isotropy the outcome of the Copernican Principle is that the Universe
is isotropic around every point. It can be shown that the last statement
leads to the conclusion of spatial homogeneity.

Accepting the validity of the Cosmological Principle we can find the
metric for the homogeneous and isotropic Universe. This can be done
using only geometric considerations \cite{Trodden_Carroll(2004)intro}
giving the proper time interval as

\begin{equation}
\d{s^{2}}{}=\d t{}^{2}-a^{2}\left(t\right)\left[\frac{\d{x^{2}}{}}{1-Kx^{2}}+x^{2}\left(\d{\theta^{2}}{}+\sin^{2}\theta\d{\phi^{2}}{}\right)\right].\label{eq:FRW-metric-Spherical}\end{equation}
This metric is expressed in spherical coordinates and is called the
Friedmann-Robertson-Walker (FRW) metric. $t$ in Eq.~\eqref{eq:FRW-metric-Spherical}
is the coordinate time and spatial coordinates $l\left(t\right)\equiv a\left(t\right)x$
are decomposed into the comoving coordinates $x$ which are constant
in time and the time dependent scale factor $a\left(t\right)$ which
parametrizes the evolution of the Universe, i.e. its expansion or
contraction. In this metric, $K$ parametrizes the curvature of space-time:
if $K<0$, the Universe is spatially open, if $K>0$ it is closed
and if $K=0$ it is flat. As will be seen later, the inflationary
paradigm predicts $K\approx0$, which is in a very good agreement
with observations. Therefore, in Chapters~\ref{cha:Scalars} and
\ref{cha:Vectors} we consider only the flat Universe in order to
dispense with the unnecessary complications related with the curvature
term. Furthermore, instead of using the spherical coordinate system
in Eq.~\eqref{eq:FRW-metric-Spherical} in many situations it will
be more convenient to use the Cartesian coordinate system. Then the
flat ($K=0$) FRW metric in Eq.~\eqref{eq:FRW-metric-Spherical}
takes a simple form

\begin{equation}
\mathrm{d}s^{2}=\mathrm{d}t^{2}-a^{2}\left(t\right)\mathrm{d}x^{i}\mathrm{d}x^{j}.\label{eq:FRW-metric-flat-Cartesian}\end{equation}

\section{Dynamics of the HBB\label{sec:HBB-dynamics}}

From Eq.~\eqref{eq:FRW-metric-Spherical} we have seen that the evolution
of the isotropic and homogeneous Universe may be described by only
one parameter, the scale factor $a\left(t\right)$. To determine the
dynamics of $a\left(t\right)$ we have to specify the energy content
of the Universe. In most situations it may be well approximated by
an ideal fluid whose energy-momentum tensor is \begin{equation}
T_{\nu\mu}=\left(\rho+p\right)u_{\mu}u_{\nu}+pg_{\mu\nu},\end{equation}
where $u_{\mu}$ is the velocity four-vector of the fluid, $g_{\mu\nu}$
is the metric and $\rho$, $p$ are the energy density and pressure
of the fluid respectively. 

Using the FRW metric in Eq.~\eqref{eq:FRW-metric-Spherical} and
the energy momentum conservation law $\nabla_{\nu}T^{\mu\nu}=0$,
where $\nabla_{\nu}$ is the covariant derivative, we find

\begin{equation}
\dot{\rho}=-3H\left(\rho+p\right),\label{eq:FRW-rho-continuity-t}\end{equation}
where $H$ is the Hubble parameter defined as\begin{equation}
H\equiv\frac{\dot{a}}{a}\end{equation}
and the dot denotes the derivative with respect to the coordinate
time $t$. Eq.~\eqref{eq:FRW-rho-continuity-t} can also be rewritten
as\begin{equation}
a\frac{\d{\rho}{}}{\d a{}}=3\left(\rho+p\right).\label{eq:rho-continuiti-a}\end{equation}
For the perfect fluid the pressure is uniquely related to the energy
density which is conveniently parametrized by the equation of state.
Assuming a barotropic fluid we have\begin{equation}
p=w\rho,\label{eq:equation-of-state-general}\end{equation}
where $w$ is called the barotropic parameter. For different kinds
of perfect fluid $w$ will have different values, for example, for
non-relativistic pressureless matter (sometimes called `dust') $w=0$,
for radiation (relativistic particles) $w=1/3$ or $w=-1$ for the
vacuum energy. Using the equation of state, from the continuity equation~\eqref{eq:rho-continuiti-a}
it is easy to find the evolution of the energy density of the perfect
fluid by integration:\begin{equation}
\rho=\rho_{0}\left(\frac{a}{a_{0}}\right)^{-3\left(1+w\right)},\label{eq:FRW-rho-scaling-general}\end{equation}
where '$0$' denotes initial values. Hence, it is clear that the energy
density scales as $\rho\propto a^{-3}$ for the pressureless matter,
$\rho\propto a^{-4}$ for the relativistic matter and $\rho=\mathrm{constant}$
for the vacuum energy.

To find how the content of the Universe determines the time evolution
of the scale factor $a\left(t\right)$ a theory of gravity must be
assumed. For the purpose of this thesis it will be enough to consider
only Einstein's theory of General Relativity (GR) with the field equation\begin{equation}
R_{\mu\nu}-\frac{1}{2}g_{\mu\nu}R=\mpl^{2}T_{\mu\nu},\label{eq:Einstein-eqs}\end{equation}
where $R_{\mu\nu}$ and $R$ are the Ricci tensor and scalar respectively.
Using temporary component of Einstein's field equation we can find
the Friedmann equation. With the spatially homogeneous and isotropic
metric in Eq.~\eqref{eq:FRW-metric-Spherical} it becomes\begin{equation}
H^{2}=\frac{\rho}{3\mpl^{2}}-\frac{K}{a^{2}}.\label{eq:FRW-Friedmann-eq}\end{equation}
Furthermore, the acceleration of the Universe is obtained using the
spatial components of Eq.~\eqref{eq:Einstein-eqs}. Together with
Eq.~\eqref{eq:FRW-Friedmann-eq} they give

\begin{equation}
\frac{\ddot{a}}{a}=-\frac{\rho+3p}{6\mpl^{2}},\label{eq:FRW-a-ddot}\end{equation}
which can be expressed in terms of the Hubble parameter as\begin{equation}
\dot{H}+H^{2}=-\frac{\rho+3p}{6\mpl^{2}}.\label{eq:HBB-dotH}\end{equation}

It is often useful to introduce the parameter $\Omega$, which is
related to the curvature of the space time \cite{Lyth_Liddle(2009)book}:\begin{equation}
\Omega-1\equiv\frac{K}{a^{2}H^{2}}.\label{eq:Omega-def}\end{equation}
In the Einstein gravity this quantity measures the energy density
of the Universe $\rho$ relative to the energy density of the flat
Universe $\rho_{\mathrm{c}}$ for given $H$, called the critical
energy density. This can be seen from the Friedmann equation~\eqref{eq:FRW-Friedmann-eq}:
for zero curvature $K=0$, the critical energy density is \begin{equation}
\rho_{\mathrm{c}}=3\mpl^{2}H^{2}.\end{equation}
Plugging this back into the Friedmann equation and using Eq.~\eqref{eq:Omega-def},
we find that in the Einstein's gravity\begin{equation}
\Omega=\frac{\rho}{\rho_{\mathrm{c}}}.\label{eq:FRW-Omega-from-Friedmann-eq}\end{equation}
If the energy density of the Universe is critical, $\Omega=1$, the
Friedmann equation becomes\begin{equation}
H^{2}=\frac{\rho}{3\mpl^{2}}.\label{eq:FRW-Friedmann-eq-flat}\end{equation}
In accord with the comment above the equation~\eqref{eq:FRW-metric-flat-Cartesian}
for the most of this thesis we consider only $\Omega=1$ and the Friedmann
equation of the form in Eq.~\eqref{eq:FRW-Friedmann-eq-flat}.

The early Universe is dominated by radiation, as can be seen from
the scaling laws below Eq.~\eqref{eq:FRW-rho-scaling-general}. In
this era it is useful to express Eq.~\eqref{eq:FRW-Friedmann-eq-flat}
in terms of the temperature of relativistic particle species. To do
this let us remember from thermodynamics that the energy density $\rho$
of the weakly interacting gas of particles is given in terms of the
internal degrees of freedom $g_{\mathrm{dof}}$ and its phase space
distribution function $f\left(p\right)$ as \cite{KolbTurner(book)}\begin{equation}
\rho=\frac{g_{\mathrm{dof}}}{\left(2\pi\right)^{3}}\int E\left(p\right)f\left(p\right)\d p3,\label{eq:FRW-rho-distribution}\end{equation}
where $p$ is the magnitude of the momentum of the particle and $E$
is its total energy $E^{2}=p^{2}+m^{2}$. For particles in thermal
equilibrium the distribution function $f\left(p\right)$ is\begin{equation}
f\left(p\right)=\frac{1}{\exp\left(\frac{E-\mu}{T}\right)\pm1},\label{eq:FRW-f-phase-distribution}\end{equation}
where $\mu$ is the chemical potential. The `$+$' sign here corresponds
to the Fermi-Dirac species and `$-$' sign to the Bose-Einstein species. 

In the early Universe when it is dominated by the relativistic particles,
i.e. $T\gg m$, we may take the limit $T\gg\mu$. Inserting Eq.~\eqref{eq:FRW-f-phase-distribution}
into Eq.~\eqref{eq:FRW-rho-distribution} and integrating it we obtain\begin{equation}
\rho_{\gamma}=\frac{\pi^{2}}{30}g_{*}\left(T\right)T^{4},\label{eq:FRW-rho-gamma-rad-dom}\end{equation}
where $\rho_{\gamma}$ denotes the energy density of the relativistic
particles and $g_{*}$ are the number of effectively massless degrees
of freedom:\begin{equation}
g_{*}\left(T\right)=\sum_{i=\mathrm{bosons}}g_{i}+\frac{7}{8}\sum_{i=\mathrm{fermions}}g_{i}.\label{eq:FRW-g-star}\end{equation}
Note that $g_{*}$ is a function of the temperature because in this
sum we included only relativistic species, i.e. particles with the
mass $m\ll T$. For example, at temperatures $T\ll\mathrm{MeV}$ only
photons and three neutrino species are relativistic giving $g_{*}=3.36$.
At temperatures $T>300\,\mathrm{GeV}$ all particles of the Standard
Model are relativistic resulting in $g_{*}=106.75$ \cite{KolbTurner(book)}. 

Finally inserting Eq.~\eqref{eq:FRW-rho-gamma-rad-dom} into Eq.~\eqref{eq:FRW-Friedmann-eq-flat}
we find that in the flat, radiation dominated Universe the Hubble
parameter is related to the temperature as \begin{equation}
H=\pi\sqrt{\frac{g_{*}}{90}}\frac{T^{2}}{\mpl}.\label{eq:FRW-H-in-rad-dom}\end{equation}

\section{Big Bang Nucleosynthesis\label{sec:BBN}}

Arguably the biggest success of the HBB theory is the explanation
for the origin of chemical elements in the Universe. According to
this theory the lightest of them were created during the first three
minutes after the Big Bang, when the Universe content was in a state
of a very hot plasma. This process is called the Big Bang Nucleosynthesis
(BBN). Due to its immense importance for the modern cosmology in predicting
the abundances of light chemical elements and being a very sensitive
method to constraint new theories of particle physics, in this section
we give a summary of BBN.

To describe the creation of light elements in the early Universe the
crucial parameter is the reaction rate $\Gamma$ of some process under
consideration. For illustrative purposes let us consider, for example,
interactions of particles. Then $\Gamma$ would represent the interaction
rate per particle. The crucial quantity here is the ratio $\Gamma/H$,
where $H$ is the Hubble parameter. At the epoch of BBN the Universe
is dominated by the matter which satisfies the strong energy condition,
so that $H^{-1}$ represents the age of the Universe. In this case
$\Gamma/H<1$ means that on average less than one particle interacted
throughout the history of the Universe. In other words, we can say
that particles are decoupled. If, on the other hand $\Gamma/H>1$,
particles have interacted many times and it is safe to assume that
they are in thermal equilibrium. During the radiation dominated stage
the expansion rate of the Universe is proportional to the temperature
squared, $H\propto T^{2}$ (see Eq.~\eqref{eq:FRW-H-in-rad-dom}),
while the reaction rates are typically proportional to $\Gamma\propto T^{s}$.
In the adiabatically expanding Universe the temperature decreases
as $T\propto a^{-1}$. Hence, we can write $\Gamma/H\propto a^{2-s}$,
from which we see that if $s>2$ the process which was in equilibrium
at some initial time, i.e. $\Gamma/H>1$, it will fall out of equilibrium
at later times. If the process we are interested is the interaction
of particles, then we can say that after being in equilibrium, particles
{}``freeze-out'' when $\Gamma/H$ becomes smaller than one, i.e.
the number density is not affected by interactions. Then BBN can be
roughly divided into three stages depending on which processes are
in thermal equilibrium.

When the temperature of the Universe was around $10\,\mathrm{MeV}$,
which corresponds to the age of $10^{-2}\,\mathrm{s}$, the ratio
of neutrons and protons is controlled by the weak interactions:\begin{equation}
n\longleftrightarrow p+e^{-}+\bar{\nu},\quad n+\nu_{e}\longleftrightarrow p+e^{-},\quad n+e^{+}\longleftrightarrow p+\bar{\nu}_{e}.\label{eq:BBN-reactions-n-p}\end{equation}
where $\nu_{e}$ and $\bar{\nu}_{e}$ are the electron neutrino and
antineutrino, and $e^{-}$, $e^{+}$ are the electron and the positron.
If the rate of these interactions are much more rapid than the expansion
of the Universe, i.e. $\Gamma_{n\leftrightarrow p}/H>1$, the species
involved in these interactions are in a thermal equilibrium, which
means that the neutron to proton ratio evolves according to\begin{equation}
\frac{n}{p}=\mathrm{e}^{-Q/T},\label{eq:BBN-n/p-equilibrium}\end{equation}
where $Q\equiv m_{n}-m_{p}=1.29\,\mathrm{MeV}$ is the mass difference
of neutrons and protons. As we can see for the energies high above
$\mathrm{MeV}$, the number of neutrons and protons are almost the
same, $\left(n/p\right)\approx1$.

\begin{figure}
\begin{centering}
\includegraphics[width=9.5cm]{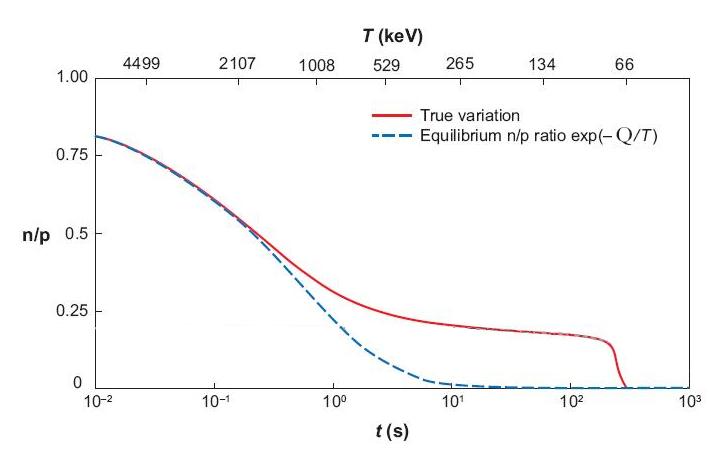}
\par\end{centering}

\caption{\label{fig:np-ratio-evolution}Evolution of the $n/p$ ratio. The
solid red line represents the true variation while the dashed blue
one represents an equilibrium evolution. BBN starts at $T\sim0.1\,\mathrm{MeV}$
which results in the steep decline of the red line at these energies.
(Figure adapted from Ref.~\cite{Steigman(2007)_BBN})}

\end{figure}

The reaction rates for the processes in Eq.~(\ref{eq:BBN-reactions-n-p})
can be calculated using Fermi theory for the weak interactions, which
gives \cite{KolbTurner(book)}\begin{equation}
\Gamma_{n\leftrightarrow p}=\begin{cases}
\tau_{n}^{-1}\left(T/m_{e}\right)^{3}\mathrm{e}^{-Q/T} & T\ll Q,\: m_{e}\\
\simeq2G_{F}^{2}T^{5} & T\gg Q,\: m_{e},\end{cases}\end{equation}
where $\tau_{n}$ is the neutron halflife and $G_{\mathrm{F}}=1.1664\times10^{-5}\,\mathrm{GeV}^{-2}$
is the Fermi constant. When the reaction rate $\Gamma_{n\leftrightarrow p}$
falls bellow the Hubble expansion rate, i.e. when $\mbox{\ensuremath{\Gamma_{n\leftrightarrow p}}/H\,\ensuremath{\lesssim}\,1}$,
processes in Eq.~(\ref{eq:BBN-reactions-n-p}) depart from the equilibrium
and the number of neutrons and protons {}``freezes-out''. The approximate
temperature of the freeze-out can be calculated using Eq.~\eqref{eq:FRW-H-in-rad-dom}
and considering that $T\gtrsim m_{e}$:\begin{equation}
\frac{\Gamma_{n\leftrightarrow p}}{H}\sim\left(\frac{1}{g_{*}^{1/6}}\frac{T}{0.8\,\mathrm{MeV}}\right)^{3}.\label{eq:BBN-Tfr-def}\end{equation}
 In the Standard Model of particle physics with the three (almost)
massless neutrino species the number of relativistic degrees of freedom
is $g_{*}=10.75$. Thus, the freeze-out temperature is found to be
\begin{equation}
T_{\mathrm{fr}}\sim1\,\mathrm{MeV},\end{equation}
which corresponds to about $1\,\mathrm{\mathrm{s}}$. The ratio $n/p$
at this moment can be calculated from Eq.~\eqref{eq:BBN-n/p-equilibrium}\begin{equation}
\frac{n}{p}\approx\frac{1}{6}.\label{eq:BBN-n/p-1/6}\end{equation}
However, this is not a true {}``freeze-out'' because the $n/p$
ratio is not constant but decreases slowly. This happens because of
occasional weak interactions among neutrons, protons, $e^{\pm}$ and
$\nu_{e}$, $\bar{\nu}_{e}$ eventually dominated by the free neutron
$\beta$ decay. However, this decrease is much slower than the equilibrium
value given in Eq.~\eqref{eq:BBN-n/p-equilibrium} (see Figure~\ref{fig:np-ratio-evolution}).

When the nucleosynthesis starts at about $0.1\,\mathrm{MeV}$ (corresponding
to about 180 s), the neutron to proton ration had been decreased to\begin{equation}
\frac{n}{p}\approx\frac{1}{7}.\end{equation}
The main product of BBN is the helium-4, $^{4}\mathrm{He}$. The production
of heavier elements is very subdominant because there are no stable
nuclei with the mass number 5 or 8 and hence no elements form through
reactions such as $n+^{4}\mathrm{He}$, $p+^{4}\mathrm{He}$ or $^{4}\mathrm{He}+^{4}\mathrm{He}$.
In addition reactions such as $\mathrm{T}+^{4}\mathrm{He}\longleftrightarrow\gamma+^{7}\mathrm{Li}$
and $^{3}\mathrm{He}+^{4}\mathrm{He}\longleftrightarrow\gamma+^{7}\mathrm{Be}$
are suppressed because of the large Coulomb barriers. The formation
of $^{4}\mathrm{He}$ in principle could proceed directly through
the four body collision. But the very low number densities of neutrons
and protons renders this type of reactions negligible. Hence, the
element formation must start with the production of deuterium through
the two-body collision:\begin{equation}
p+n\longleftrightarrow\mathrm{D}+\gamma.\end{equation}
 Although the binding energy of the deuterium is $\Delta_{\mathrm{D}}=2.23\,\mathrm{MeV}$,
the formation of this element becomes effective only at much smaller
temperatures. This is because of a large number of energetic photons
which destroy deuterium. So $\mathrm{D}$ nuclei can start forming
without being immediately photo-dissociated only when the number of
such photons per baryon falls below unity, which occurs at the temperature
$T<0.1\,\mathrm{MeV}$ \cite{Iocco_etal(2009)_BBN}. Therefore, this
period is called the \emph{deuterium bottleneck}. But once deuterium
starts forming, the whole set of reactions sets in producing other
heavier elements.

\begin{table}
\begin{centering}
\begin{tabular}{|cc|cc|}
\hline 
Number & Reaction & Number & Reaction\tabularnewline
\hline
\hline 
1 & $\tau_{n}$ & 9 & $^{3}\mathrm{He}\left(\mathrm{T},\gamma\right)\mathrm{^{7}Be}$\tabularnewline
\hline 
2 & $p\left(n,\gamma\right)d$ & 10 & $\mathrm{T}\left(\mathrm{T},\gamma\right)\mathrm{^{7}Li}$\tabularnewline
\hline 
3 & $\mathrm{D}\left(p,\gamma\right)\mathrm{^{3}He}$ & 11 & $^{7}\mathrm{Be}\left(n,p\right)\mathrm{^{7}Li}$\tabularnewline
\hline 
4 & $\mathrm{D}\left(\mathrm{D},n\right)\mathrm{^{3}He}$ & 12 & $^{7}\mathrm{Li}\left(p,\mathrm{T}\right)\mathrm{^{4}He}$\tabularnewline
\hline 
5 & $\mathrm{D}\left(\mathrm{D},p\right)\mathrm{T}$ & 13 & $^{4}\mathrm{He}\left(\mathrm{D},\gamma\right)\mathrm{^{6}Li}$\tabularnewline
\hline 
6 & $^{3}\mathrm{He}\left(n,p\right)\mathrm{T}$ & 14 & $^{6}\mathrm{Li}\left(p,\mathrm{T}\right)\mathrm{^{3}He}$\tabularnewline
\hline 
7 & $\mathrm{T}\left(\mathrm{D},n\right)\mathrm{^{4}He}$ & 15 & $^{7}\mathrm{Be}\left(n,\mathrm{T}\right)\mathrm{^{4}He}$\tabularnewline
\hline 
8 & $^{3}\mathrm{He}\left(\mathrm{D},p\right)\mathrm{^{4}He}$ & 16 & $^{7}\mathrm{Be}\left(\mathrm{D},p\right)2\mathrm{^{4}He}$\tabularnewline
\hline
\end{tabular}
\par\end{centering}

\caption{\label{tab:BBN-reactions}The most relevant reactions of BBN. Here,
numbers of reactions correspond to the ones in Figure~\ref{fig:BBN-reactions}
(adapted from \cite{Iocco_etal(2009)_BBN}).}

\end{table}

\begin{figure}
\begin{centering}
\includegraphics{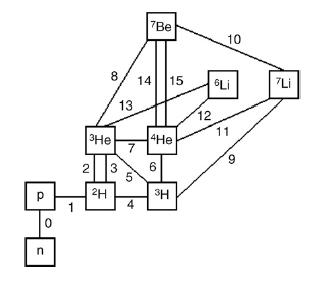}
\par\end{centering}

\caption{\label{fig:BBN-reactions}The network of most relevant reactions of
BBN. The numbers represent reactions in Table~\ref{tab:BBN-reactions}
\cite{Iocco_etal(2009)_BBN}.}

\end{figure}

The final number density of $^{4}\mathrm{He}$ depends on the whole
nuclear network only very weakly. And it is a very good approximation
to assume that all neutrons which didn't $\beta$ decay will end up
being bound into the $^{4}\mathrm{He}$ atoms. Hence, the helium mass
fraction $\mathrm{Y}_{p}$ can be calculated very easily just by power
counting:\begin{equation}
\mathrm{Y}_{p}\simeq\frac{2n}{n+p}=\frac{2\left(n/p\right)}{1+\left(n/p\right)}\simeq0.25.\label{eq:BBN-Yp}\end{equation}

The essential parameter for the processes of BBN is the number density
of baryons. To quantify it, one usually uses the ratio of baryons
over photons defined as\begin{equation}
\eta_{10}\equiv10^{10}\frac{n_{\mathrm{B}}}{n_{\gamma}}.\end{equation}
At temperatures somewhat below $T\lesssim0.3\,\mathrm{MeV}$, all
the positrons have annihilated with the electrons and hence the number
of baryons and photons in a comoving volume does not change. Therefore,
$\eta_{10}$ must stay constant from BBN through recombination until
today. And one can relate this value to the energy density parameter
for the baryons $\Omega_{B}$ today \cite{Serpio(2004)BBN}:\begin{equation}
\eta_{10}=\frac{273.45\Omega_{B}h^{2}}{1-0.007\mathrm{Y}_{p}}\left(\frac{2.725\,\mathrm{K}}{T_{\mathrm{CMB}}}\right)^{3}\left(\frac{6.708\times10^{-45}\,\mathrm{MeV}^{-2}}{G}\right),\end{equation}
where $T_{\mathrm{CMB}}$ is the photon temperature today and $G$
is Newton's gravitational constant.

\begin{figure}
\begin{centering}
\includegraphics[width=12cm]{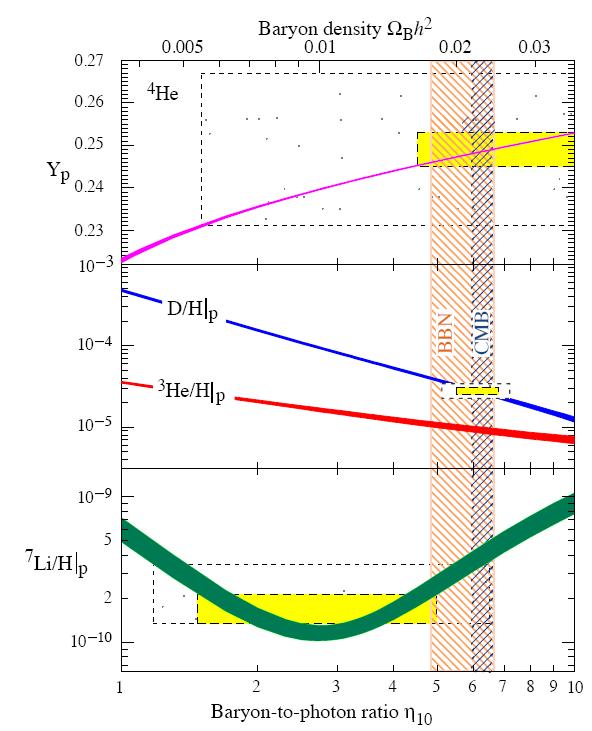}
\par\end{centering}

\caption{\label{fig:BBN-abundances}Abundances of light elements from the standard
model of BBN. The bands represent 95\% CL, the boxes represent observed
values (smaller - $\pm2\sigma$ only statistical errors; larger -
$\pm2\sigma$ statistical and systematic errors), the narrow vertical
column represents $\eta_{10}$ value inferred from CMB observations
and the wider column indicates the BBN concordance range (both at
95\% CL) \cite{Rev_of_PP(2008)}.}

\end{figure}

The most relevant reactions for the BBN are summarized in Table~\ref{tab:BBN-reactions}
and Figure~\ref{fig:BBN-reactions}. The precise final abundances
of each element, including $^{4}\mathrm{He}$, are calculated numerically,
solving a system of coupled kinetic equations for each element as
well as  Einstein equations, including the covariant conservation
of total energy momentum tensor and conservation of baryon number
and electric charge. Results are given in Figure~\ref{fig:BBN-abundances}
together with observationally inferred values for some elements.

\section{The Problem of Initial Conditions of the Hot Big Bang\label{sec:HBB-problems}}

The HBB cosmology is very successful in explaining the structure and
evolution of the Universe after $1\,\mathrm{s}$. However, in order
to agree with observations the initial conditions of the HBB model
have to be fine tuned. In this section we review briefly the problem
of this fine tuning.

\subsection{The Flatness Problem}

\begin{figure}
\begin{centering}
\includegraphics[width=8.5cm]{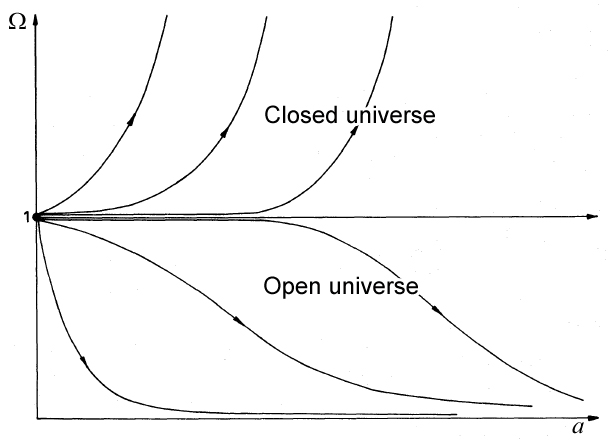}
\par\end{centering}

\caption{\label{fig:Omega-evolution-MD-RD}The evolution of the density parameter
of the Universe filled with an ideal fluid whose barotropic parameter
is $w>-1/3$. From the figure it is clear that any initial departure
from critical density will grow in time, hence the spatially flat
Universe is an unstable fixed point in the phase diagram (adapted
from Ref.~\cite{Madsen_Ellis(1988)}).}

\end{figure}

Current observations agree very well with the density parameter $\Omega$
of the Universe being very close to one, i.e. the Universe is spatially
flat. However, in the phase diagram the value $\Omega=1$ is the unstable
fixed point. In other words, any initially tiny departure from flatness
will become larger and larger as the Universe evolves. This can be
easily seen by using the Friedmann equation~\eqref{eq:rho-continuiti-a}
and the definition of the $\Omega$ in Eq.~\eqref{eq:Omega-def}:\begin{equation}
\Omega=\frac{1}{1-\frac{\rho}{3\mpl^{2}}\frac{K}{a^{2}}}.\end{equation}
Assuming $\rho$ corresponds to the energy density of a perfect fluid
and using Eq.~\eqref{eq:FRW-rho-scaling-general} we arrive at\begin{equation}
\Omega=\frac{1}{1-\left[\left(\oi-1\right)/\oi\right]y^{3w+1}},\label{eq:HBB-O-evolution}\end{equation}
where $\oi$ denotes the initial value of the density parameter and
$y\equiv a/a_{0}$. From this equation it is already clear that if
$w>-1/3$ any initial departure from the flat Universe with $\oi\neq1$
will grow in time. For example if initially the Universe is open,
$\oi<1$, the energy density at any later time $y>1$ will decrease
monotonically towards zero $\Omega\rightarrow0$, i.e. towards the
empty Milne Universe. On the other hand, if initially the Universe
is closed, $\oi>1$, its energy density will increase rapidly and
reach a singularity in a finite time. This behavior of the density
parameter is illustrated in the phase diagram in Figure~\ref{fig:Omega-evolution-MD-RD}.
Therefore, for the present Universe to be flat, $\Omega_{0}\approx1$,
its initial energy density had to be extremely close to the critical
value. For example, in order to reproduce the present Universe, the
energy density at the time of BBN had to be\begin{equation}
\left|\Omega_{\mathrm{BBN}}-1\right|\lesssim10^{-16}.\end{equation}
It is extremely unlikely for $\Omega_{\mathrm{BBN}}$ to be so close
to unity by accident.

The flatness problem of the HBB is sometimes rephrased as an age problem.
This can be seen from Eq.~\eqref{eq:HBB-O-evolution} and Figure~\ref{fig:Omega-evolution-MD-RD}.
If initially the Universe is closed $\oi>1$, very soon after its
birth it recollapses without having time to form any galaxies and
stars. If, on the other hand, initially the Universe is open $\oi<1$,
it soon becomes the empty Milne Universe before the formation of any
structure. In both cases the Universe does not have time to become
the one we observe today.

\subsection{The Horizon Problem}

In section~\ref{sec:Kinematics-of-HHB} we have demonstrated that
the isotropy and (with mild assumptions) homogeneity of the Universe
at present as well as at the Last Scattering Surface (LSS) is an observationally
established fact, which justifies the use of FRW metric. However,
in the HBB model this fact is highly non-trivial due to the finite
age of the Universe. We expect the space-time regions to be homogeneous
and isotropic on scales which could have been in a causal contact,
i.e. which could have {}``communicated'' with each other. But because
the maximum velocity of the signal is finite (equal to the speed of
light in vacuum) and because the age of the Universe is finite too,
there is only a limited distance at which two regions could have had
a causal contact.

To illustrate the horizon problem in HBB let us consider the epoch
of the last scattering. From observations of the CMB we know that
the fractional temperature variations at that time was of order $10^{-5}$.
The distance of maximal causal contact at LSS is \begin{equation}
l_{\mathrm{causal}}\sim c\, t_{\mathrm{LSS}},\end{equation}
where $c$ is the speed of light and $t_{\mathrm{LSS}}$ is the age
of the Universe at LSS. At this epoch the size of the present horizon,
given by $l_{\mathrm{today}}\sim c\, t_{\mathrm{today}}$, was $a_{\mathrm{LSS}}/a_{\mathrm{today}}$
times smaller\begin{equation}
l_{\mathrm{LSS}}\sim c\, t_{\mathrm{today}}\frac{a_{\mathrm{LSS}}}{a_{\mathrm{today}}}.\end{equation}
Taking into account that the Universe was matter dominated at the
last scattering and remained so until very recently, i.e. $a\propto t^{2/3}$,
we may compare $l_{\mathrm{LSS}}$ with the size of the causality
length\begin{equation}
\frac{l_{\mathrm{LSS}}}{l_{\mathrm{causal}}}\sim\left(\frac{a_{\mathrm{today}}}{a_{\mathrm{LSS}}}\right)^{1/3}.\end{equation}
Hydrogen recombined at the temperature $T_{\mathrm{LSS}}\approx1.6\times10^{5}\,\mathrm{K}$.
Assuming adiabatic expansion for the Universe $T\propto a^{-1}$,
and using $T_{\mathrm{CMB}}\approx2.7\,\mathrm{K}$ at $a=a_{\mathrm{today}}$
we find\begin{equation}
\left(\frac{l_{\mathrm{LSS}}}{l_{\mathrm{causal}}}\right)^{3}\sim10^{5}.\end{equation}
Therefore, at the time of recombination the observable Universe consisted
of at least $10^{5}$ causally disconnected regions with the fractional
temperature variation of only $10^{-5}$. No physical process could
have caused such extreme smoothness in so many causally disconnected
regions. This constitutes the horizon problem of HBB cosmology.

\subsection{The Origin of Primordial Perturbations}

\begin{figure}
\begin{centering}
\includegraphics[width=12.5cm]{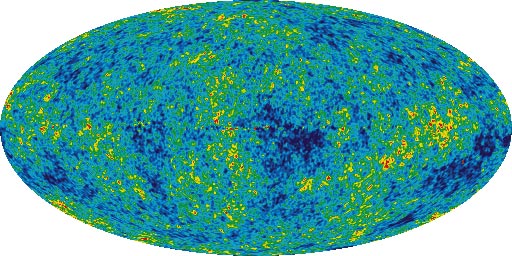}
\par\end{centering}

\caption{\label{fig:CMB-map}Cosmic Microwave Background temperature variation
map observed by the WMAP satellite \cite{WMAP_CMB_picture}.}

\end{figure}

As we have seen in Figure~\ref{fig:smoothing_and_perturbations}
the Universe can be considered isotropic and homogeneous only on smoothing
scales larger that a few hundreds of Mpcs. On smaller scales it is
highly inhomogeneous due to the presence of structures such as stars,
galaxies and galaxy clusters. It is already established that this
structure in the Universe formed due to gravitational instability,
when slightly denser regions collapsed onto themselves forming a complicated
web distribution of galaxies. However, for this process to be initiated
the existence of some primordial seed density perturbations must be
postulated. Indeed, the first observational proof of such perturbations
was provided by the COBE satellite \cite{Smoot_etal(1992)COBE} (see
Figure~\ref{fig:CMB-map} for the high resolution CMB map from WMAP
measurements). Unfortunately, the properties of these seed perturbations
cannot be explained within the framework of HBB cosmology.

\begin{figure}
\begin{centering}
\includegraphics[width=14cm]{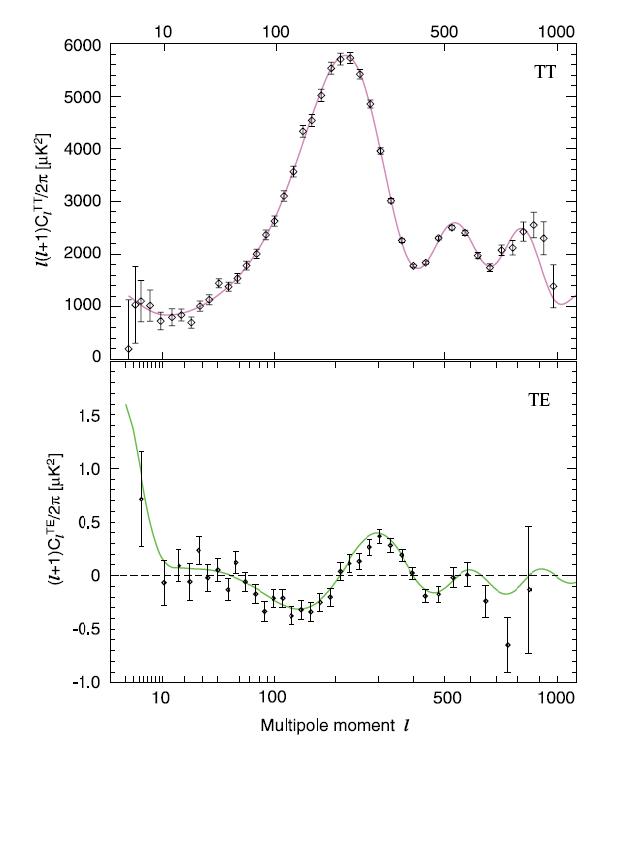}
\par\end{centering}

\caption{\label{fig:CMB-TT-TE}The angular power spectrum of the CMB temperature-temperature
(TT) and temperature-polarization (TE) anisotropies. Solid lines represent
the best fit $\Lambda$CDM model \cite{Nolta_etal(2008)WMAP_spectr}.}

\end{figure}

Causality constraints require that seed perturbations could have formed
only due to processes inside the causal horizon, which in the HBB
model is monotonically decreasing as we go back in time. However,
it was already realized in 60's and 70's that random displacements
and movements of particles inside the horizon cannot produce the necessary
perturbation spectrum. Zel'dovich \cite{Zeldovic(1965),Zeldovich&Novikov(1983)vol2}
and Peebles \cite{Peebles(1974)} were the first to realize that such
processes would produce matter density perturbations with the power
spectrum $P\left(k\right)\propto k^{4}$. Such perturbations would
result in an excessive overproduction of black holes on small length
scales. Indeed, assuming a smooth power law spectrum of primordial
perturbations, $P\left(k\right)\propto k^{n}$, from the CMB observations
it was found that $n\approx0.96$ \cite{Komatsu_etal_WMAP5(2008)},
very different from the $n=4$ case.

In addition, with more precise measurements of primordial perturbations
other properties became clear which cannot be explained by the standard
HBB cosmology \cite{Dodelson(2003)}. As seen in the upper graph of
Figure~\ref{fig:CMB-TT-TE} the CMB temperature power spectrum features
so called {}``acoustic peaks''. But more importantly this pattern
is caused by adiabatic density perturbations. Such perturbations in
the baryon-photon fluid upon horizon entry start oscillating only
with excited cosine modes and with the same phase. Therefore, this
is a strong indication that seed perturbations are present on scales
larger than the horizon, i.e. they could not be created by causal
processes during HBB.

Although no viable model exists, in principle one could construct
a model in which causal processes mimic the pattern of adiabatic acoustic
peaks \cite{Turok(1996a),Turok(1996b),Hu_etal(1996)}. However, even
stronger proof for the superhorizon origin of primordial perturbations
is provided by the temperature-polarization cross correlation function
\cite{Spergel_Zaldarriaga(1997)}. The polarization signal is not
affected on its path from LSS towards us. Therefore, by measuring
polarization perturbations we can be certain to be probing the era
of recombination. But as clearly seen in the lower graph of Figure~\ref{fig:CMB-TT-TE}
on angular scales $50<l<200$ the temperature and polarization anticorrelates.
Since these low multipoles represent superhorizon scales at LSS it
is certain that primordial density perturbations were already present
before entering the horizon.

From this discussion, one can see that the problem of seed perturbations
in the HBB cosmology in essence is a restatement of the horizon problem.
The superhorizon origin of primordial perturbations is the strongest
support for the inflationary scenario.

\section{Inflation\label{sec:Inflation}}

\subsection{The Accelerated Expansion\label{sub:The-Accelerated-Expansion}}

The initial condition problems of standard HBB cosmology named in
section~\ref{sec:HBB-problems} may be substantially alleviated if
we postulate an accelerated expansion of the Universe at it's earliest
stages. This epoch is called inflation. When we say {}``accelerated
expansion'' we mean that the distance between any two comoving points
in the Universe is increasing with a positive acceleration. In other
words the scale factor in Eqs.~\eqref{eq:FRW-metric-Spherical} or
\eqref{eq:FRW-metric-flat-Cartesian} obeys \begin{equation}
\ddot{a}>0.\label{eq:Inflation-scale-factor}\end{equation}
This might be considered as the definition of the era when gravity
is repulsive. 

Instead of Eq.~\eqref{eq:Inflation-scale-factor} we may rewrite
it in the form which gives more physical interpretation. As was discussed
in section~\ref{sec:HBB-dynamics}, $H^{-1}$ defines the Hubble
length. Then $\left(aH\right)^{-1}$ is the comoving Hubble length.
From Eq.~\eqref{eq:Inflation-scale-factor} we find that\begin{equation}
\frac{\d{}{}}{\d t{}}\left(aH\right)^{-1}<0.\label{eq:Inflation-comoving-horizon}\end{equation}
It shows that the comoving Hubble length during inflation decreases.
Therefore, two points which initially were inside the Hubble radius
(their comoving distance smaller than $\left(aH\right)^{-1}$) at
some moment goes outside this radius. This moment is called {}``the
horizon exit''.

The condition in Eq.~\eqref{eq:Inflation-scale-factor} can be written
even in another form, which will be very useful in later sections.
Substituting the derivative of the Hubble parameter $\dot{H}$ into
Eq.~\eqref{eq:Inflation-scale-factor} after some calculations we
find $-\dot{H}<H^{2}$. When \begin{equation}
\left|\dot{H}\right|\ll H^{2},\label{eq:Inflation-Hdot-ll-Hsq}\end{equation}
the expansion is almost exponential. Even more so, if $\dot{H}\rightarrow0$,
it is exactly exponential, i.e. $a\propto\exp\left(Ht\right)$, and
we call this de Sitter expansion.

By postulating the early phase of accelerated expansion (Eq.~\eqref{eq:Inflation-scale-factor})
of the early Universe, the fine tuning problem of initial conditions
of the HBB model discussed in section~\ref{sec:HBB-problems} are
substantially alleviated. The crucial condition for this is Eq.~\eqref{eq:Inflation-comoving-horizon}. 

\begin{figure}
\begin{centering}
\includegraphics[width=8.5cm]{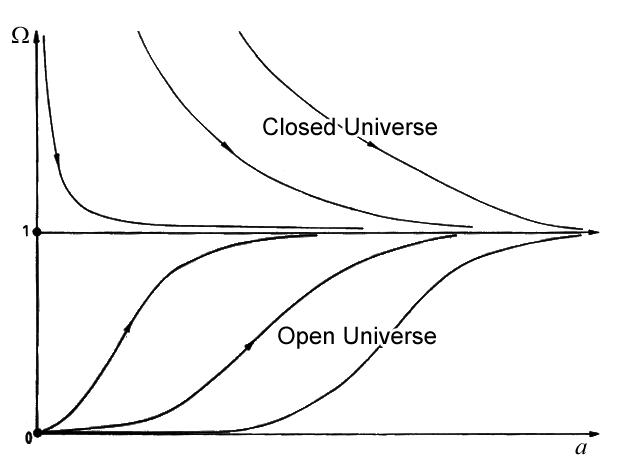}
\par\end{centering}

\caption{\label{fig:Omega-evolution-inflation}The evolution of the density
parameter for the Universe dominated by the vacuum energy with $w=-1$.
Instead of being an unstable point in HBB model, $\Omega=1$ becomes
an attractor (adapted from Ref.~\cite{Madsen_Ellis(1988)}).}

\end{figure}

How the period of accelerated expansion solves the flatness problem
can be seen by inserting Eq.~\eqref{eq:Inflation-comoving-horizon}
into Eq.~\eqref{eq:Omega-def}. Because during inflation the comoving
horizon is decreasing, $a^{2}H^{2}$ grows with time and $\left|\Omega-1\right|$
is driven towards zero. Therefore, $\Omega=1$ instead of being an
unstable fixed point in the HBB model, becomes an attractor during
the inflationary stage. To illustrate this let us use Eq.~\eqref{eq:HBB-O-evolution}.
As we will see shortly in the next section, the accelerated expansion
of the Universe may be achieved if it is dominated by the vacuum energy
for which $\rho=-p$, i.e. $w=-1$. Substituting the value $w=-1$
into Eq.~\eqref{eq:HBB-O-evolution} we find that the density parameter
approaches $\Omega\rightarrow1$ and the phase diagram in the Figure~\ref{fig:Omega-evolution-MD-RD}
changes into the Figure~\ref{fig:Omega-evolution-inflation}. %

The decreasing comoving horizon size during inflation has an effect
that initially any two causally connected points within the Hubble
radius at some later moment leaves the horizon. This makes the causally
connected regions to be larger than the horizon size and the postulate
of the inflationary period solves a second problem of the HBB model.

However, the most important achievement of inflation is that it explains
the origin of superhorizon seed perturbations. According to the inflationary
paradigm during accelerated expansion of the Universe vacuum quantum
fluctuations are converted into classical perturbations. Details of
this process with scalar field quantum fluctuations will be discussed
in Chapter~\ref{cha:Scalars} and extended to the quantum fluctuations
of vector fields in Chapter~\ref{cha:Vectors}. But before that let
us discuss what may cause an accelerated expansion of the Universe.

\subsection{The Scalar Field Driven Inflation\label{sub:sFd-inflation}}

There are several reasons why the Universe could have expanded exponentially.
It might be that, at the relevant energy scales, the Einstein gravity
is not a viable theory of Nature and it must be modified. Modification
is such that it gives almost exponential expansion of space-time.
The very first proposed model of inflation was due to this kind of
modified gravity theory \cite{Starobinsky(1980)}. 

Another possibility is that inflation happens at the energy scales
where Einstein gravity is still a viable theory of nature. Then Eq.~\eqref{eq:Inflation-scale-factor}
puts constraints on properties of matter which may be responsible
for the inflationary expansion. This can be found using Eq.~\eqref{eq:FRW-a-ddot}:\begin{equation}
\rho+3p<0.\end{equation}
Because the energy density $\rho$ must always be positive, it follows
that the Universe undergoes accelerated expansion if the pressure
is negative enough, $p<-\rho/3$. The lower bound for pressure is
determined by the dominant energy condition, which requires that $p\ge-\rho$.
Should this bound be violated the propagation of energy outside the
lightcone becomes possible and one cannot guarantee the stability
of the vacuum \cite{Carroll_etal(2003)ghosts}. Taking the extreme
case $p=-\rho$ and from Eq.~\eqref{eq:HBB-dotH} we find that $\dot{H}\rightarrow0$
and the Universe is expanding exponentially (de Sitter Universe).

The equation of state $p\approx-\rho$ may be realized in the framework
of GR if the Universe is assumed to be dominated by classical scalar
fields. For simplicity we will assume that only one such field is
relevant, which is then called the inflaton. The most general Lagrangian
which is consistent with the GR for a single field inflation is given
by \cite{Garriga&Mukhanov(1999)_kInflation}\begin{equation}
\mathcal{L}=P\left(X,\phi\right),\label{eq:Infl-kInflationary-Lagrangian}\end{equation}
where $\phi$ is the scalar field, $X\equiv\frac{1}{2}\partial_{\mu}\phi\partial^{\mu}\phi$
and $P$ is some function. Inflationary models which study the evolution
of the Universe under the influence of such fields are called k-inflation.
But to make essential properties of inflationary models more transparent
let us concentrate on a particular case where the field is canonically
normalized. Then Eq.~\eqref{eq:Infl-kInflationary-Lagrangian} becomes\begin{equation}
\mathcal{L}=\frac{1}{2}\partial_{\mu}\phi\partial^{\mu}\phi-V\left(\phi\right),\label{eq:Infl-Lagrangian}\end{equation}
where $V\left(\phi\right)$ is the potential. The equation of motion
of the field is obtained by requiring that the variation of the action
with respect to the field vanishes. For the homogeneous component
this gives\begin{equation}
\ddot{\phi}+3H\dot{\phi}+V_{\phi}=0,\label{eq:Infl-EoM}\end{equation}
where $V_{\phi}\equiv\partial V\left(\phi\right)/\partial\phi$ is
the derivative of the potential with respect to the field and because
we are interested in the homogeneous part of the field we have neglected
gradient terms.

The energy-momentum tensor of a scalar field may be obtained using
the variation of the action with respect to $g_{\mu\nu}$ \cite{Weinberg_Gravitation(1972)}\begin{equation}
T_{\mu\nu}=-2\frac{\partial\mathcal{L}}{\partial g^{\mu\nu}}+g_{\mu\nu}\mathcal{L}.\end{equation}
For the scalar field with the Lagrangian of the form in Eq.~\eqref{eq:Infl-Lagrangian}
it becomes\begin{equation}
T_{\nu}^{\mu}=\partial^{\mu}\phi\partial_{\nu}\phi-\delta_{\nu}^{\mu}\left[\frac{1}{2}\partial^{\sigma}\phi\partial_{\sigma}\phi-V\left(\phi\right)\right].\label{eq:Infl-Tmunu}\end{equation}
From this equation one can notice that the energy-momentum tensor
for the homogeneous scalar field (homogenized by inflation) becomes
as that of the perfect fluid so that the energy density $\rho$ and
pressure $p$ can be defined as

\begin{eqnarray}
\rho & = & \frac{1}{2}\dot{\phi}^{2}+V\left(\phi\right),\label{eq:Infl-rho}\\
p & = & \frac{1}{2}\dot{\phi}^{2}-V\left(\phi\right).\label{eq:Infl-p}\end{eqnarray}
Using the first of these relations the Friedmann equation in Eq.~\eqref{eq:FRW-Friedmann-eq}
becomes\begin{equation}
3\mpl^{2}H^{2}=\frac{1}{2}\dot{\phi}^{2}+V.\end{equation}
Differentiating it, we derive another useful expression\begin{equation}
3\mpl^{2}\dot{H}=-\dot{\phi}^{2},\end{equation}
were we have used the equation of motion in Eq.~\eqref{eq:Infl-EoM}
as well.

From expressions \eqref{eq:Infl-rho} and \eqref{eq:Infl-p} it is
clear that the condition $p\approx-\rho$ is satisfied if the kinetic
energy of the field is negligible compared to the potential one, i.e.
\begin{equation}
\frac{\left|\dot{H}\right|}{H^{2}}\ll1\;\Leftrightarrow\;\dot{\phi}^{2}\ll V\left(\phi\right).\label{eq:Infl-dotPhi-ll-V}\end{equation}
 This requirement is called {}``the slow-roll condition'' and is
fulfilled if the potential of the field is sufficiently shallow. These
conditions may be conveniently rewritten in terms of slow-roll parameters
$\epsilon$ and $\eta$ in the following way. Because the field is
slowly rolling we might also expect that the second derivative is
also small, $\ddot{\phi}/H\ll\dot{\phi}$. Then, in the equation of
motion \eqref{eq:Infl-EoM}, the first term is negligible \begin{equation}
3H\dot{\phi}\simeq-V_{\phi}.\label{eq:Infl-EoM-slow-roll}\end{equation}
The first slow-roll parameter becomes\begin{equation}
\epsilon\equiv\frac{\mpl^{2}}{2}\left(\frac{V_{\phi}}{V}\right)^{2}\ll1.\label{eq:Infl-e-slow-roll-parameter}\end{equation}
On the other hand, differentiating Eq.~\eqref{eq:Infl-EoM-slow-roll}
and using conditions in Eq.~\eqref{eq:Infl-dotPhi-ll-V} we define
the second slow-roll parameter\begin{equation}
\eta\equiv\mpl^{2}\frac{V_{\phi\phi}}{V}\ll1.\label{eq:Infl-eta-slow-roll-condition}\end{equation}

Conditions in Eq.~\eqref{eq:Infl-e-slow-roll-parameter} and \eqref{eq:Infl-eta-slow-roll-condition}
are called flatness conditions for the shape of the potential of the
scalar field. The quasi exponential expansion during inflation lasts
as long as those conditions are satisfied. When any of the parameters
$\epsilon$ of $\eta$ becomes of order one, inflation ends.

It is convenient to define a number of e-folds until the end of inflation\begin{equation}
N\equiv\ln\frac{a\left(t_{\mathrm{end}}\right)}{a\left(t\right)}=\int_{t}^{t_{\mathrm{end}}}H\d t{}=\int_{\phi}^{\phi_{\mathrm{end}}}\frac{H}{\dot{\phi}}\d{\phi}{}=\mpl^{-2}\int_{\phi_{\mathrm{end}}}^{\phi}\frac{V}{V_{\phi}}\d{\phi}{}.\label{eq:Infl-N-FRW-def}\end{equation}
If the initial time in this equation is chosen to be when cosmological
scales leaves the horizon, we may calculate how many numbers of e-folds
of inflation is needed to solve the horizon and flatness problems
of the HBB model. If the energy scale of inflation and the reheating
temperature in Eq.~\eqref{eq:Infl-T-reh} is at the supersymmetry
energy scales, then $N\approx60$ \cite{Lyth_Liddle(2009)book}. In
some models, with very low reheating temperature it can go down to
$N\approx40$. If we assume the validity of Einstein's gravity after
inflation and fields with canonical kinetic terms the maximum number
of e-folds is $N\approx70$.

\subsection{The End of Inflation and Reheating\label{sub:Reheating}}

Inflation ends when the slow-roll parameters defined in Eqs.~\eqref{eq:Infl-e-slow-roll-parameter}
and \eqref{eq:Infl-eta-slow-roll-condition} become of order one.
Soon after this happens, the inflaton field fast-rolls towards its
VEV and starts oscillating around the minimum of its effective potential.
Expanding this potential around the minimum, the leading term in the
series is $V\left(\phi\right)=\frac{1}{2}m^{2}\left(\phi-\left\langle \phi\right\rangle \right)^{2}$,
where $\left\langle \phi\right\rangle $ is the VEV and $m$ is the
mass of the inflaton field. Inserting this into Eq.~\eqref{eq:Infl-EoM}
we see that for $m\gg H$ the field acts as the underdamped harmonic
oscillator with the frequency $\omega=m$, much larger than the Hubble
time. Therefore, in accord with the equation of motion of the harmonic
oscillator we may write $\overline{\dot{\phi}^{2}}=2\overline{V}\left(\phi\right)$,
were the average values are defined over one Hubble time. Inserting
this into Eq.~\eqref{eq:Infl-p} we find that the average pressure
of the oscillating scalar field is $\overline{p}=0$ and the energy
density from Eq.~\eqref{eq:Infl-rho} is $\rho=\overline{\dot{\phi}^{2}}$.
Using this, Eq.~\eqref{eq:Infl-EoM} may be rewritten as\begin{equation}
\dot{\rho}+3H\rho=0.\label{eq:Infl-rho-reheat}\end{equation}
Taking into account that the Universe is dominated by the oscillating
inflaton field with the zero pressure from this equation we obtain\begin{equation}
\overline{p}=0\quad\mathrm{and}\quad\rho\propto a^{-3}.\label{eq:Infl-oscillating-scaling-law}\end{equation}
Therefore, the oscillating inflaton field acts as the pressureless
matter and the Universe evolves as dominated by the non-relativistic
dust particles (inflatons) \cite{Turner(1983)osc_infl}.

Because the field is oscillating it might be interpreted as the collection
of massive inflaton particles with zero momentum. Before inflaton
oscillations the temperature of the Universe is effectively zero.
However, for the successful BBN, discussed in section~\ref{sec:BBN},
the Universe must be radiation dominated with the temperature above
$10\,\mathrm{MeV}$. Therefore, to recover the successes of the HBB
cosmology, the energy stored in the inflaton field must be released
to effectively massless particles. This process is known as `reheating'.
The first proposals for the mechanism to reheat the Universe were
based on the single-body decays \cite{Abbott_etal(1982)reheating,Dolgov_Linde(1982)reheating}.
During inflation such decays may be neglected because the field is
not oscillating and cannot be interpreted as a collection of particles.
But during the phase of coherent oscillations inflaton particles may
decay into other scalar particles $\chi$ or fermions $\psi$ through
the terms in the Lagrangian such as $g\phi\chi^{2}$ and $h\phi\bar{\psi}\psi$,
where $g$ is the coupling constant with the dimension of mass and
$h$ is a dimensionless coupling constant. Due to these couplings
the equation~\eqref{eq:Infl-rho-reheat} must include an additional
friction term $\Gamma$ which parametrizes the inflaton decay into
these particles\begin{equation}
\dot{\rho}+\left(3H+\Gamma\right)\rho=0,\end{equation}
where $\Gamma\equiv\Gamma_{\phi\rightarrow\chi\chi}+\Gamma_{\phi\rightarrow\psi\bar{\psi}}$.
When the mass of the inflaton is much larger than those of $\chi$
and $\psi$, i.e. $m\gg m_{\chi},m_{\psi}$, the decay rates are known
to be \cite{Dolgov_Linde(1982)reheating,Linde_book(1990)}\begin{equation}
\Gamma_{\phi\rightarrow\chi\chi}=\frac{g^{2}}{8\pi m}\quad\mathrm{and}\quad\Gamma_{\phi\rightarrow\psi\overline{\psi}}=\frac{h^{2}m}{8\pi}.\end{equation}

When $H>\Gamma$ the number of produced particles is very small (see
section~\ref{sec:BBN}) and they do not influence the dynamics of
the Universe. However, these particles may still thermalise and their
temperature becomes much larger than the temperature at reheating
(given in Eq.~\eqref{eq:Infl-T-reh}) \cite{Scherrer_Turner(1984)}.
At time $t_{\mathrm{reh}}$, when the Hubble parameter becomes $H\sim\Gamma$,
the decay processes become significant and practically all inflaton
energy is transferred to the newly created particles. The temperature
of the Universe at this moment may be calculated using the flat Friedman
equation in Eq.~\eqref{eq:FRW-Friedmann-eq-flat} and assuming that
new particles are relativistic, then from Eq.~\eqref{eq:FRW-rho-gamma-rad-dom}
we get\begin{equation}
T_{\mathrm{reh}}\simeq g_{*}^{-1/4}\sqrt{\Gamma\mpl},\label{eq:Infl-T-reh}\end{equation}
where $g_{*}=10^{2}-10^{3}$ \cite{Bassett_etal(2005)review} is the
number of effective relativistic degrees of freedom defined in Eq.~\eqref{eq:FRW-g-star}.

The mechanism of reheating described above is based on perturbative
particle decay. However, in some inflationary scenarios the energy
transfer from the inflaton field may be preceded by another, much
more efficient process. To distinguish it from the conventional reheating,
it is called `preheating'. In the first such proposal, the parametric
preheating, the inflaton field decays into relativistic particles
of other fields very rapidly in short, explosive bursts due to the
parametric resonance effects \cite{Kofman_etal(1994)PreheatingPRL,Kofman_etal(1997)PreheatingPRD}.
At the second stage, these particles decay into relativistic species
which finally thermalise. It should be noted, however, that it is
not possible to transfer the total energy stored in the inflaton field
by this process. When the amplitude of inflaton oscillations decreases
below some critical value, the parametric resonance becomes inefficient.
The residual oscillating inflaton field must decay through the perturbative
reheating processes described above. If these processes are not efficient
enough, due to the scaling law in Eq.~\eqref{eq:Infl-oscillating-scaling-law},
the residual oscillating inflaton field comes to dominate the relativistic
decay products of preheating. In this situation the transfer of the
inflaton energy into radiation is still dominated by the perturbative
reheating processes.

\chapter{The Origin of the Primordial Curvature Perturbation\label{cha:Scalars}}

\section{Statistical Properties of the Curvature Perturbation\label{sec:Statistical-Properties}}

\subsection{Random Fields\label{sub:Random-Fields}}

As it will become clear in section~\ref{sec:Scalar-Field-Quantization}
the origin of cosmological perturbations is quantum mechanical. But
quantum mechanical processes are non-deterministic: one can only predict
the probability of experimental outcome. Therefore, to make quantitative
descriptions of these processes one needs to use statistical methods.
The same is true for cosmological perturbations. One cannot calculate
exact values of perturbations at each space point, only the statistical
properties may be predicted by theories and compared with observations.
To quantify the properties of cosmological perturbations a very useful
method is to describe them as random fields.

Let us introduce some random field $\beta$. It is assumed that our
Universe is just one realization of many (hypothetical) possible universes.
Then, to each of these universes one can assign a particular realization
$\beta_{n}$ from the whole ensemble $\beta$.%
\footnote{As is usual in the literature the notation $\beta_{n}\left(\mathbf{x}\right)$
is used to denote two things: a function itself and the value of that
function at the point $\mathbf{x}$. We will adopt the same notation
here hoping that the meaning will be clear from the context and no
confusion will arise. In addition, to denote a function itself (not
it's value) we will use $\beta_{n}$ too, keeping in mind that it
is a function of the spatial argument $\mathbf{x}$.%
}%
\begin{comment}
\begin{itemize}
\item In the literature very often the notation $\beta_{n}\left(\mathbf{x}\right)$
is used to denote a function itself and the value of that function
at the point $\mathbf{x}$. To avoid confusion in this section I will
use $\beta_{n}$ when talking about the realization of the random
field (keeping in mind that it is a function of the argument $\mathbf{x}$)
and $\beta_{n}\left(\mathbf{x}\right)$ when talking about the value
of that realization at a particular point. However, when talking about
Fourier expansion I will use notations $\beta_{n}\left(\mathbf{x}\right)$
and $\beta_{n}\left(\mathbf{k}\right)$ to stress that they are functions
of position and wave-vector respectively.
\end{itemize}

\end{comment}
{} Depending on the problem to be solved, functions $\beta_{n}$ may
parametrize, for example, the spatial distribution of the density,
velocity or other fields. Each of the functions $\beta_{n}$ are realized
with the probability $p\left(\beta_{n}\right)\d n{}$, where $n$
is a continuous index and $p$ is the probability distribution function
(PDF). 

Properties of the random field $\beta$ are specified by the form
of PDF. It is said that the random field $\beta$ is statistically
homogeneous if the probability $p\left(\beta_{n}\right)$ of the realization
$\beta_{n}$ is the same as that of realization $\beta_{m}$, where
$\beta_{n}\left(\mathbf{x}\right)=\beta_{m}\left(\mathbf{x}+\mathbf{X}\right)\;\forall\;\mathbf{X}$.
In other words, probabilities are equal for realizations which differ
only by the spatial translation. And $\beta$ is said to be statistically
isotropic at a point $\mathbf{x}$ if probabilities are equal for
realizations which differs only by rotation, i.e. $p\left(\beta_{n}\right)=p\left(\beta_{m}\right)$
for $\beta_{n}\left(\mathbf{x}\right)=\beta_{m}\left(\mathcal{R}\mathbf{x}\right)$,
where $\mathcal{R}$ is the rotation matrix, $\mathcal{R}\in SO\left(3\right)$.
Analogously, $\beta$ is parity conserving if $p\left(\beta_{n}\right)=p\left(\beta_{m}\right)$
for $\beta_{n}\left(\mathbf{x}\right)=\beta_{m}\left(-\mathbf{x}\right)\;\forall\;\mathbf{x}$.
For the following discussion we will consider only statistically homogeneous
and parity conserving fields. Usually in cosmology it is assumed that
the field of the primordial density perturbation is statistically
isotropic as well. But, as we will show in Chapter~\ref{cha:Vectors}
this might not necessarily be so.

Instead of working with PDF of the random field directly more convenient
and observationally more relevant quantities are $N$-point correlation
functions. For example the two-point correlation function is related
to the PDF as\begin{equation}
\left\langle \beta\left(\mathbf{x}_{1}\right)\beta\left(\mathbf{x}_{2}\right)\right\rangle \equiv\int p\left(\beta_{n}\right)\beta_{n}\left(\mathbf{x}_{1}\right)\beta_{n}\left(\mathbf{x}_{2}\right)\d n{}.\label{eq:ensemble-average}\end{equation}
Integration over $n$ shows that it is the average over all the ensemble.
In general the two-point correlation function does not specify the
PDF uniquely, one needs to calculate higher order correlators which
are defined analogously.

A very powerful way to analyze correlation functions is by decomposing
them into the eigenvectors of the translation operator. In flat space
this corresponds to the decomposition into Fourier series. But to
perform this decomposition it is necessary to chose the box of a certain
size with periodic boundary conditions. In the cosmological context
the choice of the box size is a very important issue. One requires
that the box is large enough so that wave-vectors $\mathbf{k}$ could
be treated as continuous and the Fourier series could be replaced
by an integral. On the other hand, it is undesirable that the box
is infinitely large. It might be that at very large distances the
Universe becomes very anisotropic and inhomogeneous. This for example
happens in chaotic inflationary models. Therefore, choosing too large
a box one would have to take into account unknown physics. Usually
it is enough for the box size to be only several orders of magnitude
larger than the horizon of the observable Universe, so that $\ln\left(H_{0}L\right)\sim\mathcal{O}\left(1\right)$,
where $L$ is the comoving box size and $H_{0}$ is the Hubble parameter
today. Such a box is called a minimal box \cite{Lyth_Liddle(2009)book,Lyth(2007)}.
This choice is sufficient to approximate Fourier series as integrals.
And we normalize Fourier modes such that \begin{equation}
\beta_{n}\left(\mathbf{x}\right)=\int\beta_{n}\left(\mathbf{k}\right)\mathrm{e}^{i\mathbf{k}\cdot\mathbf{x}}\frac{\d k3}{\left(2\pi\right)^{3}}.\label{eq:Fourier-decomposition}\end{equation}

Because $\beta_{n}\left(\mathbf{x}\right)$ describes the distribution
of real quantities in the Universe, they must be real functions themselves.
This translates into the requirement that imaginary Fourier modes
$\beta_{n}\left(\mathbf{k}\right)$ must satisfy the reality condition
$\beta_{n}\left(-\mathbf{k}\right)=\beta_{n}^{*}\left(\mathbf{k}\right)$.
We note as well, that if the random field \textbf{$\beta$ }is statistically
isotropic, then $\beta_{n}\left(\mathbf{k}\right)$ does not depend
on the direction of the wave-vector \textbf{$\mathbf{k}$,} only on
it's modulus $k$, i.e. $\beta_{n}\left(\mathbf{k}\right)=\beta_{n}\left(k\right)$,
where $k\equiv\left|\mathbf{k}\right|$.

If the random field is invariant under spatial translations, i.e.
if it is statistically homogeneous, then the Fourier transform of
the two-point correlator in Eq.~\eqref{eq:ensemble-average} is determined
by the reality condition\begin{equation}
\left\langle \beta_{n}\left(\mathbf{k}_{1}\right)\beta_{n}^{*}\left(\mathbf{k}_{2}\right)\right\rangle =\left(2\pi\right)^{3}\delta\left(\mathbf{k}_{1}-\mathbf{k}_{2}\right)P_{\beta}\left(\mathbf{k}\right),\label{eq:two-point-power-spec-deltaminus}\end{equation}
where $P_{\beta}\left(\mathbf{k}\right)\equiv\left\langle \left|\beta_{n}\left(\mathbf{k}\right)\right|^{2}\right\rangle $
is called the power spectrum (remember that $\left\langle \ldots\right\rangle $
means the ensemble average). Note that the presence of the delta function
in this expression is the result of statistical homogeneity of the
random field. This relation can be rewritten using $\beta_{n}\left(-\mathbf{k}\right)=\beta_{n}^{*}\left(\mathbf{k}\right)$
as\begin{equation}
\left\langle \beta_{n}\left(\mathbf{k}_{1}\right)\beta_{n}\left(\mathbf{k}_{2}\right)\right\rangle =\left(2\pi\right)^{3}\delta\left(\mathbf{k}_{1}+\mathbf{k}_{2}\right)P_{\beta}\left(\mathbf{k}_{1}\right).\label{eq:two-point-power-spec-deltaplus}\end{equation}

The power spectrum $P_{\beta}\left(\mathbf{k}\right)$ is related
to the two-point correlation function in the position space by the
Wiener-Khinchin theorem. This theorem states that $P_{\beta}\left(\mathbf{k}\right)$
is the Fourier transform of the latter\begin{equation}
P_{\beta}\left(\mathbf{k}\right)=\int\left\langle \beta_{n}\left(\mathbf{x}\right)\beta_{n}\left(\mathbf{x}+\mathbf{r}\right)\right\rangle \mathrm{e}^{-i\mathbf{k}\cdot\mathbf{r}}\d{\mathbf{r}}{}.\label{eq:power-spec-two-point-corr}\end{equation}

It is often convenient to use another definition of the power spectrum
which differs from the first one just by normalization\begin{equation}
\mathcal{P}_{\beta}\left(\mathbf{k}\right)\equiv\frac{k^{3}}{2\pi^{2}}P_{\beta}\left(\mathbf{k}\right).\label{eq:curly-P-def}\end{equation}
Both of these definitions have to satisfy the reality condition, i.e.
$\mathcal{P}_{\beta}\left(-\mathbf{k}\right)=\mathcal{P}_{\beta}\left(\mathbf{k}\right)$.
For the future convenience we will parametrize the directional dependence
of the power spectrum as \cite{Ackerman_etal(2007)} \begin{equation}
\mathcal{P}_{\beta}\left(\mathbf{k}\right)=\mathcal{P}_{\beta}^{\mathrm{iso}}\left(k\right)\left[1+g\left(\hat{\mathbf{d}}\cdot\hat{\mathbf{k}}\right)^{2}+\ldots\right],\label{eq:anisotropic-power-spec-def}\end{equation}
where $\mathcal{P}_{\beta}^{\mathrm{iso}}$ is the average over all
directions, $\hat{\mathbf{d}}$ is some unit vector, $\hat{\mathbf{k}}$
is the unit vector along $\mathbf{k}$ and $k\equiv\left|\mathbf{k}\right|$
is the modulus of $\mathbf{k}$.

The meaning of the power spectrum $\mathcal{P}_{\beta}$ can be easily
understood in case of statistically isotropic perturbations, i.e.
when $\mathcal{P}_{\beta}\left(\mathbf{k}\right)=\mathcal{P}_{\beta}\left(k\right)$.
Then from the inverse of Eq.~\eqref{eq:power-spec-two-point-corr}
we find that the variance of the random field $\beta$ is equal to\begin{equation}
\sigma_{\beta}^{2}\left(\mathbf{x}\right)\equiv\left\langle \beta^{2}\left(\mathbf{x}\right)\right\rangle =\frac{1}{\left(2\pi\right)^{3}}\int_{0}^{\infty}P_{\beta}\left(k\right)\,\d k3.\label{eq:mean-square-def}\end{equation}
Since for statistically isotropic perturbations $P_{\beta}\left(k\right)$
depends only on the modulus of $\mathbf{k}$ it is convenient to express
this integral in spherical coordinates. Then the definition in Eq.~\eqref{eq:curly-P-def}
can be rewritten as\begin{equation}
\sigma_{\beta}^{2}=\int_{0}^{\infty}\mathcal{P}_{\beta}\left(k\right)\:\d{\ln k}{}.\label{eq:mean-square-dlnk}\end{equation}
Therefore, $\mathcal{P}_{\beta}\left(k\right)$ corresponds to the
contribution to the variance $\sigma_{\beta}^{2}$ per logarithmic
interval in $k$. And because we assumed statistical homogeneity of
$\beta$, the variance does not depend on position.

If the power spectrum $\mathcal{P}_{\beta}$ is scale independent
then the integral in Eq.~\eqref{eq:mean-square-dlnk} is logarithmically
divergent. Divergences for large and small $k$ in this integral are
avoided by introducing cutoff scales. For large $k$ the cutoff scale
$R_{\mathrm{s}}$ corresponds to the smoothing scale and for small
$k$ (large spatial distances) $R_{\mathrm{box}}$ corresponds to
the maximum size of the box in which we perform calculations\begin{equation}
\sigma_{\beta}^{2}=\int_{R_{\mathrm{box}}^{-1}}^{R_{\mathrm{s}}^{-1}}\mathcal{P}_{\beta}\:\d{\ln k}{}=\mathcal{P}_{\beta}\ln\frac{R_{\mathrm{box}}}{R_{\mathrm{s}}}.\label{eq:mean-square-dlnk-scale-inv}\end{equation}
With the minimal box size, such that $\ln\left(R_{\mathrm{box}}/R_{\mathrm{s}}\right)$
is of order one, the mean-square is roughly of the order of the spectrum.

If Eq.~\eqref{eq:power-spec-two-point-corr} is to be applied in
the cosmological perturbation theory it requires an additional assumption.
In practice we can observe and make measurements only of one Universe.
Hence, the ensemble average over one Universe does not make sense
and we cannot use this equation directly. To connect theoretical predictions
with observations we have to assume the validity of ergodicity for
our Universe. This assumption states that the average over the whole
ensemble of universes is equivalent to the spatial average over one
universe. To see what this means in mathematical language let us write
the spatial average of the product of two points over the universe
of realization $\beta_{n}$\begin{equation}
\overline{\beta_{n}\left(\mathbf{x}\right)\beta_{n}\left(\mathbf{x}+\mathbf{r}\right)}=L^{-3}\int\beta_{n}\left(\mathbf{x}\right)\beta_{n}\left(\mathbf{x}+\mathbf{r}\right)\d{\mathbf{x}}{},\label{eq:two-point-averate}\end{equation}
where $L^{-3}$ is the box over which the averaging is performed.
Then ergodic assumption states that in the limit $L\rightarrow\infty$
\begin{equation}
\left\langle \beta\left(\mathbf{x}\right)\beta\left(\mathbf{x}+\mathbf{r}\right)\right\rangle =\overline{\beta_{n}\left(\mathbf{x}\right)\beta_{n}\left(\mathbf{x}+\mathbf{r}\right)}.\label{eq:ergodic-assumption}\end{equation}

As one can see, this assumption relates averages over the all ensemble
of universes, which cannot be measured, to the average over one universe,
which can be measured. For Eq.~\eqref{eq:ergodic-assumption} to
be strictly valid we required an infinite box over which the measurement
is performed. Of course this cannot be realized practically. The effect
of the finite box introduces the so called `cosmic variance' - when
the separation between points in the correlators approaches the size
of the box, the probability that the spatial average differs from
the ensemble average increases.

Until now we have considered only the two-point correlation function
of Eq.~\eqref{eq:ensemble-average} which is demanded by the reality
condition. If the random field $\beta$ is Gaussian, this correlator
specifies PDF completely. Which means that the three-point correlator
vanishes, while the four-point correlator can be expressed as the
sum of two-point correlator products and so on. In the non-Gaussian
case, the random field has a non-vanishing three-point correlator
and the four-point correlator has additional terms which cannot be
reduced to the product of two-point correlators. Let us limit ourselves
only up to the three-point correlator. Although in cosmological context
for some models higher order correlators might be as important as
the three-point correlator, for the scope of this thesis the three-point
correlator will be sufficient. It can parametrized similarly to Eq.~\eqref{eq:two-point-power-spec-deltaplus}
as\begin{equation}
\left\langle \beta\left(\mathbf{k}_{1}\right)\beta\left(\mathbf{k}_{2}\right)\beta\left(\mathbf{k}_{3}\right)\right\rangle =\left(2\pi\right)^{3}\delta\left(\mathbf{k}_{1}+\mathbf{k}_{2}+\mathbf{k}_{3}\right)B_{\beta}\left(\mathbf{k}_{1},\mathbf{k}_{2},\mathbf{k}_{3}\right),\label{eq:bispectrum-def}\end{equation}
where $B_{\beta}$ is called the bispectrum.

\subsection{The Curvature Perturbation and Observational Constraints}

In the previous section we discussed random fields in general. Let
us now turn to the discussion of the curvature perturbation $\zeta$
which will be the main topic for the rest of this thesis. As was explained
in section~\ref{sec:Inflation} the largest achievement of the inflationary
paradigm is that it predicts the statistical properties of the curvature
perturbation which can be compared with observations. 

Usually observational constraints on the statistical properties of
$\zeta$ are obtained with the assumption of statistical isotropy.
However, one would expect that the presence of anisotropy at $10\%$
level or so would not alter the results significantly. The strongest
constraints on $\zeta$ comes from the measurements of the CMB and
large scale structure which probe the range $\Delta\ln k\sim10$ \cite{Komatsu_etal_WMAP5(2008)}.
The largest probable scale corresponds to the size of the observable
Universe, $k^{-1}\sim H_{0}^{-1}$.

\subsubsection{The Power Spectrum \label{sub:Spectrum-and-constraints}}

The shape of the power spectrum $\mathcal{P}_{\zeta}$ is the primary
tool to contrast predictions of the inflationary models with observations.
To quantify this shape the power spectrum is parametrized as \begin{equation}
\mathcal{P}_{\zeta}\left(k\right)=\mathcal{P}_{\zeta}\left(k_{0}\right)\left(\frac{k}{k_{0}}\right)^{n\left(k_{0}\right)-1+\frac{1}{2}n'},\label{eq:power-spec-observational}\end{equation}
where $k_{0}\equiv0.002\,\mathrm{Mpc}^{-1}$ is the pivot scale, $n$
is called the spectral index and parametrizes the scale dependence
of the power spectrum and $n'\equiv\d n{}/\d{\ln k}{}$ is the running
of the spectral index. Such parametrization is sufficient because
according to observations $\left.n'\ll n\right.$, thus higher derivatives
are even smaller and can be neglected. Of course with such simple
parametrization one looses sensitivity to the sharp features of the
power spectrum. But according to some investigations (e.g. Ref.~\cite{Verde&Peiris(2008)})
such features are not detected. The normalization of the power spectrum
$\mathcal{P}_{\zeta}\left(k_{0}\right)$ was first measured by the
COBE satellite and most recently by the WMAP \cite{Komatsu_etal_WMAP5(2008)}.
The present value is\begin{equation}
\mathcal{P}_{\zeta}\left(k_{0}\right)=\left(2.445\pm0.096\right)\times10^{-9},\end{equation}
where this and later intervals are given at $68\%$ CL. 

For the simplest, scale invariant case, called Harrison-Zel'dovich
or flat power spectrum, $n=1$ and $n'=0$. However, according to
current observations the spectral index is 3.1 standard deviations
away from the Harrison-Zel'dovich one. Indeed, $n$ is smaller than
1. Such power spectrum is called red. With the assumption of negligible
running, $n'=0$, and no gravitational waves the spectral index is
determined as \begin{equation}
n=0.960\pm0.013.\label{eq:spec-indx-WMAP5}\end{equation}
If the running of the spectral index is allowed then this constraint
is relaxed\begin{equation}
n=1.017\pm0.043\quad\mathrm{and}\quad n'=-0.028\pm0.020,\end{equation}
where gravitational wave contribution is still neglected. Letting
non-negligible contribution from the gravitational waves relaxes these
bound even further. However, gravitational waves are not observed
yet, and as was mentioned earlier, in this thesis I will assume that
their contribution is negligible.

There are several reports of the detection of the angular modulation
of the power spectrum in Refs.~\cite{Groeneboom_etal(2009)anisotropy2,Hanson_Lewis(2009)CMBanisotropy}.
Ref.~\cite{Groeneboom_etal(2009)anisotropy2} determined the modulation
amplitude $g$ defined in Eq.~\eqref{eq:anisotropic-power-spec-def}
as \begin{equation}
g=0.29\pm0.031,\end{equation}
at $9\sigma$ confidence level. This is a definite proof of the existence
of the preferred direction in the power spectrum. However, these two
works also show that this direction is very close to the ecliptic
poles, with the galactic coordinates $\left(l,b\right)=\left(96,30\right)$.
This is a very strong indication that the origin of the observed anisotropy
is not cosmological but most probably caused by some systematic effects
or comes from within the solar system. Although Ref.~\cite{Groeneboom_etal(2009)anisotropy2}
have investigated the known systematic effects, including the Zodiacal
light, but they could not find any which reproduces the observed signal. 

Given the above value of $g$ we may place an upper limit on the anisotropy
in the power spectrum of the cosmological origin. In this thesis we
will assume that the upper bound on $g$ in the primordial power spectrum
is

\begin{equation}
g\lesssim0.3.\label{eq:bound-on-g}\end{equation}

\subsubsection{The Bispectrum\label{sub:fNL-observational-bounds}}

Although the shape of the two-point correlator or it's counterpart
in the Fourier space, the power spectrum, provides a very valuable
information in discriminating inflationary scenarios and constraining
physics of the early Universe, it has a limited potential. There are
plenty of different inflationary models which predict similar power
spectrum. Very powerful additional tools for distinguishing these
models are higher order correlators. The Fourier transform of the
three-point correlator is called the bispectrum and was defined in
Eq.~\eqref{eq:bispectrum-def}. While only two points are cross-correlated
to obtain the power spectrum an infinitely more configurations are
possible by cross-correlating three points. Therefore, the amount
of information stored in the bispectrum is immensely richer than in
the power spectrum, provided the curvature perturbation is non-Gaussian.

However, as will be seen in section~\ref{sub:Single-Field-Inflation}
single field, slow-roll inflationary models predict negligible non-Gaussianity
of the curvature perturbation. Observationally interesting non-Gaussianity
can be generated only if any of the single field slow-roll assumptions
or some combination of them are violated. These can be classified
into four classes \cite{Komatsu_etal_nonGauss(2009)}: 1) single free
field, 2) canonical kinetic energy, 3) slow roll and 4) initial Bunch-Davies
vacuum. In the first case large non-Gaussianity can be present if
the curvature perturbation is generated by the different field from
the one which drives inflation (two of such mechanisms are discussed
in sections~\ref{sub:End-of-Inflation-Scenario} and \ref{sub:Curvaton-Mechanism})
or in the multifield inflation where the curvature perturbation is
generated by many fields which drives inflation. In addition if the
inflaton cannot be considered as a free field, interaction terms can
produce large non-Gaussianity as well. The second condition is violated
for example in the k-inflation models \cite{Armendriz_etal(1999)kInflation}.
In these class of models the speed of sound is different from the
speed of light. The third condition might be violated if, for example,
the inflaton potential has some sharp features which result in temporally
violation of slow-roll conditions. The fourth assumption considers
the initial fluctuations of the field. Usually it is assumed that
initial quantum fluctuations correspond to the Bunch-Davies vacuum
(see section~\ref{sub:Bunch-Davies-vacuum}) which results in Gaussian
statistics. If, due to some quantum gravitational effects, the initial
quantum state does not correspond to the Bunch-Davies vacuum, field
perturbations may be non-Gaussian and this non-Gaussianity will be
translated into the statistical properties of the curvature perturbation
$\zeta$. 

Usually non-vanishing three point correlator of the curvature perturbation
is parametrized by the non-linearity parameter $\fnl$. There are
several definitions of $\fnl$ in the literature. I will use the one
which coincides with the definition used by WMAP team\begin{equation}
\frac{6}{5}\fnl\equiv\frac{B_{\zeta}\left(\mathbf{k}_{1},\mathbf{k}_{2},\mathbf{k}_{3}\right)}{P_{\zeta}\left(\mathbf{k}_{1}\right)P_{\zeta}\left(\mathbf{k}_{2}\right)+\cp},\label{eq:fNL-def}\end{equation}
where '$\cp$' stands for `cyclic permutations'.%
\footnote{The factor $6/5$ comes from the fact that during matter domination,
which is the case at the era of decoupling, the Newtonian potential
$\Phi$ is related to the curvature perturbation by $\Phi=\frac{3}{5}\zeta$.%
} 

The simplest form of non-Gaussianity is of the local type which can
be written as\begin{eqnarray}
\zeta\left(\mathbf{x}\right) & = & \zeta_{\mathrm{g}}\left(\mathbf{x}\right)+\zeta_{\mathrm{ng}}\left(\mathbf{x}\right)\label{eq:non-Gauss-local-def}\\
 & = & \zeta_{g}\left(\mathbf{x}\right)+\frac{3}{5}\fnl\left(\zeta_{\mathrm{g}}^{2}\left(\mathbf{x}\right)-\left\langle \zeta_{\mathrm{g}}^{2}\right\rangle \right),\nonumber \end{eqnarray}
where $\zeta_{\mathrm{g}}$ is the Gaussian part with zero mean, $\left\langle \zeta_{g}\right\rangle =0$.

The strongest constraints on $\fnl$ comes from the measurements of
the CMB sky. If the non-Gaussianity is of the local type in Eq.~\eqref{eq:non-Gauss-local-def},
then from WMAP5 data the constraint with $95\%$ confidence level
(CL) is (Ref.~\cite{Komatsu_etal_WMAP5(2008)}) \begin{equation}
-9<\fnll<111.\label{eq:fNLl-observational}\end{equation}
In this expression '$\mathrm{local}$' means the `squeezed' configuration
where one momentum is much smaller that the other two, $k_{1}\simeq k_{2}\gg k_{3}$.
In the equilateral configuration with all three momenta of the same
size, $k_{1}=k_{2}=k_{3}$, the constraint on $\fnl$ is weaker\begin{equation}
-151<\fnle<253\label{eq:fNLe-oservational}\end{equation}
at the same CL.

The bounds on the magnitude of $\fnl$ will improve substantially
in the very near future. If it is not detected by the Planck satellite,
the constraints will reduce to $\left|\fnl\right|\lesssim5$ at $95\%$
CL, which is very close to the limit of an ideal experiment of $\left|\fnl\right|\approx3$
at $95\%$ CL, limited by the cosmic variance \cite{Komatsu_Spergel(2001)fNL_bounds}.
The above bounds are given with the assumption that $\fnl$ is isotropic.
Ref.~\cite{Rudjord_etal(2009)anisotropic_fNL} analyzed the WMAP5
data for the angular modulation of $\fnll$. However, due to the large
measurement errors no conclusive statement can be made.

\section{Scalar Field Quantization\label{sec:Scalar-Field-Quantization}}

In this section we discuss the quantization procedure of quantum field
theory (QFT) in flat space-time (FST) and then generalize this formalism
to curved space-time (CST). The discussion is solely about quantization
of scalar fields, because they are the most simple ones and help to
highlight the underlying principles. The extension to vector fields
will be given in Chapter~\ref{cha:Vectors}.

Quantum mechanics was firstly formulated in the so called Schrödinger
picture in which operators are time independent and state vectors
evolve according to the Schrödinger equation. Equally well one can
formulate this theory in the Heisenberg picture, where state vectors
are constant but operators are changing with time. Quantum field theory
can be formulated in both of these pictures as well, but this is much
easier done in the Heisenberg picture, where operators are time dependent
and satisfying field equations. Hence, the name quantum \emph{field}
theory.

\subsection{Quantization in Flat Space-Time\label{sub:Quantization-in-FST}}

Field quantization in FST may be presented in two ways \cite{Teller(1995)book_InterpretiveQFT}.
In the first one we consider a classical field theory. Expand the
field in Fourier modes and find that Fourier coefficients obey the
equation of harmonic oscillator. With every harmonic oscillator we
associate a position variable and the conjugate momentum in field
space. The classical harmonic oscillator is then first quantized.
This is done by substituting c-numbers (classical numbers) of the
position and momentum to the q-numbers (quantum numbers) and imposing
commutation relations which are the result of the Heisenberg uncertainty
principle. Then one finds that the Fourier coefficients (which are
now operators) correspond to the raising and lowering operators in
the Fock space, which are commonly called creation and annihilation
operators respectively.

Another approach is to quantize the degrees of freedom of the classical
field directly. In this approach one identifies the degrees of freedom
of the field, finds their conjugate pairs of variables, changes them
into q-numbers and imposes the same commutation relations as in the
previous case. Only after quantization do we resort to the Fourier
series. Again, we find that Fourier coefficients correspond to creation
and annihilation operators. 

Results of both methods are the same. Although the first method is
more intuitive and easier interpretable, the second method is more
directly generalizable to CST. Hence, in this section we will take
a standpoint of the second method in order to present the FST formalism
in a way which is directly generalizable to the CST case.

In the classical field theory equations of motion (EoM) for fields
are obtained using the least action principle. Forming the action
as\begin{equation}
S\left(\phi\right)=\int\mathcal{L}\left(\phi_{I},\partial_{\mu}\phi_{I}\right)\d x4,\label{eq:FST-action-def}\end{equation}
the classical field equations are calculated by requirement that the
variation of the action should vanish\begin{equation}
\frac{\delta S}{\delta\phi_{I}\left(x\right)}=0,\end{equation}
where $x=\tx$ and $\phi_{I}\left(x\right)$ are classical fields.
In principle these fields could be complex and after quantization
we would find that they describe pairs of particles and antiparticles,
i.e. the field would have a charge. But in context of producing the
curvature perturbation in the Universe we are interested only in the
neutral particles, which agrees with observations of the neutrality
of the Universe. Therefore we will be interested only in real fields
$\phi_{I}\left(x\right)$. 

Let us start with a free, massive, real scalar field. The relativistically
invariant Lagrangian for such field is written as\begin{equation}
\mathcal{L}=\frac{1}{2}\partial_{\mu}\phi\partial^{\mu}\phi-\frac{1}{2}m^{2}\phi^{2},\label{eq:FST-massive-sFd-Lagrangian}\end{equation}
where $m$ is the mass term. The variation of the action in Eq.~\eqref{eq:FST-action-def}
with this Lagrangian gives the familiar Klein-Gordon field equation
for the relativistic field\begin{equation}
\left[\partial_{\mu}\partial^{\mu}+m^{2}\right]\phi=0.\label{eq:FST-Klein-Gordon-eq}\end{equation}

The general solution for this equation can be written as the superposition
of the complete set of orthonormal solutions, $\left\{ u_{\alpha}\left(x\right)\right\} $.
Where orthonormality is defined through the scalar product. For the
Klein-Gordon equation in FST the scalar product of two wave functions
is 

\begin{equation}
\left(u_{m},u_{n}\right)=i\int u_{n}^{*}\overset{\leftrightarrow}{\partial_{0}}u_{m}\d{\mathbf{x}}{}\equiv i\int\left(u_{m}^{*}\partial_{0}u_{n}-u_{n}\partial_{0}u_{m}^{*}\right)\d{\mathbf{x}}{}.\label{eq:FST-scalar-product}\end{equation}
Then the complete set of orthonormal solutions $\left\{ u_{\alpha}\left(x\right)\right\} $
must satisfy \begin{equation}
\left(u_{m},u_{n}\right)=\delta_{mn},\quad\left(u_{m}^{*},u_{n}^{*}\right)=-\delta_{mn}\quad\mathrm{and}\quad\left(u_{m},u_{n}^{*}\right)=\left(u_{m}^{*},u_{n}\right)=0.\label{eq:FST-orthonormality-of-mode-fns}\end{equation}
Indices $m$ and $n$ can be discrete or continuous. In the latter
case Kronecker symbols $\delta_{mn}$ should be replaced by Dirac
delta functions. In these expressions $\left\{ u_{m}\left(x\right)\right\} $
and $\left\{ u_{m}^{*}\left(x\right)\right\} $ denotes a complete
set of positive and negative frequency solutions respectively. Using
these sets of solutions the general solution of Eq.~\eqref{eq:FST-Klein-Gordon-eq}
may be written as the sum of $\left\{ u_{m},u_{m}^{*}\right\} $:

\begin{equation}
\phi\left(x\right)=\sum_{m}\left[a_{m}u_{m}\left(x\right)+a_{m}^{\dagger}u_{m}^{*}\left(x\right)\right],\label{eq:FST-Operator-expansion}\end{equation}
where coefficients $a_{m}$ are given by \begin{equation}
a_{m}=\left(\phi,u_{m}\right)\end{equation}

In the classical field theory $\phi\left(x\right)$ actually describes
an infinite number of degrees of freedom at each space point $\mathbf{x}$.
One can find a conjugate momentum for each of these degrees of freedom
by using equation

\begin{equation}
\pi=\frac{\delta\mathcal{L}}{\delta\dot{\phi}}=\dot{\phi}.\end{equation}

In this way for each spatial point $\mathbf{x}$ we prescribe a generalized
coordinate variable, $\phi\left(x\right)$, and a conjugate momentum,
$\pi\left(x\right)$. Field quantization proceeds by analogy with
quantum mechanics, which is changing c-numbers $\phi\left(x\right)$
and $\pi\left(x\right)$ into q-numbers $\hat{\phi}\left(x\right)$
and $\hat{\pi}\left(x\right)$ and imposing commutation relations
for them\begin{equation}
\left[\hat{\phi}\left(t,\mathbf{x}\right),\hat{\pi}\left(t,\mathbf{x}'\right)\right]=i\delta^{3}\left(\mathbf{x}-\mathbf{x}'\right),\;\left[\hat{\phi}\left(t,\mathbf{x}\right),\hat{\phi}\left(t,\mathbf{x}'\right)\right]=\left[\hat{\pi}\left(t,\mathbf{x}\right),\hat{\pi}\left(t,\mathbf{x}'\right)\right]=0.\label{eq:commutations-for-fields}\end{equation}

Because the field variable $\phi\left(x\right)$ was promoted into
the operator $\hat{\phi}\left(x\right)$, the expansion coefficients
in Eq.~\eqref{eq:FST-Operator-expansion} have to be operators as
well, i.e. the substitution $a_{m}\rightarrow\hat{a}_{m}$ must be
made. And commutation relations for these coefficients may be calculated
from Eqs.~\eqref{eq:commutations-for-fields} \cite{Birrell_Davies(1982),Ford(1997)QFTinCST,Narlikar_Padmanabhan_QuantCosm}:\begin{equation}
\left[\hat{a}_{m},\hat{a}_{n}^{\dagger}\right]=\delta_{mn}\quad\mathrm{and}\quad\left[\hat{a}_{m},\hat{a}_{n}\right]=\left[\hat{a}_{m}^{\dagger},\hat{a}_{n}^{\dagger}\right]=0.\label{eq:commutations-for-a}\end{equation}

Operators $\hat{a}_{m}$ and $\hat{a}_{m}^{\dagger}$ are interpreted
as the rising and lowering operators in Fock space, or creation and
annihilation operators respectively, \begin{equation}
\hat{a}_{m}^{\dagger}\left|n_{m},\left\{ u\right\} \right\rangle =\sqrt{n_{m}+1}\left|n_{m}+1,\left\{ u\right\} \right\rangle \quad\mathrm{and}\quad\hat{a}_{m}\ket{n_{m},\left\{ u\right\} }=\sqrt{n_{m}}\ket{n_{m}-1,\left\{ u\right\} },\end{equation}
 where $n_{m}$ is the number of particles in a state $m$, notation
of $\left\{ u\right\} $ in the ket reminds us that the definition
is for particular complete set of orthonormal mode functions $\left\{ u_{m}\right\} $.
In these equations coefficients $\sqrt{n_{m}+1}$ and $\sqrt{n_{m}}$
are chosen for the correct normalization of the vacuum state $\bra{0,\left\{ u\right\} }\left.0,\left\{ u\right\} \right\rangle =1$,
where the vacuum of this Fock space is defined as\begin{equation}
\hat{a}_{m}\left|0,\left\{ u\right\} \right\rangle =0.\end{equation}

For the following discussion it will be useful to introduce an operator
$\hat{N}$ such that\begin{equation}
\hat{N}_{m}\equiv\hat{a}_{m}^{\dagger}\hat{a}_{m}.\label{eq:FST-number-operator-def}\end{equation}
The meaning of this operator becomes clear when we take the expectation
value \linebreak{}
$\left\langle n_{m}\right|\hat{N}_{m}\ket{n_{m}}$,
\begin{equation}
\left\langle n_{m}\right|\hat{N}_{m}\ket{n_{m}}=\left\langle n_{m}\right|\hat{a}_{m}^{\dagger}\hat{a}_{m}\ket{n_{m}}=n_{m}.\end{equation}
Hence $\hat{N}_{m}$ can be interpreted as the number operator of
$m$ particles.

\subsubsection{\boldmath Interpretation of $\hat{a}_{m}$ and $\hat{a}_{m}^{\dagger}$\label{sub:FST-interpretation-of-a}}

We have mentioned that operators $\hat{a}_{m}$ and $\hat{a}_{m}^{\dagger}$
are interpreted as creation and annihilation operators. What justifies
such interpretation? To show this let us find mode functions $\left\{ u_{m}\right\} $
explicitly. 

First of all $\left\{ u_{m}\right\} $ must represent particles with
positive energy, therefore these functions must be positive frequency
solutions of Eq.~\eqref{eq:FST-Klein-Gordon-eq}, where positive
frequency is defined along some time-like Killing vector satisfying
Lie equation\begin{equation}
\pounds_{\xi}u_{m}=-i\omega_{m}u_{m},\quad\omega_{m}>0.\label{eq:Lie-equation}\end{equation}
To find such solution in the case of FST is a straightforward task
related to the fact that the Poincaré group is the symmetry group
of Minkowski space-time. Therefore, FST possesses the global time-like
Killing vector $\xi^{\mu}=\left(1,0,0,0\right)$. With this Killing
vector Eq.~\eqref{eq:Lie-equation} becomes\begin{equation}
\frac{\partial u_{m}}{\partial t}=-i\omega_{m}u_{m},\end{equation}
 from which it is clear that functions $u_{m}$ must be proportional
to \begin{equation}
u_{m}\propto\mathrm{e}^{-i\omega_{m}t}.\label{eq:FST-u-time-dependance}\end{equation}

Let us conjecture that the full set of orthonormal mode functions
have the form of plane waves \begin{equation}
u_{\mathbf{k}}=\frac{A_{\mathbf{k}}}{\left(2\pi\right)^{3/2}}\mathrm{e}^{i\left(\mathbf{k}\cdot\mathbf{x}-\omega_{\mathbf{k}}t\right)},\label{eq:FST-plane-w-mode-fns}\end{equation}
where instead of the discrete indices $m$ now we have continuous
indices $\mathbf{k}$, corresponding to the wave number of the plane
wave. At the moment $\mathbf{k}$ is just a parameter of the mode
function not related to the momentum. $A_{\mathbf{k}}$ is the normalization
constant which will be fixed later. We can easily check that these
mode functions satisfy the orthonormality conditions in Eqs.~\eqref{eq:FST-orthonormality-of-mode-fns}.
The frequency $\omega_{\mathbf{k}}$ is defined using the Klein-Gordon
equation~\eqref{eq:FST-Klein-Gordon-eq} to be\begin{equation}
\omega_{\mathbf{k}}=+\sqrt{\mathbf{k}^{2}+m^{2}}.\label{eq:FST-frequency-def}\end{equation}
This equation is called relativistic dispersion relation. Note that
we have chosen $\omega_{\mathbf{k}}>0$ in accordance with Eq.~\eqref{eq:Lie-equation}.

With these mode functions the expansion of the field operator in Eq.~\eqref{eq:FST-Operator-expansion}
becomes\begin{equation}
\hat{\phi}\left(x\right)=\int A_{\mathbf{k}}\left[\hat{a}_{\mathbf{k}}\mathrm{e}^{i\left(\mathbf{k}\cdot\mathbf{x}-\omega_{\mathbf{k}}t\right)}+\hat{a}_{\mathbf{k}}^{\dagger}\mathrm{e}^{-i\left(\mathbf{k}\cdot\mathbf{x}-\omega_{\mathbf{k}}t\right)}\right]\frac{\d{\mathbf{k}}{}}{\left(2\pi\right)^{3/2}}.\label{eq:field-operator-expansion}\end{equation}
And because $\mathbf{k}$ is the continuous index, the Kronecker delta
in commutation relations of Eq.~\eqref{eq:commutations-for-a} must
be changed into Dirac delta\begin{equation}
\left[\hat{a}_{\mathbf{k}},\hat{a}_{\mathbf{k}'}^{\dagger}\right]=\delta^{3}\left(\mathbf{k}-\mathbf{k}'\right)\label{eq:commutations-for-a-continuous}\end{equation}
with other commutators being zero. Using these conditions and commutation
relations for the field operator in Eq.~\eqref{eq:commutations-for-fields}
we find\begin{equation}
\left[\hat{\phi}\left(t,\mathbf{x}\right),\hat{\pi}\left(t,\mathbf{x}'\right)\right]=i\int A_{\mathbf{k}}^{2}\omega_{\mathbf{k}}\left(\mathrm{e}^{i\mathbf{k}\cdot\left(\mathbf{x}-\mathbf{x}'\right)}+\mathrm{e}^{-i\mathbf{k}\left(\mathbf{x}-\mathbf{x}'\right)}\right)\frac{\d{\mathbf{k}}{}}{\left(2\pi\right)^{3}}=i\delta^{3}\left(\mathbf{x}-\mathbf{x}'\right),\end{equation}
which fixes the normalization constant \begin{equation}
A_{\mathbf{k}}=\frac{1}{\sqrt{2\omega_{\mathbf{k}}}}.\label{eq:FST-mode-Fn-amplitude}\end{equation}
A more physical motivation for the normalization constant being proportional
to $\left.A_{\mathbf{k}}\propto\omega_{\mathbf{k}}^{-1/2}\right.$
is that the expansion of the operator $\hat{\phi}$ in Eq.~\eqref{eq:field-operator-expansion}
with this normalization becomes relativistically invariant.

To motivate the interpretation of $\hat{a}_{\mathbf{k}}$ and $\hat{a}_{\mathbf{k}}^{\dagger}$
as annihilation and creation operators let us exploit the space translational
symmetry of the FST due to which the following relation must hold:\begin{equation}
\hat{\phi}\left(t,\mathbf{x}\right)=\hat{\phi}\left(t,\mathbf{x}+\delta\mathbf{x}\right),\end{equation}
where $\delta\mathbf{x}$ is the infinitesimal translation vector.
The process of spatial translation of the system may be described
using a unitary transformation

\begin{equation}
\hat{U}=\mathrm{e}^{i\hat{\mathbf{P}}\cdot\mathbf{l}},\label{eq:translation-operator}\end{equation}
where $\mathbf{l}$ is the finite translation vector. Hence, we can
write\begin{equation}
\hat{\phi}\left(t,\mathbf{x}+\mathbf{l}\right)=\hat{U}^{-1}\hat{\phi}\left(t,\mathbf{x}\right)\hat{U}.\label{eq:finite-translation}\end{equation}
For the infinitesimal translation $\mathbf{l}=\delta\mathbf{x}$ the
exponent in Eq.~\eqref{eq:translation-operator} may be expanded
to the first order as $\exp\left(i\hat{\mathbf{P}}\cdot\delta\mathbf{x}\right)=\hat{I}+i\hat{\mathbf{P}}\cdot\delta\mathbf{x}$
and the translational transformation Eq.~\eqref{eq:finite-translation}
becomes\begin{equation}
\hat{\phi}\left(t,\mathbf{x}+\delta\mathbf{x}\right)=\left(\hat{I}-i\hat{\mathbf{P}}\cdot\delta\mathbf{x}\right)\hat{\phi}\left(t,\mathbf{x}\right)\left(\hat{I}+i\hat{\mathbf{P}}\cdot\delta\mathbf{x}\right)=\hat{\phi}\left(t,\mathbf{x}\right)+i\left[\hat{\phi}\left(t,\mathbf{x}\right),\hat{\mathbf{P}}\right]\delta\mathbf{x}.\end{equation}
On the other hand for infinitesimal $\delta\mathbf{x}$ we can also
write \begin{equation}
\hat{\phi}\left(t,\mathbf{x}+\delta\mathbf{x}\right)=\hat{\phi}\left(t,\mathbf{x}\right)+\nabla_{\mathbf{x}}\hat{\phi}\left(t,\mathbf{x}\right)\cdot\delta\mathbf{x}.\end{equation}
Combining these two equations we find that the commutator of $\hat{\phi}$
and $\hat{\mathbf{P}}$ defines the gradient of the quantum field
$\hat{\phi}$: \begin{equation}
\left[\hat{\phi}\left(t,\mathbf{x}\right),\hat{\mathbf{P}}\right]=-i\nabla_{\mathbf{x}}\hat{\phi}\left(t,\mathbf{x}\right).\label{eq:FST-operator-translation}\end{equation}

Let's concentrate now only on the positive frequency part of Eq.~\eqref{eq:field-operator-expansion}\begin{equation}
\hat{\phi}^{+}\left(x\right)\equiv\int\mathrm{e}^{i\left(\mathbf{k}\cdot\mathbf{x}-\omega_{\mathbf{k}}t\right)}\hat{a}_{\mathbf{k}}\d{\tilde{\mathbf{k}}}{},\end{equation}
where \begin{equation}
\d{\tilde{\mathbf{k}}}{}\equiv\frac{\d{\mathbf{k}}{}}{\left(2\pi\right)^{3/2}\sqrt{2\omega_{\mathbf{k}}}}.\end{equation}
Inserting this expression into Eq.~\eqref{eq:FST-operator-translation}
we find\begin{equation}
\int\mathbf{k}\mathrm{e}^{i\mathbf{k}\cdot\mathbf{x}}\hat{a}_{\mathbf{k}}\d{\tilde{\mathbf{k}}}{}=\left[\hat{\phi}^{+}\left(x\right),\hat{\mathbf{P}}\right].\end{equation}
Using commutation relations in Eq.~\eqref{eq:commutations-for-a-continuous}
one can show that this equation is satisfied if the operator $\hat{\mathbf{P}}$
is equal to\begin{equation}
\hat{\mathbf{P}}=\int\mathbf{k}\hat{a}_{\mathbf{k}}^{\dagger}\hat{a}_{\mathbf{k}}\d{\tilde{\mathbf{k}}}{}.\end{equation}

Acting with $\hat{\mathbf{P}}$ on the state $\hat{a}_{\mathbf{k}}^{\dagger}\ket 0$
gives\begin{equation}
\hat{\mathbf{P}}\hat{a}_{\mathbf{k}}^{\dagger}\ket{0,\left\{ u\right\} }=\mathbf{k}\hat{a}_{\mathbf{k}}^{\dagger}\ket{0,\left\{ u\right\} }.\label{eq:FST-momentum-operation}\end{equation}

What does this relation mean? From Nöther's theorem we know that translational
invariance corresponds to the conservation of momentum. From which
follows that the generator of the infinitesimal spatial translation
is the operator for the total momentum, i.e. the operator $\hat{\mathbf{P}}$.
From Eq.~\eqref{eq:FST-momentum-operation} it is clear that the
state $\ket{\mathbf{k},\left\{ u\right\} }=\hat{a}_{\mathbf{k}}^{\dagger}\ket{0,\left\{ u\right\} }$
is the eigenstate of the total momentum $\hat{\mathbf{P}}$ with the
eigenvalue $\mathbf{k}$. Remember, that until know $\mathbf{k}$
was just the index for the mode function. From the last relation $\mathbf{k}$
can be interpreted as the momentum and $\hat{a}_{\mathbf{k}}^{\dagger}$
acts as the momentum rising operator.

If instead of using spatial translation symmetry we would have used
time translational symmetry of the FST with the corresponding unitary
operator\begin{equation}
\hat{T}=\mathrm{e}^{-i\hat{H}t},\end{equation}
we would have found that infinitesimal time translation gives\begin{equation}
\hat{H}=\int\left(\omega_{\mathbf{k}}\hat{a}_{\mathbf{k}}^{\dagger}\hat{a}_{\mathbf{k}}+c\hat{I}\right)\d{\tilde{\mathbf{k}}}{},\label{eq:FST-Hamiltonian}\end{equation}
where $c$ is an arbitrary constant. The divergent constant term in
the above equation in FST can be subtracted by appropriate procedures,
but at the moment it must not concern us. The analogous arguments
which relates $\hat{\mathbf{P}}$ with the total momentum operator,
leads to $\hat{H}$ interpretation as the energy operator. Acting
on the $\hat{a}_{\mathbf{k}}^{\dagger}\ket{0,\left\{ u\right\} }$
state we would find that the operator $\hat{a}_{\mathbf{k}}^{\dagger}$
raises the energy of the state by one unit and that $\omega_{\mathbf{k}}$
can be interpreted as the total energy of that unit or quantum.

\subsection{Quantization in Curved Space-Time\label{sub:Quantization-in-CST}}

In the previous section we have described how fields are quantized
in FST. This procedure is sufficient for the particle physics models,
which studies only three fundamental forces of nature: electromagnetic,
weak and strong. But to give a complete description of the Universe
we need to study how all four fundamental forces, including gravity,
shape and influence each other as well as the structure of the Universe.
This requires a theory which puts all four forces on the same footing.
Unfortunately such theory is still absent - the gravitational force
resists the unification with the other three. In the presence of such
resistance the only hope is to use a semiclassical description of
the Nature, where we treat a classical gravitational background on
which other quantized fields live.

This approximation can be justified by noting that the Planck scale
is the only scale of GR. If we consider small perturbations of the
gravitational field and try to quantize them, then $\mpl^{2}$ plays
the role of the coupling constant. Hence, perturbation theory should
be a good approximations for the energies much smaller than $\mpl$.

We gain confidence in this approach from the early development stages
of the quantum electrodynamics (QED) theory, where the electromagnetic
field was considered as a classical background on which fully quantized
matter lives. And this method is fully consistent with a complete
QED theory.

\subsubsection{From FST to CST}

The quantization of the field living in CST proceeds in the same line
as the quantization in FST. First we write the action for the field.
In CST the analog of Eq.~\eqref{eq:FST-action-def} would be\begin{equation}
S=\int\sqrt{-\dg}\,\mathcal{L}\left(\phi_{I},\nabla_{\mu}\phi_{I}\right)\d x4,\label{eq:CST-action-def}\end{equation}
where $\dg\equiv\det\left(g_{\mu\nu}\right)$ is the determinant of
the metric and $\nabla_{\mu}$ is the covariant derivative. With the
massive free scalar field Lagrangian, which is written in Eq.~\eqref{eq:FST-massive-sFd-Lagrangian},
this action becomes \begin{equation}
S=\int\sqrt{-\dg}\left(\frac{1}{2}\partial_{\mu}\phi\partial^{\mu}\phi-\frac{1}{2}m^{2}\phi^{2}\right)\d x4.\label{eq:CST-action-massive-sFd}\end{equation}
Taking the variation with respect to the scalar field, $\delta S/\delta\phi=0$,
we arrive at the field equation (cf. Eq.~\eqref{eq:FST-Klein-Gordon-eq})\begin{equation}
\left(\square+m^{2}\right)\phi=0,\label{eq:CST-Klein-Gordon-eq}\end{equation}
where the $\square$ operator is defined by\begin{equation}
\square\phi\equiv g^{\mu\nu}\nabla_{\mu}\nabla_{\nu}=\frac{1}{\sqrt{-\dg}}\partial_{\mu}\left(\sqrt{-\dg}g^{\mu\nu}\partial_{\nu}\phi\right).\end{equation}
The scalar product of Eq.~\eqref{eq:FST-scalar-product} for the
Klein-Gordon equation in CST (Eq.~\eqref{eq:CST-Klein-Gordon-eq})
must be generalized as well (e.g. Ref.~\cite{Ford(1997)QFTinCST})\begin{equation}
\left(u_{m},u_{n}\right)=i\int_{\Sigma}u_{n}^{*}\overset{\leftrightarrow}{\partial_{\mu}}u_{m}\d{\Sigma^{\mu}}{},\label{eq:CST-scalar-product}\end{equation}
where $\d{\Sigma}{}^{\mu}\equiv n^{\mu}\d{\Sigma}{}$ and $\d{\Sigma}{}$
is the volume element in a given space-like hypersurface while $n^{\mu}$
is the orthogonal to this hypersurface time-like unit vector. It may
be shown that the value of $\left(u_{m},u_{n}\right)$ is independent
on the choice of the space-like hypersurface $\Sigma$, i.e. $\left(u_{m},u_{n}\right)_{\Sigma_{1}}=\left(u_{m},u_{n}\right)_{\Sigma_{2}}$. 

As in the FST, functions $u_{m}$ must satisfy the orthonormality
conditions in Eq.~\eqref{eq:FST-orthonormality-of-mode-fns} and
then we can write a general solution of Eq.~\eqref{eq:CST-Klein-Gordon-eq}
as the superposition of a complete set of positive frequency $\left\{ u_{m}\right\} $
and negative frequency $\left\{ u_{m}^{*}\right\} $ solutions\begin{equation}
\phi\left(x\right)=\sum_{m}\left[a_{m}u_{m}\left(x\right)+a_{m}^{\dagger}u_{m}^{*}\left(x\right)\right].\label{eq:CST-Fd-expansion-u}\end{equation}

The quantization of the field proceeds exactly as in the FST: change
c-numbers $\phi$ and $\pi$ into q-numbers $\hat{\phi}$ and $\hat{\pi}$
and impose canonical commutation relations of Eq.~\eqref{eq:commutations-for-fields}.
Then operators $\hat{a}_{m}$ and $\hat{a}_{m}^{\dagger}$ are interpreted
as lowering and rising operators in the Fock space\begin{equation}
\hat{a}_{m}^{\dagger}\ket{n_{m},\left\{ u\right\} }=\sqrt{n_{m}+1}\ket{n_{m}+1,\left\{ u\right\} },\end{equation}
with the vacuum defined as \begin{equation}
\hat{a}_{m}\ket{0,\left\{ u\right\} }=0.\label{eq:CST-vacuum-u-def}\end{equation}
As in the previous section $\left\{ u\right\} $ inside the ket reminds
us that we are dealing with the vacuum defined by the complete set
of orthonormal mode functions $\left\{ u_{m}\right\} $. This emphasis
on the choice of mode functions becomes very important in CST as will
be seen in a moment.

\subsubsection{Bogolubov Transformations}

In section~\ref{sub:FST-interpretation-of-a} it was shown that in
FST a natural choice for the complete, orthonormal set of mode functions
exist, which are plane waves of Eq.~\eqref{eq:FST-plane-w-mode-fns}.
And it was emphasized that this happens because the Poincaré group
is the symmetry group of the Minkowski space-time. Hence, using a
global time-like Killing vector, $\partial/\partial t$, of this symmetry
group we could pick-out positive frequency solutions $u_{\mathbf{k}}\propto\exp\left(-i\omega_{\mathbf{k}}t\right)$.
And in all Lorentz frames, where $t$ is the time coordinate, these
mode functions define the same vacuum state. But in CST the Poincaré
group is no longer a symmetry group and in general there will be no
global time-like Killing vectors in respect to which one could define
positive frequency solutions. Therefore, the field expansion in mode
functions $\left\{ u_{m}\right\} $ in Eq.~\eqref{eq:CST-Fd-expansion-u}
is as good as in any other complete set of orthonormal functions:\begin{equation}
\phi\left(x\right)=\sum_{m}\left[b_{m}v_{m}\left(x\right)+b_{m}^{\dagger}v_{m}^{*}\left(x\right)\right].\label{eq:CST-Fd-expansion-v}\end{equation}
After quantization the vacuum state for this expansion is defined
by\begin{equation}
\hat{b}_{m}\ket{0,\left\{ v\right\} }=0.\label{eq:CST-vacuum-v-def}\end{equation}

The definition of the vacuum state in Eq.~\eqref{eq:CST-vacuum-u-def}
with mode functions $\left\{ u_{m}\right\} $ at least formally differs
from the definition with the mode functions $\left\{ v_{m}\right\} $
in Eq.~\eqref{eq:CST-vacuum-v-def}. Shortly it will be clear that
this difference is not only formal but indeed both states $\ket{0,\left\{ u\right\} }$
and $\ket{0,\left\{ v\right\} }$ correspond to a different physical
vacuum. Which means that there is no way to define uniquely a state
without particles: what for one is a vacuum state, for the other this
state contains particles. In such situation the notion of {}``the
physical particle'' becomes ambiguous.

Since both sets of mode functions are complete orthonormal sets of
solutions, each function in one set can be expanded in terms of the
another set, i.e.\begin{equation}
v_{n}=\sum_{m}\left(\alpha_{nm}u_{m}+\beta_{nm}u_{m}^{*}\right)\quad\mathrm{or}\quad u_{m}=\sum_{n}\left(\alpha_{nm}^{*}v_{n}+\beta_{nm}v_{n}^{*}\right).\label{eq:Bogolubov-transformations}\end{equation}
These are the so called Bogolubov transformations, and matrices $\alpha_{nm}$
and $\beta_{nm}$ are called Bogolubov coefficients. It can be easily
checked that these coefficients satisfy the relations\begin{eqnarray}
\sum_{l}\left(\alpha_{nl}\alpha_{ml}^{*}-\beta_{ml}\beta_{nl}^{*}\right) & = & \delta_{nm},\\
\sum_{l}\left(\alpha_{ml}\beta_{nl}-\beta_{ml}\alpha_{nl}\right) & = & 0.\end{eqnarray}

Comparing Eqs.~\eqref{eq:CST-Fd-expansion-u} and \eqref{eq:CST-Fd-expansion-v}
and using Bogolubov transformations in Eq.~\eqref{eq:Bogolubov-transformations}
we find the relation between creation and annihilation operators of
one set of mode functions and the other:\begin{equation}
\hat{a}_{m}=\sum_{n}\left(\alpha_{nm}\hat{b}_{n}+\beta_{nm}^{*}\hat{b}_{n}^{\dagger}\right)\quad\mathrm{or}\quad\hat{b}_{n}=\sum_{m}\left(\alpha_{nm}^{*}\hat{a}_{m}-\beta_{nm}^{*}\hat{a}_{m}^{\dagger}\right).\end{equation}

Using these relations we can calculate the expectation value of the
number operator defined by $\hat{N}_{m}^{\left\{ u\right\} }\equiv\hat{a}_{m}^{\dagger}\hat{a}_{m}$
(cf. Eq.~\eqref{eq:FST-number-operator-def}). Acting with $\hat{N}_{m}^{\left\{ u\right\} }$
on the vacuum defined by the mode functions $\left\{ v_{m}\right\} $
we find\begin{eqnarray}
\bra{0,\left\{ v\right\} }\hat{N}_{m}^{\left\{ u\right\} }\ket{0,\left\{ v\right\} } & = & \bra{0,\left\{ v\right\} }\sum_{n,n'}\left(\alpha_{nm}^{*}\hat{b}_{n}^{\dagger}+\beta_{nm}\hat{b}_{n}\right)\left(\alpha_{n'm}\hat{b}_{n'}+\beta_{n'm}^{*}\hat{b}_{n'}^{\dagger}\right)\ket{0,\left\{ v\right\} }\nonumber \\
 & = & \bra{0,\left\{ v\right\} }\sum_{n,n'}\beta_{nm}\beta_{n'm}^{*}\hat{b}_{n}\hat{b}_{n'}^{\dagger}\ket{0,\left\{ v\right\} }\label{eq:particle-number-EV}\\
 & = & \sum_{n}\left|\beta_{mn}\right|^{2}.\nonumber \end{eqnarray}
This shows that the vacuum defined by the complete set $\left\{ v_{m}\right\} $
contains particles of the mode functions $\left\{ u_{m}\right\} $.

The freedom of the choice of mode functions and the related ambiguity
of the vacuum state constitutes the main problem of quantum field
theory in curved space-time. One is naturally led to ask, which is
{}``the physical vacuum'' and what are the observables of such theory.
In general, there is no way to pick out one particular set of mode
functions. But in some space-times, which have a high degree of symmetry,
this might be possible. As will be seen in the following subsection,
this for example happens in a space-time with maximal spatial symmetry
such as FRW and (quasi) de Sitter universes. The phenomena described
in Eq.~\eqref{eq:particle-number-EV} are very important in inflationary
particle creation.

\subsubsection{Quantization in Spatially Homogeneous and Isotropic Backgrounds}

In this section we describe the process of the scalar field particle
creation in the exponentially expanding Universe. Let us consider
spatially homogeneous and isotropic FRW metric of Eq.~\eqref{eq:FRW-metric-flat-Cartesian}.
But instead of using the cosmic time $t$ we rewrite this metric in
terms of the conformal time $\tau$ defined as\begin{equation}
\tau\left(t\right)\equiv\int^{t}\frac{\d t{}'}{a\left(t'\right)}.\end{equation}
Then the line element with FRW metric becomes manifestly conformal
to Minkowski space-time\begin{equation}
\d s{}^{2}=a^{2}\left(\tau\right)\left(\d{\tau}{}^{2}-\d{\mathbf{x}}{}^{2}\right).\label{eq:FRW-conformal-metric}\end{equation}

In the FRW metric the action of the free massive scalar field written
Eq.~\eqref{eq:CST-action-massive-sFd} becomes\begin{equation}
S=\frac{1}{2}\int a^{2}\left(\phi'^{2}-\left(\nabla\phi\right)^{2}-a^{2}m^{2}\phi^{2}\right)\d{\mathbf{x}}{}\d{\tau}{},\end{equation}
where the prime denotes the derivative with respect to the conformal
time, $'\equiv\frac{\d{}{}}{\d{\tau}{}}$ and $\nabla\equiv\partial_{i}$
is the spatial gradient. As was already performed several times, taking
the variation of this action gives the field equation\begin{equation}
\phi''+2\frac{a'}{a}\phi'-\nabla^{2}\phi+a^{2}m^{2}\phi=0.\label{eq:FRW-massive-sFd-eq}\end{equation}

This equation is very similar to the Klein-Gordon equation in FST
given in Eq.~\eqref{eq:FST-Klein-Gordon-eq}, except that it has
a friction term $2a'/a\cdot\phi'$. Let us transform this equation
in such a way that it does become like the Klein-Gordon equation in
FST. This can be achieved using the following mathematical trick,
which brings any second order linear differential equation to it's
normal form. If the equation is given as\begin{equation}
\frac{\d y2}{\d x{}^{2}}+P\left(x\right)\frac{\d y{}}{\d x{}}+Q\left(x\right)y=0,\label{eq:diff-eq}\end{equation}
then the transformation\begin{equation}
u=y\mathrm{e}^{\frac{1}{2}\int^{x}P\left(x'\right)\d{x'}{}}\label{eq:diff-eq-transformation}\end{equation}
brings it into the form of the harmonic oscillator \begin{equation}
\frac{\d u2}{\d x{}^{2}}+\left(Q-\frac{1}{2}\frac{\d P{}}{\d x{}}-\frac{1}{4}P^{2}\right)u=0.\end{equation}

For the equation~\eqref{eq:FRW-massive-sFd-eq} the analogous transformation
would be\begin{equation}
\chi\equiv\phi\mathrm{e}^{\frac{1}{2}\int2\frac{a'}{a}\d{\tau}{}}=a\left(\tau\right)\phi,\label{eq:chi-def}\end{equation}
which transforms Eq.~\eqref{eq:FRW-massive-sFd-eq} into the form\begin{equation}
\chi''-\nabla^{2}\chi+\left(a^{2}m^{2}-\frac{a''}{a}\right)\chi=0.\label{eq:FRW-Klein-Gordon-eq}\end{equation}
This equation does look like the Klein-Gordon one in FST except the
time varying mass.

The quantization of the scalar field $\chi$ again proceeds as in
the previous section: find the conjugate momentum of the field, which
in conformal FRW space-time is $\pi\equiv\delta\mathcal{L}/\delta\chi'=\chi'-\frac{a'}{a}\chi$,
make changes of c-numbers into q-numbers, i.e. $\chi\rightarrow\hat{\chi}$
and $\pi\rightarrow\hat{\pi}$ and impose canonical commutation relations
(cf. Eq.~\eqref{eq:commutations-for-fields})\begin{equation}
\left[\hat{\chi}\left(\tau,\mathbf{x}\right),\hat{\pi}\left(\tau,\mathbf{x}'\right)\right]=i\delta\left(\mathbf{x}-\mathbf{x}'\right),\;\left[\hat{\chi}\left(\tau,\mathbf{x}\right),\hat{\chi}\left(\tau,\mathbf{x}'\right)\right]=\left[\hat{\pi}\left(\tau,\mathbf{x}\right),\hat{\pi}\left(\tau,\mathbf{x}'\right)\right]=0.\label{eq:FRW-chi-commutators}\end{equation}

The field operator $\hat{\chi}$ expanded into the creation and annihilation
operators is written as (c.f. Eqs.~\eqref{eq:CST-Fd-expansion-u}
or \eqref{eq:CST-Fd-expansion-v})\begin{equation}
\hat{\chi}\left(x\right)=\sum_{m}\left[\hat{a}_{m}\chi_{m}\left(x\right)+\hat{a}_{m}^{\dagger}\chi_{m}^{*}\left(x\right)\right],\label{eq:FRW-chi-formal-expansion}\end{equation}
where $x=\left(\tau,\mathbf{x}\right)$ from the metric in Eq.~\eqref{eq:FRW-conformal-metric}.

Mode functions $\chi_{m}\left(x\right)$ must satisfy the orthonormality
conditions in Eq.~\eqref{eq:FST-orthonormality-of-mode-fns}. The
general scalar product of Eq.~\eqref{eq:CST-scalar-product} in the
FRW metric becomes \begin{equation}
\left(\chi_{m},\chi_{n}\right)=i\int\left(\chi_{n}^{*}\chi_{m}'-\chi_{m}\chi_{n}^{*}\overset{}{}'\right)\d{\mathbf{x}}{}.\label{eq:FRW-scalar-product}\end{equation}

Because the $\left(0,i\right)$ and $\left(i,0\right)$ components
of the FRW metric in Eq.~\eqref{eq:FRW-conformal-metric} are zero,
the mode functions $\chi_{m}\left(x\right)$ can be chosen in such
a way that the temporal and spatial parts are separated \begin{equation}
\chi_{m}\left(\tau,\mathbf{x}\right)\equiv\left(2\pi\right)^{-3/2}\chi_{\mathbf{k}}\left(\tau\right)\mathrm{e}^{i\mathbf{k}\cdot\mathbf{x}},\label{eq:FRW-chi-ansatz}\end{equation}
where $\mathbf{k}$ is now the continuous expansion coefficient and
the factor $\left(2\pi\right)^{-3/2}$ is pulled out in order for
the normalization of $\chi_{\mathbf{k}}\left(\tau\right)$ (see Eq.~\eqref{eq:FST-orthonormality-of-mode-fns})
to give the Wronskian of the form\begin{equation}
\chi_{\mathbf{k}}\chi_{\mathbf{k}}^{*}\overset{}{}'-\chi_{\mathbf{k}}^{*}\chi_{\mathbf{k}}'=i,\label{eq:FRW-Wronskian}\end{equation}
which is required from the orthonormality condition and is obtained
using the scalar product in Eq.~\eqref{eq:FRW-scalar-product}.

With the ansatz in Eq.~\eqref{eq:FRW-chi-ansatz} the expansion of
the operator $\hat{\chi}$ in Eq.~\eqref{eq:FRW-chi-formal-expansion}
becomes\begin{equation}
\hat{\chi}\left(x\right)=\int\left(\hat{a}_{\mathbf{k}}\chi_{\mathbf{k}}\left(\tau\right)\mathrm{e}^{i\mathbf{k}\cdot\mathbf{x}}+\hat{a}_{\mathbf{k}}^{\dagger}\chi_{\mathbf{k}}^{*}\left(\tau\right)\mathrm{e}^{-i\mathbf{k}\cdot\mathbf{x}}\right)\frac{\d{\mathbf{k}}{}}{\left(2\pi\right)^{3/2}}.\label{eq:FRW-chi-Fourier}\end{equation}

Substituting $\hat{\chi}\left(x\right)$ into Eq.~\eqref{eq:FRW-Klein-Gordon-eq}
we find that functions $\chi_{\mathbf{k}}\left(\tau\right)$ must
satisfy the equation of motion\begin{equation}
\chi_{\mathbf{k}}\left(\tau\right)''+\left(aH\right)^{2}\left[\left(\frac{m}{H}\right)^{2}+\left(\frac{k}{aH}\right)^{2}-\frac{\dot{H}}{H^{2}}-2\right]\chi_{\mathbf{k}}\left(\tau\right)=0,\label{eq:FRW-chik-FdEq}\end{equation}
where we used $a''/a^{3}=\dot{H}+2H^{2}$. One can see that the equation
of motion for $\chi_{\mathbf{k}}$ is formally the same as of the
harmonic oscillator with the time dependent frequency\begin{equation}
\omega_{\mathbf{k}}^{2}\left(\tau\right)\equiv a^{2}m^{2}+k^{2}-\frac{a''}{a}=\left(aH\right)^{2}\left[\left(\frac{m}{H}\right)^{2}+\left(\frac{k}{aH}\right)^{2}-\frac{\dot{H}}{H^{2}}-2\right].\label{eq:FRW-frequency-def}\end{equation}

Although Eq.~\eqref{eq:FRW-chik-FdEq} constraints the time dependence
of functions $\chi_{\mathbf{k}}\left(\tau\right)$ it does not determine
the function uniquely. In fact, any function which satisfies Eq.~\eqref{eq:FRW-Wronskian}
will be as good a choice as $\chi_{\mathbf{k}}\left(\tau\right)$.
As was explained in the previous subsection, this fact deprive us
of possibility to determine a vacuum state, defined as $\hat{a}_{\mathbf{k}}\ket{0,\left\{ \chi_{\mathbf{k}}\right\} }$,
which would be seen as absent of particles by any observer. On the
other hand, the quantum field theory in FST is a very successful theory
although we do live in the expanding Universe, i.e. curved space-time.
Therefore, we can expect that it is possible to pick out some special
definition of the vacuum which would give correct predictions for
laboratory experiments. The main reason why flat space-time QFT is
so successful from this point of view is that it describes phenomena
which take place in a very weak gravitational field, or in other words,
very weakly curved space-time, which may be neglected. 

This can be easily seen from Eq.~\eqref{eq:FRW-chik-FdEq}. If we
neglect the effect of gravity, which corresponds to taking $a=1$
and therefore $H=0$, the frequency term in Eq.~\eqref{eq:FRW-frequency-def}
becomes constant, $\omega_{\mathbf{k}}^{2}=m^{2}+k^{2}=\mathrm{const}$,
the same as in FST in Eq.~\eqref{eq:FST-frequency-def}.%
\footnote{This motivates us to interpret the expansion coefficient $\mathbf{k}$
as the comoving momentum of the particle (cf. section~\ref{sub:FST-interpretation-of-a})
and $\mathbf{k}/a$ as the physical momentum.%
} With a constant frequency, functions $\chi_{\mathbf{k}}\left(\tau\right)$
(or functions $u_{m}\left(t\right)$ in Eq.~\eqref{eq:FST-u-time-dependance})
have the time dependent part $\exp\left(-i\omega_{\mathbf{k}}\tau\right)$.
The vacuum defined in this way will be the same for all inertial observers
at all times. But the frequency term in the expanding Universe in
Eq.~\eqref{eq:FRW-chik-FdEq} is time dependent. Hence, the vacuum
defined at time $\tau_{i}$ will contain particles as seen by the
observer at some later time. This is the main reason why particles
get produced during inflation, but let us postpone this discussion
until a bit later. At the moment the important thing is the choice
of initial conditions which would fix the form of mode functions and
therefore the initial vacuum state.

\subsubsection{The Vacuum State in FRW Background\label{sub:Bunch-Davies-vacuum}}

Although, as was mentioned earlier, in general, the particle concept
in CST is ambiguous, in some special cases it is possible to define
an approximate particle concept which would be as close as possible
to the one known from QFT in FST. This is the case, for example, in
space-times described by the FRW metric or anisotropic Bianchi universes.
In the case of the present interest we may look for the solution of
Eq.~\eqref{eq:FRW-chik-FdEq} with the ansatz\begin{equation}
\chi_{\mathbf{k}}\left(\tau\right)=\frac{1}{\sqrt{2W_{\mathbf{k}}\left(\tau\right)}}\mathrm{e}^{-i\int_{\eta_{i}}^{\eta}W_{\mathbf{k}}\left(\tau'\right)\d{\tau'}{}}.\label{eq:WKB-ansatz}\end{equation}
The factor $1/\sqrt{2}$ is chosen so that $\chi_{\mathbf{k}}\left(\tau\right)$
would satisfy Eq.~\eqref{eq:FRW-Wronskian} and the function $W_{\mathbf{k}}\left(\tau\right)$
satisfies\begin{equation}
W_{\mathbf{k}}^{2}\left(\tau\right)=\omega_{\mathbf{k}}^{2}\left(\tau\right)-\left[\frac{1}{2}\frac{W''_{\mathbf{k}}}{W_{\mathbf{k}}}-\frac{3}{4}\left(\frac{W_{\mathbf{k}}'}{W_{\mathbf{k}}}\right)^{2}\right],\label{eq:W-inWKB-def}\end{equation}
which can be found by substituting the ansatz in Eq.~\eqref{eq:WKB-ansatz}
into Eq.~\eqref{eq:FRW-chik-FdEq}. If the time variation of $\omega_{\mathbf{k}}\left(\tau\right)$
is very slow, it is said that it satisfies the adiabatic condition
and the vacuum defined when this condition is valid is called the
adiabatic vacuum. By {}``slow'' we mean that $\omega_{\mathbf{k}}\left(\tau\right)$
and all its derivatives change substantially, $\Delta\omega_{\mathbf{k}}/\omega_{\mathbf{k}}\sim\mathcal{O}\left(1\right)$,
only during the time interval $T\gg\omega_{\mathbf{k}}^{-1}$ (Ref.~\cite{Mukhanov_Winitzki_book}).
In the adiabatic case, derivative terms in Eq.~\eqref{eq:W-inWKB-def}
will be small and this equation can be solved using the recursive
method. For example, to the zeroth order we can take\begin{equation}
W_{\mathbf{k}}^{\left(0\right)}\left(\tau\right)=\omega_{\mathbf{k}}\left(\tau\right).\end{equation}
Note that for the constant frequency, $\omega_{\mathbf{k}}=\mathrm{const}$,
the mode functions in CST (Eq.~\eqref{eq:FRW-chi-ansatz}) with $\chi_{\mathbf{k}}\left(\tau\right)$
given by Eq.~\eqref{eq:WKB-ansatz} reduce to the mode functions
in FST (cf. Eqs.~\eqref{eq:FST-plane-w-mode-fns} and \eqref{eq:FST-mode-Fn-amplitude}).

In Eq.~\eqref{eq:FRW-chik-FdEq} the adiabatic vacuum can be defined
for light particles ($m/H\ll1$) whose Compton wavelength is much
smaller than the curvature scale $H^{-1}$, or in other words whose
physical momentum is much grater than the Hubble expansion rate, $\left.k/a\gg H\right.$.%
\footnote{For the quasi de Sitter expansion $\dot{H}/H^{2}\ll1$ (cf. Eq.~\eqref{eq:Inflation-Hdot-ll-Hsq})
and for the FRW Universe $\dot{H}/H^{2}\sim\mathcal{O}\left(1\right)$,
so that, when $k/a\gg H$, terms of order one or less are subdominant
in Eq.~\eqref{eq:FRW-frequency-def}.%
} We may say that such particles do not {}``feel'' the gravitational
field. Therefore, by substituting $W_{\mathbf{k}}^{\left(0\right)}\left(\tau\right)$
into Eq.~\eqref{eq:WKB-ansatz} and taking that \begin{equation}
\omega_{\mathbf{k}}\approx k,\end{equation}
we find the initial condition for the mode function $\chi_{\mathbf{k}}\left(\tau\right)$\begin{equation}
\chi_{\mathbf{k}}\left(\tau\right)=\frac{1}{\sqrt{2k}}\mathrm{e}^{-ik\tau},\label{eq:initial-vacuum-mode-Fn}\end{equation}
which is the same as that of the massless field in FST (cf. Eqs.~\eqref{eq:FST-plane-w-mode-fns}
and \eqref{eq:FST-mode-Fn-amplitude}). This vacuum state is often
called the Bunch-Davies vacuum.

The result of Eq.~\eqref{eq:initial-vacuum-mode-Fn} was obtained
by assuming that the space-time can be considered flat at the zero
order approximation for subhorizon modes. This must be always valid
for the Einstein gravity in accordance to the equivalence principle.
But in many models the inflationary energy scale is just couple of
orders of magnitude below the Planck scale. At such energy scales
it might be that Einstein's theory of gravity is not precise enough
to describe Nature. In this case, one may expect that the equivalence
principle does not hold anymore. But this failure might be only at
the level of $\left(H/\mpl\right)^{2}\lesssim10^{-10}$, where $H$
is the inflationary Hubble parameter \cite{Lyth_Liddle(2009)book}.

\subsubsection{The Field Perturbation in the Inflationary Universe\label{sub:Fd-Perturbation-Inflation}}

It was already mentioned in section~\ref{sec:Inflation} that inflation
provides a natural mechanism to explain the origin of the curvature
perturbation in the early Universe. Upon entering the horizon this
perturbation sets the initial conditions for the tiny density inhomogeneities
which seeded the subsequent growth of large scale structure such as
galaxies and galaxy clusters. In this subsection we describe how field
perturbations are generated in the inflationary Universe and in the
next section how they are transformed into the curvature perturbation. 

The generation of the field perturbation can be computed from Eq.~\eqref{eq:FRW-chik-FdEq}
with appropriate initial conditions and assumptions relevant for the
inflationary expansion. During inflation the Universe undergoes quasi
de Sitter expansion for which the condition $\left|\dot{H}\right|/H^{2}\ll1$
is satisfied (see Eq.~\eqref{eq:Inflation-Hdot-ll-Hsq}). But for
our purpose in this section and for later discussions in Chapter~\ref{cha:Vectors}
it is enough to take the approximation of a quasi de Sitter Universe.
Hence, we will set $\dot{H}=0$ which is equivalent to considering
exact de Sitter space-time.%
\footnote{More precisely only a part of de Sitter space-time is considered because
inflation lasts only for a finite time. %
}

Another assumption we make is that initially the state corresponds
to its vacuum, i.e. the average occupation number of particles with
momentum $k$ is much less than 1, $\bar{n}_{k}\ll1$. This assumption
is easily justified if enough amount of inflation occurred before
the horizon exit of the scales of interest (see e.g. Ref.~\cite{Lyth_Liddle(2009)book}).
In this case the initial conditions for the mode functions is determined
by the Bunch-Davies vacuum in Eq.~\eqref{eq:initial-vacuum-mode-Fn}.

Perturbations of scalar fields with the mass comparable to the Hubble
parameter do not grow significantly as will be seen in Eq.~\eqref{eq:sFd-massive-perturb-spec}.
While in section~\ref{sub:Quantum-to-Classical} it will be shown
that perturbations of heavy fields, with the mass $m>\frac{3}{2}H$,
do not become classical after horizon crossing. Therefore, only light
scalar fields are considered for the generation of the curvature perturbation.
Assuming de Sitter expansion the general solution of Eq.~\eqref{eq:FRW-chik-FdEq}
becomes\begin{equation}
\chi_{\mathbf{k}}\left(\tau\right)=\sqrt{\frac{-\tau\pi}{2}}\mathrm{e}^{i\frac{\pi}{4}\left(2\nu+1\right)}H_{\nu}^{\left(1\right)}\left(-k\tau\right),\end{equation}
where the initial state was matched to the Bunch-Davies vacuum in
Eq.~\eqref{eq:initial-vacuum-mode-Fn}, and $H_{\nu}^{\left(1\right)}\left(-k\tau\right)$
denotes the Hankel function and we used in the de Sitter space-time
$\tau=-\left(aH\right)^{-1}$. The order of $H_{\nu}^{\left(1\right)}$
is defined as \begin{equation}
\nu\equiv\sqrt{\frac{9}{4}-\left(\frac{m}{H}\right)^{2}}.\label{eq:niu-def}\end{equation}
Well after horizon exit, when $\left|k\tau\right|\ll1$, this solution
approaches to\begin{equation}
\chi_{\mathbf{k}}\left(\tau\right)\simeq\frac{\mathrm{e}^{i\frac{\pi}{2}\left(\nu-\frac{1}{2}\right)}}{\sqrt{2k}}\frac{\Gamma\left(\nu\right)}{\sqrt{\pi}}\left(\frac{-k\tau}{2}\right)^{\frac{1}{2}-\nu}.\label{eq:sFd-superhorizon-perturbations}\end{equation}

For the light field $m<\frac{2}{3}H$ the parameter $\nu$ is real
and this solution is not oscillatory, therefore interpretation of
corresponding states $\ket{\psi}$ as physical particle states in
the Fock space is problematic. The reason for this, as can be seen
from Eq.~\eqref{eq:FRW-frequency-def}, is that the dispersion relation
for a light field, with $m\ll H$, becomes imaginary and the mode
function does not oscillate.

However, the amplitude of quantum fluctuation in the state $\ket{\psi}$
is always well defined and we can calculate the expectation value
for the vacuum state $\ket 0$ as\begin{equation}
\bra 0\hat{\chi}\left(\tau,\mathbf{x}\right)\hat{\chi}\left(\tau,\mathbf{y}\right)\ket 0=\frac{1}{2\pi^{2}}\int_{0}^{\infty}k^{3}\left|\chi_{\mathbf{k}}\left(\tau\right)\right|^{2}\frac{\sin kL}{kL}\frac{\d k{}}{k}\equiv\int_{0}^{\infty}\mathcal{P}_{\chi}\left(k\right)\frac{\sin kL}{kL}\frac{\d k{}}{k},\end{equation}
where $L\equiv\left|\mathbf{x}-\mathbf{y}\right|$ and $\mathcal{P}_{\chi}\left(k\right)\equiv\left(k^{3}/2\pi^{2}\right)\left|\chi_{\mathbf{k}}\right|^{2}$
is the power spectrum.

The power spectrum of the superhorizon massive scalar field perturbations
can be easily calculated using Eq.~\eqref{eq:sFd-superhorizon-perturbations}
\begin{equation}
\mathcal{P}_{\chi}=\frac{4\Gamma^{2}\left(\nu\right)}{\pi}\left(\frac{aH}{2\pi}\right)^{2}\left(\frac{k}{2aH}\right)^{3-2\nu}.\end{equation}
However, this expression is derived for the comoving field $\chi$.
Going back to the physical field $\phi=\chi/a$ (see Eq.~\eqref{eq:chi-def})
and considering the massless limit $\nu=3/2$, this expression reduces
to \begin{equation}
\mathcal{P}_{\phi}=\left(\frac{H}{2\pi}\right)^{2}.\label{eq:sFd-scale-inv-perturb-spect}\end{equation}
Or more generally, for a light field, $m\lesssim\frac{3}{2}H$, we
can express Eq.~\eqref{eq:niu-def} as\begin{equation}
\nu\simeq\frac{3}{2}-\frac{m^{2}}{3H^{2}},\end{equation}
and the power spectrum becomes\begin{equation}
\mathcal{P}_{\phi}\simeq\left(\frac{H}{2\pi}\right)^{2}\left(\frac{k}{2aH}\right)^{\frac{2}{3}\left(\frac{m}{H}\right)^{2}}.\label{eq:sFd-massive-perturb-spec}\end{equation}

Although Eq.~\eqref{eq:sFd-scale-inv-perturb-spect} was calculated
for the exact de Sitter expansion, one can easily include a very slow
variation of the Hubble parameter, $\dot{H}\ne0$. Due to this small
variation, the horizon size changes very slightly during inflation
and therefore each mode $k$ exits a horizon of slightly different
size. But from Eq.~\eqref{eq:sFd-scale-inv-perturb-spect} it is
clear that the amplitude of the field perturbation is proportional
to the horizon size. Therefore, different modes will have slightly
different amplitudes. This mild dependence of the power spectrum on
$k$ may be accounted for in Eq.~\eqref{eq:sFd-scale-inv-perturb-spect}
by writing\begin{equation}
\mathcal{P}_{\phi}=\left(\frac{H_{k}}{2\pi}\right)^{2},\label{eq:slow-roll-Fd-perturb-spect}\end{equation}
where the Hubble parameter $H_{k}$ in this expression has to be evaluated
at the horizon exit for each mode, i.e. when $aH_{k}=k$, and $H_{k}$
is slowly varying with $k$.

It is important to note that because we have assumed the initial state
for each mode to start in the Bunch-Davies vacuum and considered a
free field, perturbations of the field are Gaussian. This is the result
of the equivalence principle valid for Einstein's gravity. But if
at inflationary energies the equivalence principle does not hold,
for example due to modified gravity, the perturbations of the field
may be significantly non-Gaussian. In addition, because we were concerned
in this section about scalar fields which are rotationally invariant,
the perturbations of the field are statistically isotropic. This will
not be the case in Chapter~\ref{cha:Vectors} where we discuss perturbations
of vector fields.

\subsubsection{Quantum to Classical Transition\label{sub:Quantum-to-Classical}}

As it is clear from the discussion so far, the origin of the field
perturbation is quantum mechanical. But as was claimed in section~\ref{sub:The-Accelerated-Expansion}
the greatest success of inflationary paradigm is that it can explain
how these quantum mechanical perturbations give rise to the initial
density perturbations in the Universe which are observed as CMB temperature
anisotropies and which seed the formation of galaxies. But CMB temperature
anisotropies and galaxies are not quantum but classical objects. Hence,
there must be some transition period where quantum mechanical perturbations
are transformed into classical ones. This process is analogous to
the decoherence in quantum mechanics. The result of it is that the
quantum mechanical superposition principle is violated and the wavefunction
collapses to a particular state obeying the classical evolution. The
coherence between different states is lost, since after the collapse
only one state can be observed, although quantum mechanically all
states should be allowed. In usual applications of quantum mechanics
this happens due to the wavefunction interaction with the degrees
of freedom of the environment. But in cosmological context the transition
from quantum-to-classical does not require environment, therefore
in Ref.~\cite{Polarski_Starobinski(1995)} it was named {}``decoherence
without decoherence''. In the exposition of quantum-to-classical
transition below we will follow Refs.~\cite{Polarski_Starobinski(1995),Lesgourgues_etal(1997),Kiefer_etal(1998a),Kiefer_etal(1998b)}.

What does it mean, that perturbations in the Universe are classical?
As was discussed in section~\ref{sec:Statistical-Properties} these
perturbations can be described by random fields $\beta$. The classicality
of perturbations means that $\beta$ is described by classical stochastic
variables. But to make the discussion easier, instead of treating
some general variable $\beta$ let us specialize to a field perturbation
$\chi\left(\mathbf{x}\right)$ defined in Eq.~\eqref{eq:chi-def}.
Then $\chi\left(\mathbf{x}\right)$ will be classical if it is described
as classical stochastic variable with the probability distribution
function $p\left(\left|\chi\right|,\left|\pi\right|\right)$, where
$\pi$ is the canonical conjugate of $\chi$.

In quantum mechanics a classical limit is achieved when the state
collapses to the definite numerical value. But in the cosmological
context we cannot assign the definite numerical value to a collapsed
state. So we say that the quantum state becomes classical if the field
modes become equivalent to the classical stochastic functions with
the probability distribution $p\left(\text{\ensuremath{\left|\chi\right|},\ensuremath{\left|\pi\right|}}\right)$
\cite{Polarski_Starobinski(1995)}. This can be written as\begin{eqnarray}
 &  & \bra 0G\left(\hat{X}_{m},\hat{\pi}_{m}\right)G^{\dagger}\left(\hat{X}_{m},\hat{\pi}_{m}\right)\ket 0=\nonumber \\
 &  & \qquad\qquad=\int\d{X_{m1}}{}\d{X_{m2}}{}\d{\pi_{m1}}{}\d{\pi_{m2}}{}p\left(\left|X_{m}\right|,\left|\pi_{m}\right|\right)\left|G\left(X_{m},\pi_{m}\right)\right|^{2},\qquad\label{eq:quantum-classical-equivalence-gen}\end{eqnarray}
where from Eq.~\eqref{eq:FRW-chi-formal-expansion} \begin{equation}
\hat{\chi}\left(x\right)=\int\hat{X}_{m}\left(x\right)\d m{}\equiv\int\left[\hat{a}_{m}\chi_{m}\left(x\right)+\hat{a}_{m}^{\dagger}\chi_{m}^{*}\left(x\right)\right]\d m{},\label{eq:q-to-c-quantum-expansion}\end{equation}
$m$ is a continuous index and $X_{m1}\equiv\mathrm{Re}\left(X_{m}\right)$,
$X_{m2}\equiv\mathrm{Im}\left(X_{m}\right)$. Note that in this equation
operators are denoted by hats and classical fields without hats.

Of course Eq.~\eqref{eq:quantum-classical-equivalence-gen} is not
valid in general. But this equality is valid when quantum fields can
be treated as classical, i.e. when conjugate variables commute. To
show this let us use a quantum field $\hat{\chi}$. For the moment
restoring physical units, the non-zero commutation relation for this
operator and its conjugate pair in Eq.~\eqref{eq:FRW-chi-commutators}
becomes \begin{equation}
\left[\hat{\chi}\left(\tau,\mathbf{x}\right),\hat{\pi}\left(\tau,\mathbf{x}'\right)\right]=i\hbar\delta\left(\mathbf{x}-\mathbf{x}'\right).\label{eq:q-to-c-commutators}\end{equation}
The classical limit of a quantum description must be achieved when
the Planck constant becomes negligibly small, $\hbar\rightarrow0$.
In this limit the commutator in Eq.~\eqref{eq:q-to-c-commutators}
becomes zero. If the operator $\hat{\chi}$ is expanded into the complete
set of orthonormal mode functions as in Eq.~\eqref{eq:FRW-chi-formal-expansion},
then in the limit $\hbar\rightarrow0$ the orthonormality condition
in Eq.~\eqref{eq:FST-orthonormality-of-mode-fns} for the mode functions
$\left\{ \chi_{m}\right\} $ becomes $\left(\chi_{m},\chi_{m}\right)\rightarrow0$.
Using Eq.~\eqref{eq:FST-scalar-product} we find that this condition
results in the mode function $\chi_{m}$ and its complex conjugate
$\chi_{m}^{*}$ being different only by the \emph{time independent}
phase factor \begin{equation}
\chi_{m}^{*}=c_{m}\chi_{m}.\label{eq:q-to-c-phase-rotation}\end{equation}
 But the phase of $\chi_{m}$ is completely arbitrary. Therefore we
are free to choose it in a way that makes $\chi_{m}$ real. And because
$c_{m}$ is time independent, $\chi_{m}$ is real at all times and
$\hat{X}_{m}$ in Eq.~\eqref{eq:q-to-c-quantum-expansion} can be
rewritten as\begin{equation}
\hat{X}_{m}\left(x\right)=\chi_{m}\left(x\right)\left[\hat{a}_{m}+\hat{a}_{m}^{\dagger}\right].\label{eq:q-to-c-classical-expansion}\end{equation}
One can easily check that with this expression the commutator in Eq.~\eqref{eq:q-to-c-commutators}
is zero. In Refs.~\cite{Polarski_Starobinski(1995),Lesgourgues_etal(1997)}
it was calculated explicitly that an operator of the form in Eq.~\eqref{eq:q-to-c-classical-expansion}
satisfies the equivalence equation~\eqref{eq:quantum-classical-equivalence-gen}
for the quantum field and stochastic classical field. Which shows
as well that the classical stochastic field can be expressed as \begin{equation}
X_{m}\left(x\right)=\chi_{m}\left(x\right)e_{m},\label{eq:q-to-c-classical-stochastic-Fd-def}\end{equation}
 where $e_{m}$ are time independent, complex, stochastic c-number
functions with zero average and unit dispersion: $\left\langle e_{m}\right\rangle =0$
and $\left\langle e_{m},e_{n}^{*}\right\rangle =\delta\left(m-n\right)$.
And $e_{m}$ obeys the same statistics as $\hat{X}_{m}\left(\tau_{0},\mathbf{x}\right)$.

Note that the time dependent part $\chi_{m}\left(\tau,\mathbf{x}\right)$
can be factored out from both: the quantum field $\hat{X}_{m}$ in
Eq.~\eqref{eq:q-to-c-classical-expansion} and from the classical
field $X_{m}$ in Eq.~\eqref{eq:q-to-c-classical-stochastic-Fd-def}.
As the result the evolution of a given mode function is completely
deterministic after the realization of some stochastic amplitude have
occurred. In other words, if we measure the amplitude of the field
perturbation some time after the horizon exit, it will continue to
have a definite value. 

To show how a quantum field becomes of the form in Eq.~\eqref{eq:q-to-c-classical-expansion}
in the accelerating Universe, let us consider such a field in the
de Sitter background. In subsection~\ref{sub:Fd-Perturbation-Inflation}
it was shown that if we choose mode functions $\chi_{m}\left(x\right)$
to be Fourier modes (see Eq.~\eqref{eq:FRW-chi-ansatz}) on the superhorizon
scales they will have the solution \begin{equation}
\chi_{m}\left(\tau,\mathbf{x}\right)\equiv\sqrt{2}\frac{\mathrm{e}^{i\frac{\pi}{2}\left(\nu-\frac{5}{2}\right)}}{\sqrt{2k}}\frac{\Gamma\left(\nu\right)}{4\pi^{2}}\left(\frac{-k\tau}{2}\right)^{\frac{1}{2}-\nu}\mathrm{e}^{i\mathbf{k}\cdot\mathbf{x}},\label{eq:q-to-c-sFd-mode-fn}\end{equation}
where we have used Eq.~\eqref{eq:sFd-superhorizon-perturbations}
and $\nu$ is defined in Eq.~\eqref{eq:niu-def}. If $\nu$ is not
imaginary, corresponding to $m^{2}<9H^{2}/4$, this mode function
does not have a time dependent phase and can be made real by choosing
a time independent phase rotation $c_{\mathbf{k}}$ (defined in Eq.~\eqref{eq:q-to-c-phase-rotation}).
Note that this transformation makes $\chi_{m}$ real at all times
on the superhorizon scales. Therefore, quantum mechanical operator
$\hat{X}_{m}$ satisfies Eq.~\eqref{eq:q-to-c-classical-expansion}
and consequently is equivalent to the classical stochastic field $X_{m}$.

The process of quantum-to-classical transition described above suffers
from the usual interpretational problem of measurement in quantum
mechanics. The first question is why did Nature choose this particular
value for the realization of the field amplitude when other infinite
possibilities were available? Another is a cosmological variant of
the Schr\"{o}dinger's Cat problem related to the question of when
the state collapsed into its observed value. According to the usual
Copenhagen interpretation, this happens at the time of measurement.
But does that mean that CMB perturbation pattern did not exist before
we have measured it for the first time?

\section{The Primordial Curvature Perturbation\label{sec:zeta} }

Quantum field perturbations described in the last section on superhorizon
scales give rise to the classical cosmological perturbations. These
perturbations are most conveniently described by the intrinsic spatial
curvature $\zeta$, which is commonly called the curvature perturbation.
It is defined on the hypersurfaces of constant energy density. We
will postpone the discussion how quantum field perturbations may generate
$\zeta$ until section~\ref{sec:Mechanisms-for-zeta-generation}.
In this section we discuss the properties of $\zeta$ and the formalism
which relates perturbations of quantum fields with the curvature perturbation.

\subsection{Gauge Freedom in General Relativity}

When discussing the FRW Universe in section~\ref{sec:Kinematics-of-HHB}
we implicitly chose the coordinate system in which the metric in Eq.~\eqref{eq:FRW-metric-flat-Cartesian}
attained it's elegant form. Although the physical results in GR should
not depend on the coordinate system, the homogeneity and isotropy
of the Universe singles out a preferred reference frame in which equations
reduces to their simplest form. However, in space-times without symmetries
such preferred coordinate system does not exist. Therefore, the choice
of coordinates is purely arbitrary and may be selected depending on
the problem at hand. 

Fixing the coordinate system in GR specifies how space-time is threaded
by the lines of constant spatial coordinate $\mathbf{x}$ (\emph{threading})
and how it is divided into the hypersurfaces of constant coordinate
time $t$ (\emph{slicing}). And because the coordinate system is arbitrary,
so are the threading and slicing.

The arbitrariness of the coordinate system becomes especially problematic
when describing tiny departures from homogeneity of the actual Universe.
Since the real space-time with these departures does not posses any
symmetry, it would be impossible to solve the exact relativistic evolution
equations because GR is a non-linear theory. But luckily on large
enough scales these departures from ideal homogeneity are very small,
only of the order $10^{-5}$ (see the discussion on the cosmological
principle in section~\ref{sec:Kinematics-of-HHB}). Therefore, a
very good approximation to the actual Universe is to treat these inhomogeneities
as tiny perturbations of the otherwise homogeneous and isotropic background.
And using perturbation theory, non-linear equations of GR may be linearized.

However, separating physical quantities into background value and
perturbations are not so trivial. Due to the freedom for the choice
of the coordinate system this separation is not unique. Such complication
is due to the fact that by separation we mean that for each space-time
point in the background or reference manifold we associate a perturbation,
corresponding to the actual or physical space-time. But because these
are two different manifolds with different curvature we must specify
how the mapping from one manifold to the other is performed. This
may be done by choosing the specific threading and slicing of the
space-time. Or in perturbation theory it is called by fixing the \emph{gauge}. 

By changing the gauge we must redefine what is the background value
and what is the perturbation. Let us consider an infinitesimal gauge
transformation in which new coordinates $\tilde{x}^{\mu}$ are related
to the old ones by\begin{equation}
\tilde{x}^{\mu}=x^{\mu}+\delta x^{\mu}\left(x\right),\label{eq:infinit-coord-transform}\end{equation}
and see how the perturbation of some scalar quantity $\delta f\left(x\right)$
changes by this transformation, where the perturbation is defined
as the difference between the actual value $f\left(x\right)$ and
the background value $f_{0}\left(x\right)$, $\delta f\left(x\right)\equiv f\left(x\right)-f_{0}\left(x\right)$.
In the new gauge this perturbation will be $\widetilde{\delta f}\left(\tilde{x}\right)=\tilde{f}\left(\tilde{x}\right)-\tilde{f}_{0}\left(\tilde{x}\right)$.
But because the physical value of the quantity does not change under
the gauge transformation, only the separation into the background
and perturbation does, we can write $\tilde{f}\left(\tilde{x}\right)=f\left(\tilde{x}\right)$.
Further, because $f$ is a scalar, it should be invariant under the
coordinate change, thus $f\left(\tilde{x}\right)=f\left(x\right)$.
On the other hand, we are keeping fixed the point in the background
manifold and investigate the change in the mapping to the perturbed
manifold, therefore $\tilde{f}_{0}\left(\tilde{x}\right)=f_{0}\left(\tilde{x}\right)$.
However, although it is the same point on the background manifold,
in a new coordinate system it will have a different value. With the
infinitesimal transformation in Eq.~\eqref{eq:infinit-coord-transform}
this may be written as $f_{0}\left(\tilde{x}\right)=f_{0}\left(x\right)+\left(\partial f_{0}\left(x\right)/\partial x\right)\cdot\delta x$.
Putting all this discussion together, we may write \begin{eqnarray}
\widetilde{\delta f}\left(\tilde{x}\right) & = & \tilde{f}\left(\tilde{x}\right)-\tilde{f}_{0}\left(\tilde{x}\right)\nonumber \\
 & = & f\left(x\right)-f_{0}\left(\tilde{x}\right)\label{eq:scalar-gauge-transformation}\\
 & = & \delta f\left(x\right)+f_{0}\left(x\right)-\frac{\partial f_{0}\left(x\right)}{\partial x^{\mu}}\delta x^{\mu}-f_{0}\left(x\right)\nonumber \\
 & = & \delta f\left(x\right)-\frac{\partial f_{0}\left(x\right)}{\partial x^{\mu}}\delta x^{\mu}.\nonumber \end{eqnarray}

In the later discussion, of special importance will be the transformation
of the scalar quantity when only the slicing is changed, keeping the
threading constant. This corresponds to a time shift, for which Eq.~\eqref{eq:infinit-coord-transform}
reduces to\begin{equation}
\tilde{t}=t+\delta t\left(x\right).\end{equation}
And Eq.~\eqref{eq:scalar-gauge-transformation} becomes\begin{equation}
\delta\tilde{f}-\delta f=-\dot{f}_{0}\delta t.\label{eq:scalar-slicing-transformation}\end{equation}

\subsection{Smoothing and The Separate Universe Assumption }

Calculations of the curvature perturbation in this thesis are performed
using the separate Universe assumption. And the central concept for
this assumption is that of smoothing.

Let us assume that we are interested in the perturbation of the energy
density $\delta\rho$. Then the smoothed value $\delta\rho\left(t,\mathbf{x},L\right)$
will correspond to the value of the energy density at the space-time
point $\left(t,\mathbf{x}\right)$ after averaging over the sphere
of comoving size $L$. If we expand $\delta\rho\left(t,\mathbf{x},L\right)$
in a Fourier series, smoothing would correspond to dropping off all
modes bigger than $k\sim L^{-1}$. The dynamics of this smoothed quantity
is assumed to be determined by the averaged Einstein equations. However,
GR is a non-linear theory which makes the issue of smoothing non trivial.
For example, it is not clear how small scale fluctuations with $k>L^{-1}$
can influence the evolution of the quantity smoothed on scales $L^{-1}$.
At present there is no satisfactory conclusion to this issue, but
we will assume that such length scale does exist above which the smoothed
Universe is a good approximation to the actual one.

\begin{figure}
\begin{centering}
\includegraphics[width=5.5cm]{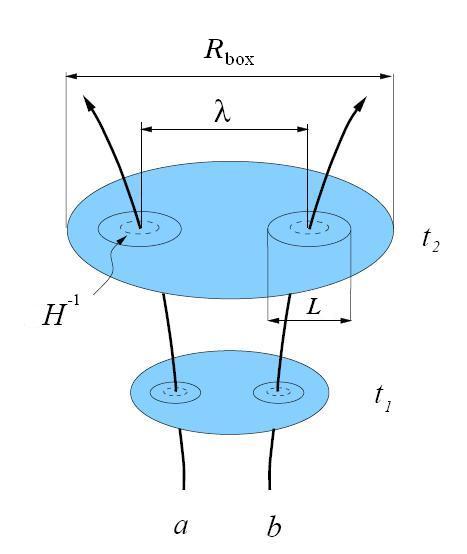}
\par\end{centering}

\caption{\label{fig:Separate-Universes}The schematic illustration of the separate
Universe assumption. According to this assumption all threadings coincide
with the comoving one. Hence, curves (a) and (b) represent comoving
trajectories of any of these threadings (adapted from Ref.~\cite{Wands_etal(2000)zeta_conservation})}

\end{figure}

The separate Universe approach assumes that each region, smoothed
on scales larger than the horizon size, locally evolves as a separate
unperturbed Universe. The basic idea is presented in Figure~\ref{fig:Separate-Universes}.
In this figure $L$ corresponds to the smoothing length scale which
is larger than the horizon size $H^{-1}$ but smaller than the largest
box size $R_{\mathrm{box}}$, within which we perform our calculations
(see Eq.~\ref{eq:mean-square-dlnk-scale-inv}). Ideally $R_{\mathrm{box}}\rightarrow\infty$,
but, as was discussed in section~\ref{sub:Random-Fields}, one should
keep a box size finite in order to avoid unknown physics and keep
calculations under control. The lines (a) and (b) represent two comoving
worldlines for two different space points.

Another assumption made in the separate Universe approach is that
all length scales introduced by the energy momentum tensor are much
smaller than the smoothing length scale $k^{-1}$. Then $k^{-1}$
is the only relevant superhorizon length scale and all spatial gradients
of order $k/a$ are negligible. When this assumption is satisfied,
the locally measurable parts of the metric should reduce to those
of the FRW \cite{Lyth_etal(2005)}. In other words, every comoving
location smoothed on superhorizon distances, evolves as the unperturbed
Universe with the FRW metric of Eqs.~\eqref{eq:FRW-metric-Spherical}
or \eqref{eq:FRW-metric-flat-Cartesian}.

\subsection{Conservation of the Curvature Perturbation}

Every smooth space-time metric can be decomposed into 3+1 components
as \cite{Arnowitt_etal(2004)}:\begin{equation}
\d s{}^{2}=\mathcal{N}^{2}\d t{}^{2}-\gamma_{ij}\left(\d x{}^{i}+\beta^{i}\d t{}\right)\left(\d x{}^{j}+\beta^{j}\d t{}\right),\label{eq:line-element-smooth-ST}\end{equation}
where $\mathcal{N}$ is the lapse function, $\beta^{i}$ is the shift
vector and $-\gamma_{ij}$ is the spatial three metric tensor.

With this decomposition one can define a unit time-like vector, $n^{\mu}$,
normal to the hypersurface of constant coordinate time $t$. The components
of this vector are\begin{equation}
n_{\mu}=\left(\mathcal{N},\mathbf{0}\right);\quad n^{\mu}=\left(-\frac{1}{\mathcal{N}},\frac{\beta^{i}}{\mathcal{N}}\right).\end{equation}
Then the volume expansion rate of the hypersurface along some integral
curve $\gamma\left(\tau\right)$ of $n^{\mu}$ will be given as\begin{equation}
\vartheta=\nabla_{\mu}n^{\mu},\label{eq:expan-rate-genrl}\end{equation}
where $\nabla_{\mu}$ is the covariant derivative and $\tau$ is the
proper time, which can be found from Eq.~\eqref{eq:line-element-smooth-ST},
$\d{\tau}{}=\mathcal{N}\d t{}$. Along each of these integral curves
we may define the number of e-folds of expansion\begin{equation}
N\left(t_{1},t_{2};\mathbf{x}\right)\equiv\frac{1}{3}\int_{\gamma\left(\tau\right)}\vartheta\d{\tau}{}=\frac{1}{3}\int_{t_{1}}^{t_{2}}\vartheta\mathcal{N}\d t{},\label{eq:N-sepU-def}\end{equation}
where the vector $\mathbf{x}$ is chosen to be the comoving spatial
coordinate.

The spatial metric of Eq.~\eqref{eq:line-element-smooth-ST} can
be further decomposed as\begin{equation}
\gamma_{ij}=a^{2}\tx\tilde{\gamma}_{ij}.\end{equation}
With the requirement $\det\left(\tilde{\gamma}_{ij}\right)=1$, $a\tx$
becomes a local scale factor. Note, that $t$ in this equation is
not necessarily a proper time, it is just the coordinate time labeling
the slices. Since we are interested in the non-homogeneity of the
scale factor, we may further decompose $a\tx$ into some global scale
factor $a\left(t\right)$, which is independent of position, and the
local deviation $\psi\left(t,\mathbf{x}\right)$ \begin{equation}
a\left(t,\mathbf{x}\right)=a\left(t\right)\mathrm{e}^{\psi\tx}.\label{eq:metric-decomposition-gen}\end{equation}
This decomposition into the global quantity and its perturbation is
completely arbitrary. Hence, we may choose $a\left(t\right)$ in such
a way that $\psi\tx$ vanishes somewhere inside the observable Universe.
Then $\psi\tx$ becomes small everywhere inside this Universe \cite{Lyth_etal(2005)}.

A similar decomposition may be done for the $\tilde{\gamma}_{ij}$
part of the metric\begin{equation}
\tilde{\gamma}_{ij}=\left(I\mathrm{e}^{h}\right)_{ij},\end{equation}
where $I$ is the unit matrix and $h_{ij}$ must be a traceless matrix
due to the requirement $\det\left(\tilde{\gamma}\right)=1$. It can
be shown that $h_{ij}$ corresponds to the primordial tensor perturbation,
i.e. gravitational waves. But, for the time being, we assume that
GWs are negligible so that we can set $h=0$.

According to the separate Universe assumption, if the metric is smoothed
on superhorizon scales, at each space-time point we should be able
to find such coordinates which reduce the metric into the form of
FRW:\begin{equation}
\d s{}^{2}=\d t{}^{2}-a^{2}\left(t\right)\delta_{ij}\d x{}^{i}\d x{}^{j}.\end{equation}
This metric is chosen to be flat in agreement with observations, but
as noted in Ref.~\cite{Lyth_etal(2005)} a small homogeneous curvature
should not make much difference.

In accord with this assumption and with the appropriate coordinate
choice, the metric in Eq.~\eqref{eq:line-element-smooth-ST} should
reduce to the form\begin{equation}
\d s{}^{2}=\mathcal{N}^{2}\d t{}^{2}-a^{2}\left(t\right)\mathrm{e}^{2\psi\tx}\delta_{ij}\d x{}^{i}\d x{}^{j}.\label{eq:metric-sepU}\end{equation}
The separate Universe assumption does not pose any constraints on
$\mathcal{N}$ and $\psi$ since they are not locally observable quantities.
And in view of this assumption we have neglected all terms of order
$\mathcal{O}\left(k/aH\right)$, which on superhorizon scales approach
zero, $k/aH\rightarrow0$. In particular, in Ref.~\cite{Lyth_etal(2005)}
it was shown that\begin{equation}
\beta_{i}=\mathcal{O}\left(k/aH\right)\qquad\mathrm{and}\qquad\dot{\tilde{\gamma}}_{ij}=\mathcal{O}\left[\left(k/aH\right)^{2}\right].\end{equation}
In the following discussions we will keep in mind that the separate
Universe assumption is valid up to this order, but will omit terms
$\mathcal{O}\left(k/aH\right)$ from equations. Note, however, that
the smallness of $\beta_{i}$ just corresponds to our choice of the
coordinate system. The generalization of the formalism to a threading
with non negligible $\beta_{i}$ is straightforward \cite{Lyth_etal(2005)}.

With the line element of Eq.~\eqref{eq:metric-sepU} the local expansion
rate $\vartheta$, defined in Eq.~\eqref{eq:expan-rate-genrl}, takes
the form\begin{equation}
\vartheta=\frac{3}{\mathcal{N}}\left(\frac{\dot{a}\left(t\right)}{a\left(t\right)}+\dot{\psi}\tx\right).\label{eq:expan-rate-local}\end{equation}
For later convenience we define the local Hubble parameter $\tilde{H}\tx\equiv\frac{1}{3}\vartheta$:\begin{equation}
\tilde{H}\tx=\frac{1}{\mathcal{N}}\left(\frac{\dot{a}\left(t\right)}{a\left(t\right)}+\dot{\psi}\tx\right).\label{eq:Hubble-param-local}\end{equation}

In what follows, an important quantity is the number of e-folds of
the local expansion, which is defined in Eq.~\eqref{eq:N-sepU-def}.
With the line element in Eq.~\eqref{eq:metric-sepU} it becomes\begin{equation}
N\left(t_{1},t_{2};\mathbf{x}\right)=\int_{\gamma\left(\tau\right)}\tilde{H}\tx\d{\tau}{}=\int_{t_{1}}^{t_{2}}\left(\frac{\dot{a}\left(t\right)}{a\left(t\right)}+\dot{\psi}\tx\right)\d t{}.\label{eq:N-sepU}\end{equation}

According to the separate Universe assumption each space point evolves
as the unperturbed Universe with the locally defined expansion rate
in Eq.~\eqref{eq:expan-rate-local} (or equivalently local Hubble
parameter in Eq.~\eqref{eq:Hubble-param-local}). Therefore, at each
point we can write the energy-momentum conservation law, $\nabla_{\nu}T^{\mu\nu}=0$,
from which it follows\begin{equation}
\frac{\d{\rho\tx}{}}{\d t{}}=-3\tilde{H}\tx\left[\rho\tx+p\tx\right].\label{eq:Friedmann-eq-local-gen}\end{equation}
It has the same form in the FRW Universe (c.f. Eq.~\eqref{eq:FRW-rho-continuity-t}).

This equation is valid independently of the slicing. But let us specialize
further to the slicing on which energy density is uniform, i.e. independent
on space coordinate at each given time. Such slicing is called comoving
or uniform density slicing and the value of $\psi$ on this slicing
is usually denoted by $\zeta$. It determines the perturbation in
the intrinsic curvature of the slices. Then, Eq.~\eqref{eq:Friedmann-eq-local-gen}
can be rewritten as\begin{equation}
\dot{\rho}\left(t\right)=-3\left[\frac{\dot{a}\left(t\right)}{a\left(t\right)}+\dot{\zeta}\tx\right]\left[\rho\left(t\right)+p\tx\right],\end{equation}
where we have used Eq.~\eqref{eq:Hubble-param-local} as well.

Now let us limit ourselves to the case where pressure is adiabatic,
which is equivalent to saying that pressure is a unique function of
the energy density, i.e. $p=p\left(\rho\right)$. In this case, because
$\rho$ is independent of position, the pressure must be independent
of position as well, $p=p\left(t\right)$ only. Therefore, the same
must be true for $\dot{\zeta}$, i.e. $\dot{\zeta}=\dot{\zeta}\left(t\right)$
only.

On the other hand, the decomposition of the spatial part of the metric
in Eq.~\eqref{eq:metric-decomposition-gen} into the background value
$a\left(t\right)$ and deviation from that value $\psi\tx$ was purely
arbitrary. So we may choose the normalization of $a\tx$ such that
$a\left(t\right)$ corresponds to the scale factor at our location
(or any other location). In other words, we choose $a\left(t\right)$
in such a way that $\zeta$ vanishes at our location at all times.
Hence, at this location\begin{equation}
\dot{\zeta}=0.\label{eq:zeta-constant}\end{equation}
But because $\dot{\zeta}$ is independent on position (when $p=p\left(\rho\right)$)
it must vanish everywhere.

In this way we have found a very important quantity, the curvature
perturbation $\zeta$, which determines the intrinsic curvature of
constant time spatial hypersurfaces. As was shown above, on superhorizon
scales $\zeta$ stays constant whenever pressure is the unique function
of the energy density. In the history of the Universe this happens
when the latter is dominated by radiation or matter. More generally,
the pressure of the multicomponent fluid is adiabatic if each component
of the fluid satisfies the relation $\rho_{a}=\rho_{a}\left(\rho\right)$,
where $\rho_{a}$ and $\rho$ are the energy densities of each component
and of the total fluid respectively. Thus, around the matter-radiation
equality era, $\zeta$ is constant too if perturbations are adiabatic.
By adiabatic perturbations we mean that on uniform total energy density
slices, perturbations of each component are independent of position.
A more rigorous proof of the constancy of the curvature perturbation
$\zeta$ can be found for example in Refs.~\cite{Lyth_etal(2005),Wands_etal(2000)zeta_conservation}.
In Ref.~\cite{Wands_etal(2000)zeta_conservation}, the constancy
of $\zeta$ was proved using perturbation theory, without the assumption
of separate universes. As shown in these works, the change in the
curvature perturbation to the first order is proportional to the non-adiabatic
part of the pressure $\delta p_{\mathrm{nad}}$ as\begin{equation}
\dot{\zeta}=-\frac{H}{\rho+p}\delta p_{\mathrm{nad}},\label{eq:zeta-t-derivative}\end{equation}
where $\delta p_{\mathrm{nad}}$ is defined as the pressure perturbation
on the uniform density slicing.

\subsection{\boldmath The $\delta N$ Formalism\label{sub:dN-formalism}}

On superhorizon scales all threadings are equivalent to the comoving
threading up to the order $\mathcal{O}\left[\left(k/aH\right)^{2}\right]$
\cite{Lyth_etal(2005)}. Hence, only the slicing is arbitrary. In
this section we show how using this fact and the separate Universe
assumption one can relate the curvature perturbation $\zeta$ with
the energy density perturbation $\delta\rho$ without invoking cosmological
perturbation theory.

Let us consider one of the comoving threads drawn in Figure~\ref{fig:Separate-Universes}.
Going from the coordinate time $t_{1}$ to $t_{2}$ along this thread
we can calculate the change in the value of $\psi$ for some slicing
using Eq.~\eqref{eq:N-sepU}\begin{equation}
\psi\left(t_{2},\mathbf{x}\right)-\psi\left(t_{1},\mathbf{x}\right)=N\left(t_{1},t_{2};\mathbf{x}\right)-N_{0}\left(t_{1},t_{2}\right),\label{eq:slice-evolution-gen}\end{equation}
where $N_{0}\left(t_{1},t_{2}\right)\equiv\ln\left[a\left(t_{2}\right)/a\left(t_{1}\right)\right]$
is the number of e-folds for the background expansion. For the flat
slicing we have $\psi_{\mathrm{flat}}=0$, and thus the number of
e-folds for the local expansion coincides with the background one,
$N_{\mathrm{flat}}\left(t_{1},t_{2};\mathbf{x}\right)=N_{0}\left(t_{1},t_{2}\right)$. 

Let us apply now Eq.~\eqref{eq:slice-evolution-gen} for two different
choices $\psi_{\mathrm{A}}$ and $\psi_{\mathrm{B}}$ corresponding
to two different slicings, which coincide at time $t_{1}$. Then at
time $t_{2}$ the difference between $\psi_{\mathrm{A}}$ and $\psi_{\mathrm{B}}$
will be\begin{equation}
\psi_{\mathrm{A}}\left(t_{2},\mathbf{x}\right)-\psi_{\mathrm{B}}\left(t_{2},\mathbf{x}\right)=N_{\mathrm{A}}\left(t_{1},t_{2};\mathbf{x}\right)-N_{\mathrm{B}}\left(t_{1},t_{2};\mathbf{x}\right).\label{eq:slice-difference}\end{equation}
Let us further specify slicings $\mathrm{A}$ and $\mathrm{B}$ in
the following way. The slicing $\mathrm{B}$ will be the flat slicing,
giving $N_{\mathrm{B}}=N_{0}$. And let the slicing $\mathrm{A}$
be such that at time $t_{1}$ it coincides with the flat slicing,
$\psi_{\mathrm{A}}\left(t_{1},\mathbf{x}\right)=0$, while at time
$t_{2}$ it coincides with the uniform density slicing, $\psi_{\mathrm{A}}\left(t_{2},\mathbf{x}\right)=\zeta\left(t_{2},\mathbf{x}\right)$.
Then Eq.~\eqref{eq:slice-difference} takes the form\begin{equation}
\zeta\left(t_{2},\mathbf{x}\right)=N_{A}\left(t_{2};\mathbf{x}\right)-N_{0}\left(t_{2}\right).\label{eq:zeta-N-diff}\end{equation}
 Due to our choice of $\mathrm{A}$ such that $\psi_{\mathrm{A}}\left(t_{1},\mathbf{x}\right)=0$,
the number of e-folds of the local expansion on this slicing becomes
$N_{\mathrm{A}}\left(t_{2},\mathbf{x}\right)=\ln\left[a\left(t_{2},\mathbf{x}\right)/a\left(t_{1}\right)\right]$.
This means that Eq.~\eqref{eq:zeta-N-diff} is independent on the
initial time and this is why we omitted the notation of $t_{1}$.
In other words, the calculation of $\zeta$ is independent on the
initial epoch, because when going from one flat slice to the other
the expansion is uniform.

From this equation it is clear that the curvature perturbation $\zeta\tx$
specifies the perturbation in the number of e-folds of the local expansion
starting from any flat slice and ending on the uniform density slice
at time $t$:\begin{equation}
\zeta\tx=\delta N\tx.\label{eq:zeta-deltaN}\end{equation}

Until now our discussion didn't require that perturbations should
be small. Hence, they are valid to any order in the perturbation expansion.
To relate $\delta N\tx$ with the field perturbation (discussed in
section~\ref{sub:Quantization-in-CST}) we will need to specialize
further in small perturbations.

Let us assume that the local expansion of the Universe is determined
solely by the value of a classical scalar field, $N\tx=N\left(\phi\tx\right)$.
By this assumption we neglect the contribution, for example, by the
kinetic term of the field, $\dot{\phi}\tx$. This is valid in most
cosmologically interesting cases, for example during inflation with
the slowly varying field. Then Eq.~\eqref{eq:zeta-N-diff} can be
written as\begin{equation}
\zeta\tx=N\left(\phi\tx\right)-N\left(\phi\left(t\right)\right).\end{equation}
Taking the field perturbation to be small $\delta\phi\tx\ll\phi\left(t\right)$,
where $\left.\phi\tx\equiv\phi\left(t\right)+\delta\phi\tx\right.$,
this equation becomes\begin{equation}
\zeta\tx=N_{\phi}\delta\phi+\frac{1}{2}N_{\phi\phi}\left(\delta\phi\right)^{2}+\ldots,\label{eq:zeta-dN-single-Fd}\end{equation}
where $N_{\phi}\equiv\partial N\left(\phi\left(t\right)\right)/\partial\phi$
and $N_{\phi\phi}\equiv\partial^{2}N\left(\phi\left(t\right)\right)/\partial\phi^{2}$.
Note that derivatives are taken of the unperturbed value of $N$,
and the field perturbation is evaluated on the initial flat slice.
This equation can be easily generalized to the many field case, when
$N\tx=N\left(\phi_{1}\tx,\phi_{2}\tx,\ldots\right)$, in which case
Eq.~\eqref{eq:zeta-dN-single-Fd} becomes\begin{equation}
\zeta\tx=\sum_{I}N_{I}\delta\phi_{I}+\frac{1}{2}\sum_{IJ}N_{IJ}\delta\phi_{I}\delta\phi_{J}+\ldots.\label{eq:zeta-dN-multi-Fd}\end{equation}

In the rest of this thesis it is sufficient to consider the curvature
perturbation only to the second order in the field perturbations,
i.e we will drop out terms denoted by the ellipsis.

\subsection{\boldmath The Power Spectrum and Non-Gaussianity of $\zeta$}

To calculate the power spectrum and bispectrum of the curvature perturbation,
$\zeta$ must be transformed to the Fourier space by Eq.~\eqref{eq:Fourier-decomposition}.
Then Eq.~\eqref{eq:zeta-dN-multi-Fd} becomes \begin{equation}
\zeta_{k}=N_{\phi}\delta\phi_{k}+\frac{1}{2}N_{\phi\phi}\left(\delta\phi_{k}\right)^{2}.\label{eq:zeta-k-space}\end{equation}
Note, that in this expression $\zeta_{k}$ is dependent only on the
modulus of $k\equiv\left|\mathbf{k}\right|$. This is because perturbations
of the scalar field are rotationally invariant. We will drop this
assumption in Chapter~\ref{cha:Vectors} when discussing perturbations
of vector fields.

The two point correlation function for the curvature perturbation
in Eq.~\eqref{eq:zeta-k-space} is\begin{equation}
\left\langle \zeta_{k}\left(t\right),\zeta_{k'}\left(t\right)\right\rangle =N_{\phi}^{2}\left\langle \delta\phi_{k},\delta\phi_{k'}\right\rangle +\frac{1}{4}N_{\phi\phi}^{2}\left\langle \left(\delta\phi_{k}\right)^{2}\left(\delta\phi_{k'}\right)^{2}\right\rangle .\label{eq:two-point-corr-singl-Fd-infl}\end{equation}
Because we have assumed that field perturbations are Gaussian, the
first term of this equation is the Gaussian contribution. The second
term gives a non-Gaussian contribution and according to observations
this contribution must be subdominant. 

Taking only the dominant part in Eq.~\eqref{eq:two-point-corr-singl-Fd-infl}
we find that the power spectrum of the curvature perturbation is related
to the power spectrum of the field perturbation by\begin{equation}
\Pz{}=N_{\phi}^{2}\mathcal{P}_{\phi},\label{eq:zeta-power-sp-single-Fd}\end{equation}
where the power spectrum of the field perturbations $\mathcal{P}_{\phi}$
is the one in Eq.~\eqref{eq:sFd-scale-inv-perturb-spect} for de
Sitter inflation or in Eq.~\eqref{eq:slow-roll-Fd-perturb-spect}
for the slow-roll inflation.

As was discussed in section~\ref{sub:Random-Fields} if perturbations
are Gaussian the two point correlator is the only non-zero correlator.
The non-Gaussianity manifest itself in the non-vanishing higher order
correlators. For the aim of the present thesis it is enough to consider
only the three point correlator, although in some models it might
be that, for example the four point correlator is even larger than
the three point (see e.g. Ref.~\cite{Engel_etal(2009)Tri_and_Bispectrum}).
The bispectrum (defined in Eq.~\eqref{eq:bispectrum-def}) of the
curvature perturbation usually is parametrized by the non-linearity
(or non-Gaussianity) parameter $\fnl$ defined in Eq.~\eqref{eq:fNL-def}.

If the field perturbation $\delta\phi$ is Gaussian, then $\fnl$
becomes practically independent of $k$. In Ref.~\cite{Lyth_Rodriguez(2005)}
it was calculated that if the first term of Eq.~\eqref{eq:zeta-k-space}
is dominant the $\fnl$ parameter becomes\begin{equation}
\frac{6}{5}\fnl=-\frac{N_{\phi\phi}}{N_{\phi}^{2}}.\label{eq:fNL-single-Fd}\end{equation}

\subsection{Density Perturbations}

In the previous subsection we have shown how to calculate the curvature
perturbation $\zeta$, which is conserved on superhorizon scales whenever
the pressure of the cosmic fluid is adiabatic. However, our primary
interest is in small perturbations of the energy density, which upon
horizon entry form the seeds for the subsequent structure formation
in the Universe. To make a connection between the curvature perturbation
$\zeta$, which we calculated so far, and inhomogeneities of the energy
density, we will use a limit of small perturbations up to the first
order.

As was discussed earlier, on superhorizon scales the threading is
defined uniquely, changing the slicing corresponds only to a shift
in the coordinate time. So let us consider a change from the uniform
density slicing to some generic one. At any given position this will
correspond to a time change $\delta t\tx$, so that on a new slicing
$\tilde{t}=t+\delta t\tx$. Then, the local scale factor on the new
slicing can be found using Eq.~\textbf{\eqref{eq:scalar-slicing-transformation}}
and considering that the background value is the same for both slicings\begin{equation}
a\left(\tilde{t},\mathbf{x}\right)=a\tx-\dot{a}\left(t\right)\delta t.\end{equation}
Separating local scale factors into the background value and the perturbation
as was done in Eq.~\eqref{eq:metric-decomposition-gen} and considering
that the perturbation is small, we find to first order\begin{equation}
\psi=\zeta-H\delta t.\end{equation}

A similar reasoning applies to the energy density giving \begin{equation}
\delta\rho_{\psi}\tx=-\dot{\rho}\left(t\right)\delta t\tx,\label{eq:rho-change-of-gauge}\end{equation}
where $\delta\rho=0$ on the uniform density slicing. For the time
being we use the index '$\psi$' to remind ourselves that density
perturbation is defined on an arbitrary slicing. While $\delta\rho$
without this index will correspond to the density perturbation in
a flat slicing.

Combining the last two equations we arrive at\begin{equation}
\zeta=\psi-H\frac{\delta\rho_{\psi}\tx}{\dot{\rho}\left(t\right)}=\psi+\frac{1}{3}\frac{\delta\rho_{\psi}\tx}{\rho+p},\label{eq:zeta-rho-gen-slicing}\end{equation}
where in the last equation the continuity equation for the FRW Universe
(Eq.~\eqref{eq:FRW-rho-continuity-t}) was applied. Thus we derived
the equation for the transformation going from the uniform density
slicing to arbitrary slicing. An important choice of the latter is
the flat slicing, $\psi=0$, for which we get\begin{equation}
\zeta=-H\frac{\delta\rho}{\dot{\rho}}=\frac{1}{3}\frac{\delta\rho}{\rho+p}.\label{eq:zeta-rho}\end{equation}

To derive this equation the uniform density slicing was defined with
respect to the total energy density of the cosmic fluid. But for the
fluid with several components, we can equally well define the uniform
density slicing for each component. Then if there is no total energy
exchange between these components, Eq.~\eqref{eq:zeta-rho} can be
rewritten as\begin{equation}
\zeta_{n}=-H\frac{\delta\rho_{n}}{\dot{\rho}_{n}}=\frac{1}{3}\frac{\delta\rho_{n}}{\rho_{n}+p_{n}},\label{eq:zeta-rho-many-Fds}\end{equation}
where $n$ is the index for a particular component of the fluid and
$\delta\rho_{n}$ is the energy density perturbation of that component
on a flat slicing. Using $\delta\rho=\sum_{n}\delta\rho_{n}$ we can
calculate the total curvature perturbation from Eq.~\eqref{eq:zeta-rho}\begin{equation}
\zeta=\frac{\sum_{n}\left(\rho_{n}+p_{n}\right)\zeta_{n}}{\rho+p}.\label{eq:zeta-composite}\end{equation}
This equation will be important when we consider curvaton models where
the primordial perturbation can be generated by several fluids.

\section{\label{sec:Mechanisms-for-zeta-generation}Mechanisms for the Generation
of the Curvature Perturbation}

Sections~\ref{sub:Fd-Perturbation-Inflation} and \ref{sub:Quantum-to-Classical}
described how, in the inflationary Universe, quantum fluctuations
are amplified and converted into classical field perturbations $\delta\phi$.
Then in section~\ref{sec:zeta} we have shown how to calculate the
intrinsic curvature perturbation of the space-time which is measured
after the horizon entry. In this section we will connect those two
parts and show three mechanisms by which fluctuations of quantum fields
during inflation can generate the curvature perturbation $\zeta$.
These three models of the generation of the curvature perturbation
by no means are the only possible. However, only these three are necessary
for our purpose when we discuss vector fields in Chapter~\ref{cha:Vectors}.

\subsection{Single Field Inflation\label{sub:Single-Field-Inflation}}

Let us assume in this section that the single field which drives the
slow-roll inflation, as discussed in section~\ref{sec:Inflation},
is the same field which is responsible for the total curvature perturbation
in the Universe. 

During single field inflation the value of the field $\phi\tx$ at
any given instant determines the energy density $\rho\tx$. Therefore,
we can calculate the curvature perturbation $\zeta$ directly from
Eq.~\eqref{eq:zeta-rho} by using the expression for the energy density
and pressure of the scalar field in Eqs.~\eqref{eq:Infl-rho} and
\eqref{eq:Infl-p}. Imposing the slow-roll condition for which $\rho\simeq V\left(\phi\right)$
and $3H\dot{\phi}\simeq-V_{\phi}$ we find to the first order\begin{equation}
\zeta=\frac{1}{3}\frac{V_{\phi}}{\dot{\phi}^{2}}\delta\phi=\frac{1}{\mpl^{2}}\frac{V}{V_{\phi}}\delta\phi.\label{eq:zeta-rho-single-Fd-infl}\end{equation}

Alternatively we can use the $\delta N$ formula directly. This method
renders the second order calculations more straightforward and comparison
with other methods for generation of $\zeta$ easier. 

In Eq.~\eqref{eq:zeta-dN-single-Fd} $N_{\phi}$ and $N_{\phi\phi}$
are derivatives of the number of e-folds of expansion of the unperturbed
Universe. During single field inflation the local evolution of the
Universe is determined only by the value of a single scalar field
$\phi$. Therefore, the change of $\phi$ corresponds to the shift
in time along the same unperturbed trajectory, $N_{\phi}$ and $N_{\phi\phi}$
are independent on the final epoch, which makes $\zeta$ independent
on time%
\footnote{The situation is different for the multifield inflation, where the
change in the field space, $\vec{\phi}\equiv\left(\phi_{1},\phi_{2},\ldots\right)$,
does not only correspond to the shift in time along the unperturbed
trajectory, but also the rotation in this space. Then $\zeta$ becomes
time dependent until the end of inflation or until the trajectories
of fields $\phi_{1},\phi_{2},\ldots$ become straight lines.%
}\begin{equation}
\zeta\left(\mathbf{x}\right)=N_{\phi}\delta\phi\left(\mathbf{x}\right)+N_{\phi\phi}\left[\delta\phi\left(\mathbf{x}\right)\right]^{2}.\label{eq:zeta-dN-single-Fd-infl}\end{equation}

The definition of the number of unperturbed e-folds of expansion is
given in Eq.~\eqref{eq:Infl-N-FRW-def}, from which it follows that
$\dot{N}=-H$, or alternatively\begin{equation}
N_{\phi}=-\frac{H}{\dot{\phi}}=\frac{1}{\mpl^{2}}\frac{V}{V_{\phi}},\label{eq:N'-single-Fd-infl}\end{equation}
where for the last equality the slow-roll equation of motion in Eq.~\eqref{eq:Infl-EoM-slow-roll}
was used. Inserting this expression into Eq.~\eqref{eq:zeta-dN-single-Fd-infl}
at the first order one recovers the same equation as in Eq.~\eqref{eq:zeta-rho-single-Fd-infl}.

The field perturbations $\delta\phi_{k}$ was discussed in section~\ref{sub:Fd-Perturbation-Inflation},
and the power spectrum for a light scalar field $\mathcal{P}_{\phi}$
was calculated in Eq.~\eqref{eq:slow-roll-Fd-perturb-spect}: \begin{equation}
\mathcal{P}_{\phi}\left(k\right)\simeq\left(\frac{H_{k}}{2\pi}\right)^{2}.\end{equation}
A few Hubble times after horizon exit $\zeta$ becomes constant as
shown in Eq.~\eqref{eq:zeta-constant}, hence it is enough to evaluate
\eqref{eq:zeta-power-sp-single-Fd} at this time, which gives\begin{equation}
\mathcal{P}_{\zeta}\left(k\right)=N_{\phi}^{2}\left(\frac{H_{k}}{2\pi}\right)^{2}.\label{eq:Power-spec-single-Fd-infl-2}\end{equation}

Using this expression for the power spectrum of the scalar field and
Eq.~\eqref{eq:N'-single-Fd-infl} with $3\mpl^{2}H_{k}^{2}\simeq\left.V\left(\phi\right)\right|_{k}$
gives us the power spectrum of the curvature perturbation\begin{equation}
\mathcal{P}_{\zeta}\left(k\right)=\frac{1}{24\mpl^{4}\pi^{2}}\left.\frac{V}{\epsilon}\right|_{k},\label{eq:power-spec-flow-roll-infl}\end{equation}
where $\epsilon$ is a slow-roll parameter defined in Eq.~\eqref{eq:Infl-e-slow-roll-parameter}
and the right hand side has to be evaluate at horizon exit, $aH_{k}=k$.

For exponential inflation, when $\dot{H}=0$, $\mathcal{P}_{\zeta}$
is scale independent. But in the slow-roll inflation $H$ is only
approximately constant, which makes the power spectrum slowly varying
with $k$. This variation usually is approximated by a power law (c.f.
Eq.~\eqref{eq:power-spec-observational}) and is parametrized by
the spectral index $n$ such that $\mathcal{P}_{\zeta}\left(k\right)\propto k^{n-1}$.
One can take the definition of the spectral index to be\begin{equation}
n-1\equiv\frac{\d{\ln\mathcal{P}_{\zeta}}{}}{\d{\ln k}{}}.\label{eq:spec-indx-def}\end{equation}

To evaluate this equation it is useful to derive the following relations\begin{equation}
\d{\ln k}{}=\d{\ln\left(a_{k}H_{k}\right)}{}\simeq\d{\ln a_{k}}{}=H_{k}\d t{},\label{eq:dlnk-Hdt-relation}\end{equation}
from which we can write \begin{equation}
\frac{\d{}{}}{\d{\ln k}{}}=\frac{\dot{\phi}}{H}\frac{\d{}{}}{\d{\phi}{}}=-\frac{1}{N_{\phi}}\frac{\d{}{}}{\d{\phi}{}},\label{eq:d/dlnk-d/dphi-relation}\end{equation}
where Eq.~\eqref{eq:N'-single-Fd-infl} was used. Then one can readily
calculate\begin{equation}
\frac{\d{\ln V}{}}{\d{\ln k}{}}=-\frac{1}{N_{\phi}}\frac{V_{\phi}}{V}=-\mpl^{2}\left(\frac{V_{\phi}}{V}\right)^{2}=-2\epsilon,\label{eq:1-for-sp-indx}\end{equation}
where we have used Eq.~\eqref{eq:N'-single-Fd-infl} and the definition
of the slow-roll parameter in Eq.~\eqref{eq:Infl-e-slow-roll-parameter}.
Now, we also have\begin{equation}
\frac{\d{\ln\epsilon}{}}{\d{\ln k}{}}=2\left(\frac{V}{V_{\phi}}\right)\frac{\d{}{}}{\d{\ln k}{}}\left(\frac{V_{\phi}}{V}\right)=2\mpl^{2}\left[\left(\frac{V_{\phi}}{V}\right)^{2}-\frac{V_{\phi\phi}}{V}\right]=4\epsilon-2\eta.\end{equation}
Using the last two relations and the power spectrum in Eq.~\eqref{eq:power-spec-flow-roll-infl}
from the definition of the spectral index in Eq.~\eqref{eq:spec-indx-def}
we find that the spectral index for single field slow-roll inflation
is \begin{equation}
n-1=2\eta-6\epsilon.\label{eq:spectral-tilt-slow-roll-infl}\end{equation}

From Eq.~\eqref{eq:N'-single-Fd-infl} we may write\begin{equation}
\frac{1}{\mpl}\left|\frac{\d{\phi}{}}{\d N{}}\right|=\mpl\left|\frac{V_{\phi}}{V}\right|=\sqrt{2\epsilon}.\label{eq:smallness-of-e-1}\end{equation}
The number of e-folds of inflationary expansion when the observable
cosmological scales leave the horizon correspond approximately to
$N\sim10$ \cite{Lyth_Liddle(2009)book}. Let us denote the change
in the field value during this period by $\Delta\phi_{10}$. Considering
that the low-roll parameter $\epsilon$ is almost constant during
this period from Eq.~\eqref{eq:smallness-of-e-1} we find\begin{equation}
\epsilon\sim10^{-2}\left(\frac{\Delta\phi_{10}}{\mpl}\right)^{2},\label{eq:smallness-of-e-2}\end{equation}
which is the value of $\epsilon$ when the cosmological scales leaves
the horizon. Observational constraints on $n$ where discussed in
section~\ref{sub:Spectrum-and-constraints} Eq.~\eqref{eq:spec-indx-WMAP5},
i.e. $n-1\approx0.04$, which gives $2\eta-6\epsilon\approx0.04$.
Therefore, from Eq.~\eqref{eq:smallness-of-e-2} we see that for
small field inflationary models, for which $\Delta\phi_{10}\ll\mpl$,
the slow-roll parameter $\epsilon$ is much smaller than $\eta$ and
Eq.~\eqref{eq:spectral-tilt-slow-roll-infl} may be written as\begin{equation}
n-1=2\eta.\label{eq:spectral-tilt-small-Fd}\end{equation}

One can go further and consider the running of the spectral index
as well, i.e. the scale dependence $n=n\left(k\right)$. The running
is defined as $n'\equiv\d n{}/\d{\ln k}{}$, from which one finds\begin{equation}
n'=-16\epsilon\eta+24\epsilon^{2}+2\xi,\end{equation}
where \begin{equation}
\xi\equiv\mpl^{4}\frac{V_{\phi}V_{\phi\phi\phi}}{V^{2}},\end{equation}
where $V_{\phi\phi\phi}\equiv\d V3/\d{\phi^{3}}{}$.

So far we have discussed the two point correlation function of Eq.~\eqref{eq:two-point-corr-singl-Fd-infl}.
If perturbations are exactly Gaussian all higher correlators must
vanish. Indeed, the non-Gaussianity of the curvature perturbation
generated by the inflaton field is very small in single field inflation
and this can be seen by calculating the three point correlators. To
show this let us calculate the second derivative of Eq.~\eqref{eq:N'-single-Fd-infl}\begin{equation}
\frac{N_{\phi\phi}}{N_{\phi}^{2}}=2\epsilon-\eta.\label{eq:N''/N'2}\end{equation}

Then using Eq.~\eqref{eq:fNL-single-Fd} we find that non-Gaussianity
is of order the slow roll parameters:\begin{equation}
\frac{6}{5}\fnl=\eta-2\epsilon,\label{eq:single-Fd-infl-fNL}\end{equation}
were we have used Eq.~\eqref{eq:N''/N'2}. From this relation it
is clear that $\left|\fnl\right|\ll1$. The cosmic variance limits
the detection of $\fnl$ to the values $\left|\fnl\right|>3$ \cite{Komatsu_Spergel(2001)fNL_bounds}.
Therefore, non-Gaussianity produced by a single field inflation is
too small to ever be observed. However, there are other scenarios
of the generation of the curvature perturbation for which non-Gaussianity
can be large enough to be observable in the near future.

One may consider even higher order correlators of the curvature perturbations.
And in some models they can be large enough to be observable as well.
But this is beyond the scope of this thesis.

\subsection{At the End of Inflation\label{sub:End-of-Inflation-Scenario}}

In the previous section we discussed the scenario when the inflation
is driven by a single scalar field $\phi$. In this scenario inflation
ends when the inflaton field reaches a critical value $\phi_{c}$,
where the slow-roll conditions in Eqs.~\eqref{eq:Infl-e-slow-roll-parameter}
and \eqref{eq:Infl-eta-slow-roll-condition} are violated, which is
solely determined by the inflaton field. Therefore, inflation ends
on the uniform energy density slice.

In Ref.~\cite{Lyth(2005a)} a scenario was suggested where the critical
value $\phi_{c}$ is modulated by some other scalar field $\sigma$,
$\phi_{c}=\phi_{c}\left(\sigma\right)$. For single field inflation,
the contribution from $\sigma$ to the inflaton dynamics must be negligible.
$\phi_{c}$ in this case must depend only on the perturbation $\delta\sigma\left(\mathbf{x}\right)$.
Then the hypersurface of constant $\phi_{c}$ does no longer coincide
with the uniform density hypersurface. 

For clarity let us assume for the moment that the perturbation $\delta\phi$
of the inflaton field generated during inflation is negligible. Then,
if the inflaton is a free field, the slice of constant $\phi$ will
coincide with the flat slice and with the constant energy density
slice even at the end of inflation, i.e there will be no curvature
perturbation. But if the end of inflation value $\phi_{c}$ is modulated
by some other perturbed field, $\phi_{c}\left(\sigma\left(\mathbf{x}\right)\right)$,
then the uniform density slice at the end of inflation no longer coincides
with the flat slice. According to the section~\ref{sub:dN-formalism},
this produces a perturbation in the amount of expansion between the
flat and uniform energy density slices which, from Eq.~\eqref{eq:zeta-deltaN},
is equal to the curvature perturbation $\zend$.

If, on the other hand, $\delta\phi$ is not negligible, then the same
argument holds, but $\zend$ will correspond to the perturbation in
the amount of expansion between the uniform energy density just before
the end of inflation and the one just after the end of inflation.
However, $\zend$ may be still large enough to dominate the curvature
perturbation which is generated during inflation, $\zend\gg\zinf$. 

For the following discussion we will assume that after inflation the
Universe undergoes prompt reheating, i.e. the inflaton field energy
is promptly converted into radiation. Then from Eq.~\eqref{eq:zeta-dN-single-Fd}
up to the second order we can write\begin{equation}
\zend=N_{c}\delta\phi_{c}+N_{cc}\left(\delta\phi_{c}\right)^{2},\end{equation}
where we have defined \begin{equation}
N_{c}\equiv\frac{\partial N}{\partial\phi_{c}}\quad\mathrm{and}\quad N_{cc}\equiv\frac{\partial^{2}N}{\partial\phi_{c}^{2}}.\end{equation}
In the former expression, $\delta\phi_{c}$ is the perturbation of
the end-of-inflation field value due to the coupling to the field
$\sigma$. Because, as was argued before, $\phi_{c}$ depends only
on the perturbation $\delta\sigma$, we can expand to the second order\begin{equation}
\delta\phi_{c}=\phi_{c}'\delta\sigma+\frac{1}{2}\phi_{c}''\left(\delta\sigma\right)^{2},\end{equation}
where $\phi_{c}'\equiv\partial\phi_{c}/\partial\sigma$ and $\phi_{c}''\equiv\partial^{2}\phi_{c}/\partial\sigma^{2}$.
Keeping terms only to the second order in $\delta\sigma$ the curvature
perturbation $\zend$ becomes\begin{equation}
\zend=N_{c}\phi_{c}'\delta\sigma+\frac{1}{2}\left[N_{c}\phi_{c}''+N_{cc}\left(\phi_{c}'\right)^{2}\right]\left(\delta\sigma\right)^{2}.\label{eq:z-end-gen}\end{equation}

To illustrate this model in a concrete example, let us consider the
hybrid inflation scenario \cite{Coplend_etal(19994)Hybrid_infl,Linde(1991),Linde(1994)}
first. In this scenario the slowly rolling inflaton field is coupled
to another scalar field, called the waterfall field. The potential
for this type of models is given by\begin{equation}
V\left(\phi,\chi\right)=V\left(\phi\right)-\frac{1}{2}m_{\chi}^{2}\chi^{2}+\frac{1}{4}\lambda\chi^{4}+\frac{1}{2}\lambda_{\phi}\phi^{2}\chi^{2},\label{eq:hybrid-potential}\end{equation}
where $\phi$ is the slow rolling inflaton field and $\chi$ is the
waterfall field. From this expression it can be seen that the effective
mass of $\chi$ is\begin{equation}
m_{\mathrm{eff}}^{2}=\lambda_{\phi}\phi^{2}-m_{\chi}^{2}.\label{eq:m-eff-hybrid-infl}\end{equation}
Initial conditions are such that $m_{\mathrm{eff}}^{2}>0$ and the
waterfall field is located at $\chi=0$ while the inflaton field $\phi$
slowly rolls towards zero. With this configuration the dominant term
of the potential in Eq.~\eqref{eq:hybrid-potential} is the term
$V\left(\phi\right)$. 

Inflation ends when the inflaton field reaches a critical value $\phi_{c}$
and the effective mass in Eq.~\eqref{eq:m-eff-hybrid-infl} becomes
negative. This destabilizes the waterfall field which very rapidly
rolls down to the minimum of the potential and acquires the vacuum
expectation value. This also very promptly changes the evolution of
the inflaton $\phi$: instead of slowly rolling it is quickly driven
towards zero. 

For the end-of-inflation scenario the critical value $\phi_{c}$ is
additionally modulated by including one more field $\sigma$, which
is coupled to the waterfall field. The potential in Eq.~\eqref{eq:hybrid-potential}
then becomes\begin{equation}
V\left(\phi,\chi,\sigma\right)=V\left(\phi\right)-\frac{1}{2}m_{\chi}^{2}\chi^{2}+\frac{1}{4}\lambda\chi^{4}+\frac{1}{2}\lambda_{\phi}\phi^{2}\chi^{2}+\frac{1}{2}\lambda_{\sigma}\sigma^{2}\chi^{2}+V\left(\sigma\right).\label{eq:end-of-infl-sFd-potential}\end{equation}
It is clear that the effective mass of $\chi$ becomes\begin{equation}
m_{\mathrm{eff}}^{2}=\lambda_{\phi}\phi^{2}+\lambda_{\sigma}\sigma^{2}-m_{\chi}^{2}.\end{equation}
In this case the waterfall field is destabilized and inflation ends
when\begin{equation}
\lambda_{\phi}\phi_{c}^{2}=m_{\chi}^{2}-\lambda_{\sigma}\sigma^{2}.\end{equation}
As we can see, the critical value $\phi_{c}$ is a function of another
field, $\phi_{c}\left(\sigma\right)$. The first and second derivative
with respect to this field are\begin{eqnarray}
\phi_{c}' & = & -\frac{\lambda_{\sigma}}{\lambda_{\phi}}\frac{\sigma}{\phi_{c}},\label{eq:phi'-end-of-infl}\\
\phi_{c}'' & = & -\frac{\lambda_{\sigma}}{\lambda_{\phi}\phi_{c}}\left[1+\frac{\lambda_{\sigma}}{\lambda_{\phi}}\left(\frac{\sigma}{\phi_{c}}\right)^{2}\right].\label{eq:phi''-end-of-infl}\end{eqnarray}

In Ref.~\cite{Lyth(2005a)} it was considered that the curvature
perturbation generated at the end of inflation dominates over the
one generated at the horizon exit. This happens if $\left.N_{c}\phi_{c}'\gg N_{\phi}\right.$,
where $N_{\phi}$ is given in Eq.~\eqref{eq:N'-single-Fd-infl}.
Using Eq.~\eqref{eq:phi'-end-of-infl} this condition can be rewritten
as\begin{equation}
\left(\frac{\lambda_{\sigma}}{\lambda_{\phi}}\frac{\sigma}{\phi_{c}}\right)^{2}\gg\frac{\epsilon_{c}}{\epsilon_{k}},\end{equation}
where $\epsilon_{c}$ is the slow-roll parameter just before the end
of inflation. $N_{c}$ in this expression was taken from Eq.~\eqref{eq:N'-single-Fd-infl}
giving $N_{c}^{2}=1/\left(2\mpl^{2}\epsilon_{c}\right)$.

With $\zend$ dominating over $\zinf$ from Eq.~\eqref{eq:z-end-gen}
we obtain the power spectrum of the produced curvature perturbation
in this model as\begin{equation}
\mathcal{P}_{\zend}=\frac{1}{2\mpl^{2}\epsilon_{c}}\left(\frac{\lambda_{\sigma}\sigma_{\mathrm{end}}}{\lambda_{\phi}\phi_{c}}\right)^{2}\left(\frac{H_{k}}{2\pi}\right)^{2},\label{eq:end-of-infl-power-spec}\end{equation}
where $\sigma_{\mathrm{end}}$ is the value of $\sigma$ at the end
of inflation. Since the potential of the field $\sigma$ is flat during
inflation, we may apply calculations in section~\ref{sub:Single-Field-Inflation}
for its perturbation spectrum. Therefore, using Eq.~\eqref{eq:spectral-tilt-small-Fd}
we find the spectral tilt as \begin{equation}
n-1=2\eta_{\sigma},\end{equation}
where $\eta_{\sigma}$ is the second slow roll parameter for the field
$\sigma$.

The non-Gaussianity parameter for this model when $\zend$ is dominant
can be calculated as follows. From the expression of $\fnl$ in Eq.~\eqref{eq:fNL-single-Fd}
and the curvature perturbation in Eq.~\eqref{eq:z-end-gen} one finds\begin{equation}
\frac{6}{5}\fnl=-\left[\frac{\phi''}{N_{c}\left(\phi_{c}'\right)^{2}}+\frac{N_{cc}}{N_{c}^{2}}\right].\end{equation}
Since we already know that $N_{cc}/N_{c}^{2}$ is of the order of
slow roll parameters, the first term dominates. And using Eqs.~\eqref{eq:phi'-end-of-infl}
and \eqref{eq:phi''-end-of-infl} we arrive at \begin{equation}
\frac{6}{5}\fnl=\eta\left[\frac{\lambda_{\phi}}{\lambda_{\sigma}}\left(\frac{\phi_{c}}{\sigma_{\mathrm{end}}}\right)^{2}\right].\label{eq:end-of-infl-fNL}\end{equation}

Expressions of the power spectrum and non-Gaussianity parameter $\fnl$
in Eqs.~\eqref{eq:end-of-infl-power-spec} and \eqref{eq:end-of-infl-fNL}
involve the homogeneous mode of the field $\sigma$. This mode is
defined as the average value of the field, where the averaging is
done over the same comoving box in which perturbations are defined.
The comoving box size $L$ must be larger than the observable Universe,
but not too large \cite{Lyth_Liddle(2009)book,Lyth(2006),Lyth(2007)}.
As discussed in subsection~\ref{sub:Random-Fields}, depending on
the accuracy required the box should be such that $\ln\left(LH_{0}\right)\sim\mathcal{O}\left(1\right)$.

In single field inflation the value of the field at horizon exit may
be calculated from the number of e-folds of remaining inflation. However,
in general it is not possible to calculate the unperturbed value of
the field and it must be specified as the free parameter of the model.
In some cases, it can be evaluated using the stochastic formalism
and assuming that our Universe is the typical realization of the whole
ensemble.

\subsection{The Curvaton Mechanism\label{sub:Curvaton-Mechanism}}

In section~\ref{sub:Single-Field-Inflation} we have demonstrated
the mechanism for the generation of the curvature perturbation at
the horizon exit during single field inflation. In section~\ref{sub:End-of-Inflation-Scenario}
the curvature perturbation was generated at the end of inflation.
Here we will discuss a mechanism by which $\zeta$ is generated some
time after inflation when the Universe is radiation dominated. Such
model is called the curvaton model and was first introduced in Refs.~\cite{Lyth_Ungarelli_Wands(2003),Mollerach(1990),Lyth_Wands(2002),Moroi_Takahashi(2001)}.
It is possible that the total curvature perturbation is generated
by the curvaton mechanism, or only a part of it. For simplicity, let
us assume first, that the curvature perturbation is generated only
by the curvaton mechanism.

In these models the field which is responsible for the curvature perturbation
is different from the field which drives inflation. In fact, it is
not even necessary to assume any particular model of inflation and
validity of Einstein gravity during that era: it might be slow roll
inflation due to scalar fields, due to modified gravity or any other
mechanism. The only assumptions necessary are that inflationary expansion
is almost exponential and that after inflation the Universe undergoes
reheating and becomes radiation dominated.

The curvaton mechanism liberates inflation models from the need for
the inflaton field to drive inflationary expansion as well as generate
the primordial curvature perturbation. Therefore, it substantially
increases the available parameter space for viable inflation models
\cite{Dimopoulos_Lyth(2004)}.

Let us denote the scalar curvaton field by $\sigma$. Although it
is a different field from the one in the previous section, in this
case too $\sigma$ during inflation is subdominant with a sufficiently
flat potential, $\left|V_{\sigma\sigma}\right|\ll H^{2}$, where $V_{\sigma\sigma}\equiv\d V2/\d{\sigma^{2}}{}$.
The unperturbed curvaton field satisfies equation of motion\begin{equation}
\ddot{\sigma}+3H\dot{\sigma}+V_{\sigma}=0.\label{eq:curvaton-EoM-unperturbed}\end{equation}
As for all light fields, quantum fluctuations of the curvaton field
during inflation are promoted to classical perturbations after horizon
exit. Then the curvaton perturbations $\delta\sigma\left(\mathbf{x}\right)$
satisfy\begin{equation}
\ddot{\delta\sigma}+3H\dot{\sigma}+V_{\sigma\sigma}\delta\sigma=0,\label{eq:curvaton-EoM-perturbed}\end{equation}
where we have taken into account that on superhorizon scales all gradients
vanish since $k/a\rightarrow0$. With vacuum initial conditions the
power spectrum for the curvaton field perturbations is given by Eq.~\textbf{\eqref{eq:slow-roll-Fd-perturb-spect}\begin{equation}
\mathcal{P}_{\sigma}=\left(\frac{H_{k}}{2\pi}\right)^{2}.\label{eq:curvaton-field-power-spec}\end{equation}
}

At some epoch after inflation when $V_{\sigma\sigma}\sim H^{2}\left(t\right)$
the curvaton field starts to oscillate around its VEV. Let us assume
that at this epoch Einstein gravity is already valid and that this
happens during the radiation domination (the energy density of the
dominant contribution, i.e. radiation, decreases as $\rho_{\gamma}\propto a^{-4}$).
Then $3\mpl^{2}H^{2}=\rho_{\mathrm{\gamma}}$ and the Hubble parameter
decreases as $H\propto a^{-2}$. 

The onset of oscillations of the curvaton field is taken to occur
much before the cosmological scales enter the horizon. The potential
of the curvaton near its VEV can be approximated as \begin{equation}
V\left(\sigma\right)=\frac{1}{2}m_{\sigma}^{2}\sigma^{2}.\label{eq:curvaton-potential}\end{equation}
Then oscillations start when $H\sim m_{\sigma}$. However, even if
the potential is not of that form at the start of oscillations, after
a few Hubble times, when the amplitude decreases, it can be approximated
to high accuracy by this quadratic form. Then, from Eqs.~\eqref{eq:curvaton-EoM-unperturbed}
and \eqref{eq:curvaton-EoM-perturbed}, it is clear that the unperturbed
and perturbed values of the curvaton field satisfy the same equation
of motion resulting in $\delta\sigma/\sigma=\mathrm{const}$. But
even if this condition is not satisfied, it will result only in a
scale independent factor which does not spoil scale invariance \cite{Lyth_Wands(2002)}.

During oscillations, the curvaton field evolves as the underdamped
harmonic oscillator with the energy density decreasing as $\rho_{\sigma}\tx\approx\frac{1}{2}m_{\sigma}^{2}\sigma_{\mathrm{A}}^{2}\tx\propto a^{-3}\tx$,
where $\sigma_{\mathrm{A}}$ is the amplitude of oscillations. This
decrease is slower than that of radiation. Therefore, the relative
energy density of the curvaton increases, $\rho_{\sigma}/\rho_{\gamma}\propto a$.
If the curvaton decay rate is small enough it can dominate the Universe,
resulting in the second reheating at its decay. Or the curvaton can
decay when it is still subdominant. In both cases during the period
when the relative energy density of the curvaton field increases the
total pressure of the Universe is not adiabatic. According to Eq.~\eqref{eq:zeta-t-derivative}
this results in a growth of the curvature perturbation $\zeta$ which
settles at its constant value soon after the curvaton decay. 

To calculate the power spectrum and non-Gaussianity of the curvature
perturbation we use the $\delta N$ formalism (see Ref.~\cite{Lyth_Rodriguez(2005)}).
This can be applied with the sudden decay approximation. The curvature
perturbation without this approximation can be calculated using Eq.~\eqref{eq:zeta-t-derivative}
and knowing the decay rate of the curvaton. Such calculation can only
be done numerically. However, in Refs.~\cite{Malik_etal(2003),Sasaki_etal(2006)}
it was shown that the sudden decay approximation agrees with the numerical
results within $10\%$. 

As we have assumed that the curvature perturbation in the radiation
dominated Universe is negligible before the curvaton starts oscillating,
the number of e-folds from the end of inflation until oscillations
is unperturbed. Therefore, the initial epoch for the $\delta N$ formula
can be taken just after oscillations commence. Let us denote the value
of the curvaton field at the onset of oscillations by $\sigma_{\mathrm{osc}}\left(\mathbf{x}\right)$.
In general this will depend on the value of the curvaton field $\sigma_{*}$
a few Hubble times after horizon exit of a given scale $k$, i.e.
$\sigma_{\mathrm{osc}}=\sigma_{\mathrm{osc}}\left(\sigma_{*}\right)$.
Then the number of e-folds from the beginning of oscillations until
the curvaton decay is \begin{equation}
N\left(\rho_{\mathrm{dec}},\rho_{\mathrm{osc}},\sigma_{*}\right)=\ln\frac{a_{\mathrm{dec}}}{a_{\mathrm{osc}}}=\frac{1}{3}\ln\frac{\rho_{\sigma,\mathrm{osc}}}{\rho_{\sigma,\mathrm{dec}}},\end{equation}
where `osc' and `dec' denotes values at the start of curvaton oscillations
and at the decay respectively. On the other hand, for the energy density
we have $\rho_{\mathrm{osc}}\left(\mathbf{x}\right)\approx\frac{1}{2}m_{\sigma}^{2}\sigma_{\mathrm{osc}}^{2}$,
which gives\begin{equation}
N\left(\rho_{\mathrm{dec}},\rho_{\mathrm{osc}},\sigma_{*}\right)=\frac{1}{3}\ln\frac{\frac{1}{2}m_{\sigma}^{2}\sigma_{\mathrm{osc}}^{2}}{\rho_{\sigma,\mathrm{dec}}}.\label{eq:curvaton-N-expr}\end{equation}
Let us denote the total energy density at the start of oscillations
$\rho_{\mathrm{osc}}$. Because the curvaton energy density at this
epoch is negligible $\rho_{\mathrm{osc}}$ corresponds primarily to
the radiation energy density $\rho_{\mathrm{osc}}\simeq\rho_{\gamma,\mathrm{osc}}$.
But the curvaton is not negligible just before its decay, making the
total energy density at this epoch $\rho_{\mathrm{dec}}=\rho_{\gamma,\mathrm{dec}}+\rho_{\sigma,\mathrm{dec}}$.
From the scaling laws of matter and radiation (see Eq.~\eqref{eq:FRW-rho-scaling-general})
we find $\rho_{\sigma,\mathrm{dec}}/\rho_{\sigma,\mathrm{osc}}=\left(\rho_{\gamma,\mathrm{dec}}/\rho_{\gamma,\mathrm{osc}}\right)^{3/4}$.
Putting everything together we find \begin{equation}
\rho_{\sigma,\mathrm{dec}}=\frac{1}{2}m_{\sigma}^{2}\sigma_{\mathrm{osc}}^{2}\left(\frac{\rho_{\mathrm{dec}}-\rho_{\sigma,\mathrm{dec}}}{\rho_{\mathrm{osc}}}\right)^{3/4}.\label{eq:rho-sigma-decay}\end{equation}

The derivative of $N$ with respect to the curvaton field may be found
using the chain rule $\partial/\partial\sigma_{*}=\sigma_{\mathrm{osc}}'\cdot\partial/\partial\sigma_{\mathrm{osc}}$
and keeping $\rho_{\mathrm{osc}}$ and $\rho_{\mathrm{dec}}$ fixed
(the prime in this relation denotes differentiation with respect to
$\sigma_{*}$). Then from Eq.~\eqref{eq:rho-sigma-decay} we find\begin{equation}
\frac{\partial_{\sigma_{\mathrm{osc}}}\rho_{\sigma,\mathrm{dec}}}{\rho_{\sigma,\mathrm{dec}}}=\frac{8}{\sigma_{\mathrm{osc}}}\frac{\rho_{\mathrm{dec}}-\rho_{\sigma,\mathrm{dec}}}{4\rho_{\mathrm{dec}}-\rho_{\sigma,\mathrm{dec}}}.\end{equation}
Using this relation, from Eq.~\eqref{eq:curvaton-N-expr} we calculate\begin{equation}
N_{\sigma_{*}}=\frac{2}{3}\hat{\Omega}_{\sigma}\frac{\sigma_{\mathrm{osc}}'}{\sigma_{\mathrm{osc}}},\label{eq:curvaton-dN-dsigma}\end{equation}
where \begin{equation}
\hat{\Omega}_{\sigma}\equiv\frac{3\rho_{\sigma,\mathrm{dec}}}{3\rho_{\sigma,\mathrm{dec}}+4\rho_{\gamma,\mathrm{dec}}}.\label{eq:curvaton-Omega-hat-def}\end{equation}
If the curvaton energy density is subdominant at the decay, then $\rho_{\sigma,\mathrm{dec}}\ll\rho_{\gamma,\mathrm{dec}}$,
giving $\hat{\Omega}_{\sigma}=\frac{3}{4}\Omega_{\sigma}$, where
$\Omega_{\sigma}\equiv\rho_{\sigma,\mathrm{dec}}/\rho_{\mathrm{dec}}$
is the density parameter of $\sigma$ at decay. If, on the other hand,
the curvaton is dominant at that epoch, then $\hat{\Omega}_{\sigma}=\Omega_{\sigma}=1$.
In both cases it is a good approximation to write $\hat{\Omega}_{\sigma}\approx\Omega_{\sigma}$.
The error introduced by such approximation is not bigger than that
of the sudden decay approximation \cite{Lyth_Liddle(2009)book}.

Inserting Eq.~\eqref{eq:curvaton-dN-dsigma} into the equation~\eqref{eq:zeta-dN-single-Fd}
we find that the curvature perturbation generated by the curvaton
mechanism to first order is \begin{equation}
\zeta_{\sigma}=\frac{2}{3}\Omega_{\sigma}\frac{\sigma_{\mathrm{osc}}'}{\sigma_{\mathrm{osc}}}\delta\sigma_{*}.\end{equation}
The power spectrum becomes\begin{equation}
\mathcal{P}_{\zeta_{\sigma}}=N_{\sigma_{*}}\left(\frac{H_{k}}{2\pi}\right)^{2}=\frac{4}{9}\Omega_{\sigma}^{2}\left(\frac{\sigma_{\mathrm{osc}}'}{\sigma_{\mathrm{osc}}}\right)^{2}\left(\frac{H_{k}}{2\pi}\right)^{2},\label{eq:curvaton-zeta-power-spec-gen}\end{equation}
where for the curvaton field perturbation the power spectrum was given
in Eq.~\eqref{eq:curvaton-field-power-spec}.

To find the non-Gaussianity parameter $\fnl$ for this model we have
to know $\zeta_{\sigma}$ at least up to second order. Calculating
the second derivative of Eq.~\eqref{eq:curvaton-dN-dsigma} we find\begin{equation}
\frac{6}{5}\fnl=-\frac{N_{\sigma_{*}\sigma_{*}}}{N_{\sigma_{*}}^{2}}=-2+\Omega_{\sigma}+\frac{3}{2\Omega_{\sigma}}\left(1+\frac{\sigma_{\mathrm{osc}}\sigma_{\mathrm{osc}}''}{\sigma_{\mathrm{osc}}'^{2}}\right).\label{eq:curvaton-fNL-gen}\end{equation}

Eqs.~\eqref{eq:curvaton-zeta-power-spec-gen} and \eqref{eq:curvaton-fNL-gen}
were calculated using the $\delta N$ formalism and they agree very
well with the calculations performed using the first and second order
perturbation theory in Refs.~\cite{Lyth_Rodriguez(2005),Lyth_Ungarelli_Wands(2003)}.

In the previous calculations we have $\sigma_{\mathrm{osc}}\left(\sigma_{*}\right)$
as a general function. In some models it might happen that this is
a highly non-trivial function (for example Ref.~\cite{Dimopoulos_etal(2003)PNG})
but for the future reference let us consider the case when $\sigma_{\mathrm{osc}}\simeq\sigma_{*}$.
Then the power spectrum in Eq.~\eqref{eq:curvaton-zeta-power-spec-gen}
becomes\begin{equation}
\mathcal{P}_{\zeta_{\sigma}}\simeq\frac{4}{9}\Omega_{\sigma}^{2}\left(\frac{H_{k}}{2\pi\sigma_{*}}\right)^{2}.\label{eq:curvaton-zeta-spec}\end{equation}
And the non-Gaussianity from Eq.~\eqref{eq:curvaton-fNL-gen} is\begin{equation}
\frac{6}{5}\fnl=\frac{3}{2\Omega_{\sigma}},\end{equation}
where we have considered the case when $\fnl\gg1$.

The result shows that in the curvaton scenario the non-Gaussianity
can be very large, $\fnl\gg1$. This is in contrast to single field
inflation, where $\fnl$ was of order of the slow-roll parameters
(see Eq.~\eqref{eq:single-Fd-infl-fNL}). In the curvaton case the
non-Gaussianity can be large because $N_{\sigma}$ and $N_{\sigma\sigma}$
have nothing to do with the slow roll parameters.

There is another notable difference from single field inflation -
the power spectrum of the curvature perturbation in Eq.~\eqref{eq:curvaton-zeta-spec}
depends on the homogeneous mode of the field. This situation is analogous
to the one discussed in subsection~\ref{sub:End-of-Inflation-Scenario}.
The value of $\sigma_{*}\left(\mathbf{x}\right)$ should be taken
as an average within the observable Universe. Then it can be calculated
assuming that our Universe is typical.

Until now we have considered only the case when the curvaton is the
only source of the curvature perturbation $\zeta$. In other words,
we have assumed that the perturbation in the Universe is negligible
prior to the domination (or near domination) of the curvaton. But
if we drop this assumption then Eq.~\eqref{eq:zeta-composite} can
be used to calculate the resulting curvature perturbation from both
components:\begin{equation}
\zeta=\frac{4\rho_{\gamma,\mathrm{dec}}}{3\rho_{\sigma,\mathrm{dec}}+4\rho_{\gamma,\mathrm{dec}}}\zeta_{\gamma}+\frac{3\rho_{\sigma,\mathrm{dec}}}{3\rho_{\sigma,\mathrm{dec}}+4\rho_{\gamma,\mathrm{dec}}}\zeta_{\sigma},\end{equation}
where $\zeta_{\gamma}$ is the curvature perturbation in the radiation
dominated background. Using the definition in Eq.~\eqref{eq:curvaton-Omega-hat-def}
this equation becomes\begin{equation}
\zeta=\left(1-\hat{\Omega}_{\sigma}\right)\zeta_{\gamma}+\hat{\Omega}_{\sigma}\zeta_{\sigma}.\label{eq:curvaton-zeta-composite}\end{equation}
Such curvaton models with a two component contribution to the total
curvature perturbation were considered in Refs.~\cite{Lazarides_etal(2004),Lyth(2006),Ichikawa_etal(2008)}.
While for the negligible curvature perturbation from the inflaton
this equation reduces to\begin{equation}
\zeta\approx\hat{\Omega}_{\sigma}\zeta_{\sigma}.\label{eq:zeta-curvaton}\end{equation}
\pagebreak{}

\thispagestyle{empty}\onehalfspacing~

\pagebreak{}

\chapter{The Primordial Curvature Perturbation from Vector Fields\label{cha:Vectors}}

\section{Vector Fields in Cosmology\label{sec:vFds-in-Cosmology}}

In the previous chapter we have discussed the generation of the primordial
curvature perturbation in the Universe. It was shown that during inflation
the quantum mechanical fluctuations of light scalar fields are transformed
into the classical curvature perturbation. Long after inflation upon
horizon entry this perturbation seeds the formation of large scale
structure. Until very recently the generation of the curvature perturbation
by this mechanism was assigned solely to scalar fields. The main reason
for this is that scalar degrees of freedom are the simplest ones.
The preference for scalar fields is further supported by the observational
fact that the Universe on large scales is predominantly isotropic
and statistical properties of the temperature perturbation in the
CMB sky are predominantly isotropic too. In addition, particle physics
theories beyond the Standard Model are abundant with scalar fields. 

In the rest of this thesis we will show that quantum fluctuations
of vector fields may influence or even generate the total curvature
perturbation in the Universe. But why should we consider something
else than the scalar field? The motivation comes from both sides:
theoretical as well as observational.

From the theoretical side, for the inflationary model building only
scalar fields have been used to generate the curvature perturbation,
even if no fundamental scalar field is discovered yet. Although it
is widely accepted that all elementary particles possess masses due
to the Higgs scalar field, this might be explained by other mechanisms,
without invoking fundamental scalar fields (see e.g. the technicolor
model in Ref.~\cite{Sannino_technicolor(2008)}). It is expected
that in the near future the Large Hadron Collider (LHC) in CERN will
discover the Higgs boson and prove the existence of the fundamental
scalar fields. But if the Higgs boson is not discovered, the alternative
models explaining masses of elementary particles will become more
favorable. On the other hand, in the case of inflation, the generation
of the curvature perturbation from scalar fields will become much
less attractive. Furthermore, even if scalar fields are discovered,
particle physics theories, such as supersymmetry or supergravity,
incorporate many vector bosons and fermions. However, despite this,
the possible contribution to the curvature perturbation from other
kind of fields usually has been ignored.

Observationally there is some indication that the simplest scalar
field scenario may not be sufficient to explain some features of the
CMB sky. These features, which challenge the simple homogeneous and
isotropic model of the Universe, were first discovered already after
the release of the first year WMAP satellite data. For example it
was found that the quadrupole moment of the power spectrum of the
temperature perturbation was too small compared with predictions of
the currently favored $\Lambda$CDM cosmology. The lack of power in
the quadrupole was still present in the five year data (see Ref.~\cite{Komatsu_etal_WMAP5(2008)}).
Another discovery was that 2-4-8-16 spherical harmonics of the CMB
temperature map seem to be aligned, suggesting the presence of a preferred
direction in the Universe, so called the {}``Axis of Evil'' \cite{Land_Magueijo(2005)_AxisOfEvil}.
In addition this preferred direction is aligned with the large cold
spot in the CMB \cite{Vielva_etal(2004)_ColdSpot}, with the large
void in the radio galaxy distribution \cite{Rudnick_etal(2007)_GalaxyVoid}
and with the galaxy spin directions \cite{Longo(2009)GalaxyHandedness}.
Although currently these anomalies are under intense debate about
their statistical significance, they might be an indication that the
Universe is mildly anisotropic on large scales.

If these anomalies are confirmed it will prove the existence of the
preferred direction in the Universe and this can not be explained
solely by scalar fields. On the other hand, for vector fields the
existence of the preferred direction is natural. But employing vector
fields for the generation of the curvature perturbation we encounter
with two complications: conformal invariance and excessive large scale
anisotropy of the Universe.

In this Chapter it will be enough to approximate the inflationary
expansion of the Universe to be exactly exponential, i.e. with the
constant Hubble parameter $\dot{H}=0$.

\subsection{Conformal Invariance}

The evolution of the conformally flat space-time, such as de Sitter
or matter/radiation dominated FRW universes, can be modeled as the
conformal rescaling of the Minkowski space-time, i.e. by transforming
the metric $g_{\mu\nu}\left(\mathbf{x}\right)\rightarrow a^{2}\left(\tau\right)g_{\mu\nu}\left(\mathbf{x}\right)$,
where $\tau$ is the conformal time. As we have seen in section~\ref{sub:Quantization-in-CST}
for a light, minimally coupled scalar field on superhorizon scales
this leads to the amplification of vacuum fluctuations. But this is
not the case for the conformally trivial theories, for which the field
equations are invariant under the rescaling of the metric. For such
theories the form of the field equations is time independent in the
conformally flat space-times. 

This is the case, indeed, for the massless U(1) vector field with
the Lagrangian\begin{equation}
\mathcal{L}=-\frac{1}{4}F_{\mu\nu}F^{\mu\nu},\end{equation}
where $F_{\mu\nu}$ is the field strength tensor \begin{equation}
F_{\mu\nu}\equiv\partial_{\mu}A_{\nu}-\partial_{\nu}A_{\mu},\label{eq:vFd-F-def}\end{equation}
 and $A_{\mu}$ is the four-vector. In contrast to the light scalar
field case discussed so far, quantum fluctuations of a vector field
with this Lagrangian do not undergo amplification after the horizon
exit. But if the vector field is to generate the non-negligible curvature
perturbation, its fluctuations have to undergo amplification. Therefore,
the conformal invariance of the U(1) massless vector field must be
broken.

This problem is well known in the literature on the generation of
primordial magnetic fields (PMF) during inflation \cite{TurnerWidrow1988,Widrow_PMFs_review(2002),Giovannini_PMFs_review(2004),Dimopoulos_etal(2001)PMFs,Davis_etal(2000)PMFs}.
In this literature there are numerous suggested ways of breaking the
conformal invariance for vector fields: (i) introducing a mass for
the vector field, (ii) making the kinetic term time dependent, (iii)
introducing an anomaly term (iv) coupling a vector field to another
field which is not conformally coupled to gravity, (v) or using non-Abelian
vector fields. In this thesis we use only the first two methods. In
the context of the curvature perturbation the breaking of the conformal
invariance of U(1) field by introducing the mass term was first considered
in Ref.~\cite{Dimopoulos2006} and using the time dependent kinetic
term in Ref.~\cite{Dimopoulos2007}. The first attempt to calculate
the generation of the curvature perturbation by the non-Abelian SU(2)
vector fields was reported in Refs.~\cite{Bartolo_etal(2009)_Bispectrum,Bartolo_etal(2009)Trispectrum}.
The other methods of breaking the conformal invariance were investigated
only for the generation of PMF.

\subsection{Large Scale Anisotropy\label{sub:vFd-Large-Scale-Anisotropy}}

If the vector field is to influence or generate the curvature perturbation,
its energy density must dominate or nearly dominate the Universe for
the effect to be non-negligible. But the energy-momentum tensor of
the light vector field has an anisotropic stress. For example the
energy-momentum tensor of the light Abelian vector field can be written
as \cite{Dimopoulos2006}\begin{equation}
T_{\mu\nu}=\mathrm{diag}\left(\rho,-p_{\perp},-p_{\perp},+p_{\perp}\right),\label{eq:T-massive-vFd-gen}\end{equation}
where we have chosen the coordinate axis in a way that spatial components
of the homogeneous vector field are equal to $\mathbf{A}=\left(0,0,A\right)$
and we use this choice of coordinates in the rest of this thesis.
From Eq.~\eqref{eq:T-massive-vFd-gen} it is clear that if such a
vector field is to dominate or nearly dominate the Universe, the expansion
along the vector field direction will be different from the transverse
directions. This would induce an excessive large scale anisotropy
which is ruled out by observations.

To bypass this problem there are four methods proposed in the literature.
The author of the earliest one in Ref.~\cite{Ford(1989)_Vtriad}
considered three orthogonal identical vector fields. In this model
the total energy-momentum tensor, which is the sum of all three vector
fields is isotropic. Therefore, these fields can dominate the Universe
and even drive the inflationary expansion. Another mechanism, with
vector fields responsible for the inflationary expansion, was proposed
in Ref.~\cite{Golovnev_etal(2008)}. The authors introduced a large
number of identical vector fields which are randomly oriented in space
with identical initial conditions. Due to random orientation, the
average pressure becomes almost isotropic. The residual anisotropy
is proportional to $N^{-1/2}$, where $N$ is the number of vector
fields. If this number is sufficiently large, the induced large scale
anisotropy can be small enough to agree with observational bounds.
In another model in Ref.~\cite{Yokoyama_Soda(2008)} the vector field
is always subdominant and therefore does not generate excessive large
scale anisotropy. It cannot be responsible for the inflationary expansion,
but the vector field influences the generation of the curvature perturbation
by coupling to the scalar field. In particular, the authors of this
paper consider that the vector field modulates the end of inflation.
Finally in Ref.~\cite{Dimopoulos2006} the excessive large scale
anisotropy is avoided by introducing the vector curvaton scenario.
In this scenario, a massive vector field is subdominant during inflation
and afterwards until it becomes massive. As was demonstrated in that
work, when the vector field becomes massive it starts to oscillate
with a frequency much larger than the Hubble time. The pressure components
in Eq.~\eqref{eq:T-massive-vFd-gen} induced by the oscillating vector
field oscillates rapidly themselves. Therefore, the time averaged
value of the pressure over one Hubble time is zero and the vector
field on average acts as the pressureless, isotropic matter. It can
dominate the Universe without generating excessive large scale anisotropy. 

%
\begin{comment}
The requirement of the model in Ref.~\cite{Ford(1989)_Vtriad} to
have three identical and exactly orthogonal vector fields seems to
be very unnatural. This is avoided in Ref.~\cite{Golovnev_etal(2008)}.
However, in this case one has to assume a very large number of vector
fields which are identical and, more importantly, with similar initial
conditions.
\end{comment}
{} In this thesis we will use the end-of-inflation scenario of Ref.~\cite{Yokoyama_Soda(2008)}
as an example to calculate the statistical properties of the curvature
perturbation (section~\ref{sec:vFd-SodYok-Scenario}). But mostly
we will be occupied with the vector curvaton scenario (section~\ref{sec:vFd-Vector-Curvaton}).

\subsection{The Physical Vector Field\label{sub:vFd-physical}}

Before going into the description of the vector field quantization
and the generation of the curvature perturbation let us make a comment
about the distinction between the field which appears in the Lagrangian
and the physical vector field. Consider, for example, the Lagrangian
of the massive Abelian vector field%
\footnote{To be more precise, the Lagrangian of the Abelian vector field whose
gauge symmetry is broken by an explicit mass term.%
}\begin{equation}
\mathcal{L}=-\frac{1}{4}F_{\mu\nu}F^{\mu\nu}+\frac{1}{2}m^{2}A^{\mu}A_{\mu},\label{eq:vFd-Lagrangian-massive-vFd}\end{equation}
with the field strength tensor $F_{\mu\nu}$ defined in Eq.~\eqref{eq:vFd-F-def}.
Using the FRW metric the mass term may be expanded as\begin{equation}
\mathcal{L}=-\frac{1}{4}F_{\mu\nu}F^{\mu\nu}+\frac{1}{2}m^{2}A_{t}^{2}-\frac{1}{2}m^{2}a^{-2}\mathbf{A}^{2},\end{equation}
from which one notices that the spatial term dependents on the scale
factor. This might be alarming, because the Lagrangian is the physical
quantity and cannot depend on the arbitrary choice of the normalization
of the scale factor. This shows that the vector field $\mathbf{A}$
in the Lagrangian is the comoving field defined with respect to the
comoving, Cartesian space coordinates $x^{i}$. The physical four-vector
field is\begin{equation}
W_{\mu}=\left(A_{0},\, A_{i}/a\right),\label{eq:vCurv-phys-massive-vFd}\end{equation}
which is defined in in the basis of the physical coordinate system
$a\left(t\right)x^{i}$. For the comoving and physical vector fields
the corresponding upper-index quantities are $A^{i}=-a^{-2}A_{i}$
and $W^{i}=-W_{i}$.

This is the case for the Lagrangian with the canonically normalized
field such as in Eq.~\eqref{eq:vFd-Lagrangian-massive-vFd}. In section~\ref{sub:vCurvaton-fF2}
we will consider a vector field with the time dependent kinetic term.
In this case $W_{\mu}$ will denote a canonically normalized physical
vector field (see Eq.~\eqref{eq:vCurv-fF-W-def}).

\section{Vector Field Quantization and the Curvature Perturbation\label{sec:vFd-Perturbations-and-z}}

\subsection{\boldmath$\delta N$ Formula with the Vector Field }

In this section we will generalize the $\delta N$ formalism introduced
in section~\ref{sub:dN-formalism} to include perturbations of the
vector field. For simplicity we will assume that only one perturbed
vector field affects the local expansion rate in the Universe. And
keeping one scalar field we can write to the second order (c.f. Eq.~\eqref{eq:zeta-dN-multi-Fd})\begin{eqnarray}
\zeta\left(t,\mathbf{x}\right) & = & \delta N\left(\phi\left(\mathbf{x}\right),W_{i}\left(\mathbf{x}\right),t\right)\nonumber \\
 & = & N_{\phi}\delta\phi+N_{W}^{i}\delta W_{i}+\frac{1}{2}N_{\phi\phi}\left(\delta\phi\right)^{2}+N_{\phi W}^{i}\delta\phi\delta W_{i}+\frac{1}{2}N_{W}^{ij}\delta W_{i}\delta W_{j}+...,\qquad\label{eq:vFd-dN-gen}\end{eqnarray}
where\begin{equation}
N_{\phi}\equiv\frac{\partial N}{\partial\phi},\; N_{W}^{i}\equiv\frac{\partial N}{\partial W_{i}},\; N_{\phi\phi}\equiv\frac{\partial^{2}N}{\partial\phi^{2}},\; N_{W}^{ij}\equiv\frac{\partial^{2}N}{\partial W_{i}\partial W_{j}},\; N_{\phi W}^{i}\equiv\frac{\partial^{2}N}{\partial W_{i}\partial\phi},\end{equation}
with $i$ and $j$ denoting spatial indices running from 1 to 3. As
with scalar fields, the unperturbed vector field values are defined
as averages within the chosen box (see the discussion below Eq.~\eqref{eq:end-of-infl-fNL}).

For this expression there is no need to define $W_{i}$ as components
of the vector field. Even more, this expression is valid not only
for the isotropic background expansion, but for anisotropic as well.
Although for the aim of this thesis it will be enough to consider
spatially flat isotropic geometry with the line element in the conformal
time\begin{equation}
\d s2=a^{2}\left(\tau\right)\left(\d{\tau}2-\d{\mathbf{x}}2\right).\end{equation}

\subsection{The Vector Field Quantization}

To quantize the vector field, let us expand perturbations of the field
in Fourier modes, similarly to the case of the scalar field\begin{equation}
\delta W_{i}\left(\tau,\mathbf{x}\right)=\int\delta\mathcal{W}_{i}\left(\tau,\mathbf{k}\right)\mathrm{e}^{i\mathbf{k}\cdot\mathbf{x}}\d{\mathbf{k}}{}.\label{eq:vFd-q-Fourier-exp-gen}\end{equation}
The massive vector field has three degrees of freedom, and the massless
vector field has two, in contrast to the scalar field case which has
only one degree of freedom. In Eq.~\eqref{eq:vFd-q-Fourier-exp-gen}
we have included only spatial components of the vector field, because
the temporal component is non-dynamical, i.e. it is not a degree of
freedom and is related to the spatial components through the equation
of motion (for the massive vector field). The perturbation of each
degree of freedom may be parametrized using polarization vectors as\begin{equation}
\delta\mathcal{W}_{i}\left(\tau,\mathbf{k}\right)=\sum_{\lambda}e_{i}^{\lambda}\left(\hat{\mathbf{k}}\right)w_{\lambda}\left(\tau,k\right),\label{eq:vFd-q-polariz-decompose}\end{equation}
where $e_{i}^{\lambda}$ are polarization vectors, $\hat{\mathbf{k}}\equiv\mathbf{k}/k$
is the unit vector in the direction of $\mathbf{k}$ and $k$ is the
modulus $k\equiv\left|\mathbf{k}\right|$. The most convenient choice
is the circular polarization for which two transverse vectors have
different handedness. Because both of them transform differently under
rotations, the rotational invariance of the Lagrangian prevents any
coupling between them. Choosing the coordinate $z$ axis to point
into the direction of $\mathbf{k}$, the circular polarization vectors
$e_{i}^{\lambda}$ take the form

\begin{equation}
\el i{}=\left(1,i,0\right)/\sqrt{2},\quad e_{i}^{\mathrm{R}}=\left(1,-i,0\right)/\sqrt{2}\quad\mathrm{and}\quad\eln i=\left(0,0,1\right).\label{eq:vFd-q-Pol-vecs-z-along-k}\end{equation}
In these expressions superscripts `$\mathrm{L}$', `R' and `$||$'
indicate the left-handed, right-handed and longitudinal polarizations
respectively. For the massive vector field all three polarizations
are present, but for the massless one $w_{||}=0$ in Eq.~\eqref{eq:vFd-q-polariz-decompose}.
These expressions define polarization vectors only up to a rotation
about the $\mathbf{k}$ direction, but this is enough for the present
purpose. Under the transformation $\mathbf{k}\rightarrow-\mathbf{k}$
one of the axis $x$ or $y$ change the sign as well. We will choose
that $x$ changes the sign and $y$ stays the same. Then $e_{i}^{\lambda}\left(-\hat{\mathbf{k}}\right)=-e_{i}^{\lambda*}\left(\hat{\mathbf{k}}\right)$
and because $\delta W_{i}\left(\tau,\mathbf{x}\right)$ is real, imposing
the reality condition onto Eqs.~\eqref{eq:vFd-q-Fourier-exp-gen}
and \eqref{eq:vFd-q-polariz-decompose} gives $w_{\lambda}^{*}\left(-\mathbf{k}\right)=-w_{\lambda}\left(\mathbf{k}\right)$.

Later it will be useful to rewrite polarization vectors in Eq.~\eqref{eq:vFd-q-Pol-vecs-z-along-k}
when $\mathbf{k}$ points to an arbitrary direction, not only along
the $z$ axis. Using the Cartesian coordinate system they become\begin{eqnarray}
\eln i\khp & = & \kh=\left(k_{x},k_{y},k_{z}\right),\nonumber \\
\el i{}\khp & = & \frac{1}{\sqrt{2\left(k_{x}^{2}+k_{y}^{2}\right)}}\left(-k_{y}+ik_{x}k_{z},\: k_{x}+ik_{y}k_{z},\:-i\left(k_{x}^{2}+k_{y}^{2}\right)\right),\label{eq:vFd-q-pol-vecs-arbitr-k}\\
\er i{}\khp & = & \el i*\khp.\nonumber \end{eqnarray}

Quantization of the vector field proceeds in the same way as for the
scalar field: we expand each degree of freedom in a complete set of
orthonormal mode functions and promote the expansion coefficients
to operators with appropriate commutation relations. Vector field
components satisfy the Klein-Gordon equation and the field lives in
the homogeneous and isotropic FRW space-time, therefore, the complete
set of orthonormal mode functions was already chosen to be $\mathrm{e}^{i\mathbf{k}\cdot\mathbf{x}}$
in Eq.~\eqref{eq:vFd-q-Fourier-exp-gen}. The quantized vector field
then takes the form%
\footnote{Note that in this chapter we have changed our normalization of Fourier
modes from $\left(2\pi\right)^{-3/2}$ to $\left(2\pi\right)^{-3}$.
This resulted in the $\left(2\pi\right)^{3}$ factor in the commutation
relations for creation and annihilation operators.%
}\begin{equation}
\delta\hat{W}_{i}=\sum_{\lambda}\int\left[e_{i}^{\lambda}\khp w_{\lambda}\left(\tau,k\right)\hat{a}_{\lambda}\left(\mathbf{k}\right)\mathrm{e}^{i\mathbf{k}\cdot\mathbf{x}}+e_{i}^{\lambda*}\khp w_{\lambda}^{*}\left(\tau,k\right)\hat{a}_{\lambda}^{\dagger}\left(\mathbf{k}\right)\mathrm{e}^{-i\mathbf{k}\cdot\mathbf{x}}\right]\frac{\d{\mathbf{k}}{}}{\left(2\pi\right)^{3}},\end{equation}
where \begin{equation}
\left[\hat{a}_{\lambda}\left(\mathbf{k}\right),\hat{a}_{\lambda'}\left(\mathbf{k}'\right)\right]=\left(2\pi\right)^{3}\delta\left(\mathbf{k}-\mathbf{k}'\right)\delta_{\lambda\lambda'}\end{equation}
and other commutators being zero. As in the scalar field case, after
the horizon exit vector field perturbations become classical in the
sense that the commutator $\left[\delta\hat{W}_{i}\left(\mathbf{x}\right),\partial_{\tau}\delta\hat{W}_{j}\left(\mathbf{x}'\right)\right]$
approaches zero. 

In later sections we will discuss several mechanisms to generate scale
invariant perturbations of the vector field. In all of these mechanisms
perturbations will be Gaussian, with no correlation between different
polarizations $\lambda$ or between perturbations of scalar and vector
fields. With these conditions it is sufficient to consider only the
spectra of vector field perturbations $\mathcal{P}_{\lambda}\equiv\mathcal{P}_{w_{\lambda}}$.
They are defined by the analogue to Eqs.~\eqref{eq:two-point-power-spec-deltaminus}
and \eqref{eq:two-point-power-spec-deltaplus} as\begin{equation}
\left\langle w_{\lambda}\left(\mathbf{k}\right)w_{\lambda'}^{*}\left(\mathbf{k}'\right)\right\rangle =\left(2\pi\right)^{3}\delta_{\lambda\lambda'}\delta\left(\mathbf{k}-\mathbf{k}'\right)\frac{2\pi^{2}}{k^{3}}\mathcal{P}_{\lambda}\left(\mathbf{k}\right),\label{eq:vFd-q-power-sp}\end{equation}
\begin{equation}
\left\langle w_{\lambda}\left(\mathbf{k}\right)w_{\lambda'}\left(\mathbf{k}'\right)\right\rangle =-\left(2\pi\right)^{3}\delta_{\lambda\lambda'}\delta\left(\mathbf{k}+\mathbf{k}'\right)\frac{2\pi^{2}}{k^{3}}\mathcal{P}_{\lambda}\left(\mathbf{k}\right),\end{equation}
where $\mathcal{P}_{\lambda}\left(\mathbf{k}\right)\equiv P_{\lambda}\left(\mathbf{k}\right)k^{3}/\left(2\pi^{2}\right)$
as defined in Eq.~\eqref{eq:curly-P-def}, and we have suppressed
the notation of time, i.e. $w_{\lambda}\left(\tau,\mathbf{k}\right)\equiv w_{\lambda}\left(\mathbf{k}\right)$.

If the Lagrangian is parity conserving then $\mathcal{P}_{\mathrm{L}}=\mathcal{P}_{\mathrm{R}}$,
which will be the case in all models considered in this thesis. Parity
violation is introduced by terms involving the dual of $F_{\mu\nu}$,
i.e. $\tilde{F}^{\mu\nu}=\epsilon^{\mu\nu\rho\sigma}F_{\rho\sigma}$,
where $\epsilon^{\mu\nu\rho\sigma}$ is the totally antisymmetric
tensor. Examples of such theories may be found in Refs.~\cite{Bamba_etal(2008)_PMF_ChernSim,Campanelli_Cea(2008)PMFs,Carroll_Field(1991)}.
The difference between the left-handed and right-handed power spectra
would indicate a parity violation. Therefore, it is convenient to
define parity conserving $\mathcal{P}_{+}$ and parity violating $\mathcal{P}_{-}$
power spectra by\begin{equation}
\mathcal{P}_{\pm}\equiv\frac{\mathcal{P}_{\mathrm{R}}\pm\mathcal{P}_{\mathrm{L}}}{2}.\label{eq:vFd-q-Ppm-def}\end{equation}
We also define two parameters which quantify the anisotropy in the
particle production of the vector field\begin{equation}
p\left(k\right)\equiv\frac{\Pl\left(k\right)-\Pp\left(k\right)}{\Pp\left(k\right)}\quad\mathrm{and}\quad q\left(k\right)\equiv\frac{\Pm\left(k\right)}{\Pp\left(k\right)},\label{eq:vFd-p-q-def}\end{equation}
\begin{figure}
\begin{centering}
\includegraphics[width=4.4cm]{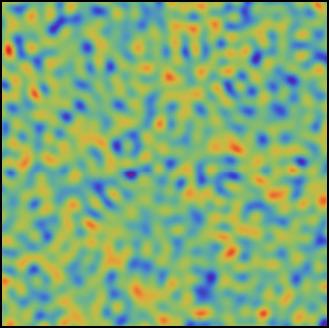}~~~~\includegraphics[width=4.4cm]{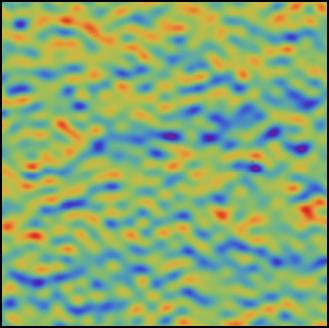}~~~~\includegraphics[width=4.4cm]{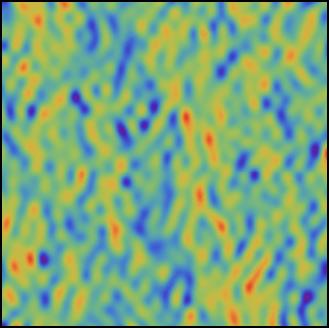}
\par\end{centering}

\caption{\label{fig:anisotropic-random-fields}Random field maps. The first
is statistically isotropic random field. If the vector field undergoes
isotropic particle production with $p=0$ and $q=0$, the curvature
perturbation generated by such a vector resembles the first map. If,
on the other hand, $p\ne0$ and/or $q\ne0$ the curvature perturbation
generated by the vector field is statistically anisotropic and resembles
the second or the third map. The second map is of the statistically
anisotropic random field with the preferred direction pointing vertically,
while the preferred direction of the third map is horizontal.}

\end{figure}
where we have also assumed that the expansion during inflation is
isotropic making $\Pl$ and $\mathcal{P}_{\pm}$ dependent only on
the modulus of $\mathbf{k}$. By the isotropic particle production
it is meant that the perturbation spectrum for all three degrees of
freedom is the same, i.e. $p=0$ and $q=0$. In this case the curvature
perturbation generated by the vector field does not differ from the
scalar field. But if $p\ne0$ and/or $q\ne0$ the particle production
of the vector field is anisotropic. If such a vector field generates
the curvature perturbation, its statistical properties are not invariant
under the rotations, i.e. it is statistically anisotropic (see Figure~\ref{fig:anisotropic-random-fields})

Calculating the two-point correlators of $\delta W_{i}\left(\mathbf{k}\right)$
we find\begin{equation}
\left\langle \delta W_{i}\left(\mathbf{k}\right)\delta W_{j}\left(\mathbf{k}'\right)\right\rangle =\left(2\pi\right)^{3}\delta\left(\mathbf{k}+\mathbf{k}'\right)\frac{2\pi^{2}}{k^{3}}\left[\el i{}\el j*\:\mathcal{P}_{\mathrm{L}}+\er i{}\er j*\:\mathcal{P}_{\mathrm{R}}+\eln i\eln j\:\Pl\right],\end{equation}
where we have suppressed the notation of $\kh$ for $e_{i}^{\lambda}\khp$
and of $k$ for $\mathcal{P}_{\lambda}\left(k\right)$. This expression
may be rewritten in terms of $\mathcal{P}_{\pm}$ in Eq.~\eqref{eq:vFd-q-Ppm-def}
as\begin{eqnarray}
\left\langle \delta W_{i}\left(\mathbf{k}\right)\delta W_{j}\left(\mathbf{k}'\right)\right\rangle  & = & \left(2\pi\right)^{3}\delta\left(\mathbf{k}+\mathbf{k}'\right)\frac{2\pi^{2}}{k^{3}}\times\nonumber \\
 & \times & \left[\Te\khp\Pp\left(k\right)+i\To\khp\Pm\left(k\right)+\Tl\khp\Pl\left(k\right)\right],\qquad\end{eqnarray}
where we have introduced tensors\begin{eqnarray}
\Te\khp & \equiv & \el i{}\khp\el j*\khp+\er i{}\khp\er j*\khp=\el i{}\khp\er j{}\khp+\er i{}\khp\el j{}\khp,\nonumber \\
\To\khp & \equiv & i\left[\el i{}\khp\el j*\khp-\er i{}\khp\er j*\khp\right]=i\left[\el i{}\khp\er j{}\khp-\er i{}\khp\el j{}\khp\right],\qquad\\
\Tl\khp & \equiv & \eln i\khp\eln j\khp.\nonumber \end{eqnarray}
With the circular polarization vectors derived in Eq.~\eqref{eq:vFd-q-pol-vecs-arbitr-k}
these tensors take a simple form

\begin{eqnarray}
\Te\khp & = & \delta_{ij}-\hat{k}_{i}\hat{k}_{j},\\
\To\khp & = & \epsilon_{ijk}\hat{k}_{k},\\
\Tl\khp & = & \hat{k}_{i}\hat{k}_{j}.\end{eqnarray}

\subsection{The Power Spectrum}

Since $\zeta$ is Gaussian to high accuracy, it seems reasonable to
expect that $\zeta$ will be dominated by one or more of the linear
terms in Eq.~\eqref{eq:vFd-dN-gen}. Keeping only them (corresponding
to what is called the tree-level contribution) we find\begin{eqnarray}
\Pz{}\kp & = & N_{\phi}^{2}\mathcal{P}_{\phi}\left(k\right)+N_{W}^{i}N_{W}^{j}\left[\Te\khp\Pp\left(k\right)+\Tl\Pl\left(k\right)\right]\label{eq:vFd-z-Power-sp-1}\\
 & = & N_{\phi}^{2}\mathcal{P}_{\phi}\left(k\right)+N_{W}^{2}\Pp\left(k\right)+\left(\mathbf{N}_{W}\cdot\kh\right)^{2}\left(\Pl-\Pp\right).\label{eq:vFd-z-Power-sp-2}\end{eqnarray}
Note that the power spectrum of $\zeta$ is dependent on the direction
of $\mathbf{k}$. In the upcoming discussion we will frequently use
the modulus of $\mathbf{N}_{W}$ and the unit vector along its direction
defined by\begin{equation}
N_{W}\equiv\left|\mathbf{N}_{W}\right|=\sqrt{N_{W}^{i}N_{W}^{i}}\quad\mathrm{and}\quad\hat{\mathbf{N}}_{W}\equiv\frac{\mathbf{N}_{W}}{N_{W}}.\end{equation}
 The curvature perturbation power spectrum $\Pz{}\kp$ may be further
separated into isotropic and anisotropic parts\begin{equation}
\Pz{}\kp=\Pz{iso}\left(k\right)\left[1+g\left(k\right)\left(\hat{\mathbf{N}}_{W}\cdot\kh\right)^{2}\right],\label{eq:vFd-z-Power-sp-3}\end{equation}
which has the same form as anisotropic power spectrum in Eq.~\eqref{eq:anisotropic-power-spec-def}
keeping only up to the quadratic term. Comparing this expression with
Eq.~\eqref{eq:vFd-z-Power-sp-2} we find that the isotropic part
of the spectrum is\begin{equation}
\Pz{iso}\left(k\right)=N_{\phi}^{2}\mathcal{P}_{\phi}\left(k\right)+N_{W}^{2}\Pp\left(k\right)=N_{\phi}^{2}\mathcal{P}_{\phi}\left(k\right)\left[1+\xi\frac{\Pp\left(k\right)}{\mathcal{P}_{\phi}\left(k\right)}\right],\label{eq:vFd-z-Piso-def}\end{equation}
where we have introduced the parameter $\xi$\begin{equation}
\xi\equiv\left(\frac{N_{W}}{N_{\phi}}\right)^{2}.\label{eq:vFd-ksi-def}\end{equation}
This parameter specifies the relative contribution from the vector
field to the statistically isotropic part of the curvature perturbation. 

From Eq.~\eqref{eq:vFd-z-Power-sp-2} we can also find that the anisotropy
in the curvature perturbation power spectrum $\Pz{}$ is equal to\begin{equation}
g\left(k\right)=N_{W}^{2}\frac{\Pl\left(k\right)-\Pp\left(k\right)}{\Pz{iso}\left(k\right)}=\frac{\xi}{\left[\mathcal{P}_{\phi}\left(k\right)/\Pp\left(k\right)\right]+\xi}\, p\left(k\right).\label{eq:vFd-z-g-def}\end{equation}

If the vector field perturbation dominates $\zeta$, i.e. $\xi\gg1$,
the anisotropy in the power spectrum of the curvature perturbation
is equal to the anisotropy in the particle production of the vector
field $g\approx p$. As was mentioned in section~\ref{sub:Spectrum-and-constraints}
the observational bound for the anisotropy in the power spectrum of
the curvature perturbation is $g\lesssim0.3$. Therefore, if the anisotropy
in the particle production of the vector field is larger than this
bound and there is no other vector field contribution, the produced
statistical anisotropy would violate observational constraints. To
prevent this, the dominant contribution to $\zeta$ must come from
one or more statistically isotropic scalar field perturbations.

Equations~\eqref{eq:vFd-z-Power-sp-1} and \eqref{eq:vFd-z-Power-sp-2}
for the power spectrum of the curvature perturbation were calculated
only at the tree level. Analogous calculation for the one-loop contribution
may be found in Ref.~\cite{Dimopoulos_etal_anisotropy(2008)}.

\subsection{The Bispectrum}

Working to the leading order in the quadratic terms of the $\delta N$
formula, we calculate the tree-level contribution to the bispectrum
defined in Eq.~\eqref{eq:bispectrum-def}\begin{equation}
B_{\zeta}\left(\kbi 1,\kbi 2,\kbi 3\right)=B_{\phi}\left(\kbi 1,\kbi 2,\kbi 3\right)+B_{\phi W}\left(\kbi 1,\kbi 2,\kbi 3\right)+B_{W}\left(\kbi 1,\kbi 2,\kbi 3\right),\end{equation}
where we have separated the bispectrum into three parts: one due to
perturbations in the scalar field, another part due to the vector
field perturbations and the mixed term. These terms are given by\begin{eqnarray}
B_{\phi}\left(\mathbf{k}_{1},\mathbf{k}_{2},\mathbf{k}_{3}\right) & = & N_{\phi}^{2}N_{\phi\phi}\left[\frac{4\pi^{4}}{k_{1}^{3}k_{2}^{3}}\mathcal{P}_{\phi}^{2}+\mathrm{c.p.}\right],\nonumber \\
B_{\phi W}\left(\mathbf{k}_{1},\mathbf{k}_{2},\mathbf{k}_{3}\right) & = & -N_{\phi}N_{\phi W}^{i}\left[\frac{4\pi^{4}}{k_{1}^{3}k_{2}^{3}}\mathcal{P}_{\phi}\mathcal{M}_{i}\left(\mathbf{k}_{2}\right)+\mathrm{c.p.}\right],\label{eq:vFd-bispectrum-gen}\\
B_{W}\left(\mathbf{k}_{1},\mathbf{k}_{2},\mathbf{k}_{3}\right) & = & \frac{4\pi^{4}}{k_{1}^{3}k_{2}^{3}}\mathcal{M}_{i}\left(\mathbf{k}_{1}\right)N_{W}^{ij}\mathcal{M}_{j}\left(\mathbf{k}_{2}\right)+\mathrm{c.p.}\nonumber \end{eqnarray}
The power spectrum $\mathcal{P}_{\phi}\left(k\right)$ in the above
equations depends only on the modulus of $\mathbf{k}$ because we
assumed that the expansion during inflation is isotropic, and the
vector $\mathcal{M}_{i}\left(\mathbf{k}\right)$ characterizes perturbations
of the vector field:\begin{equation}
\mathcal{M}_{i}\left(\mathbf{k}\right)\equiv\mathcal{P}_{+}N_{W}\left[\hat{N}_{W}^{i}+\hat{k}_{i}\left(\hat{\mathbf{k}}\cdot\hat{\mathbf{N}}_{W}\right)p+i\left(\hat{\mathbf{k}}\times\hat{\mathbf{N}}_{W}\right)_{i}q\right].\label{eq:vFd-M-def}\end{equation}

Reversal of the three wave-vectors in Eq.~\eqref{eq:vFd-bispectrum-gen}
corresponds to the parity transformation, and using the reality condition
$\zeta\left(-\mathbf{k}\right)=\zeta^{*}\left(\mathbf{k}\right)$
we find that it changes each correlator into its complex conjugate.
This does not affect the power spectrum because the reality condition
also makes the spectrum real. This may not affect the isotropic bispectrum
as well, because the reality condition and statistical isotropy make
the bispectrum real. In our case, the bispectrum is anisotropic, and
is guaranteed to be real only if the theory is parity conserving,
i.e. if $q=0$.

The second order contribution of the quadratic terms in the $\delta N$
formula gives the one-loop contribution to the bispectrum. It could
be significant or even dominant. It has been calculated for the scalar
case in Ref.~\cite{Boubekeur_Lyth(2006)_nonG}, and has been investigated
for the case of multifield inflation in Refs.~\cite{Cogollo_etal(2008),Rodriguez_Toledo(2008b)}
for example. The one-loop contribution from the vector perturbation
is calculated in Ref.~\cite{Toledo_etal(2009)}.

\subsection{\boldmath The Non-Linearity Parameter $\fnl$}

In calculating the non-linearity parameter defined in Eq.~\eqref{eq:fNL-def}
we will be interested in two configurations: equilateral, with $\ki 1=\ki 2=\ki 3$,
and squeezed, with $\ki 1\simeq\ki 2\gg\ki 3$. In the equilateral
configuration the bispectra from Eqs.~\eqref{eq:vFd-bispectrum-gen}
become\begin{eqnarray}
\mathcal{B}_{\phi}^{\mathrm{equil}}\left(\mathbf{k}_{1},\mathbf{k}_{2},\mathbf{k}_{3}\right) & = & 3N_{\phi}^{2}N_{\phi\phi}\mathcal{P}_{\phi}^{2},\nonumber \\
\mathcal{B}_{W\phi}^{\mathrm{equil}}\left(\mathbf{k}_{1},\mathbf{k}_{2},\mathbf{k}_{3}\right) & = & -N_{\phi}N_{\phi W}^{i}\mathcal{P}_{\phi}\left[\mathcal{M}_{i}\left(\mathbf{k}_{1}\right)+\mathcal{M}_{i}\left(\mathbf{k}_{2}\right)+\mathcal{M}_{i}\left(\mathbf{k}_{3}\right)\right],\label{eq:vFd-BA-equil-def}\\
\mathcal{B}_{W}^{\mathrm{equil}}\left(\mathbf{k}_{1},\mathbf{k}_{2},\mathbf{k}_{3}\right) & = & \mathcal{M}_{i}\left(\mathbf{k}_{1}\right)N_{W}^{ij}\mathcal{M}_{j}\left(\mathbf{k}_{2}\right)+\mathrm{c.p.},\nonumber \end{eqnarray}
where we have defined for the equilateral configuration \begin{equation}
\mathcal{B}_{\zeta}^{\mathrm{equil}}\left(\mathbf{k}_{1},\mathbf{k}_{2},\mathbf{k}_{3}\right)\equiv\left(\frac{k_{1}^{3}}{2\pi^{2}}\right)^{2}B_{\zeta}^{\mathrm{equil}}\left(\mathbf{k}_{1},\mathbf{k}_{2},\mathbf{k}_{3}\right),\label{eq:vFd-curlyB-equil-def}\end{equation}
and $\mathcal{B}_{\zeta}^{\mathrm{equil}}=\mathcal{B}_{\phi}^{\mathrm{equil}}+\mathcal{B}_{W\phi}^{\mathrm{equil}}+\mathcal{B}_{W}^{\mathrm{equil}}$.
In this case the non-linearity parameter $\fnle$ is expressed using
the power spectrum and the bispectrum as:\begin{equation}
\frac{6}{5}\fnle=\frac{\mathcal{B}_{\zeta}^{\mathrm{equil}}\left(\mathbf{k}_{1},\mathbf{k}_{2},\mathbf{k}_{3}\right)}{3\left(\mathcal{P}_{\zeta}^{\mathrm{iso}}\right)^{2}}.\label{eq:vFd-fNL-equil-def}\end{equation}
Observations give a limit on the anisotropy $g\lesssim0.3$ (see the
discussion above Eq.~\eqref{eq:bound-on-g}). Therefore, since the
anisotropic contribution to the curvature perturbation is subdominant
compared to the isotropic one, we have included only $\Pz{iso}$ into
the expression of $\fnle$.

In the squeezed configuration we have for the two vectors $\mathbf{k}_{1}\simeq-\mathbf{k}_{2}$,
but the third vector $\mathbf{k}_{3}$ is of much smaller modulus
than the other two and almost perpendicular to them. For this configuration
Eqs.~\eqref{eq:vFd-bispectrum-gen} take the form\begin{eqnarray}
\mathcal{B}_{\phi}^{\mathrm{local}}\left(\mathbf{k}_{1},\mathbf{k}_{2},\mathbf{k}_{3}\right) & = & 2\, N_{\phi}^{2}N_{\phi\phi}\mathcal{P}_{\phi}\left(\ki 1\right)\mathcal{P}_{\phi}\left(\ki 3\right),\nonumber \\
\mathcal{B}_{W\phi}^{\mathrm{local}}\left(\mathbf{k}_{1},\mathbf{k}_{2},\mathbf{k}_{3}\right) & = & -N_{\phi}N_{\phi W}^{i}\left\{ \mathcal{P}_{\phi}\left(\ki 1\right)\mathcal{M}_{i}\left(\mathbf{k}_{3}\right)+\mathcal{P}_{\phi}\left(\ki 3\right)\mathrm{Re}\left[\mathcal{M}_{i}\left(\mathbf{k}_{1}\right)\right]\right\} ,\label{eq:vFd-BA-local-def}\\
\mathcal{B}_{W}^{\mathrm{local}}\left(\mathbf{k}_{1},\mathbf{k}_{2},\mathbf{k}_{3}\right) & = & 2\,\mathrm{Re}\left[\mathcal{M}_{i}\left(\mathbf{k}_{1}\right)\right]N_{W}^{ij}\mathrm{Re}\left[\mathcal{M}_{j}\left(\mathbf{k}_{3}\right)\right],\nonumber \end{eqnarray}
where $\mathrm{Re}\left[\ldots\right]$ denotes the real part and
$\mathcal{B}_{\zeta}^{\mathrm{local}}\left(\mathbf{k}_{1},\mathbf{k}_{2},\mathbf{k}_{3}\right)$
is defined similarly to Eq.~\eqref{eq:vFd-curlyB-equil-def}\begin{equation}
\mathcal{B}_{\zeta}^{\mathrm{local}}\left(\mathbf{k}_{1},\mathbf{k}_{2},\mathbf{k}_{3}\right)\equiv\frac{k_{1}^{3}k_{3}^{3}}{4\pi^{4}}B_{\zeta}^{\mathrm{local}}\left(\mathbf{k}_{1},\mathbf{k}_{2},\mathbf{k}_{3}\right).\end{equation}
Then, the non-linearity parameter $\fnll$ in the squeezed configuration
becomes\begin{equation}
\frac{6}{5}f_{\mathrm{NL}}^{\mathrm{local}}=\frac{\mathcal{B}_{\zeta}^{\mathrm{local}}\left(\mathbf{k}_{1},\mathbf{k}_{2},\mathbf{k}_{3}\right)}{2\Pz{iso}\left(\ki 1\right)\Pz{iso}\left(\ki 3\right)}.\end{equation}

In the treatment so far we have calculated the curvature perturbation
generated by the vector field as well as the resulting anisotropic
power spectrum and the non-linearity parameter $\fnl$ in equilateral
and squeezed configurations. However, we didn't discuss how the perturbation
of the vector field is generated and which mechanism transformed the
field perturbation into the curvature perturbation. In the rest of
the thesis we consider several examples of the conformal invariance
breaking for the vector field, which generates scale invariant perturbation
spectrum and determines the values of $p$ and $q$ parameters. We
also consider two scenarios in which the vector field perturbation
influences or generates the curvature perturbation $\zeta$.

\section{The Vector Curvaton Scenario\label{sec:vFd-Vector-Curvaton}}

In section~\ref{sub:vFd-Large-Scale-Anisotropy} it was discussed
that a dominant light vector field would generate large scale anisotropy
in the Universe which violates observational bounds. But to generate
or influence the curvature perturbation by a vector field it has to
dominate or nearly dominate. One of the ways to overcome this difficulty
is the curvaton scenario. This scenario was summarized in section~\ref{sub:Curvaton-Mechanism}
with the curvaton acted by a scalar field. In this section we consider
scenarios with the curvaton acted by a vector field, which was first
introduced in Ref.~\cite{Dimopoulos2006}.

\subsection{The Vector Curvaton Dynamics\label{sub:vCurv-dynamics}}

Let us consider a massive Abelian vector field in the Universe dominated
by matter (radiation) with the barotropic parameter $w=0$ ($w=1/3$).
The Lagrangian of the vector field is\begin{equation}
\mathcal{L}=-\frac{1}{4}F_{\mu\nu}F^{\mu\nu}+\frac{1}{2}m^{2}A^{\mu}A_{\mu}.\label{eq:vCurv-Intro-Lagrangian}\end{equation}
And let us choose a coordinate system in such a way that spatial part
of the homogeneous mode of the vector field has components $A_{i}=\left(0,0,A\right)$.
In Ref.~\cite{Dimopoulos2006} it was shown that the energy momentum
tensor for this field may be written as\begin{equation}
T_{\mu}^{\nu}=\mathrm{diag}\left(\rho_{W},-p_{\perp},-p_{\perp},+p_{\perp}\right),\label{eq:vCurv-Intro-T-massive-vFd}\end{equation}
where\begin{equation}
\rho_{W}\equiv\rho_{\mathrm{kin}}+V_{W},\qquad p_{\perp}\equiv\rho_{\mathrm{kin}}-V_{W}\label{eq:vCurv-Intro-r-p-def}\end{equation}
with\begin{eqnarray}
\rho_{\mathrm{kin}} & \equiv & -\frac{1}{4}F_{\mu\nu}F^{\mu\nu}=\frac{1}{2}\left(\frac{\dot{A}}{a}\right)^{2},\label{eq:vCurv-Intro-rkin-massive-vFd}\\
V_{W} & \equiv & -\frac{1}{2}m^{2}A_{\mu}A^{\mu}=\frac{1}{2}m^{2}\left(\frac{A}{a}\right)^{2}.\label{eq:vCurv-Intro-V-massive-vFd}\end{eqnarray}
From which we see that in general the energy momentum tensor will
have anisotropic stress due to the opposite sign of the pressure components
in the direction parallel to the field and the perpendicular one.
Indeed, the equation of motion for the homogeneous mode of the vector
field with the Lagrangian in Eq.~\eqref{eq:vCurv-Intro-Lagrangian}
is given by \cite{Dimopoulos2006}\begin{equation}
\ddot{\mathbf{A}}+H\dot{\mathbf{A}}+m^{2}\mathbf{A}=0.\label{eq:vCurv-Intro-EoM}\end{equation}
In the matter or radiation dominated Universe the solution of this
equation is \begin{equation}
A\left(t\right)=t^{v}\left[C_{1}J_{v}\left(mt\right)+C_{2}J_{-v}\left(mt\right)\right],\end{equation}
and the time derivative of the vector field is given by\begin{equation}
\dot{A}\left(t\right)=\frac{t^{v}}{m}\left[C_{1}J_{v-1}\left(mt\right)-C_{2}J_{1-v}\left(mt\right)\right],\end{equation}
where $C_{1}$ and $C_{2}$ are constants of integration, $J_{v}$
is the Bessel function of the first kind and \begin{equation}
v\equiv\frac{1+3w}{6\left(1+w\right)}.\end{equation}
In the FRW Universe the Hubble parameter is $H\sim t^{-1}$ and the
light field corresponds to $mt\ll1$. In this limit Bessel functions
can be approximated by power law functions and the growing mode of
the vector field changes with time as \begin{equation}
A\left(t\right)\propto t^{2v}.\end{equation}
Inserting this into Eqs.~\eqref{eq:vCurv-Intro-rkin-massive-vFd},
\eqref{eq:vCurv-Intro-V-massive-vFd} and using Eq.~\eqref{eq:vCurv-Intro-T-massive-vFd}
it is clear that the energy-momentum tensor of the light vector field
has anisotropic stress. If such a vector field dominated the Universe,
the expansion rate in the direction of the field would be different
from the transverse directions and the Universe would become predominantly
anisotropic. Such excessive \emph{large scale} anisotropy is forbidden
by observations, therefore a light vector field cannot dominate the
Universe.

In the opposite regime, when the vector field is heavy, $mt\gg1$,
the Bessel functions may be approximated by trigonometric functions
giving\begin{eqnarray}
A\left(t\right) & = & t^{v}\sqrt{\frac{2}{\pi mt}}\left[C_{1}\sin\left(mt+\frac{1-2v}{4}\pi\right)+C_{2}\cos\left(mt-\frac{1-2v}{4}\pi\right)\right],\qquad\label{eq:vCurv-Intro-A-oscillating}\\
\dot{A}\left(t\right) & = & -mt^{v}\sqrt{\frac{2}{\pi mt}}\left[C_{1}\cos\left(mt+\frac{1-2v}{4}\pi\right)-C_{2}\sin\left(mt-\frac{1-2v}{4}\pi\right)\right].\qquad\label{eq:vCurv-Intro-Adot-oscillating}\end{eqnarray}
From these equations it is clear that the evolution of the heavy vector
field resembles the evolution of the underdamped harmonic oscillator
with decreasing amplitude of oscillations. To see this let us rewrite
Eqs.~\eqref{eq:vCurv-Intro-A-oscillating} and \eqref{eq:vCurv-Intro-Adot-oscillating}
in the form\begin{eqnarray}
A\left(t\right) & = & \Lambda t^{v}\sqrt{\frac{2}{\pi mt}}\sin\left(mt+\varphi\right),\label{eq:vCurv-Intro-A-oscillating-sin}\\
\dot{A}\left(t\right) & = & -\Lambda mt^{v}\sqrt{\frac{2}{\pi mt}}\cos\left(mt+\varphi\right),\label{eq:vCurv-Intro-Adot-oscillating-cos}\end{eqnarray}
where constants $\Lambda$ and $\varphi$ are related to the original
constants by\begin{equation}
\Lambda\equiv\sqrt{C_{1}^{2}+C_{2}^{2}+2C_{1}C_{2}\cos\left(\pi v\right)},\end{equation}
and\begin{equation}
\varphi\equiv\arccos\left[\frac{C_{1}\cos\left(\frac{1-2v}{4}\pi\right)+C_{2}\sin\left(\frac{1-2v}{4}\pi\right)}{\Lambda}\right].\end{equation}
Calculating the energy density and pressure from Eq.~\eqref{eq:vCurv-Intro-r-p-def}
we find\begin{eqnarray}
\rho_{W} & = & a^{-3}\frac{m}{\pi}\Lambda^{2},\label{eq:vCurv-Intro-rho-massive-vFd}\\
p_{\perp} & = & a^{-3}\frac{m}{\pi}\Lambda^{2}\cos\left(2mt+2\varphi\right).\label{eq:vCurv-Intro-p-massive-vFd}\end{eqnarray}
Since the vector field is heavy, i.e. $m\gg H$, the frequency of
oscillating functions in Eq.~\eqref{eq:vCurv-Intro-p-massive-vFd}
is much larger than the Hubble parameter. Therefore, during one Hubble
time the average pressure of the heavy vector field is zero and we
can write \begin{equation}
\rho_{W}\propto a^{-3}\quad\mathrm{and}\quad\overline{p_{\perp}}\approx0.\label{eq:vCurv-Intro-isotr-pressureless-cond}\end{equation}
Thus the heavy vector field acts as the pressureless isotropic matter
and it can dominate the Universe without generating excessive large
scale anisotropy.

\begin{figure}
\begin{centering}
\includegraphics[width=11cm]{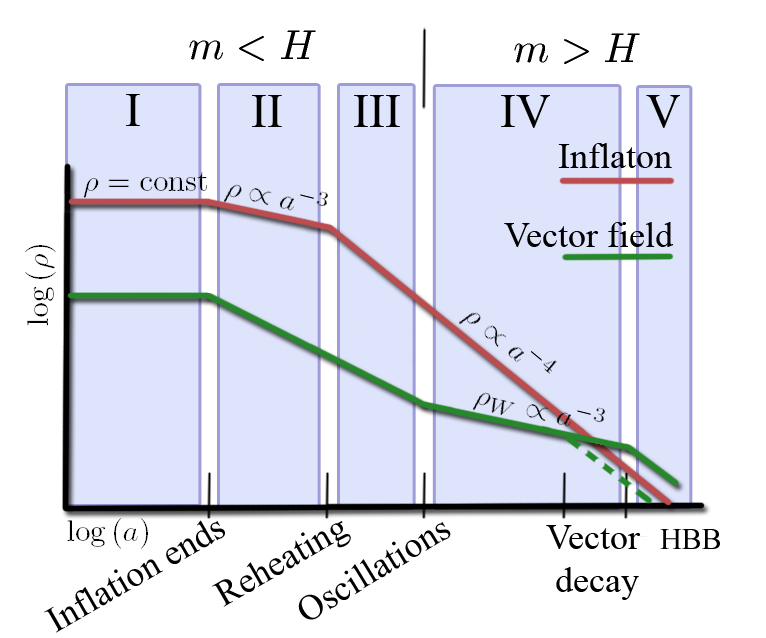}
\par\end{centering}

\caption{\label{fig:vCurvaton-graph}The schematic graph of the vector curvaton
scenario: I. during inflation the vector field is subdominant but
it acquires a scale invariant perturbation spectrum; II. inflation
ends and the inflaton field starts oscillating around its minimum,
while the energy density of the Universe scales as $\rho\propto a^{-3}$
(see section~\ref{sub:Reheating}); III. after reheating the Universe
is radiation dominated and the vector field is still subdominant until
it becomes heavy; IV. the heavy vector field oscillates and acts as
the pressureless isotropic matter with the energy density decreasing
as $\rho_{W}\propto a^{-3}$; V. the vector curvaton dominates (solid
line) or nearly dominates (dashed line) the Universe, imprints its
perturbation spectrum and decays, recovering the standard HBB cosmology.
Note, that in this graph we depicted the situation when the vector
field becomes heavy after reheating. However, this might happen before
reheating as well.}

\end{figure}

This property of the vector field is utilized in the vector curvaton
scenario (see Figure~\ref{fig:vCurvaton-graph}). During inflation
the vector field is light, and although its energy-momentum tensor
has non-vanishing stress in accordance with the curvaton scenario,
it is subdominant, allowing the expansion of the Universe to be isotropic.
During this period the vector field with broken conformal invariance
undergoes particle production. (In sections~\ref{sub:vCurvaton-RA2}
and \ref{sub:vCurvaton-fF2} we consider two mechanisms of breaking
the conformal invariance of the vector field and producing a flat
perturbation spectrum.) After inflation, the Hubble parameter decreases
as $H\propto t^{-1}$. When it becomes smaller than the mass of the
vector field, the latter starts to oscillate and, as was shown in
Eq.~\eqref{eq:vCurv-Intro-isotr-pressureless-cond}, acts as the
pressureless isotropic matter. After reheating the relative energy
density of the vector field increases as $\rho_{W}/\rho_{\gamma}\propto a$,
where $\rho_{\gamma}$ is the energy density of the radiation dominated
Universe. As in the original curvaton scenario (see section~\ref{sub:Curvaton-Mechanism})
the curvaton field during the radiation dominated period dominates
(or nearly dominates) the Universe, imprints its perturbation spectrum
and decays, recovering the standard Hot Big Bang cosmology. If the
curvaton decays when it is dominant, the Universe undergoes a second
reheating.

\subsection{\boldmath The Generic Treatment of $\fnl$\label{sub:vCurv-generic-fNL}}

In this section we obtain analytic expressions for the non-linearity
parameter $f_{{\rm NL}}$ without assuming a specific vector curvaton
model. In contrast to the original curvaton idea we include perturbations
already present in the radiation dominated Universe when the vector
field energy density is still negligible (for a similar study in the
scalar curvaton case see Refs.~\cite{Lazarides_etal(2004),Ichikawa_etal(2008)}).

Some time after inflation the mass of the vector field becomes bigger
than the Hubble parameter. Then the field starts to oscillate and
acts as the pressureless isotropic matter. The total contribution
to the curvature perturbation by the vector field before its decay
can be found using Eqs.~\eqref{eq:zeta-rho-many-Fds} and \eqref{eq:curvaton-zeta-composite}
\begin{equation}
\zeta_{\mathrm{vec}}\equiv\ohw\zeta_{W}=\frac{1}{3}\ohw\frac{\delta\rho_{W}}{\rho_{W}},\end{equation}
where $\ohw$ is defined similarly to the Eq.~\eqref{eq:curvaton-Omega-hat-def}
\begin{equation}
\ohw\equiv\frac{3\rho_{W}}{3\rho_{W}+4\rho_{\gamma}}=\frac{3\ow}{4-\ow}\label{eq:vCurv-Omega-hat-def}\end{equation}
and energy densities $\rho_{W}$ and $\rho_{\gamma}$ are evaluated
at the curvaton decay with $\ow\equiv\rho_{W}/\rho$ and $\rho=\rho_{W}+\rho_{\gamma}$.
This expression is valid to the first order in $\delta\rho_{W}$,
which is evaluated on a flat slice, where $a\left(t,\mathbf{x}\right)$
is unperturbed.

We assume that the curvaton decays instantly (sudden-decay approximation)
and evaluate $\zeta_{W}$ just before the curvaton decays, leaving
$\zeta$ constant thereafter. Evaluating $\delta\rho_{W}$ to the
second order we have \cite{Lyth_Rodriguez(2005b)}\begin{equation}
\zeta_{\mathrm{vec}}=\frac{2}{3}\ohw\frac{W_{i}}{W^{2}}\delta W_{i}+\frac{1}{3}\ohw\frac{1}{W^{2}}\delta W_{i}\delta W_{i},\label{eq:vCurv-z-gen}\end{equation}
where $W\equiv\left|\mathbf{W}\right|$ is evaluated just before the
vector field decays. This is valid only for $\ohw\ll1$. To calculate
the same expression when $\ohw\simeq1$ one could evaluate $N$ and
hence $\delta N$ directly. All of this is the same as for a scalar
field contribution, where the calculation of $N$ was done in Ref.~\cite{Lyth_Rodriguez(2005b)}.

Comparing Eq.~\eqref{eq:vCurv-z-gen} with \eqref{eq:vFd-dN-gen}
we find $N_{W}^{i}$ and $N_{W}^{ij}$ to be equal to \begin{eqnarray}
N_{W}^{i} & = & \frac{2}{3}\ohw\frac{W_{i}}{W^{2}},\label{eq:vCurv-NAi-def}\\
N_{W}^{ij} & = & \frac{2}{3}\ohw\frac{\delta_{ij}}{W^{2}}.\label{eq:vCurv-NAij-def}\end{eqnarray}
Using Eq.~\eqref{eq:vCurv-NAi-def} the isotropic part of the total
power spectrum in Eq.~\eqref{eq:vFd-z-Piso-def} becomes \begin{equation}
\mathcal{P}_{\zeta}^{\mathrm{iso}}\left(k\right)=N_{\phi}^{2}\mathcal{P}_{\phi}\left(k\right)\left(1+\xi\,\frac{\mathcal{P}_{+}\left(k\right)}{\mathcal{P}_{\phi}\left(k\right)}\right),\label{eq:P_iso_delta}\end{equation}
and from Eq.~\eqref{eq:vCurv-NAi-def} the preferred direction in
the power spectrum in Eq.~\eqref{eq:vFd-z-Power-sp-3} is \begin{equation}
\hat{\mathbf{N}}_{W}=\hat{\mathbf{W}},\label{eq:vCurv-pref-direction}\end{equation}
where $\hat{\mathbf{W}}\equiv\mathbf{W}/W$. 

Then the vector part of the bispectrum for equilateral configuration
in Eq.~\eqref{eq:vFd-BA-equil-def} reduces to \begin{eqnarray}
\mathcal{B}_{W}^{\mathrm{equil}}\left(\mathbf{k}_{1},\mathbf{k}_{2},\mathbf{k}_{3}\right) & = & \left(\frac{2}{3}\frac{\ohw}{W}\right)^{3}\frac{1}{W}\mathcal{P}_{+}\left(k_{1}\right)\mathcal{P}_{+}\left(k_{2}\right)\left\{ 1+p\left(k_{1}\right)W_{1}^{2}+p\left(k_{2}\right)W_{2}^{2}+\right.\nonumber \\
 &  & +W_{1}W_{2}\left[q\left(k_{1}\right)q\left(k_{2}\right)-\frac{1}{2}p\left(k_{1}\right)p\left(k_{2}\right)\right]+\nonumber \\
 &  & +i\sqrt{\frac{3}{4}-\left(W_{1}^{2}+W_{1}W_{2}+W_{2}^{2}\right)}\left[W_{1}p\left(k_{1}\right)q\left(k_{2}\right)-\right.\nonumber \\
 &  & \left.\left.-W_{2}p\left(k_{2}\right)q\left(k_{1}\right)\right]+\frac{1}{2}q\left(k_{1}\right)q\left(k_{2}\right)\right\} +\mathrm{c.p.}\label{eq:M^2_equilateral}\end{eqnarray}
 In the above we used the notation $W_{1}\equiv\hat{\mathbf{W}}\cdot\hat{\mathbf{k}}_{1}$
etc. Because the configuration of wave vectors $\hat{\mathbf{k}}_{1}$,
$\hat{\mathbf{k}}_{2}$ and $\hat{\mathbf{k}}_{3}$ is equilateral,
with the angle between any two of them being $2\pi/3$, we find $\hat{\mathbf{k}}_{1}\cdot\hat{\mathbf{k}}_{2}=\hat{\mathbf{k}}_{1}\cdot\hat{\mathbf{k}}_{3}=\hat{\mathbf{k}}_{2}\cdot\hat{\mathbf{k}}_{3}=-\frac{1}{2}$.
Eq.(\ref{eq:M^2_equilateral}) simplifies further if we consider a
scale invariant power spectrum, then the expression for $f_{\mathrm{NL}}^{\mathrm{equil}}$
becomes:\begin{equation}
\frac{6}{5}f_{\mathrm{NL}}^{\mathrm{equil}}=\xi^{2}\mathcal{P}_{+}^{2}\,\frac{3}{2\ohw}\,\frac{\left(1+\frac{1}{2}q^{2}\right)+\left[p+\frac{1}{8}\left(p^{2}-2q^{2}\right)\right]W_{\bot}^{2}}{\left(\mathcal{P}_{\phi}+\xi\,\mathcal{P}_{+}\right)^{2}},\label{eq:vCurv-fnl-eql-M^2-scaleInv}\end{equation}
 where we have taken into account that the non-Gaussianity generated
during the single field inflation is negligible (see Eq.~\eqref{eq:single-Fd-infl-fNL}).
The quantity $W_{\bot}\leq1$ is the modulus of the projection of
the unit vector $\hat{\mathbf{W}}$ onto the plane containing the
three vectors $\hat{\mathbf{k}}_{1}$, $\hat{\mathbf{k}}_{2}$ and
$\hat{\mathbf{k}}_{3}$. The calculation of $W_{\bot}$ in the equilateral
configuration is explained in more detail in the Appendix~\ref{cha:Appendix-A-calculation}.

Similarly to the definition of the anisotropic power spectrum in Eq.~\eqref{eq:vFd-z-Power-sp-3}
we may separate the non-linearity parameter $\fnl$ in Eq.~\eqref{eq:vCurv-fnl-eql-M^2-scaleInv}
into the isotropic and anisotropic parts\begin{equation}
\frac{6}{5}\fnl=\fnli\left(1+\mathcal{G}\cdot W_{\perp}^{2}\right),\label{eq:vCurv-fNL-aniso-parametrization}\end{equation}
where $\mathcal{G}$ parametrizes the anisotropy in $\fnl$. Comparing
this equation with Eq.~\eqref{eq:vCurv-fnl-eql-M^2-scaleInv} we
find that in the equilateral configuration\begin{equation}
\fnlei=\xi^{2}\mathcal{P}_{+}^{2}\,\frac{3}{2\ohw}\,\frac{\left(1+\frac{1}{2}q^{2}\right)}{\left(\mathcal{P}_{\phi}+\xi\,\mathcal{P}_{+}\right)^{2}},\label{eq:vCurv-fNL-equil-iso-gen}\end{equation}
and\begin{equation}
\Ge=\frac{p+\frac{1}{8}\left(p^{2}-2q^{2}\right)}{1+\frac{1}{2}q^{2}}.\label{eq:vCurv-fNL-curG-equil}\end{equation}

For the squeezed configuration the bispectrum from the vector field
perturbation in Eqs.~\eqref{eq:vFd-BA-local-def} becomes \begin{eqnarray}
\mathcal{B}_{W}^{\mathrm{local}}\left(\mathbf{k}_{1},\mathbf{k}_{2},\mathbf{k}_{3}\right) & = & 2\left(\frac{2}{3}\frac{\ohw}{W}\right)^{3}\frac{1}{W}\mathcal{P}_{+}\left(k_{1}\right)\mathcal{P}_{+}\left(k_{3}\right)\times\nonumber \\
 &  & \times\left[1+p\left(k_{1}\right)W_{1}^{2}+p\left(k_{3}\right)W_{3}^{2}\right].\end{eqnarray}
 Working as in the equilateral case, we find that the non-linearity
parameter for the scale invariant power spectra is \begin{equation}
\frac{6}{5}f_{\mathrm{NL}}^{\mathrm{local}}=\xi^{2}\mathcal{P}_{+}^{2}\,\frac{3}{2\ohw}\,\frac{1+pW_{\bot}^{2}}{\left(\mathcal{P}_{\phi}+\xi\,\mathcal{P}_{+}\right)^{2}}.\label{eq:vCurv-fNL-scaleInv-local}\end{equation}
Using the parametrization of Eq.~\eqref{eq:vCurv-fNL-aniso-parametrization}
we write for the squeezed configuration\begin{equation}
\fnlli=\xi^{2}\mathcal{P}_{+}^{2}\,\frac{3}{2\ohw}\,\frac{1}{\left(\mathcal{P}_{\phi}+\xi\,\mathcal{P}_{+}\right)^{2}},\label{eq:vCurv-fNL-local-iso-gen}\end{equation}
and \begin{equation}
\Gl=p.\label{eq:vCurv-fNL-curG-local}\end{equation}

As one can see from the above equations $f_{\mathrm{NL}}$ in general
depends on $W_{\bot}$ in both configurations, i.e. $\Ge\ne0$ and
$\Gl\ne0$. This means that $f_{\mathrm{NL}}$ is anisotropic, with
the same preferred direction as in the power spectrum (c.f. Eqs.~\eqref{eq:vFd-z-Power-sp-3}
and \eqref{eq:vCurv-pref-direction}). The isotropic parts of $\fnl$
in both configurations may be rewritten as\begin{equation}
\fnlei=\fnlli\left(1+\frac{1}{2}q^{2}\right)=g^{2}\frac{3}{2\ohw}\cdot\frac{1+\frac{1}{2}q^{2}}{p^{2}}.\label{eq:vCurv-fNLe-fNLl}\end{equation}
Given the (quasi) exponential expansion of the Universe during inflation,
the value of $p$ depends only on the Lagrangian of the vector field.
As Eq.~\eqref{eq:vFd-z-g-def} suggests it relates to $g$ only indirectly,
through parameters determining the generation of anisotropic as well
as isotropic parts of $\zeta$. In other words, specifying the value
of $p$ does not determine $g$. In view of this, from Eq.~\eqref{eq:vCurv-fNLe-fNLl}
we see that the amount of non-Gaussianity is correlated with the statistical
anisotropy in the spectrum, $\fnl\propto g^{2}$. If, instead, the
particle production is isotropic (i.e. $\mathcal{P}_{\parallel}=\mathcal{P}_{+}$
and $\mathcal{P}_{-}=0$) Eqs.~\eqref{eq:vFd-q-Ppm-def} and \eqref{eq:vFd-p-q-def}
give $p=q=0$ and therefore $\mathcal{G}^{\mathrm{equil}}=\mathcal{G}^{\mathrm{local}}=0$.
In this case $f_{\mathrm{NL}}^{\mathrm{equil}}$ and $f_{\mathrm{NL}}^{\mathrm{local}}$
become isotropic too and both reduce to $f_{\mathrm{NL}}=5/4\ohw$
as in the scalar curvaton scenario, where we used Eq.~\eqref{eq:vFd-z-g-def}
with the assumption $\mathcal{P}_{\phi}\ll\mathcal{P}_{+}$, i.e.
that the dominant contribution to the curvature perturbation is due
to the vector curvaton field only.

In addition to the $\fnl$ being anisotropic, having the same preferred
direction as the spectrum and its magnitude being correlated with
the anisotropy in the spectrum $g$ from Eqs.~\eqref{eq:vCurv-fNL-equil-iso-gen},
\eqref{eq:vCurv-fNL-curG-equil} and \eqref{eq:vCurv-fNL-local-iso-gen},
\eqref{eq:vCurv-fNL-curG-local} we find more observational signatures.
From Eq.~\eqref{eq:vCurv-fNLe-fNLl} it is clear that for parity
conserving vector fields with $q=0$, the isotropic parts of $\fnl$
are identical, i.e. $\fnlei=\fnlli$. Any departure from this equality
would indicate parity violating terms in the Lagrangian of the vector
field. But the anisotropy in the non-linearity parameter is configuration
dependent with $\Ge\ne\Gl$ in both - parity conserving as well as
parity violating - theories. In the squeezed configuration, $\Gl$
is sensitive only to the anisotropy in the parity conserving perturbations
of the vector field. But in the equilateral configuration, $\Ge$
is also correlated with the amount of parity violation of the field.
In both cases the anisotropy in $\fnl$ is proportional to the anisotropy
in the particle production of the vector field, $p$ and $q$. Therefore,
if the anisotropy in particle production is of order one or larger,
anisotropic parts of $\fnl$ in both configurations \emph{are not
subdominant}. At the moment, observations do not provide any information
about the values of $\Ge$ and $\Gl$. However, as can be seen from
Eqs.~\eqref{eq:vCurv-fNL-curG-equil} and \eqref{eq:vCurv-fNL-curG-local}
the observational detection of $\Ge$ and $\Gl$ would allow to determine
uniquely the values of parameters $p$ and $q$, therefore, allowing
to constraint very narrowly the possible range of conformal invariance
breaking Lagrangians for the vector field.

\subsection{\boldmath Generation of $\zeta$\label{sub:vCurv-Intro-z-generation}}

To calculate the curvature perturbation generated by the vector curvaton
field consider an era after reheating when the Universe is radiation
dominated, with the energy density decreasing as $\rho_{\gamma}\propto a^{-4}$.
As was discussed at the end of section~\ref{sub:vCurv-dynamics}
the relative energy density of the heavy vector field during this
era increases as $\rho_{W}/\rho_{\gamma}\propto a$ (see Eq.~\eqref{eq:vCurv-Intro-isotr-pressureless-cond}).
When the vector field becomes dominant (or nearly dominant) it imprints
its perturbation spectrum onto the Universe.

The curvature perturbation $\zw$ generated by the vector field is
calculated as follows. On the spatially flat slicing of space-time
using Eq.~\eqref{eq:zeta-rho-many-Fds} we can write for the vector
field \begin{equation}
\zw=\left.\frac{\delta\rw}{3\rw}\right|_{\mathrm{dec}},\end{equation}
where we considered that the decay of the vector field (labeled by
`dec') occurs after the onset of its oscillations so that it is pressureless,
as shown in Eq.~\eqref{eq:vCurv-Intro-isotr-pressureless-cond}.
Note that, since $\zw$ is determined by the fractional perturbation
of the field's density, which is a scalar quantity, the perturbation
$\zw$ is scalar and not vector in nature.

Since Eq.~\eqref{eq:vCurv-Intro-EoM} is a linear differential equation,
$A$ and its perturbation $\delta A$ satisfy the same equation of
motion. Therefore, they evolve in the same way, which means that $\delta A/A$
remains constant, before and after the onset of oscillations. As was
discussed in section~\ref{sub:vCurv-dynamics} the massive vector
field acts as an underdamped harmonic oscillator. The energy of such
oscillator is determined by the amplitude of oscillations. Therefore,
we may write $\rho_{W}=\frac{1}{2}m^{2}\left|\left|W\right|\right|^{2}$,
where we used the physical vector field $W=A/a$ defined in Eq.~\eqref{eq:vCurv-phys-massive-vFd}
and $\left|\left|W\right|\right|$ is the amplitude of the oscillating
physical vector field. From the above we obtain

\begin{equation}
\zw=\left.\frac{\delta\rw}{3\rw}\right|_{\mathrm{dec}}\approx\frac{2}{3}\left.\frac{||\delta W||}{||W||}\right|_{\mathrm{dec}}\simeq\frac{2}{3}\left.\frac{\delta W}{W}\right|_{\mathrm{osc}}\simeq\frac{2}{3}\left.\frac{\delta W}{W}\right|_{\mathrm{end}},\label{eq:vCurv-Intro-zetaW-gen}\end{equation}
where `osc' denotes the onset of oscillations and `end' denotes the
time at the end of inflation. Therefore, from this equation we may
write\begin{equation}
\zw\sim\frac{\delta W_{\mathrm{end}}}{W_{\mathrm{end}}}.\label{eq:vCurv-Intro-zetaW}\end{equation}

In the usual scalar curvaton scenario the curvaton field generates
the total curvature perturbation in the Universe. This may be realized
in the vector curvaton scenario as well if the curvature perturbation
$\zw$, generated by the vector field, is statistically isotropic,
or if its statistical anisotropy is within the observationally allowed
region (see Eq.~\eqref{eq:bound-on-g}). However, if $\zw$ is predominantly
anisotropic it can only be a subdominant contribution to the total
curvature perturbation $\zeta$, while the dominant part must be generated
by some statistically isotropic source. In analogy to the Eq.~\eqref{eq:curvaton-zeta-composite}
in this case the total curvature perturbation in the curvaton scenario
may be written as \begin{equation}
\zeta=\left(1-\ohw\right)\zg+\ohw\zw,\label{eq:vCurv-Intro-zeta-composite}\end{equation}
where $\zg$ is the dominant and statistically isotropic curvature
perturbation which is present in the radiation dominated Universe
before $\zw$ is generated. In this equation $\ohw\approx\frac{3}{4}\ow<1$
because the vector field must decay before it starts dominating (the
dashed line in Figure~\ref{fig:vCurvaton-graph}). Assuming that
$\zg$ is generated by the scalar field $\phi$, the anisotropy in
the power spectrum $g$ from Eq.~\eqref{eq:vFd-z-g-def} becomes\begin{equation}
g\approx\xi\frac{\Pp}{\mathcal{P}_{\phi}}p,\label{eq:vCurv-Intro-g-subdomW}\end{equation}
were we also assumed scale invariant power spectra for the scalar
and vector field perturbations. 

Because the isotropic part of the curvature perturbation is dominant,
from Eq.~\eqref{eq:vFd-z-Power-sp-3} we may write (see \eqref{eq:mean-square-dlnk-scale-inv})\begin{equation}
\zeta\approx\sqrt{\Pz{iso}}=g^{-1/2}N_{W}\sqrt{\Pp\left(g+p\right)},\end{equation}
were we used Eqs.~\eqref{eq:vFd-z-Piso-def} and \eqref{eq:vCurv-Intro-g-subdomW}.

Using the definition of $p$ in Eq.~\eqref{eq:vFd-p-q-def} we find
that the typical amplitude of the vector field perturbation is\begin{equation}
\delta W\approx\sqrt{\Pl+2\Pp}=\sqrt{\Pp\left(p+3\right)}.\end{equation}
Combining the last two equations we obtain\begin{equation}
\zeta\approx g^{-1/2}N_{W}\delta W\sqrt{\frac{g+p}{3+p}},\end{equation}
where $\delta W$ is evaluated at the vector field decay. For the
vector curvaton scenario the parameter $N_{W}$ can be found from
Eq.~\eqref{eq:vCurv-NAi-def} as $N_{W}\approx\ow/\left(2W\right)$,
where $W$ is evaluated at the field decay. Therefore, using Eq.~\eqref{eq:vCurv-Intro-zetaW-gen}
we find that the total curvature perturbation given in Eq.~\eqref{eq:vCurv-Intro-zeta-composite}
is of order\begin{equation}
\zeta\sim g^{-1/2}\ow\zw,\label{eq:vCurv-Intro-zeta-anisotropic}\end{equation}
were we have taken $\sqrt{\left(g+p\right)/\left(3+p\right)}\sim\mathcal{O}\left(1\right)$.

After the vector field decays, the curvature perturbation $\zeta$
stays constant. This happens at the time $\Gamma_{W}^{-1}$, where
$\Gamma_{W}\sim h^{2}m$ is the field decay rate and $h$ is the vector
field coupling to its decay products. Due to gravitational decay the
lower bound for $h$ is \begin{equation}
h\gsim\frac{m}{\mpl}.\label{eq:vCurv-h-lower-limit}\end{equation}
However, during its oscillations the vector field is subject to thermal
evaporation. Were this to occur, all memory of the superhorizon perturbation
spectrum would be erased; therefore, no $\zw$ would be generated.
Considering that the scattering rate of the massive vector boson with
the thermal bath is $\Gamma_{\mathrm{sc}}\sim h^{4}T$ we can obtain
a bound such that the condensate does not evaporate before the vector
field decays, i.e. \begin{equation}
\Gamma_{\mathrm{sc}}<\Gamma_{W},\label{eq:vCurv-no-evaporation}\end{equation}
which is evaluated at $H\sim\Gamma_{W}$. The temperature of the Universe
at the vector field decay is $T\sim\sqrt{\mpl\Gamma_{W}}$, giving
$\Gamma_{\mathrm{sc}}\sim h^{5}\sqrt{\mpl m}$. Substituting this
into Eq.~\eqref{eq:vCurv-no-evaporation} and combining with Eq.~\eqref{eq:vCurv-h-lower-limit}
the range for $h$ becomes

\begin{equation}
\frac{m}{\mpl}\lsim h\lsim\left(\frac{m}{\mpl}\right)^{1/6}.\label{eq:vCurv-Intro-bound-h}\end{equation}
The lower bound in the above is due to decay through gravitational
interactions, while the upper bound is relaxed if the vector field
dominates the Universe before it decays. This happens if $\Gamma_{W}<H_{\mathrm{dom}}$,
where $H_{\mathrm{dom}}$ is the Hubble parameter when the vector
field starts to dominate. Then the energy density of the thermal bath
is exponentially smaller than $\rw$ and the vector field condensate
does not evaporate. Thus it is enough to ensure that the vector field
condensate does not evaporate before it dominates, i.e. $\Gamma_{\mathrm{sc}}<H_{\mathrm{dom}}$.
For the dominant vector curvaton this bound can be satisfied even
if the one in Eq.~\eqref{eq:vCurv-no-evaporation} is violated.

Having discussed the general predictions of the vector curvaton scenario
for the power spectrum $\Pz{}$ and non-linearity parameter $\fnl$
let us turn now to the realization of this scenario in two concrete
examples. In sections~\ref{sub:vCurvaton-RA2} and \ref{sub:vCurvaton-fF2}
we present two mechanisms for breaking the conformal invariance of
the vector field and find under which conditions the field perturbation
power spectrum is scale invariant. This allows us to calculate parameters
$p$ and $q$ as well. Then we implement these models into the vector
curvaton scenario and compute the parameter space for these models.

\section{Non-minimally Coupled Vector Curvaton\label{sub:vCurvaton-RA2}}

In Ref.~\cite{Dimopoulos2006} it was shown that a massive vector
field may acquire a scale invariant perturbation spectrum if its effective
mass during inflation is $-2H^{2}$. In this section we consider the
realization of this scenario. The negative mass squared can be achieved
by non-minimally coupling the vector field to gravity through the
Ricci scalar term. The vector field Lagrangian for this model is written
as\begin{equation}
\mathcal{L}=-\frac{1}{4}F_{\mu\nu}F^{\mu\nu}+\frac{1}{2}\left(m^{2}+\nm R\right)A_{\mu}A^{\mu},\label{eq:vCurv-RA-Lagran-gen}\end{equation}
where\begin{equation}
F_{\mu\nu}\equiv\partial_{\mu}A_{\nu}-\partial_{\nu}A_{\mu}\end{equation}
and $R$ is the Ricci scalar, with $\nm$ being a real coupling constant.
For the further discussion let us define the effective mass of the
vector field as\begin{equation}
M^{2}\equiv m^{2}+\nm R.\label{eq:vCurv-RA-M-eff}\end{equation}

Starting from Ref.~\cite{Widrow_PMFs_review(2002)} this action with
$m=0$ was invoked by many authors for the generation of primordial
magnetic fields, where $A_{\mu}$ is identified with the electromagnetic
field. Note, that due to the non-minimal coupling, this field is no
longer gauge invariant. This is in contrast to the electromagnetic
field in the Standard Model. But because in the present Universe $R$
is very small, it is thought that at present the electromagnetic field
is approximately gauge invariant. 

In our case we don't have to worry about the gauge invariance because
we don't associate $A_{\mu}$ with the electromagnetic field. Even
more so, we don't assume that the vector field $A_{\mu}$ couples
to any scalar field through the covariant derivative of the form $\mathcal{D}_{\mu}\phi\left(\mathcal{D}^{\mu}\phi\right)^{*}$.

\subsection{Equations of Motion}

During inflationary stage the spatial curvature of the Universe is
inflated away. And in accordance with the curvaton scenario, the vector
field during inflation is subdominant and does not influence the expansion
of the Universe. Therefore, we can assume to a good approximation
that inflationary expansion is homogeneous and isotropic with the
flat space-time metric in Cartesian coordinates given by\begin{equation}
\d s2=\d t2-a^{2}\left(t\right)\d{\mathbf{x}}2.\label{eq:vCurv-RA-FRW-metric}\end{equation}
In this case the Ricci scalar takes the form\begin{equation}
R=-6\left(\frac{\ddot{a}}{a}+\frac{\dot{a}^{2}}{a^{2}}\right)=-6\left(\dot{H}+2H^{2}\right)=3\left(3w-1\right)H^{2},\label{eq:vCurv-RA-R-gen}\end{equation}
where $w\approx-1$, $w=1/3$ and $w=0$ during (quasi) de Sitter
inflation, radiation and matter dominated epochs respectively. We
will further assume that inflationary expansion is of the (quasi)
de Sitter type with $H\simeq\mathrm{constant}$, making the Ricci
scalar $R\simeq-12H^{2}$. With this condition, the effective mass
of the vector field during inflation becomes $\me 2\simeq m^{2}-12\nm H^{2}\simeq\mathrm{constant}$.

Calculating equations of motion for the vector field components we
will mainly follow Ref.~\cite{Dimopoulos2006}. Using Eq.~\eqref{eq:vCurv-RA-Lagran-gen}
and the variation principle\begin{equation}
\frac{\partial\left(\sqrt{-\mathcal{D}_{g}}\mathcal{L}\right)}{\partial A_{\nu}}-\partial_{\mu}\frac{\partial\left(\sqrt{-\mathcal{D}_{g}}\mathcal{L}\right)}{\partial\left(\partial_{\mu}A_{\nu}\right)}=0\end{equation}
we find the field equation for the vector field as\begin{equation}
\left[\partial_{\mu}+\left(\partial_{\mu}\ln\sqrt{-\mathcal{D}_{g}}\right)\right]F^{\mu\nu}+\me 2A^{\nu}=0,\label{eq:vCurv-RA-Fd-eq}\end{equation}
where $\mathcal{D}_{g}\equiv\mathrm{det}\left(g_{\mu\nu}\right)$.
With the FRW metric in Eq.~\eqref{eq:vCurv-RA-FRW-metric} and the
field equation in Eq.~\eqref{eq:vCurv-RA-Fd-eq} the equation of
motion for the temporal component ($\nu=0$) is found to be \begin{equation}
\mathbf{\nabla}\cdot\dot{\mathbf{A}}-\nabla^{2}A_{t}+\left(a\me{}\right)^{2}A_{t}=0,\label{eq:vCurv-RA-EoM-0-comp}\end{equation}
where $\mathbf{\nabla}$ is the divergence and $\nabla^{2}\equiv\partial_{i}\partial_{i}$
is the Laplacian. In the same way we may find the equation of motion
for the temporal component ($\nu=i$):\begin{equation}
\ddot{\mathbf{A}}+H\dot{\mathbf{A}}-a^{-2}\left[\nabla^{2}\mathbf{A}-\mathbf{\nabla}\left(\mathbf{\nabla}\cdot\mathbf{A}\right)\right]+\me 2\mathbf{A}=\mathbf{\nabla}\left(\dot{A}_{t}+HA_{t}\right).\label{eq:vCurv-RA-Eom-i-comp}\end{equation}
A third useful relation is the integrability condition, which is obtained
by contracting Eq.~\eqref{eq:vCurv-RA-Fd-eq} with $\partial_{\nu}$:\begin{equation}
\left(a\me{}\right)^{2}\dot{A}_{t}+2\left(a\me{}\right)^{2}\frac{\dot{\me{}}}{\me{}}A_{t}-\me 2\mathbf{\nabla}\cdot\mathbf{A}+3H\left(\nabla^{2}A_{t}-\mathbf{\nabla}\cdot\dot{\mathbf{A}}\right)=0.\end{equation}
Combining the integrability condition with Eq.~\eqref{eq:vCurv-RA-EoM-0-comp}
we find\begin{equation}
\dot{A}_{t}+\left(3H+2\frac{\dot{\me{}}}{\me{}}\right)A_{t}-a^{-2}\mathbf{\nabla}\cdot\mathbf{A}=0.\label{eq:vCurv-RA-eq-At}\end{equation}
From this equation we can see that the temporal component of the vector
field is non-dynamical. Taking the gradient of Eq.~\eqref{eq:vCurv-RA-eq-At}
and plugging it into Eq.~\eqref{eq:vCurv-RA-Eom-i-comp} we arrive
at\begin{equation}
\ddot{\mathbf{A}}+H\dot{\mathbf{A}}+\me 2\mathbf{A}-a^{-2}\nabla^{2}\mathbf{A}=-2\left(H+\frac{\dot{\me{}}}{\me{}}\right)\mathbf{\nabla}A_{t}.\label{eq:vCurv-RA-eqn-for-spatial-comp}\end{equation}
Classical inhomogeneities of the vector field are diluted by inflation.
Therefore, we can neglect all gradient terms\begin{equation}
\partial_{i}A_{\mu}=0\quad\forall\,\mu.\end{equation}
Using this condition in Eqs.~\eqref{eq:vCurv-RA-EoM-0-comp} and
\eqref{eq:vCurv-RA-Eom-i-comp} we find that for the homogeneous mode
the temporal and spatial components of the vector field obey\begin{eqnarray}
A_{t} & = & 0,\label{eq:vCurv-RA-A0-0}\\
\ddot{A}+H\dot{A}+\me 2A & = & 0,\label{eq:vCurv-RA-EoM-A-homog}\end{eqnarray}
where we have used the choice of the coordinates such that the homogeneous
vector field has components $A_{\mu}=\left(A_{t},0,0,A\right)$. We
see that a temporal component of the homogeneous massive vector field
in the FRW Universe is zero. The equation of motion for the spatial
component in Eq.~\eqref{eq:vCurv-RA-EoM-A-homog} may be rewritten
in terms of the physical vector field. For this model it is $W=A/a$.
In the (quasi)de Sitter space-time ($\dot{H}\approx0$) Eq.~\eqref{eq:vCurv-RA-EoM-A-homog}
becomes\begin{equation}
\ddot{W}+3H\dot{W}+\left(\me 2+2H^{2}\right)W=0.\label{eq:vCurv-RA-EoM-Whom}\end{equation}

However, to quantize the vector field we need to perturb the field\begin{equation}
A_{\mu}\tx=A_{\mu}\left(t\right)+\delta A_{\mu}\tx\Rightarrow\begin{cases}
\mathbf{A}\tx=\mathbf{A}\left(t\right)+\delta\mathbf{A}\tx,\\
A_{t}\tx=\delta A_{t}\tx,\end{cases}\label{eq:vCurv-RA-v-perturb}\end{equation}
where we have used Eq.~\eqref{eq:vCurv-RA-A0-0}. From Eqs.~\eqref{eq:vCurv-RA-EoM-0-comp}
and \eqref{eq:vCurv-RA-eqn-for-spatial-comp} we find that the evolution
of perturbations of the vector field in (quasi)de Sitter space-time
($\dot{H}\approx0$ and $\dot{\me{}}\approx0$) follow equations\begin{eqnarray}
\mathbf{\nabla}\cdot\dot{\delta\mathbf{A}}-\nabla^{2}\delta A_{t}+\left(a\me{}\right)^{2}\delta A_{t} & = & 0,\\
\ddot{\delta\mathbf{A}}+H\,\dot{\delta\mathbf{A}}+\me 2\delta\mathbf{A}-a^{-2}\nabla^{2}\delta\mathbf{A} & = & -2H\mathbf{\nabla}\delta A_{t}.\label{eq:vCurv-RA-EoM-spatial-pert}\end{eqnarray}
Going to the Fourier space (see Eq.~\eqref{eq:vFd-q-Fourier-exp-gen})
the first equation for the temporal component becomes\begin{equation}
\delta A_{kt}+\frac{i\partial_{t}\left(\mathbf{k}\cdot\delta\mathbf{A}_{k}\right)}{k^{2}+\left(a\me{}\right)^{2}}=0,\label{eq:vCurv-RA-EoM-dA-k-temporal}\end{equation}
where $k^{2}\equiv\mathbf{k}\cdot\mathbf{k}$ and the subscript '$k$'
in $\delta A_{k\mu}$ denotes the Fourier mode of the vector field
perturbation. Using the Fourier transform of Eq.~\eqref{eq:vCurv-RA-EoM-spatial-pert}
and plugging it in Eq.~\eqref{eq:vCurv-RA-EoM-dA-k-temporal} we
find\begin{equation}
\ddot{\delta\mathbf{A}}_{k}+H\,\dot{\delta\mathbf{A}}_{k}+\me 2\delta\mathbf{A}_{k}+\left(\frac{k}{a}\right)^{2}\delta\mathbf{A}_{k}+2H\frac{\mathbf{k}\partial_{t}\left(\mathbf{k}\cdot\delta\mathbf{A}_{k}\right)}{k^{2}+\left(a\me{}\right)^{2}}=0.\label{eq:vCurv-RA-EoM-da-k-spatial}\end{equation}
The massive vector field has three degrees of freedom and all three
of them have to be quantized. Similarly to Eq.~\eqref{eq:vFd-q-polariz-decompose}
we decompose $\delta\mathbf{A}_{k}$ into three polarizations\begin{equation}
\delta A_{ki}=\sum_{\lambda}e_{i}^{\lambda}\left(\hat{\mathbf{k}}\right)\delta A_{\lambda},\label{eq:vCurv-RA-polar-decomposition}\end{equation}
and choose $e_{i}^{\lambda}$ to denote three vectors of the circular
polarization in Eq.~\eqref{eq:vFd-q-Pol-vecs-z-along-k}. Two transverse
ones are perpendicular to the wave-vector $\mathbf{k}$ giving $e_{i}^{+}k_{i}=0$,
where '$+$' stands for the left-handed `L' or right-handed `R' polarizations.
Substituting this into Eq.~\eqref{eq:vCurv-RA-EoM-da-k-spatial}
we find\begin{equation}
\left[\partial_{t}^{2}+H\partial_{t}+\me 2+\left(\frac{k}{a}\right)^{2}\right]\delta A_{+}=0.\label{eq:vCurv-RA-EoM-Atrans}\end{equation}
For the longitudinal polarization $e_{i}^{||}k_{i}=k$, and taking
into account that $e^{||}=\hat{\mathbf{k}}=\mathbf{k}/k$ from Eq.~\eqref{eq:vCurv-RA-EoM-da-k-spatial}
we find \begin{equation}
\left[\partial_{t}^{2}+\left(1+\frac{2k^{2}}{k^{2}+\left(a\me{}\right)^{2}}\right)H\partial_{t}+\me 2+\left(\frac{k}{a}\right)^{2}\right]\delta A_{||}=0.\label{eq:vCurv-RA-EoM-Along}\end{equation}

In the following sections we quantize the transverse and longitudinal
degrees of freedom separately.

\subsection{Transverse Modes}

Let us rewrite the equation of motion of the transverse polarizations
in Eq.~\eqref{eq:vCurv-RA-EoM-Atrans} in terms of the physical vector
field $W_{\mu}$ and conformal time $\tau\equiv\int\d t{}/a$:\begin{equation}
\left[\partial_{\tau}^{2}+2\frac{a'}{a}\partial_{\tau}+\left(a\me{}\right)^{2}+k^{2}+\frac{a''}{a}\right]w_{+}=0,\label{eq:vCurv-RA-EoM-wp}\end{equation}
where prime denotes the derivative with respect to the conformal time
$\tau$ and $w_{+}\left(\tau,k\right)=\delta A_{+}\left(\tau,k\right)/a\left(\tau\right)$
is defined in Eq.~\eqref{eq:vFd-q-polariz-decompose}. 

To find initial conditions for this field let us make a transformation
in Eq.~\eqref{eq:diff-eq-transformation} \begin{equation}
\varphi_{+}\equiv w_{+}\mathrm{e}^{\frac{1}{2}\int^{\tau}\frac{a'}{a}\d{\tau}{}}=aw_{+}\end{equation}
and bring the equation of motion into the form of the harmonic oscillator\begin{equation}
\left[\partial_{\tau}^{2}+\left(a\me{}\right)^{2}+k^{2}\right]\varphi_{+}=0.\label{eq:vCurv-RA-EoM-phi-trans}\end{equation}
In the subhorizon limit, for the modes with $k^{2}\gg\left|a\me{}\right|^{2}$,
Eq.~\eqref{eq:vCurv-RA-EoM-phi-trans} reduces to the flat space-time
harmonic oscillator. In other words, if we write the action for functions
$\varphi_{+}\left(k\right)$ in the limit $k/aH\rightarrow\infty$
it would correspond to the collection of harmonic oscillators. Choosing
the initial state for $\varphi_{+}$ to correspond to a vacuum (no
particles or minimum energy), from Eq.~\eqref{eq:initial-vacuum-mode-Fn}
it becomes $\varphi_{+}\left(k\right)=\exp\left(ik/aH\right)/\sqrt{2k}$,
or going back to the physical field\begin{equation}
\lim_{\frac{k}{aH}\rightarrow+\infty}w_{+}=\frac{a^{-1}}{\sqrt{2k}}\mathrm{e}^{ik/aH}.\label{eq:vCurv-RA-vacuum-wp}\end{equation}

We are interested in the power spectrum of classical perturbations
of the vector field on the superhorizon scales when $k\ll aH$, which
from Eq.~\eqref{eq:vFd-q-power-sp} can be calculated as\begin{equation}
\mathcal{P}_{\lambda}=\frac{k^{3}}{2\pi^{2}}\lim_{\frac{k}{aH}\rightarrow0}\left|w_{\lambda}\right|^{2}.\label{eq:vCurv-RA-spec-w-lambda-def}\end{equation}
To find the power spectrum we need to solve the equation of motion
in Eq.~\eqref{eq:vCurv-RA-EoM-wp} in the limit $k/aH\rightarrow0$
with the vacuum initial conditions in Eq.~\eqref{eq:vCurv-RA-vacuum-wp}.
For this purpose it is convenient to rewrite this equation in the
form\begin{equation}
\left[\partial_{x}^{2}-\frac{2}{x}\partial_{x}+1+\frac{2+\left(\me{}/H\right)^{2}}{x^{2}}\right]w_{+}=0,\end{equation}
where\begin{equation}
x\equiv\frac{k}{aH}.\label{eq:vCurv-RA-x-def}\end{equation}
This is a Bessel equation with a general solution of the form\begin{equation}
w_{+}=x^{3/2}\left[C_{1}\mathcal{H}_{\nu}^{1}\left(x\right)+C_{2}\mathcal{H}_{\nu}^{2}\left(x\right)\right],\end{equation}
where $C_{1}$, $C_{2}$ are constants of integration and $\mathcal{H}_{\nu}^{1}$,
$\mathcal{H}_{\nu}^{2}$ are the Hankel functions of the first and
second kind respectively with\begin{equation}
\nu\equiv\sqrt{\frac{1}{4}-\left(\frac{\me{}}{H}\right)^{2}}=\sqrt{\frac{1}{4}+12\nm-\left(\frac{m}{H}\right)^{2}}\label{eq:vCurv-RA-ni-def}\end{equation}
 Taking the limit $x\rightarrow\infty$ and matching to the vacuum
value in Eq.~\eqref{eq:vCurv-RA-vacuum-wp} we get\begin{equation}
w_{+}=a^{-1}\sqrt{\frac{\pi}{2k}}\,\mathrm{e}^{i\left(2\nu+1\right)\pi/4}\left(\frac{x}{2}\right)^{\frac{1}{2}}\mathcal{H}_{\nu}^{1}\left(x\right).\end{equation}
For the superhorizon perturbations ($x\rightarrow0$) this solution
becomes\begin{equation}
w_{+}=a^{-1}\frac{\Gamma\left(\nu\right)}{\sqrt{2\pi k}}\,\mathrm{e}^{i\left(2\nu-1\right)\pi/4}\left(\frac{x}{2}\right)^{\frac{1}{2}-\nu},\label{eq:vCurv-RA-wp}\end{equation}
where $\Gamma\left(\nu\right)$ is the Gamma function. Plugging this
result into Eq.~\eqref{eq:vCurv-RA-spec-w-lambda-def} we find the
power spectrum for the transverse polarization to be equal to \begin{equation}
\Pp=\frac{4}{\pi}\Gamma^{2}\left(\nu\right)\left(\frac{H}{2\pi}\right)^{2}\left(\frac{k}{2aH}\right)^{3-2\nu}.\label{eq:vCurv-RA-spec-wp-gen}\end{equation}
Note that for the light vector field, when $\me{}\rightarrow0$, the
power spectrum becomes\begin{equation}
\Pp^{\mathrm{vac}}=\left(\frac{k}{2\pi a}\right)^{2}.\end{equation}
Comparing with Eq.~\eqref{eq:vCurv-RA-vacuum-wp} we see that it
is simply the vacuum value. This is in accord with the expectation
that the massless vector field is conformally invariant and does not
undergo particle production.

The scale dependence of the power spectrum can be parametrized in
the usual way as $\Pp\propto k^{n_{\mathrm{v}}-1}$, so that $n_{\mathrm{v}}=1$
corresponds to a flat spectrum. Comparing this with Eq.~\eqref{eq:vCurv-RA-spec-wp-gen}
we find that the spectral index is $n_{\mathrm{v}}-1=3-2\nu$ and
the scale invariant spectrum of the vector field perturbation is achieved
if \begin{equation}
n_{\mathrm{v}}=1\quad\Rightarrow\quad\nu=\frac{3}{2}\quad\Rightarrow\quad\me 2=-2H^{2},\label{eq:vCurv-RA-flat-cond-gen}\end{equation}
which agrees with the findings of Ref.~\cite{Dimopoulos2006}. With
this condition the power spectrum becomes\begin{equation}
\Pp=\left(\frac{H}{2\pi}\right)^{2},\label{eq:vCurv-RA-Pp}\end{equation}
the same as for the massless scalar field. 

The condition in Eq.~\eqref{eq:vCurv-RA-flat-cond-gen} is satisfied
if the coupling constant of the vector field to gravity is \begin{equation}
\nm\approx\frac{1}{6}\left[1+\frac{1}{2}\left(\frac{m}{H}\right)^{2}\right],\label{eq:vCurv-RA-e-for-scale-inv}\end{equation}
from which it is clear that for scale invariance we need $\nm\gtrsim1/6$.
If $m\gtrsim H$ then scale invariance is attained only when $\nm$
is tuned according to Eq.~\eqref{eq:vCurv-RA-e-for-scale-inv}. However,
if $m\ll H$ then scale-invariance simply requires $\nm\approx1/6$.
In the latter case $m$ and $H$ do not have to balance each other
through the condition in Eq.~\eqref{eq:vCurv-RA-e-for-scale-inv}
and can be treated as free parameters. We feel that this is a more
natural setup.

With $\nm=1/6$ the $\nu$ parameter in Eq.~\eqref{eq:vCurv-RA-ni-def}
becomes\begin{equation}
\nu=\sqrt{\frac{9}{4}-\left(\frac{m}{H}\right)^{2}},\label{eq:vCurv-RA-ni}\end{equation}
which is reminiscent of the scalar field case, where perturbations
become classical in the superhorizon limit if $\nu$ is real, corresponding
to $m^{2}<9H^{2}/4$ (see the discussion below Eq.~\eqref{eq:q-to-c-sFd-mode-fn}).

\subsection{The Longitudinal Mode}

Let us first rewrite the equation for longitudinal perturbations of
the vector field in Eq.~\eqref{eq:vCurv-RA-EoM-Along} in terms of
the conformal time $\tau$\begin{equation}
\left[\partial_{\tau}^{2}+\frac{2k^{2}aH}{k^{2}+a^{2}\me 2}\partial_{\tau}+\left(k^{2}+a^{2}\me 2\right)\right]\delta A_{||}=0.\label{eq:vCurv-RA-EoM-Along-tau}\end{equation}
In the previous discussion on the perturbations of the transverse
components, we found that the scale invariant spectrum is achieved
if the effective mass squared of the field is negative and equal to
$\me 2=-2H^{2}$. But in this case the second term in the above equation
becomes singular when $\left(k/a\right)^{2}=2H^{2}$. This might indicate
that the longitudinal vector field perturbation becomes unstable when
approaching the horizon exit. But the two independent solutions of
this equation\begin{equation}
\delta A_{||}^{\pm}\propto\left(-k\tau+\frac{2}{k\tau}\pm2i\right)\mathrm{e}^{\mp ik\tau}\label{eq:vCurv-RA-Along-prelim-solution}\end{equation}
show that this is not the case.%
\footnote{To find this solution we have used the relation $\tau=-\left(aH\right)^{-1}$,
which is valid in de Sitter space-time.%
} 

$\me 2=-2H^{2}=\mathrm{constant}$ corresponds to the flat perturbation
power spectrum of transverse modes and, as will be seen later, of
the longitudinal mode too. However, the exactly flat spectrum is excluded
by observations (Eq.~\eqref{eq:spec-indx-WMAP5}). Therefore, one
would expect that in the realistic theory the condition $\me 2=-2H^{2}$
is violated by a small amount to give the correct spectral tilt and
the Hubble parameter is not exactly constant during inflation. In
this case the solution in Eq.~\eqref{eq:vCurv-RA-Along-prelim-solution}
is not valid and one may be worried that for general effective negative
mass $\me 2<0$ and $\dot{H}\ne0$ the solution of Eq.~\eqref{eq:vCurv-RA-EoM-Along-tau}
is still singular at $\left(k/a\right)^{2}\rightarrow\left|\me 2\right|$.
We can prove that this is not the case using the Frobenius method
for differential equations with regular singular points (see for example
Ref.~\cite{Rabenstein(1966)}). 

Using Eqs.~\eqref{eq:vCurv-RA-eq-At} and \eqref{eq:vCurv-RA-eqn-for-spatial-comp}
with $\dot{M}\ne0$ the equation of motion for the longitudinal mode
$\delta A_{||}$ becomes\begin{equation}
\left[\partial_{\tau}^{2}-\left(3w+1\right)\frac{k^{2}aH}{k^{2}+a^{2}\me 2}\partial_{\tau}+\left(k^{2}+a^{2}\me 2\right)\right]\delta A_{||}=0,\label{eq:vCurv-RA-EoM-Along-genH}\end{equation}
where $w$ is the barotropic parameter of the dominant component of
the Universe which drives inflation. For de Sitter expansion $w=-1$
and we recover Eq.~\eqref{eq:vCurv-RA-EoM-Along-tau}. However, for
this calculation we do not assume de Sitter inflation and consider
a constant $w$ in the range $-1<w<-\frac{1}{3}$, which is necessary
for the accelerated expansion of the Universe (see the discussion
in section~\ref{sub:sFd-inflation}). This equation is valid for
a general non-minimal coupling constant $\alpha$ defined in Eq.~\eqref{eq:vCurv-RA-M-eff}
and we used that $\alpha R\gg m^{2}$ during inflation.

Let us first we make a change of variables\begin{equation}
y\equiv\left(\frac{k}{a\left|\me{}\right|}\right)^{2}-1,\end{equation}
with $y$ varying in the region $-1<y<\infty$. Eq.~\eqref{eq:vCurv-RA-EoM-Along-tau}
with this transformation translates into the form\begin{equation}
\left[\partial_{y}^{2}-\frac{1}{2}\frac{\left(y+2\right)}{y\left(y+1\right)}\partial_{y}+\frac{\left|\me 2\right|}{H^{2}}\frac{y}{\left(3w+1\right)^{2}\left(y+1\right)^{2}}\right]\delta A_{||}=0,\label{eq:vCurv-RA-EoM-Along-y}\end{equation}
with $\me 2<0$ and the regular singular point at $y\rightarrow0$,
corresponding to $\left(k/a\right)^{2}\rightarrow\left|\me 2\right|$.
The general solution of this equation can be found using the ansatz\begin{equation}
\delta A_{||}=\sum_{n=0}^{\infty}D_{n}y^{s+n},\label{eq:vCurv-RA-Along-series-anstaz}\end{equation}
where $D_{0}\ne0$. In this case the series in Eq.~\eqref{eq:vCurv-RA-Along-series-anstaz}
is convergent at least in the region $-1<y<1$ (corresponding to $\left|\me 2\right|<\left(k/a\right)^{2}<2\left|\me 2\right|$)
without a singular point at $y=0$. Our aim is to prove that the solution
in Eq.~\eqref{eq:vCurv-RA-Along-series-anstaz} is not singular even
at $\left(k/a\right)^{2}=\left|\me 2\right|$, i.e. $y=0$. This will
be the case if the power series ansatz in Eq.~\eqref{eq:vCurv-RA-Along-series-anstaz}
has two independent solutions and if $s+n>0$ for all $n$, i.e. there
are no negative powers of $y$ in the series. To show this let us
substitute Eq.~\eqref{eq:vCurv-RA-Along-series-anstaz} into Eq.~\eqref{eq:vCurv-RA-EoM-Along-y}
giving\begin{eqnarray}
\sum_{n=0}^{\infty}D_{n}\left[4\left(s+n\right)\left(s+n-2\right)y^{s+n-2}+8\left(s+n\right)\left(s+n-\frac{7}{4}\right)y^{s+n-1}+\right.\nonumber \\
\left.+4\left(s+n\right)\left(s+n-\frac{3}{2}\right)y^{s+n}+12\alpha\frac{\left|3w-1\right|}{\left(3w+1\right)^{2}}y^{s+n+1}\right] & = & 0,\qquad\quad\label{eq:vCurv-RA-y-series}\end{eqnarray}
where we also used Eq.~\eqref{eq:vCurv-RA-R-gen}. In order for this
equality to be valid, coefficients in front of each $y$ with the
same power must vanish. The coefficient in front of the term with
the smallest power, i.e. $y^{s-2}$, is $4D_{0}s\left(s-2\right)$.
Because $D_{0}\ne0$, from the indicial equation $s\left(s-2\right)=0$
we find\begin{equation}
s=0\quad\mathrm{or}\quad s=2.\end{equation}
Because these two solutions differ by an integer, it might be alarming
that both series in Eq.~\eqref{eq:vCurv-RA-Along-series-anstaz}
with $s=0$ and $s=2$ do not provide two independent solutions. In
this case the second independent solution would involve the term $\ln y$,
which indeed diverges at $y\rightarrow0$. However, by closer inspection
of Eq.~\eqref{eq:vCurv-RA-y-series} we find that the coefficient
$D_{2}$ of the series with $s=0$ is arbitrary, thus the power series
in Eq.~\eqref{eq:vCurv-RA-Along-series-anstaz} with $s=0$ and $s=2$
do give two independent solutions. In addition they do not involve
negative powers of $y$, i.e. $s\geq0$, therefore, the solution with
the ansatz in Eq.~\eqref{eq:vCurv-RA-Along-series-anstaz} converges
at the singular point $y=0$. This proves that during inflation, when
$M^{2}<0$, the solution of Eq.~\eqref{eq:vCurv-RA-EoM-Along-genH}
is stable when the wavelength of the perturbation approaches $\left(k/a\right)^{2}\rightarrow\left|\me 2\right|$.

Let us turn now to the quantization of the longitudinal mode in (quasi)
de Sitter space-time. From Eq.~\eqref{eq:vCurv-RA-EoM-Along} the
equation of motion for the longitudinal physical field in the conformal
time is\begin{equation}
\left[\partial_{\tau}^{2}+2\frac{a'}{a}\left(1+\frac{k^{2}}{k^{2}+\left(a\me{}\right)^{2}}\right)\partial_{\tau}+k^{2}+\left(a\me{}\right)^{2}+2\left(\frac{a'}{a}\right)^{2}\frac{k^{2}}{k^{2}+\left(a\me{}\right)^{2}}+\frac{a''}{a}\right]w_{||}=0.\end{equation}

To quantize the longitudinal mode, let us write the Lagrangian corresponding
to the equation of motion in Eq.~\eqref{eq:vCurv-RA-EoM-Along-tau}\begin{equation}
\mathcal{L}=M^{2}\left[\frac{\left|\delta A_{\pl}'\left(\tau,\mathbf{k}\right)\right|^{2}}{\left(k/a\right)^{2}+M^{2}}-a^{2}\left|\delta A_{\pl}\left(\tau,\mathbf{k}\right)\right|^{2}\right].\label{eq:vCurv-RA-dA-long-Lagrangian}\end{equation}
This Lagrangian can also be achieved by perturbing the full Lagrangian
in Eq.~\eqref{eq:vCurv-RA-Lagran-gen} and it is unique up to the
total derivative. To set the initial conditions for the subhorizon
modes we use the transformation\begin{equation}
\varphi_{\pl}\equiv\gamma^{-1}\delta A_{\pl},\end{equation}
where $\gamma$ is the Lorentz boost factor\begin{equation}
\gamma=\frac{E}{\left|M\right|}=\frac{\sqrt{\left(k/a\right)^{2}+\left|M^{2}\right|}}{\left|M\right|}\approx\frac{k/a}{\left|M\right|},\end{equation}
and the last equality is taken in the limit $k/a\gg\left|M\right|$.
With this transformation the Lagrangian for subhorizon modes reduces
to that of the simple harmonic oscillator\begin{equation}
\mathcal{L}=\pm\left(\left|\varphi_{\pl}'\right|^{2}-k^{2}\left|\varphi_{\pl}\right|^{2}\right),\label{eq:vCurv-RA-phi-long-Lagrangian}\end{equation}
where the sign $\pm$ is that of $M^{2}$, hence negative for the
case of interest $M^{2}\simeq-2H^{2}$.

The Lagrangian in Eq.~\eqref{eq:vCurv-RA-phi-long-Lagrangian} is
the same as of the harmonic oscillator. Choosing initial conditions
to correspond to the vacuum state, we have \begin{equation}
\varphi_{\pl}=\frac{1}{\sqrt{2k}}\mathrm{e}^{-ik\tau}.\label{eq:vCurv-RA-phi-ini-cond}\end{equation}
This is similar as for the scalar field case, except that for $M^{2}<0$
the Lagrangian has a negative sign. Because of the wrong sign, initial
conditions in Eq.~\eqref{eq:vCurv-RA-phi-ini-cond} are not identical
to the scalar field case, since for the longitudinal mode occupied
initial states would have negative energy density and pressure. As
the pressure is negative it is not dangerous for inflation. Instead,
it is the negative energy density that is dangerous. As the total
energy density is required to be positive, the negative contribution
of occupied states has to be less than the total at the beginning
of inflation. This is satisfied by assuming that initially the occupation
number is much less than 1 (as in the scalar field case), justifying
both the choice of initial mode function and the assumption of the
vacuum state.

Matching the solution in Eq.~\eqref{eq:vCurv-RA-Along-prelim-solution}
to the vacuum initial conditions from Eq.~\eqref{eq:vCurv-RA-phi-ini-cond},
$\delta A_{\pl,\,\mathrm{vac}}=aw_{\pl}=\frac{\gamma}{\sqrt{2k}}\exp\left(-ik\tau\right)$,
and using Eq.~\eqref{eq:vCurv-RA-spec-w-lambda-def} we find that
the power spectrum for the longitudinal mode with $\nm=1/6$ is \begin{equation}
\Pl=2\left(\frac{H}{2\pi}\right)^{2}=2\Pp.\label{eq:vCurv-RA-Plong}\end{equation}
This corresponds to $p=1$ in Eq.~\eqref{eq:vFd-p-q-def}, meaning
that the particle production of the vector field is anisotropic.

\subsection{The Stability of the Longitudinal Mode}

As shown above, the possible instability of the longitudinal mode
when $\left.\left(k/a\right)^{2}\rightarrow\left|\me 2\right|\right.$
is absent. However, in Ref.~\cite{Himmetoglu_etal(2009d)} it has
been noted that other instabilities might be present for non-minimally
coupled vector field. The first concern is that for $m^{2}\ll\left|R\right|$
the kinetic term of the longitudinal mode is negative on the subhorizon
scales (see Eqs.~\eqref{eq:vCurv-RA-M-eff} and \eqref{eq:vCurv-RA-dA-long-Lagrangian}).
As a result one might suspect that corresponding particles carry a
negative energy and they can be created from the vacuum making it
unstable. This is indeed the case for minimally coupled scalar field
with negative kinetic term. The latter field is called a ghost and
is cosmologically unacceptable as it would create too many photons
from the present day vacuum (Ref.~\cite{Cline_etal(2003)ghosts}).
However, the flat space-time calculation of Ref.~\cite{Cline_etal(2003)ghosts}
can not be directly applied to a vector field with non-minimal coupling
to gravity and currently no such calculation exits. Moreover, the
bound on the photon creation from the vacuum at the present Universe
is irrelevant for the vector curvaton scenario as its bare mass squared
$m^{2}$ in Eq.~\eqref{eq:vCurv-RA-M-eff} dominates, making $M^{2}$
positive. But even for models with negligible $m^{2}$ one wouldn't
expect a large particle creation, as in the present day Universe $\left|R\right|\sim10^{-66}\,\mathrm{eV^{2}},$
which is extremely small compared to other energy scales.

Another concern is about the singularity when $M^{2}\rightarrow0$.
After inflation $R\propto t^{-2}$, and when both terms in Eq.~\eqref{eq:vCurv-RA-M-eff}
cancel each other out $M^{2}$ vanishes. As shown in Ref.~\cite{Himmetoglu_etal(2009d)}
this results in a singularity which invalidates linear calculations
around this point. To evaluate the effects of this, one needs to perform
a full non-linear calculation which has not been done to the present
moment. However, as the period of non-linear evolution is very brief,
one might expect that linear calculations before and after $M^{2}=0$
will match. Or one can assume that $m^{2}=0$ and consider more complicated
models to generate mass term for the vector field, in which case the
aforementioned singularity can be avoided.%
\footnote{A more thorough discussion of these issues can be found in Ref.~\cite{Karch_Lyth(2010)long_ghost}.%
}

\subsection{Statistical Anisotropy and Non-Gaussianity\label{sub:vCurv-RA-nonGauss}}

Let us calculate the statistical anisotropy and non-Gaussianity for
this model. Because $p=1$, the dominant contribution to the curvature
perturbation is assumed to be generated by the scalar field. The parity
conserving transverse power spectrum of the vector field perturbation
in Eq.~\eqref{eq:vCurv-RA-Pp} and the power spectrum generated during
the single scalar field inflation are equal, i.e. $\mathcal{P}_{+}=\mathcal{P}_{\phi}$.
Thus the isotropic part of the curvature perturbation spectrum can
be written as \begin{equation}
\mathcal{P}_{\zeta}^{\mathrm{iso}}=\mathcal{P}_{\phi}N_{\phi}^{2}\left(1+\xi\right).\end{equation}
While the anisotropy parameter from Eq.~\eqref{eq:vFd-z-g-def} becomes
\begin{equation}
g=\frac{\xi}{1+\xi}.\label{eq:vCurv-RA-g}\end{equation}
This model does not have parity violating terms, and from Eqs.~\eqref{eq:vFd-p-q-def}
and \eqref{eq:vCurv-RA-Plong} we find \begin{equation}
p=1\quad\mathrm{and}\quad q=0.\label{eq:g_h_RA^2}\end{equation}
 Thus, the anisotropy in the vector field is rather strong, which
means that it will have to remain subdominant, i.e. $\Omega_{W}\ll1$.
Using this and Eq.~\eqref{eq:vCurv-fnl-eql-M^2-scaleInv}, the $f_{\mathrm{NL}}^{\mathrm{equil}}$
for the non-minimally coupled vector curvaton is found to be \begin{eqnarray}
\frac{6}{5}f_{\mathrm{NL}}^{\mathrm{equil}}=2\frac{\xi^{2}}{\Omega_{W}}\left(1+\frac{9}{8}W_{\bot}^{2}\right),\label{eq:fNL_vCurvaton}\end{eqnarray}
Similarly, $f_{\mathrm{NL}}^{\mathrm{local}}$ for the squeezed configuration
in Eq.~\eqref{eq:vCurv-fNL-scaleInv-local} is\begin{equation}
\frac{6}{5}f_{\mathrm{NL}}^{\mathrm{local}}=2\frac{\xi^{2}}{\Omega_{W}}\left(1+W_{\bot}^{2}\right)\label{eq:fNL_vCurvaton_local}\end{equation}

Since $\Pp=\frac{1}{2}\Pl=\mathcal{P}_{\phi}=\left(H/2\pi\right)^{2}$,
for the typical values of the perturbation we have $\delta\phi\sim\delta W_{i}\sim H$.
This means that, in order for the vector field contribution to be
subdominant, we require $N_{W}\ll N_{\phi}$ (c.f. Eq.~\eqref{eq:vFd-dN-gen}),
which from Eq.~\eqref{eq:vFd-ksi-def} gives $\xi\ll1$. Using these
results and $p=1$ (see Eq.~\eqref{eq:g_h_RA^2}) from Eq.~\eqref{eq:vFd-z-g-def}
we find $g\simeq\xi$. Thus, in view of Eqs.~\eqref{eq:fNL_vCurvaton}
and \eqref{eq:fNL_vCurvaton_local}, we see that $\fnl\sim g^{2}/\Omega_{W}$.
Therefore, we find that the non-Gaussianity is determined by the magnitude
of the anisotropy in the power spectrum.

This prediction is valid in the regime $|\delta W/W|\ll1$ which corresponds
to $\Omega_{W}^{2}\gtrsim\Pz{}\xi$, which implies $\fnl\lesssim g^{3/2}/\sqrt{\Pz{}}$.
For smaller $\Omega_{W}$, the contribution of the vector field perturbation
to $\zeta$ is of order $\Omega_{W}[\delta W/(\overline{\delta W^{2}})^{1/2}]$.
In other words, it is of order $\Omega_{W}$ and is the square of
a Gaussian quantity. The resulting prediction for its contribution
to $\fnl$ would be given by the one-loop formula which is calculated
in Ref.~\cite{Toledo_etal(2009)}.

\subsection{The Energy-Momentum Tensor}

Let us now study the evolution of the vector field. For the scale
invariant perturbation spectrum with $\nm=1/6$ from Eq.~\eqref{eq:vCurv-RA-EoM-A-homog}
we find the equation of motion for the homogeneous mode of the vector
field\begin{equation}
\ddot{W}+3H\dot{W}+m^{2}W=0,\label{eq:vCurv-RA-W-EoM-R1/6}\end{equation}
which is identical to the one of a massive scalar field. When $m\ll H$
it has the solution\begin{equation}
W=W_{0}+Ca^{\frac{3}{2}\left(w-1\right)},\label{eq:vCurv-RA-W-homog-eqn}\end{equation}
where $W_{0}$ and $C$ are constants of integration. Because the
second term in Eq.~\eqref{eq:vCurv-RA-W-homog-eqn} is decaying,
as long as $m\ll H$ the physical vector field develops a condensate
which remains constant $W\simeq W_{0}$.

We can follow the evolution of the vector field condensate by considering
the energy momentum tensor, which can be written in the form\begin{equation}
T_{\mu}^{\nu}=\mathrm{diag}\left(\rho_{W},-p_{\perp},-p_{\perp},-p_{||}\right),\end{equation}
where \cite{Golovnev_etal(2008)}\begin{equation}
\rho_{W}=\frac{1}{2}\dot{W}^{2}+\frac{1}{2}m^{2}W^{2}\label{eq:vCurv-RA-rho}\end{equation}
and the transverse and longitudinal pressures are\begin{eqnarray}
p_{\perp} & = & \frac{5}{6}\left(\dot{W}^{2}-m^{2}W^{2}\right)+\frac{1}{3}\left(2H\dot{W}+\dot{H}W+3H^{2}W\right)W,\label{eq:vCurv-RA-p}\\
p_{||} & = & -\frac{1}{6}\left(\dot{W}^{2}-m^{2}W^{2}\right)-\frac{2}{3}\left(2H\dot{W}+\dot{H}W+3H^{2}W\right)W.\nonumber \end{eqnarray}
Thus, the energy-momentum tensor for the homogeneous vector field
is, in general, anisotropic because $p_{||}\ne p_{\perp}$. This is
why the vector field cannot be taken to drive inflation, for if it
did it would generate a substantial large-scale anisotropy, which
would be in conflict with the predominant isotropy in the CMB. Therefore,
we have to investigate whether, \emph{after} inflation, there is a
period in which the vector field becomes isotropic (i.e. $p_{\perp}\approx p_{||}$)
and can imprint its perturbation spectrum onto the Universe without
such problems.

Considering the growing mode in Eq.~\eqref{eq:vCurv-RA-W-homog-eqn}
and Eqs.~\eqref{eq:vCurv-RA-rho}, \eqref{eq:vCurv-RA-p} we see
that, during and after inflation, when $m\ll H$, we have\begin{equation}
\rho_{W}\simeq\frac{1}{2}m^{2}W_{0}^{2}\quad\mathrm{and}\quad p_{\perp}\simeq-\frac{1}{2}p_{||}\simeq\frac{1}{2}\left(1-w\right)H^{2}W_{0}^{2}.\end{equation}
Hence, the density of the vector field remains roughly constant, while
the vector field condensate remains anisotropic after inflation.

The above are valid under the condition $m\ll H$. However, after
the end of inflation $H\left(t\right)\propto t^{-1}$, so there will
be a moment when $m\sim H$. After this moment, due to Eq.~\eqref{eq:vCurv-RA-R-gen},
the curvature coupling becomes negligible and the vector field behaves
as a massive minimally-coupled Abelian vector field. As shown in Eq.~\eqref{eq:vCurv-Intro-A-oscillating},
when $m\gtrsim H$ a massive vector field undergoes (quasi)harmonic
oscillations of frequency $\sim m$, because the friction term in
Eq.~\eqref{eq:vCurv-RA-W-EoM-R1/6} becomes negligible. In this case,
on average over many oscillations $\overline{\dot{W}^{2}}\approx m^{2}\overline{W^{2}}$.
Hence, Eqs.~\eqref{eq:vCurv-RA-rho} and Eq.~\eqref{eq:vCurv-RA-p}
become\begin{eqnarray}
\rho_{W} & \simeq & m^{2}\overline{W^{2}},\\
p_{\perp} & \simeq & -\frac{1}{2}p_{\Vert}\simeq\frac{2}{3}mH\left[1+\frac{3}{4}\left(1-w\right)\left(\frac{H}{m}\right)\right]\overline{W^{2}}.\nonumber \end{eqnarray}
The effective barotropic parameters of the vector field are\begin{equation}
0<w_{\perp}\simeq-\frac{1}{2}w_{\Vert}=\frac{2}{3}\left[1+\frac{3}{4}\left(1-w\right)\left(\frac{H}{m}\right)\right]\left(\frac{H}{m}\right)\ll1,\end{equation}
where $w_{\perp}=p_{\perp}/\rho_{W}$ and $w_{\Vert}=p_{\Vert}/\rho_{W}$.
By virtue of the condition $m\gg H$, we see that, after the onset
of the oscillations, $w_{\perp},w_{\Vert}\rightarrow0$. This means
that the oscillating massive vector field behaves as a pressureless
isotropic matter, which can dominate the Universe without generating
an excessive large-scale anisotropy. Moreover, as was shown in Eq.~\eqref{eq:vCurv-Intro-isotr-pressureless-cond}
the energy density decreases as $\rho_{W}\propto a^{-3}$, i.e. like
dust. Thus, if the Universe is radiation dominated, $\rho_{W}/\rho\propto a$
while oscillations occur, so the field has a chance to dominate the
Universe and imprint its curvature perturbation according to the curvaton
scenario.

\subsection{Curvaton Physics\label{sub:vCurv-RA-Curvaton-Physics}}

As we have seen in section~\ref{sub:vCurv-RA-nonGauss} for the non-minimally
coupled vector field with the Lagrangian in Eq.~\eqref{eq:vCurv-RA-Lagran-gen}
the particle production is anisotropic, and the curvature perturbation
generated by such a field is statistically anisotropic. Therefore,
the non-minimally coupled vector curvaton may generate only the subdominant
contribution to the total curvature perturbation $\zeta$, while the
dominant part must be produced by a statistically isotropic source.
In such scenario the total curvature perturbation with statistically
anisotropic contribution $\zw$ was calculated in Eq.~\eqref{eq:vCurv-Intro-zeta-anisotropic}.

If $m\ll H$ during inflation the physical vector field (being non-conformally
invariant) undergoes particle production and obtains an approximately
flat superhorizon spectrum of perturbations, as shown. Indeed, if
in Eq.~\eqref{eq:vCurv-RA-ni} $\nu\approx\frac{3}{2}$ from Eq.~\eqref{eq:vCurv-RA-Plong}
we find that the typical value of the vector field perturbation is
(see Eq.~\eqref{eq:mean-square-dlnk-scale-inv}) \begin{equation}
\delta W_{\mathrm{end}}\approx\sqrt{\Pl+2\mathcal{P}_{+}}\approx\frac{H_{*}}{\pi},\end{equation}
where `end' denotes the typical value of the vector field perturbation
at the end of inflation and $H_{*}$ is the Hubble parameter during
inflation. The curvature perturbation generated by the vector curvaton
field was calculated in Eq.~\eqref{eq:vCurv-Intro-zetaW}. Using
this equation and considering that $W\approx\mathrm{constant}$ during
inflation (see Eq.~\eqref{eq:vCurv-RA-W-homog-eqn}), i.e. $W_{\mathrm{end}}=W_{0}$,
we may write\begin{equation}
\zw\sim\frac{H_{*}}{W_{0}}.\end{equation}
Thus, from this result and Eq.~\eqref{eq:vCurv-Intro-zeta-anisotropic}
we obtain \begin{equation}
\zeta\sim g^{-1/2}\ow\frac{H_{*}}{W_{0}}.\label{eq:vCurv-RA-zt-eqn}\end{equation}

At the onset of vector field oscillations the density parameter of
the vector field is\begin{equation}
\Omega_{W}\equiv\frac{\rho_{W}}{\rho}\sim\left(\frac{W_{0}}{\mpl}\right)^{2},\end{equation}
where we have used the flat Friedman equation \eqref{eq:FRW-Friedmann-eq-flat}
$\rho=3\mpl^{2}H^{2}$. To avoid excessive large scale anisotropy
the density of the vector field must be subdominant before the onset
of oscillations, which means that $W_{0}<\mpl$.

Let us assume that inflation is driven by some inflaton field, which
after inflation ends, oscillates around its VEV until its decay into
a thermal bath of relativistic particles at reheating. In this scenario
the Universe is matter dominated (by inflaton particles) until reheating.
Using the above findings we can estimate the density ratio of the
vector field at decay \begin{equation}
\od\sim\left(\frac{\mathrm{min}\left\{ m,\Gamma\right\} }{\Gamma_{W}}\right)^{1/2}\left(\frac{W_{0}}{\mpl}\right)^{2},\label{eq:vCurv-RA-Odec}\end{equation}
where $\Gamma_{W}$ is the vector field decay rate and $\Gamma$ is
the decay rate of the inflaton field. If inflation gives away directly
to a thermal bath of particles then we have prompt reheating and $\Gamma\rightarrow H_{*}$,
where $H_{*}$ is the Hubble scale of inflation. 

Using Eqs.~\eqref{eq:vCurv-RA-zt-eqn} and \eqref{eq:vCurv-RA-Odec}
and considering that at the vector field decay $\od=\ow$ we get\begin{equation}
\frac{H_{*}}{\mpl}\sim\zeta\left(\frac{g}{\od}\right)^{1/2}\left(\frac{\Gamma_{W}}{\mathrm{min}\left\{ m,\Gamma\right\} }\right)^{1/4},\label{eq:vCurv-RA-H-eqn}\end{equation}
The Hot Big Bang has to begin before nucleosynthesis (which occurs
at the temperature $T_{\mathrm{BBN}}\sim1\,\mathrm{MeV}$). Hence,
$\Gamma_{W}\gsim T_{\mathrm{BBN}}^{2}/\mpl$. Using this and also
$\mathrm{min}\left\{ m,\Gamma\right\} \lsim H_{*}$, we obtain the
bound\begin{equation}
H_{*}\gsim\zeta^{4/5}\left(\frac{g}{\od}\right)^{2/5}\left(T_{\mathrm{BBN}}^{2}\mpl^{3}\right)^{1/5}\Longleftrightarrow V_{*}^{1/4}\gsim g^{1/5}\,10^{12}\,\mathrm{GeV},\label{eq:vCurv-RA-bound-1}\end{equation}
where we used that $\od\lsim1$ and $\zeta=4.8\times10^{-5}$ from
COBE observations. For $g\lesssim0.3$ this is similar to the case
of a scalar field curvaton \cite{Lyth(2004)curvaton}.%
\footnote{The cosmological scales re-enter the horizon at temperatures $T\lsim1\,\mathrm{keV}$,
i.e. much later than nucleosynthesis and well after our vector field
condensate decays restoring local Lorentz invariance.%
}

Another bound on the inflation scale is obtained by considering that
$\Gamma_{W}\sim h^{2}m$, where $h$ is the vector field coupling
to its decay products, for which $h\gsim m/\mpl$ due to gravitational
decay. Thus, $\Gamma_{W}\gsim m^{3}/\mpl^{2}$. Combining with Eq.~\eqref{eq:vCurv-RA-H-eqn}
we obtain the bound\begin{equation}
H_{*}\gsim\zeta\left(\frac{g}{\od}\right)^{1/2}\left(\mpl m\right)^{1/2}\Longleftrightarrow V_{*}^{1/4}\gsim g^{1/4}\,10^{11}\,\mathrm{GeV},\label{eq:vCurv-RA-bound-2}\end{equation}
where we took $1\,\mathrm{TeV}\lsim m<\Gamma$.

Finally, an upper bound on inflation scale can be obtained by combining
Eq.~\eqref{eq:vCurv-RA-zt-eqn} with the requirement $W_{0}<\mpl$,
thereby finding\begin{equation}
H_{*}<g^{1/2}\zeta\od^{-1}\mpl\Longleftrightarrow V_{*}^{1/4}<g^{1/4}\,10^{16}\,\mathrm{GeV},\label{eq:vCurv-RA-bound-3}\end{equation}
where we considered that $\od\gsim10^{-3}$, in order to avoid excessive
non-Gaussianity in the CMB. This bound on $\Omega_{\mathrm{dec}}$
may be found considering that $\fnl\sim g^{2}/\Omega_{\mathrm{dec}}$
(see Eq.~\eqref{eq:vCurv-fNLe-fNLl}) and the observational constraints
on $\left|\fnl\right|\lesssim100$ (Eq.~\eqref{eq:fNLl-observational}).

As was discussed in section~\ref{sub:vCurv-Intro-z-generation} we
also need to consider the hazardous possibility of the thermal evaporation
of the vector field condensate. If it evaporates all memory of the
superhorizon perturbation spectrum is erased and no $\zw$ is generated.
This puts a bound on the allowed values of the vector field coupling
constant $h$ to its decay products which was calculated in Eq.~\eqref{eq:vCurv-Intro-bound-h}. 

The above lower bounds on $H_{*}$ can be substantially relaxed by
employing the so-called mass increment mechanism according to which,
the vector field obtains its bare mass at a phase transition (denoted
by \textquoteleft{}pt\textquoteright{}) with $m/H_{\mathrm{pt}}\gg1$.
The mechanism was firstly introduced for the scalar curvaton in Ref.~\cite{Dimopoulos_etal(2005)}
and has been already implemented in the vector curvaton case in Ref.~\cite{Dimopoulos2007}.

Let us consider now the case when $\nm\ne\frac{1}{6}$. If  $\nm=\mathcal{O}\left(1\right)$
then, according to Eq.~\eqref{eq:vCurv-RA-e-for-scale-inv}, a scale
invariant spectrum is possible only if $m\sim H_{*}$. Hence, the
oscillations begin immediately after the end of inflation. With this
in mind the previous analysis remains valid. In particular, the bound
in Eq.~\eqref{eq:vCurv-RA-bound-1} remains the same. However, the
bound in Eq.~\eqref{eq:vCurv-RA-bound-2} becomes much more stringent:\begin{equation}
H_{*}\gsim g\:10^{10}\:\mathrm{GeV}\Longleftrightarrow V_{*}^{1/4}\gsim g^{1/2}\:10^{14}\:\mathrm{GeV}.\end{equation}

\subsection{A Concrete Example}

To illustrate our findings let us consider a specific example. Let
us choose $\nm\approx\frac{1}{6}$ , $m\sim10\,\mathrm{TeV}$ and
also $\Gamma_{W}\sim10^{-10}\,\mathrm{GeV}$ such that the temperature
at the vector field decay is $T_{\mathrm{dec}}\sim10\,\mathrm{TeV}$.
Such a particle may be potentially observable in the LHC. These values
suggest $h\sim10^{-7}$, which lies comfortably within the range in
Eq.~\eqref{eq:vCurv-Intro-bound-h}. For the decay rate of the inflaton
let us chose $\Gamma\sim10^{-2}\,\mathrm{GeV}$ so that the reheating
temperature satisfies the gravitino overproduction constraint $T_{\mathrm{reh}}\sim\sqrt{\mpl\Gamma}\sim10^{8}\,\mathrm{GeV}$.
Then Eq.~\eqref{eq:vCurv-RA-H-eqn} reduces to $H_{*}/\mpl\sim10^{-4}\zeta\left(g/\od\right)^{1/2}$.
Using this and Eq.~\eqref{eq:vCurv-RA-zt-eqn} we get $W_{0}/\mpl\sim10^{-4}\sqrt{\od}$.
Hence, with the maximum observationally allowed statistical anisotropy
$g\sim0.1$ the lowest value for the inflationary Hubble scale is
$H_{*}>10^{9}\,\mathrm{GeV}$.

\subsection{\boldmath Summary of the $RA^{2}$ Model}

In Ref.~\cite{Dimopoulos2006} it was demonstrated for the first
time that the vector field may influence or generate the curvature
perturbation in the Universe. It was shown that a massive vector field
may act as a curvaton field without producing excessive large scale
anisotropy. In this reference it was also calculated that the perturbation
spectrum of a massive Abelian vector field is scale invariant if the
mass of the field is equal to $M^{2}=-2H^{2}$. Section~\ref{sub:vCurvaton-RA2}
of this thesis explored the possibility of realizing the negative
mass squared by non-minimal coupling of the vector field to gravity
through the term $\nm RA_{\mu}A^{\mu}$, where $R$ is the Ricci scalar
and $\varepsilon$ is the non-minimal coupling constant. We have calculated
the vector field perturbation spectrum for the transverse and longitudinal
degrees of freedom and found that they are scale invariant if $\nm=1/6$.
However, the magnitude of the longitudinal power spectrum is twice
the transverse ones, indicating that the particle production of the
vector field is anisotropic. If such a vector field generated the
total curvature perturbation in the Universe, the resulting magnitude
of statistical anisotropy in $\zeta$ would violate observational
bounds obtained from CMB measurements. Therefore, the vector curvaton
considered in this section may generate only a subdominant contribution
to $\zeta$.

We have also explored the parameter space of the proposed scenario.
In this thesis calculations of the constraints for the non-minimally
coupled vector curvaton model, with the statistical anisotropy taken
into account, were performed for the first time. We have shown that
there is an ample parameter space for the model to work by considering
all relevant constraints in the cosmology. 

Some of recently raised concerns \cite{Himmetoglu_etal(2009d),Himmetoglu_etal(2008)Instabilities}
about the stability of the model were also addressed. It was shown
that although the longitudinal mode is a ghost when it is subhorizon,
but it may not be dangerous during inflation if we assume no-particle
(vacuum) initial conditions (as in the scalar field case) and negligible
coupling to other fields. It was also emphasized that the equation
of motion of the longitudinal mode has a singular point at $\left(k/a\right)^{2}=\left|M^{2}\right|$,
which might indicate that the longitudinal mode becomes singular at
horizon exit ($\left|M^{2}\right|\approx H^{2}$). We have obtained
an exact solution for non-zero bare mass of the vector field, i.e.
$m\ne0$, and demonstrated that it is well behaved at all time during
inflation. However, we have not addressed the instability of the longitudinal
mode when the effective mass of the vector field becomes zero after
inflation, i.e. when $M\rightarrow0$.

\section{Vector Curvaton with a Time Varying Kinetic Function\label{sub:vCurvaton-fF2}}

In this section we consider a vector curvaton scenario with the vector
field Lagrangian during inflation\begin{equation}
\mathcal{L}=-\frac{1}{4}fF_{\mu\nu}F^{\mu\nu}+\frac{1}{2}m^{2}A_{\mu}A^{\mu},\label{eq:vCurv-fF-Lagrangian}\end{equation}
where $f=f\left(t\right)$ is the kinetic function and $m=m\left(t\right)$
is the mass and both are functions of the cosmic time $t$. $F_{\mu\nu}=\partial_{\mu}A_{\nu}-\partial_{\nu}A_{\mu}$
is the field strength tensor. If $f$ is time-independent it can be
set equal to 1 because any constant value can be absorbed into $A_{\mu}$.
Otherwise, $f$ represents a time-dependent coupling.

The above Lagrangian density can be of a massive Abelian gauge field,
in which case $f$ is the gauge kinetic function. However, we need
not restrict ourselves to gauge fields only. If no gauge symmetry
is considered the argument in support of the above Maxwell type kinetic
term is that it is one of the few (three) choices \cite{Carroll_etal(2009)Instabilities}
which avoids introducing instabilities, such as ghosts \cite{Himmetoglu_etal(2008)Instabilities}.

\subsection{Equations of Motion}

We focus, at first, on a period of cosmic inflation, during which
we assume that the contribution of the vector field to the energy
budget of the Universe is negligible. Thus, we take the inflationary
expansion to be isotropic. As in the previous model in section~\ref{sub:vCurvaton-RA2}
we also assume that inflation is of (quasi)de Sitter type, i.e. the
Hubble parameter is $H\approx\mathrm{constant}$. 

Inflation is expected to homogenize the vector field. Following the
analogous calculations as in section~\ref{sub:vCurvaton-RA2} and
Ref.~\cite{Dimopoulos2007}, we find that the temporal component
of the homogeneous vector field has to be zero, while the spatial
components satisfy the equation of motion\begin{equation}
\ddot{\mathbf{A}}+\left(H+\frac{\dot{f}}{f}\right)\dot{\mathbf{A}}+\frac{m^{2}}{f}\mathbf{A}=0,\label{eq:vCurv-fF-EoM-homog}\end{equation}
where the dot denotes derivative with respect to $t$. From the above
it is evident that the effective mass of the vector field is \begin{equation}
M\equiv\frac{m}{\sqrt{f}}\,,\label{eq:vCurv-fF-M-def}\end{equation}
 where we assumed that $m,\, f>0$.

We perturb the vector field according to Eq.~\eqref{eq:vCurv-RA-v-perturb}
and going to the Fourier space we calculate equations of motions for
the transverse and longitudinal polarizations as\begin{eqnarray}
\left\{ \partial_{t}^{2}+\left(H+\frac{\dot{f}}{f}\right)\partial_{t}+\frac{m^{2}}{f}+\left(\frac{k}{a}\right)^{2}\right\} \delta A_{+} & = & 0,\qquad\label{eq:vCurv-fF-EoM-Ap}\\
\left\{ \partial_{t}^{2}+\left[H+\frac{\dot{f}}{f}+\left(2H+2\frac{\dot{m}}{m}-\frac{\dot{f}}{f}\right)\frac{\left(\frac{k}{a}\right)^{2}}{\left(\frac{k}{a}\right)^{2}+\frac{m^{2}}{f}}\right]\partial_{t}\,+\frac{m^{2}}{f}+\left(\frac{k}{a}\right)^{2}\right\} \delta A_{\parallel} & = & 0,\qquad\qquad\label{eq:vCurv-fF-EoM-Al}\end{eqnarray}
where $\delta A_{+}$ and $\delta A_{\parallel}$ are defined in Eq.~\eqref{eq:vCurv-RA-polar-decomposition}. 

To continue we need to employ the physical (in contrast to comoving),
canonically normalized vector field\begin{equation}
\mathbf{W}=\sqrt{f}\frac{\mathbf{A}}{a}.\label{eq:vCurv-fF-W-def}\end{equation}
Note that the definition of $\mathbf{W}$ differs from the one in
Eq.~\eqref{eq:vCurv-phys-massive-vFd} because in this section $\mathbf{W}$
gets an additional factor of $\sqrt{f}$ due to canonical normalization.

Expressing Eqs.~\eqref{eq:vCurv-fF-EoM-Ap} and \eqref{eq:vCurv-fF-EoM-Al}
in terms of the physical vector field we obtain \begin{equation}
\left\{ \partial_{t}^{2}+3H\partial_{t}+\frac{1}{2}\left[\frac{1}{2}\left(\frac{\dot{f}}{f}\right)^{2}-\frac{\ddot{f}}{f}-\frac{\dot{f}}{f}H+4H^{2}\right]+M^{2}+\left(\frac{k}{a}\right)^{2}\right\} w_{+}=0\label{eq:vCurv-fF-EoM-wp-gen}\end{equation}
 and \begin{eqnarray}
\left\{ \partial_{t}^{2}+\left[3H+\left(2H+2\frac{\dot{M}}{M}\right)\frac{\left(k/a\right)^{2}}{\left(k/a\right)^{2}+M^{2}}\right]\partial_{t}+\frac{1}{2}\left[\frac{1}{2}\left(\frac{\dot{f}}{f}\right)^{2}-\frac{\ddot{f}}{f}-\frac{\dot{f}}{f}H+4H^{2}\right]+\right.\nonumber \\
\left.+\left(H-\frac{1}{2}\frac{\dot{f}}{f}\right)\left(2H+2\frac{\dot{M}}{M}\right)\frac{\left(\frac{k}{a}\right)^{2}}{\left(\frac{k}{a}\right)^{2}+M^{2}}+M^{2}+\left(\frac{k}{a}\right)^{2}\right\} w_{\parallel}=0.\qquad\label{eq:vCurv-fF-EoM-wl-gen}\end{eqnarray}
Because the theory is parity conserving the Fourier mode $w_{+}$
of $\delta\mathbf{W}\tx$ perturbations denotes both polarizations:
the left-handed and right-handed, i.e. $w_{+}=\sqrt{f}\delta A_{+}/a$. 

Let us use the following ansatz for the time dependence of the kinetic
function and the mass \begin{equation}
f\propto a^{\alpha}\quad\mathrm{and}\quad m\propto a^{\beta},\label{eq:vCurv-fF-f-m-parametriz}\end{equation}
where $\alpha$ and $\beta$ are real constants. We will also assume
that $f\rightarrow1$ at the end of inflation so that, after inflation,
the vector field is canonically normalized. Then Eqs.~\eqref{eq:vCurv-fF-EoM-wp-gen}
and \eqref{eq:vCurv-fF-EoM-wl-gen} become\begin{equation}
\ddot{w}_{+}+3H\dot{w}_{+}+\left[-\frac{1}{4}(\alpha+4)(\alpha-2)H^{2}+M^{2}+\left(\frac{k}{a}\right)^{2}\right]w_{+}=0\label{eq:vCurv-fF-EoM-wp}\end{equation}
and

\begin{eqnarray}
\ddot{w}_{\pl}+\left(3+\frac{2-\alpha+2\beta}{1+r^{2}}\right)H\dot{w}_{\pl}+\nonumber \\
+\left[\frac{1}{2}(2-\alpha)\left(\alpha+4+\frac{2-\alpha+2\beta}{1+r^{2}}\right)H^{2}+\left(\frac{k}{a}\right)^{2}(1+r^{2})\right]w_{\pl} & = & 0,\label{eq:vCurv-fF-EoM-wl}\end{eqnarray}
where $r$ is defined as \begin{equation}
r\equiv\frac{M}{k/a}.\label{eq:vCurv-fF-r-def}\end{equation}

\subsection{The Power Spectrum\label{sub:vCurv-fF-Power-Sp}}

To calculate the power spectrum one can proceed as in section~\ref{sub:vCurvaton-RA2}:
calculate general solutions of Eqs.~\eqref{eq:vCurv-fF-EoM-wp} and
\eqref{eq:vCurv-fF-EoM-wl}, determine integration constants by matching
the solution to the vacuum at the subhorizon limit, $k/a\gg H$, and
calculating the field amplitude at the superhorizon regime when $k/a\ll H$.
However, it is difficult to find general solutions for these equations,
therefore one needs to use approximate methods.

In Appendix~\ref{cha:AppendixB-Scale-Inv} it is shown that, in analogy
to the equation of motion of a scalar field during quasi de Sitter
inflation and with initial conditions in Eq.~\eqref{eq:vCurv-RA-vacuum-wp},
the scale invariant perturbation spectrum for transverse polarizations
in Eq.~\eqref{eq:vCurv-fF-EoM-wp} is achieved if\begin{equation}
\alpha=-1\pm3\label{eq:vCurv-fF-alpha}\end{equation}
(i.e. either $f\propto a^{2}$ or $f\propto a^{-4}$) and \begin{equation}
M_{*}\ll H,\label{eq:vCurv-fF-MllH}\end{equation}
where the star denotes the time when cosmological scales exit the
horizon. The latter condition simply requires that the physical vector
field $W_{\mu}$ is effectively massless at that time.%
\footnote{Note that this is not the same as having $A_{\mu}$ being effectively
massless. In the latter case the vector field is approximately conformally
invariant and does not undergo particle production. However, the conformal
invariance of the massless physical vector field $W_{\mu}$ is broken.%
}

For the longitudinal polarization the initial condition reads\begin{equation}
\lim_{\frac{k}{aH}\rightarrow+\infty}w_{\parallel}=\gamma\frac{a^{-1}}{\sqrt{2k}}\mathrm{e}^{ik/aH},\label{eq:vCurv-fF-longitudinal-vacuum}\end{equation}
where the Lorentz boost factor is\begin{equation}
\gamma=\frac{E}{M}=\frac{\sqrt{\left(\frac{k}{a}\right)^{2}+M^{2}}}{M}=\sqrt{1+\frac{1}{r^{2}}}.\label{eq:vCurv-fF-Lorentz-boost}\end{equation}
In the subhorizon limit $r\ll1$. After finding the solution of Eq.~\eqref{eq:vCurv-fF-EoM-wl}
with this condition and matching it to the vacuum solution in Eq.~\eqref{eq:vCurv-fF-longitudinal-vacuum},
one can calculate the power spectrum of $w_{\parallel}$ in the superhorizon
limit. In Ref.~\cite{Dimopoulos_us(2009)_fF2} it was shown that
the spectrum is scale invariant if \begin{equation}
\beta=-\frac{1}{2}\left(3\pm5\right).\label{eq:vCurv-fF-beta}\end{equation}
As explained in the Appendix~\ref{cha:AppendixB-Scale-Inv} the value
$\beta=-4$ must be disregarded because it implies the massive physical
vector field in the subhorizon limit. This contradicts the requirement
for the scale invariant perturbation spectrum of the transverse modes.

Having evaluated $\alpha$ and $\beta$ in Eqs.~\eqref{eq:vCurv-fF-alpha}
and \eqref{eq:vCurv-fF-beta} to give the scale invariant perturbation
spectrum of the transverse and longitudinal modes one can analyze
equations of motion in Eqs.~\eqref{eq:vCurv-fF-EoM-wp} and \eqref{eq:vCurv-fF-EoM-wl}
in more detail. In Ref.~\cite{Dimopoulos_us(2009)_fF2} these equations
were solved in different approximation regimes as well as solved numerically.
Below we provide the summary of results.

\paragraph{\boldmath Case: $f\propto a^{-4}$ \& $m\propto a$\protect \\
}

For $\alpha=-4$ and $\beta=1$, the equation of motion for the transverse
mode functions in Eq.~\eqref{eq:vCurv-fF-EoM-wp} become\begin{equation}
\ddot{w}_{+}+3H\dot{w}_{+}+\left(\frac{k}{a}\right)^{2}(1+r^{2})w_{+}=0.\label{eq:vCurv-fF-wp-a-4}\end{equation}
When the kinetic function of the vector field scales as $f\propto a^{-4}$,
from Eqs.~\eqref{eq:vCurv-fF-r-def} and \eqref{eq:vCurv-fF-M-def}
we find that $r\propto a^{4}$. In addition we assume that cosmological
scales exit the horizon when the vector field is light. Therefore,
for subhorizon perturbations when $x\gtrsim1$, where $x$ was defined
in Eq.~\eqref{eq:vCurv-RA-x-def} as $x\equiv k/\left(aH\right)$,
the $r$ parameter is very small, i.e. $r\ll1$. When the mode leaves
the horizon $x\lesssim1$ and for cosmological scales $r<1$. However,
because $r$ is a growing function, at some later time it may become
large, i.e. $r\gtrsim1$. Assuming the Bunch-Davies vacuum initial
conditions, when $x\gg1$, the solution of Eq.~\eqref{eq:vCurv-fF-wp-a-4}
in these three different regimes are given by \cite{Dimopoulos_us(2009)_fF2}

\begin{eqnarray}
w_{+}=a^{-3/2}\sqrt{\frac{\pi}{4H}}\left[J_{3/2}\left(x\right)-iJ_{-3/2}\left(x\right)\right] & \mathrm{for} & x\gtrsim1\,,\label{eq:vCurv-fF-wp-sr}\\
w_{+}=\frac{i}{\sqrt{2k}}\left(\frac{H}{k}\right)\left[1+\frac{i}{3}x^{3}\right]\simeq\frac{i}{\sqrt{2k}}\left(\frac{H}{k}\right) & \mathrm{for} & x\ll1\ll\frac{1}{z}\,,\label{eq:vCurv-fF-wp-r}\\
w_{+}=\frac{1}{\sqrt{2k}}\left(\frac{H}{k}\right)\sqrt{\frac{z\pi}{2}}\left[\frac{x^{3}}{3}J_{-1/2}\left(z\right)+iz^{-1}J_{1/2}\left(z\right)\right] & \mathrm{for} & \frac{1}{z}\lesssim1\,,\label{eq:vCurv-fF-wp-br}\end{eqnarray}
where $z$ is defined as \begin{equation}
z\equiv\frac{M}{3H}\,,\end{equation}
and $r=z/\left(3x\right)$. The solution of Eq.~\eqref{eq:vCurv-fF-wp-a-4}
was calculated using numerical methods as well and it was found that
they agree with Eqs.~\eqref{eq:vCurv-fF-wp-sr}-\eqref{eq:vCurv-fF-wp-br}
remarkably well.

The equation of motion for the longitudinal component with the same
scaling of $f$ and $m$ is\begin{equation}
\ddot{w}_{\parallel}+\left(3+\frac{8}{1+r^{2}}\right)H\dot{w}_{\parallel}+\left[\frac{24}{1+r^{2}}H^{2}+\left(\frac{k}{a}\right)^{2}(1+r^{2})\right]w_{\parallel}=0\,.\label{eq:vCurv-fF-wl-a-4}\end{equation}
And the solution of this equation in the same three regimes was found
to be\begin{eqnarray}
w_{\parallel}=-\frac{i}{6}a^{-9/2}\sqrt{\frac{\pi}{H}}x^{-2}z^{-1}\left[J_{5/2}\left(x\right)-iJ_{-5/2}\left(x\right)\right] & \mathrm{for} & x\gtrsim1\,,\label{w+0sr}\\
w_{\parallel}\simeq-a^{-1/2}\frac{3a_{k}^{4}}{\sqrt{2H}}x^{5/2}=-\frac{1}{\sqrt{2k}}\left(\frac{H}{k}\right)z^{-1} & \mathrm{for} & x\ll1\ll\frac{1}{z}\,,\label{eq:vCurv-fF-wl-r}\\
w_{\parallel}=-\frac{1}{\sqrt{2k}}\left(\frac{H}{k}\right)\sqrt{\frac{z\pi}{2}}\left[z^{-1}J_{-1/2}\left(z\right)-i\frac{x^{3}}{3}J_{1/2}\left(z\right)\right] & \mathrm{for} & \frac{1}{z}\lesssim1\,,\label{eq:vCurv-fF-wl-br}\end{eqnarray}
which agrees with the numerical solution of Eq.~\eqref{eq:vCurv-fF-wl-a-4}
very well too.

As is seen from Eqs.~\eqref{eq:vCurv-fF-wp-r}, \eqref{eq:vCurv-fF-wp-br}
and \eqref{eq:vCurv-fF-wl-r}, \eqref{eq:vCurv-fF-wl-br} on the superhorizon
scales modes $w_{+}$ and $w_{\parallel}$ evolves differently if
the vector field is light, $M\lesssim H$, or heavy, $M\gtrsim H$.
When the field is light, $w_{+}$ is constant and $w_{\parallel}\propto a^{-1}$.
Therefore using Eq.~\eqref{eq:vCurv-RA-spec-w-lambda-def} we find\begin{equation}
\mathcal{P}_{+}=\left(\frac{H}{2\pi}\right)^{2}\quad\mathrm{and}\quad\mathcal{P}_{\parallel}=\frac{1}{z^{2}}\left(\frac{H}{2\pi}\right)^{2}\propto a^{-6}\quad\mathrm{for}\quad M<H.\label{eq:vCurv-fF-Pp-a-4-light}\end{equation}
Thus the typical value of the vector field perturbation is (see Eq.~\eqref{eq:mean-square-dlnk-scale-inv})\begin{equation}
\delta W\approx\sqrt{\Pl}=\frac{3H}{M}\frac{H}{2\pi}\propto a^{-3},\label{eq:vCurv-fF-typical-pert-light}\end{equation}
where $\delta W\equiv\left|\delta\mathbf{W}\right|$ and we used $\Pl\gg\Pp$.

On the other hand, when the mass of the vector field becomes comparable
with the inflationary Hubble parameter, i.e. $M\sim H$, the transverse
and longitudinal mode functions on the superhorizon scales, with $x\ll1$,
become\begin{eqnarray}
\left.\begin{array}{c}
w_{+}=\frac{i}{\sqrt{2k}}\left(\frac{H}{k}\right)\sqrt{\frac{\pi}{2}}\frac{J_{1/2}\left(z\right)}{\sqrt{z}}\\
w_{\parallel}=\frac{-1}{\sqrt{2k}}\left(\frac{H}{k}\right)\sqrt{\frac{\pi}{2}}\frac{J_{-1/2}\left(z\right)}{\sqrt{z}}\end{array}\right\}  & \Rightarrow & \left|\left|w_{+}\right|\right|\approx\left|\left|w_{\parallel}\right|\right|,\quad z\gtrsim1.\end{eqnarray}
When the vector field becomes heavy $M\gg H$ from Eqs.~\eqref{eq:vCurv-fF-wp-br}
and \eqref{eq:vCurv-fF-wl-br} we find\begin{eqnarray}
w_{+} & = & \frac{i}{\sqrt{2k}}\left(\frac{H}{k}\right)\frac{\sin\left(z\right)}{z},\label{eq:vCurv-fF-wp-oscillating}\\
w_{\parallel} & = & -\frac{1}{\sqrt{2k}}\left(\frac{H}{k}\right)\frac{\cos\left(z\right)}{z},\label{eq:vCurv-fF-wl-oscillating}\end{eqnarray}
i.e. they oscillate with the same amplitude, $\left|\left|w_{+}\right|\right|=\left|\left|w_{\parallel}\right|\right|$,
but with the phase difference of $\pi/2$. The frequency of oscillations
is much larger than the Hubble parameter because $z\gg1$, therefore
it makes sense to use the average values of the power spectra over
many oscillations. Using Eq.~\eqref{eq:vCurv-RA-spec-w-lambda-def}
we find\begin{equation}
\overline{\Pp}=\overline{\Pl}=\frac{1}{2z^{2}}\left(\frac{H}{2\pi}\right)^{2}\quad\mathrm{for}\quad M\gtrsim H.\label{eq:vCurv-fF-Pp-a-4-heavy}\end{equation}
Thus, the typical value for the vector field perturbations in this
regime is\begin{equation}
\delta W\approx\frac{1}{\sqrt{2}}\frac{3H}{M}\frac{H}{2\pi}\propto a^{-3},\label{eq:vCurv-fF-typical-pert-heavy}\end{equation}
where for the scale invariant perturbations $\left|\delta\mathbf{W}\right|\approx\sqrt{\overline{\mathcal{P}}_{||}}$
(see Eq.~\eqref{eq:mean-square-dlnk-scale-inv}). Thus from Eq.~\eqref{eq:vCurv-fF-typical-pert-light}
we see that the typical value of the vector field perturbation is
roughly the same if the field is light or heavy.

\paragraph{\boldmath Case: $f\propto a^{2}$ \& $m\propto a$\protect \\
}

When the vector field kinetic function $f$ is increasing with time,
i.e. $\alpha=2$, and $\beta=1$ the effective mass of the field is
constant, $M=\mathrm{constant}$. The requirement that the field is
effectively massless when cosmological scales exit the horizon in
Eq.~\eqref{eq:vCurv-fF-MllH} suggests that $M/H\ll1$ at all times
when the scaling above holds. Using this condition and scaling we
can calculate the power spectra for all components of the superhorizon
vector field perturbations generated by the particle production process.

The equation of motion for the transverse mode functions\begin{equation}
\ddot{w}_{+}+3H\dot{w}_{+}+\left(\frac{k}{a}\right)^{2}(1+r^{2})w_{+}=0.\label{eq:vCurv-fF-wp-a-2}\end{equation}
is the same as for the $\alpha=-4$ case except that now $M=\mathrm{constant}$.
This condition simplifies Eq.~\eqref{eq:vCurv-fF-wp-a-2}, making
it possible to obtain the exact solution. After matching this solution
to the initial Bunch-Davies vacuum state, the power spectrum on superhorizon
scales becomes \cite{Dimopoulos_us(2009)_fF2}\begin{equation}
\Pp=\left(\frac{H}{2\pi}\right)^{2}.\label{eq:vCurv-fF-Pp-a=2}\end{equation}

The equation of motion for the longitudinal component from Eq.~\eqref{eq:vCurv-fF-EoM-wl}
and $\alpha=-4$ becomes\begin{equation}
\ddot{w}_{\parallel}+\left(3+\frac{2}{1+r^{2}}\right)H\dot{w}_{\parallel}+\left[\left(\frac{k}{a}\right)^{2}(1+r^{2})\right]w_{\parallel}=0.\end{equation}
Again, it is impossible to find an exact solution of this equation.
But using the vacuum initial conditions in Eq.~\eqref{eq:vCurv-fF-longitudinal-vacuum}
and solving it in two regimes, $r\ll1$ and $r\gg1$, and matching
those solutions at $r=1$ we find \begin{eqnarray}
w_{\parallel}=-\frac{x}{2}\sqrt{\frac{\pi}{aH}}\left[J_{-5/2}\left(x\right)+iJ_{5/2}\left(x\right)\right] & \mathrm{for} & x\gtrsim1,\\
w_{\parallel}=-\frac{z^{-1}}{\sqrt{2k}}\left(\frac{H}{k}\right) & \mathrm{for} & x\ll1\ll\frac{1}{z}.\label{eq:vCurv-fF-wl-solution-a-2}\end{eqnarray}
From Eq.~\eqref{eq:vCurv-fF-wl-solution-a-2} we calculate the power
spectrum\begin{equation}
\Pl=9\left(\frac{H}{M}\right)^{2}\left(\frac{H}{2\pi}\right)^{2},\label{eq:vCurv-fF-Pl-a=2}\end{equation}
same as in Eq.~\eqref{eq:vCurv-fF-Pp-a-4-light}. Since in this case,
we have $M/H=\mathrm{constant}\ll1$, the longitudinal power spectrum
is constant, in contrast to the $\alpha=-4$ case. Also, we see that
${\cal P}_{\parallel}\gg{\cal P}_{+}$.

\subsection{Statistical Anisotropy and Non-Gaussianity\label{sub:vCurv-fF-fNL}}

The theory studied in this section has two clear advantages. First
we can obtain a completely isotropic perturbation spectrum for the
vector field, which has previously never been achieved. As we discuss
below, this means that we may consider vector fields as dominating
the total energy density of the Universe when the curvature perturbation
is formed. The second advantage is that we can also account for a
small amount of statistical anisotropy in the curvature perturbation
spectrum depending on when inflation ends, again by considering the
vector field alone. We also demonstrate this in what follows. Finally,
statistical anisotropy can also be present in a correlated manner
in the bispectrum as well, which characterizes the non-Gaussian features
of the CMB temperature perturbations. In view of the forthcoming observations
of the recently launched Planck satellite mission this is a particularly
promising and timely result.

Let us first consider the case with $\alpha=-4$. As we have seen
in the previous section in this case the effective mass of the vector
field is time dependent during inflation, $M\propto a^{3}$. When
$M\lesssim H$ the evolution of the vector field perturbations follows
the power law on superhorizon scales (see Eqs.~\eqref{eq:vCurv-fF-wp-r}
and \eqref{eq:vCurv-fF-wl-r}) and the power spectra for the transverse
and longitudinal modes are given in Eq.~\eqref{eq:vCurv-fF-Pp-a-4-light}.
Using the definition of $p$ in Eq.~\eqref{eq:vFd-p-q-def} we find
that in this regime the anisotropy in the particle production is equal
to\begin{equation}
p=z^{-2}-1,\label{eq:vCurv-fF-p-light}\end{equation}
where $z\lesssim1$.

\begin{figure}
\begin{centering}
\includegraphics[width=7cm]{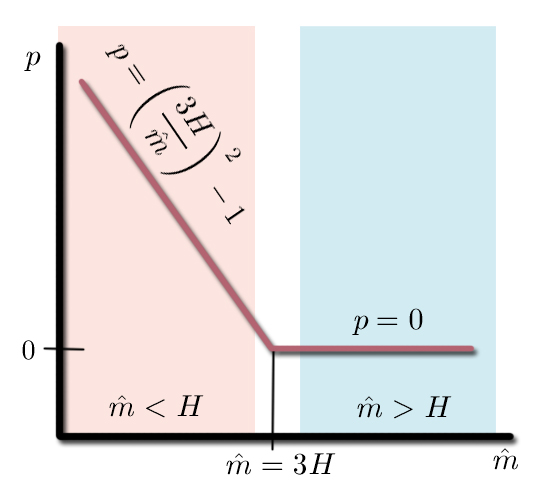}
\par\end{centering}

\caption{\label{fig:vCurv-fF-aniso-on-p}If the vector field is light at the
end of inflation, the particle production will be highly anisotropic.
If it is heavy, the transverse and longitudinal power spectra are
equal and $p=0$, i.e particle production is isotropic. In the regime
where $\hat{m}\sim H$ the anisotropy in the particle production might
be non-zero but very small, $p\lesssim1$. }

\end{figure}

In the opposite regime, when $M\gtrsim H$, the vector field perturbations
are oscillating (see Eqs.~\eqref{eq:vCurv-fF-wp-oscillating} and
\eqref{eq:vCurv-fF-wl-oscillating}) and the average power spectra
in Eq.~\eqref{eq:vCurv-fF-Pp-a-4-heavy} are equal giving\begin{equation}
p=0.\label{eq:vCurv-fF-p-heavy}\end{equation}

At the end of inflation the kinetic function and the mass of the vector
field are stabilized: $f_{\mathrm{end}}=1$ and $M_{\mathrm{end}}=\mathrm{constant}\equiv\hat{m}$.
At this epoch the vector field perturbation power spectra become constant.
Therefore, although in the curvaton scenario $\zeta$ becomes constant
only after the curvaton decay, it is enough to evaluate amplitudes
of the vector field perturbations at the end of inflation. Thus, the
value of $p$ is frozen at the end of inflation and it depends on
the ratio $\hat{m}/H$ (see Figure~\ref{fig:vCurv-fF-aniso-on-p}).
If the vector field is light at the end of inflation (or if its mass
is of the order of inflationary Hubble parameter), the particle production
is anisotropic and given in Eq.~\eqref{eq:vCurv-fF-p-light} with
$z=z_{\mathrm{end}}\equiv\hat{m}/3H$. If the field is heavy, the
transverse and longitudinal power spectra are equal, i.e. the particle
production is isotropic giving Eq.~\eqref{eq:vCurv-fF-p-heavy}. 

If $\hat{m}>H$, the particle production of the vector field is isotropic,
$p=0$, and the vector field generated curvature perturbation is statistically
isotropic\begin{equation}
g=0.\label{eq:vCurv-fF-g-zgg1}\end{equation}
Therefore, if the vector field is heavy at the end of inflation the
generated curvature perturbation is indistinguishable from the scalar
field case. Indeed, if we plug Eq.~\eqref{eq:vCurv-fF-p-heavy} into
the expression of the $\fnl$ in Eqs.~\eqref{eq:vCurv-fnl-eql-M^2-scaleInv}
and \eqref{eq:vCurv-fNL-scaleInv-local} we find\begin{equation}
\frac{6}{5}\fnle=\frac{6}{5}\fnll=\frac{3}{2\ohw},\label{eq:vCurv-fF-fNL-zgg1}\end{equation}
exactly the same as in the scalar curvaton case. In this expression
we considered that the only contribution to $\zeta$ comes from the
vector field, i.e. $\mathcal{P}_{\phi}=0$.

If, on the other hand, the vector field mass at the end of inflation
is $\hat{m}\sim H$, from Eq.~\eqref{eq:vCurv-fF-p-light} we find
$0\lesssim p<1$. Using this and Eqs.~\eqref{eq:vCurv-fNL-curG-equil}
and \eqref{eq:vCurv-fNL-curG-local} the anisotropy in $\fnl$ becomes\begin{equation}
\Ge\approx\Gl\approx p<1,\end{equation}
where again we have considered that only the vector field generates
the curvature perturbation. This regime is possible because the observational
bound on anisotropy in the spectrum (defined in Eq.~\eqref{eq:vFd-z-g-def})
is not violated\begin{equation}
g\approx p<0.3,\label{eq:vCurv-fF-p-l-0.3}\end{equation}
where $g<0.3$ is the observational constraint from CMB on the statistical
anisotropy (see the discussion above Eq.~\eqref{eq:bound-on-g}).
Using Eq.~\eqref{eq:vCurv-fF-p-l-0.3} and requiring that $z_{\mathrm{end}}\lesssim1$
from Eq.~\eqref{eq:vCurv-fF-p-light} we find the allowed range of
the mass values $\hat{m}$ for this case\begin{equation}
0.3<\frac{H}{\hat{m}}<0.4.\label{eq:vCurv-fF-m-range}\end{equation}
Unfortunately this range is very narrow and initial conditions must
be tuned accurately to achieve this possibility. 

In the vector curvaton model with the light vector field at the end
of inflation the dominant part of $\zeta$ must be generated by the
scalar field, while the vector field can generate only a subdominant
contribution. This is because if $\hat{m}\ll H$, the anisotropy in
the vector field particle production is large, i.e. $p\gg1$, and
this would violate observational constraints on $g$. Assuming a light
scalar field with perturbation power spectrum $\mathcal{P}_{\phi}=\left(H/2\pi\right)^{2}=\Pp$
from Eq.~\eqref{eq:vFd-z-g-def} we find\begin{equation}
g=\frac{\xi}{1+\xi}\, p\approx\frac{\xi}{1+\xi}\left(\frac{3H}{\hat{m}}\right)^{2},\label{eq:vCurv-fF-g-zll1}\end{equation}
where in the last expression we have used $z\ll1$. Similarly from
Eq.~\eqref{eq:vCurv-fNLe-fNLl} we find that the isotropic part of
$\fnl$ is\begin{equation}
\fnlli=\fnlei=g^{2}\frac{2}{\ow}\left(\frac{3H}{\hat{m}}\right)^{4}.\end{equation}
While amplitudes of the angular modulation of $\fnl$ in the equilateral
and squeezed configurations with $z\ll1$ are \begin{equation}
\Gl=p\approx\left(\frac{3H}{\hat{m}}\right)^{2}\label{eq:vCurv-fF-curG-zll1}\end{equation}
and\begin{equation}
\Ge\approx\frac{1}{8}p^{2}\approx\frac{1}{8}\left(\frac{3H}{\hat{m}}\right)^{4},\end{equation}
which are much larger than one in both cases.

The values of the non-linearity parameter $\fnl$ calculated in this
section correspond to the scaling of the kinetic function with $\alpha=-4$.
But in the limit $z\ll1$ Eqs.~\eqref{eq:vCurv-fF-g-zll1}-\eqref{eq:vCurv-fF-curG-zll1}
are also applicable for $\alpha=2$. In the latter case $z=\mathrm{constant}\ll1$
therefore, if $f\propto a^{2}$ the vector field may only generate
a subdominant contribution to the curvature perturbation without violating
observational bounds on statistical anisotropy, where the dominant
part is produced by a scalar field. But for $f\propto a^{-4}$, as
we have seen in Eq.~\eqref{eq:vCurv-fF-g-zgg1}, the vector field
can also produce the total curvature perturbation in the Universe.
If it is heavy at the end of inflation, i.e. $\hat{m}\gg H$, the
generated $\zeta$ will be statistically isotropic and indistinguishable
from the scalar curvaton case. If, on the other hand, $\hat{m}\sim H$,
it may still generate the total $\zeta$ which is approximately statistically
anisotropic within the observational bounds.

\subsection{Evolution of the Zero Mode\label{sub:vCurv-fF-zero-mode}}

In order to calculate the curvature perturbation associated with the
vector field one needs to study also the evolution of the homogeneous
zero mode $W$. Combining Eqs.~\eqref{eq:vCurv-fF-EoM-homog} and
\eqref{eq:vCurv-fF-W-def} and using Eq.~\eqref{eq:vCurv-fF-f-m-parametriz},
we obtain \begin{equation}
\ddot{\mathbf{W}}+3H\dot{\mathbf{W}}+\left[\left(1-\frac{1}{2}\alpha\right)\dot{H}-\frac{1}{4}(\alpha+4)(\alpha-2)H^{2}+M^{2}\right]\mathbf{W}=0\,,\label{EoMalpha}\end{equation}
 where we also used the definition of the effective mass in Eq.~\eqref{eq:vCurv-fF-M-def}.

\subsubsection{During Inflation}

As shown in Appendix~\ref{cha:AppendixB-Scale-Inv}, to obtain a
scale invariant spectrum for the transverse components of the vector
field perturbation we require $f(a)$ to scale according to Eq.~\eqref{eq:vCurv-fF-alpha},
i.e. $\alpha=-1\pm3$. Using this and considering the (quasi)de Sitter
inflation (with $\dot{H}\approx0$) the above becomes\begin{equation}
\ddot{\mathbf{W}}+3H\dot{\mathbf{W}}+M^{2}\mathbf{W}=0\,.\label{EoM}\end{equation}
We show below that, when $M\ll H$ (true at early times when $\alpha=-4$;
always true when $\alpha=2$), the solution of the above is well approximated
by \begin{equation}
W\simeq\hat{C}_{1}+\hat{C}_{2}a^{-3},\label{Wsolu0}\end{equation}
where $\hat{C}_{i}$ are constants. The dominant term to the solution
of Eq.~\eqref{Wsolu0} is determined by the initial conditions. We
choose initial conditions for the vector field zero-mode based on
energy equipartition grounds. As is demonstrated in what follows,
if the energy equipartition is assumed at the onset of inflation,
the dominant term turns out to be the decaying mode $W\propto a^{-3}$
when $\alpha=-4$, and the {}``growing'' mode $\left.W=\mathrm{constant}\right.$
when $\alpha=2$.

To apply energy equipartition in the initial conditions we need to
consider the energy-momentum tensor for this theory, which, from Eq.~\eqref{eq:vCurv-fF-Lagrangian}
is given by \cite{Dimopoulos2007} \begin{eqnarray}
T_{\mu\nu} & = & f\left(\frac{1}{4}g_{\mu\nu}F_{\rho\sigma}F^{\rho\sigma}-F_{\mu\rho}F_{\nu}^{\;\rho}\right)+m^{2}\left(A_{\mu}A_{\nu}-\frac{1}{2}g_{\mu\nu}A_{\rho}A^{\rho}\right).\label{Tmn}\end{eqnarray}
If we assume that the homogenized vector field lies along the $z$-direction,
we can write the above as \cite{Dimopoulos2007} \begin{equation}
T_{\mu}^{\,\nu}=\mathrm{diag}(\rho_{W},-p_{\perp},-p_{\perp},+p_{\perp})\,,\label{Tdiag}\end{equation}
where \begin{equation}
\rho_{W}\equiv\rho_{\mathrm{kin}}+V_{W}\;,\qquad p_{\perp}\equiv\rho_{\mathrm{kin}}-V_{W}\;,\label{rp}\end{equation}
 with \begin{eqnarray}
\rho_{\mathrm{kin}} & \equiv & -\frac{1}{4}fF_{\mu\nu}F^{\mu\nu}\;=\;\frac{1}{2}a^{-2}f\dot{A}^{2}=\frac{1}{2}\left[\dot{W}-\frac{1}{2}\left(\alpha-2\right)HW\right]^{2},\label{rkin}\\
\nonumber \\V_{W} & \equiv & -\frac{1}{2}m^{2}A_{\mu}A^{\mu}\;=\;\frac{1}{2}a^{-2}m^{2}A^{2}=\frac{1}{2}M^{2}W^{2},\label{VA}\end{eqnarray}
where $A\equiv|\mathbf{A}|$, we used Eqs.~\eqref{eq:vCurv-fF-W-def}
and \eqref{eq:vCurv-fF-f-m-parametriz}, and we assumed a negative
signature for the metric.

Energy equipartition corresponds to \begin{equation}
(\rho_{\mathrm{kin}})_{0}\simeq(V_{W})_{0}\,,\label{equip}\end{equation}
 where the subscript `0' indicates the values at some initial time,
e.g. near the onset of inflation.

\subparagraph{\boldmath Case: $f\propto a^{-4}$\protect \\
}

In this case $M\propto a^{3}$ and the solution to Eq.~\eqref{EoM}
is \begin{equation}
W=a^{-3}\left[\hat{C}_{3}\sin\left(\frac{M}{3H}\right)+\hat{C}_{2}\cos\left(\frac{M}{3H}\right)\right].\label{Wsolu}\end{equation}
When $M\gtrsim H$ the above shows that the amplitude of the oscillating
zero mode is decreasing as $\left|\left|W\right|\right|\propto a^{-3}$.
In the opposite regime, when $M\ll H$ the solution above is well
approximated by Eq.~\eqref{Wsolu0} with $\hat{C}_{1}=\hat{C}_{3}a_{0}^{-3}M_{0}/3H$,
where we considered that $a^{-3}M=a_{0}^{-3}M_{0}=\mathrm{constant}$.
Using this, the constants $\hat{C}_{2}$ and $\hat{C}_{3}$ in Eq.~\eqref{Wsolu}
can be expressed in terms of initial values of the field amplitude
$W_{0}$ and it's velocity $\dot{W}_{0}$: \begin{equation}
\hat{C}_{2}=-\frac{\dot{W}_{0}}{3H}\; a_{0}^{3}\quad\mathrm{and}\quad\hat{C}_{3}=\frac{\left(\dot{W}_{0}+3HW_{0}\right)}{M_{0}}\; a_{0}^{3}\;.\label{eq:Wconsts}\end{equation}

Assuming initial equipartition of energy we can relate $W_{0}$ with
$\dot{W}_{0}$. From Eqs.~\eqref{rkin} and \eqref{VA}, setting
$\alpha=-4$, we readily obtain \begin{equation}
\rho_{{\rm kin}}=\frac{1}{2}(\dot{W}+3HW)^{2}\quad\mathrm{and}\quad V_{W}=\frac{1}{2}M^{2}W^{2}.\label{kinV-4}\end{equation}
Then, using Eq.~\eqref{equip}, we get \begin{equation}
\dot{W}_{0}\simeq W_{0}\left(-3H\pm M_{0}\right).\end{equation}
 Substituting this relation into Eq.~\eqref{eq:Wconsts} we find
that the evolution of the vector field $W$ in Eq.~\eqref{Wsolu}
takes the simple form: \begin{equation}
W=W_{0}\left(\frac{a}{a_{0}}\right)^{-3}\sqrt{2}\cos\left(\frac{M}{3H}\pm\frac{\pi}{4}\right).\label{eq:W-EoM}\end{equation}
Note that this equation is valid for any value of $M$. However, we
can see that when $M\ll H$ the zero mode of the vector field is decreasing
as $W\propto a^{-3}$, but when $M\gg H$ it oscillates rapidly with
a decreasing amplitude proportional to $a^{-3}$. On this basis we
can assume that the typical value of the zero mode during inflation
always scales as \begin{equation}
W\propto a^{-3}.\label{Wa}\end{equation}

With the assumption of initial equipartition of energy for the vector
field at the onset of inflation we can calculate the kinetic and potential
energy densities.%
\footnote{By {}``potential'' we refer to the energy density stored in the
mass-term $V_{W}=-\frac{1}{2}m^{2}A_{\mu}A^{\mu}$.%
} Inserting Eq.~\eqref{eq:W-EoM} and its derivative into Eqs.~\eqref{rkin}
and \eqref{VA} we find \begin{equation}
\rho_{\mathrm{kin}}=\left[W_{0}M_{0}\sin\left(\frac{M}{3H}\pm\frac{\pi}{4}\right)\right]^{2}\quad\mathrm{and}\quad V_{\mathrm{W}}=\left[W_{0}M_{0}\cos\left(\frac{M}{3H}\pm\frac{\pi}{4}\right)\right]^{2}.\end{equation}
 Hence, the total energy density is constant \begin{equation}
\rho_{W}=M_{0}^{2}W_{0}^{2}.\label{eq:rhoA-duringInfl}\end{equation}
Because this relation is independent of the vector field mass $M$
it is valid in both regimes: when $M\ll H$ and $W$ follows a power
law evolution, and when $M\gg H$ and $W$ oscillates. This is valid
as long as $f(a)$ and $m(a)$ are varying with time.

In the vector curvaton scenario the vector field must be subdominant
during inflation. From Eq.~\eqref{eq:rhoA-duringInfl} we see that
assuming this to be the case at the onset of inflation, it will stay
so until the end of inflation irrespective if the field is light or
heavy.

\subparagraph{Case: $f\propto a^{2}$\protect \\
}

In this case, $M=\mathrm{constant}$, which means that the solution
of Eq.~\eqref{EoM} is \begin{equation}
W=a^{-3/2}\left[\hat{C}_{1}a^{\sqrt{\frac{9}{4}-(\frac{M}{H})^{2}}}+\hat{C}_{2}a^{-\sqrt{\frac{9}{4}-(\frac{M}{H})^{2}}}\right].\label{Wsolu2}\end{equation}
 Since in this case $M\ll H$, the above solution is always well approximated
by Eq.~\eqref{Wsolu0} and there is no oscillating regime.

Now, Eqs.~\eqref{rkin} and \eqref{VA} take the form \begin{equation}
\rho_{\mathrm{kin}}=\frac{1}{2}\dot{W}^{2}\quad\mathrm{and}\quad V_{W}=\frac{1}{2}M^{2}W^{2}.\label{kinV2}\end{equation}
 Combining Eqs.~\eqref{Wsolu0} and \eqref{kinV2}, we find \begin{equation}
\rho_{\mathrm{kin}}=\frac{9}{2}H^{2}\hat{C}_{2}^{2}a^{-6}.\label{rkin2}\end{equation}
Thus, at the onset of inflation assuming energy equipartition in Eq,~\eqref{equip}
gives \begin{equation}
\left(1+\frac{\hat{C}_{1}}{\hat{C}_{2}}a_{0}^{3}\right)^{2}=\left(\frac{3H}{M_{0}}\right)^{2}\gg1\;\Rightarrow\;\hat{C}_{1}\simeq\pm\frac{3H}{M_{0}}a_{0}^{-3}\hat{C}_{2}\;,\end{equation}
 where we used that $M_{0}=M\ll H$. Inserting the above into Eq.~(\ref{Wsolu0})
we find \begin{equation}
W=a_{0}^{-3}\hat{C}_{2}\left[\left(\frac{a_{0}}{a}\right)^{3}\pm\frac{3H}{M_{0}}\right]\simeq\mathrm{constant}\simeq W_{0}\;,\end{equation}
because, after the onset of inflation, $(a_{0}/a)^{3}\ll1\ll3H/M_{0}$.

Therefore, we have found that $W$ remains constant. Since $M=\mathrm{constant}$,
this means that $V_{W}$ also remains constant. On the other hand,
Eq.~(\ref{rkin2}) suggests that $\rho_{\mathrm{kin}}\propto a^{-6}$.
Thus, since we assumed energy equipartition at the onset of inflation,
we find that, during inflation, $\rho_{\mathrm{kin}}\ll V_{W}$. Hence,
\begin{equation}
\rho_{W}\approx V_{W}\simeq M_{0}^{2}W_{0}^{2},\label{eq:rhoA-f-prop-a}\end{equation}
 where $M=\mathrm{constant}=M_{0}$. This result is the same as in
the case $f\propto a^{-4}$ in Eq.~\eqref{eq:rhoA-duringInfl} and
the vector curvaton field is ensured to be subdominant during inflation
(as required by the curvaton mechanism) if it is subdominant at the
onset of inflation.

\subsubsection{After Inflation}

At the end of inflation we assume that the scaling of $f$ and $m$
has ended and we have \begin{equation}
f=1\quad\mathrm{and}\quad m=\hat{m}\,.\label{eq:fm-after-infl}\end{equation}
 Hence, Eqs.~(\ref{eq:rhoA-duringInfl}) and (\ref{eq:rhoA-f-prop-a})
no longer apply. The evolution of $\rho_{W}$ is determined as follows.

As mentioned already, after the end of scaling, $\alpha=0$ and $M=\hat{m}$.
Then, Eqs.~(\ref{rkin}) and (\ref{VA}) become \begin{equation}
\rho_{{\rm kin}}=\frac{1}{2}(\dot{W}+HW)^{2}\quad\mathrm{and}\quad V_{W}=\frac{1}{2}\hat{m}^{2}W^{2}.\label{kinVend}\end{equation}
The behavior of $\rho_{\mathrm{kin}}$ and $V_{W}$ depends on whether
the vector field is light or not. To see this let us calculate the
evolution of the field after inflation. With the conditions in Eq.~(\ref{eq:fm-after-infl})
the physical vector field of Eq.~\eqref{eq:vCurv-fF-W-def} is $\mathbf{W}=\mathbf{A}/a$,
while Eq.~(\ref{EoMalpha}) becomes \begin{equation}
\ddot{\mathbf{W}}+3H\dot{\mathbf{W}}+\left(\dot{H}+2H^{2}+\hat{m}^{2}\right)\mathbf{W}=0\,,\label{eq:EoM-W-after-infl}\end{equation}
where the Hubble parameter after inflation decreases as $H(t)=\frac{2}{3\left(1+w\right)t}$,
with $w\equiv p/\rho$ being the barotropic parameter of the Universe.
Solving Eq.~(\ref{eq:EoM-W-after-infl}) we find \begin{eqnarray}
W & = & t^{\frac{1}{2}\,\frac{w-1}{w+1}}\left[\tilde{C}_{1}J_{v}\left(\hat{m}t\right)+\tilde{C}_{2}J_{-v}\left(\hat{m}t\right)\right],\label{eq:W-afterInfl-gen}\\
\dot{W}+HW & = & \hat{m}\, t^{\frac{1}{2}\,\frac{w-1}{w+1}}\left[\tilde{C}_{1}J_{v-1}\left(\hat{m}t\right)-\tilde{C}_{2}J_{1-v}\left(\hat{m}t\right)\right],\label{eq:Wdot-afterInfl-gen}\end{eqnarray}
where $v=\frac{1+3w}{6\left(1+w\right)}$. One can easily see that
the vector field behaves differently if it is light, $\hat{m}t\ll1$,
or heavy, $\hat{m}t\gg1$.

Let us first see what happens if the vector field is light. Then,
Eqs.~(\ref{eq:W-afterInfl-gen}) and (\ref{eq:Wdot-afterInfl-gen})
can be approximated as \begin{eqnarray}
W & = & t^{\frac{1}{2}\,\frac{w-1}{w+1}}\left[\frac{\tilde{C}_{1}}{\Gamma\left(1+v\right)}\left(\frac{\hat{m}t}{2}\right)^{v}+\frac{\tilde{C}_{2}}{\Gamma\left(1-v\right)}\left(\frac{\hat{m}t}{2}\right)^{-v}\,\right],\\
\dot{W}+HW & = & \hat{m}\, t^{\frac{1}{2}\,\frac{w-1}{w+1}}\left[v\,\frac{\tilde{C}_{1}}{\Gamma\left(1+v\right)}\left(\frac{\hat{m}t}{2}\right)^{v-1}-\frac{1}{1-v}\frac{\tilde{C}_{2}}{\Gamma\left(1-v\right)}\left(\frac{\hat{m}t}{2}\right)^{1-v}\right].\qquad\qquad\end{eqnarray}
Although the solution has one decaying and one growing mode, it might
happen that the decaying mode stays larger than the growing mode.
To check this we calculate constants $\tilde{C}_{1}$ and $\tilde{C}_{2}$
by matching the above equations to the values $W_{\mathrm{end}}$
and $\dot{W}_{\mathrm{end}}$ at the end of inflation (denoted by
`end'). Thus, we find that\begin{eqnarray}
W & = & \frac{2}{3w+1}\left(\frac{a}{a_{\mathrm{end}}}\right)^{\frac{1}{2}\left(3w-1\right)}\left(W_{\mathrm{end}}+\frac{\dot{W}_{\mathrm{end}}}{H_{*}}\right),\\
\dot{W}+HW & = & H_{*}\left(\frac{a}{a_{\mathrm{end}}}\right)^{-2}\left(W_{\mathrm{end}}+\frac{\dot{W}_{\mathrm{end}}}{H_{*}}\right),\end{eqnarray}
where $H_{*}$ is the inflationary Hubble scale. Plugging these solutions
into Eq.~(\ref{kinVend}) (and using that $a^{3(1+w)}\propto t^{2}$)
we obtain \begin{equation}
\frac{V_{W}}{\rho_{\mathrm{kin}}}=\frac{4}{(3w+1)^{2}}\left(\frac{\hat{m}}{H_{*}}\right)^{2}\left(\frac{t}{t_{\mathrm{end}}}\right)^{2}\simeq(\hat{m}t)^{2}\ll1\,,\end{equation}
 which implies that the total energy density of the light vector field
is\begin{equation}
\rho_{W}\simeq\rho_{\mathrm{kin}}=\frac{1}{2}\left(\dot{W}_{\mathrm{end}}+W_{\mathrm{end}}H_{*}\right)^{2}\left(\frac{a}{a_{\mathrm{end}}}\right)^{-4}\;\Rightarrow\;\rho_{W}\propto a^{-4}.\label{eq:rhoA-mllH}\end{equation}
 Therefore, we see that the energy density of the light vector field
scales as that of relativistic particles. This is in striking difference
to the scalar field case, in which when the field is light its density
remains constant even after inflation.

On the other hand, if the vector field is heavy, $\hat{m}t\gg1$,
the Bessel functions in Eqs.~\eqref{eq:W-afterInfl-gen} and \eqref{eq:Wdot-afterInfl-gen}
are oscillating. Hence, as was discussed in section~\ref{sub:vCurv-dynamics},
the heavy vector field oscillates with a frequency much larger than
the Hubble parameter and with the amplitude decreasing as $t^{-1/\left(1+w\right)}\propto a^{-3/2}$.
In Eq.~\eqref{eq:vCurv-Intro-isotr-pressureless-cond} it was shown
that the energy density of such field decreases as $\rw\propto a^{-3}$
and the average pressure is zero, i.e. $\overline{p_{\perp}}\approx0$.
Therefore, on average, the oscillating vector field behaves as pressureless
isotropic matter and can dominate the Universe without generating
excessive large scale anisotropy. This is crucial for the vector curvaton
mechanism because, to produce the curvature perturbation, the field
must dominate (or nearly dominate) the Universe without inducing excessive
anisotropic expansion.

\subsection{Curvaton Physics}

In this section we calculate constraints for our vector curvaton model
assuming that the scaling behavior of $f(t)$ and $m(t)$ ends when
inflation is terminated. This implies that the scaling is controlled
by some degree of freedom which varies during inflation, e.g. the
inflaton field.

In the curvaton scenario the total curvature perturbation can be calculated
as the sum of individual curvature perturbations from the constituent
components of the Universe multiplied by the appropriate weighting
factor. In the current scenario this is written as follows

\begin{equation}
\zeta=(1-\ohw)\zg+\ohw\zw,\label{eq:vCurv-fF-zeta-general}\end{equation}
where $\ohw$ is defined in Eq.~\eqref{eq:curvaton-Omega-hat-def}.
As in the scalar curvaton paradigm, the above is to be evaluated at
the time of decay of the curvaton field.

As was discussed in section~\ref{sub:vCurv-fF-fNL}, if $\hat{m}\gg H_{*}$
at the end of inflation, then the vector field perturbation spectrum
is isotropic and may generate the total curvature perturbation in
the Universe without violating observational bounds on the statistical
anisotropy of the curvature perturbation. If this is the case, we
can assume that $\zg=0$. On the other hand, when $\hat{m}\ll H_{*}$,
the amplitude of the spectrum of the longitudinal component of the
vector field perturbations is substantially larger than the one of
the transverse perturbations. Hence, the curvature perturbation due
to the vector field is excessively anisotropic. To avoid conflict
with observational bounds, the contribution of the vector field to
the curvature perturbation has to remain subdominant. Therefore, for
this scenario, we have to consider $\zg\neq0$ and the curvature perturbation
already present in the radiation dominated Universe must dominate
the one produced by the vector curvaton field.

In Eqs.~\eqref{eq:vCurv-fF-typical-pert-light} and \eqref{eq:vCurv-fF-typical-pert-heavy}
it was shown that the typical value of the field perturbation is $\delta W\sim(3H_{*}/M)(H_{*}/2\pi)$.
If $M\ll H_{*}$ this is because the longitudinal component is dominant
over the transverse ones (see Eq.~\eqref{eq:vCurv-fF-Pp-a-4-light}).
If $M\gg H_{*}$, then the transverse and longitudinal components
are oscillating with same amplitudes (see Eq.~\eqref{eq:vCurv-fF-Pp-a-4-heavy}).

For this reason, at the end of inflation, we can write \begin{equation}
\delta W_{\mathrm{end}}\sim\frac{3H_{*}}{\hat{m}}\frac{H_{*}}{2\pi}\simeq\frac{H_{*}^{2}}{\hat{m}}\,,\label{eq:dWend}\end{equation}
 where we have taken $M=\hat{m}$ and $f=1$ at the end of inflation.
$W_{\mathrm{end}}$ can be found from Eq.~\eqref{eq:rhoA-duringInfl}
by using $(\rw)_{\mathrm{end}}\simeq W_{0}M_{0}\simeq W_{\mathrm{end}}\hat{m}$
(see Eqs.~\eqref{eq:rhoA-duringInfl} and \eqref{eq:rhoA-f-prop-a}).
Thus, \begin{equation}
W_{\mathrm{end}}\sim\frac{\sqrt{(\rw)_{\mathrm{end}}}}{\hat{m}}\,.\end{equation}
Hence, from Eq.~\eqref{eq:vCurv-Intro-zetaW} we calculate the curvature
perturbation of the vector field \begin{equation}
\zeta_{W}\sim\Omega_{\mathrm{end}}^{-1/2}\frac{H_{*}}{\mpl}\,,\label{eq:zetaW}\end{equation}
where $\Omega_{\mathrm{end}}\equiv(\rw/\rho)_{\mathrm{end}}$ is the
density parameter of the vector field at the end of inflation, $\rho_{\mathrm{end}}$
is the total energy density dominated by the inflaton field, and we
have used the Friedman equation: $3m_{P}^{2}H_{*}^{2}=\rho_{\mathrm{end}}$.
Since the vector field must be subdominant during inflation we have
$\Omega_{\mathrm{end}}\ll1$.

Eq.~\eqref{eq:zetaW} is valid in both $\alpha=-1\pm3$ cases. The
only difference is that, in the $f\propto a^{2}$ case, statistically
isotropic curvature perturbations cannot be generated. Hence, only
considerations for statistically anisotropic perturbations in Sec.~\ref{sub:vCurv-fF-fNL}
are relevant.

To calculate the parameter space for this model we note that at the
end of inflation the inflaton field starts oscillating and $w\neq-1$.
Therefore the Hubble parameter decreases as $H(t)\sim t^{-1}$. In
general, the inflaton potential is approximately quadratic around
its VEV. Thus, the coherently oscillating inflaton field corresponds
to a collection of massive particles (inflatons) whose energy density
decreases as $a^{-3}$. When the Hubble parameter falls bellow the
inflaton decay rate $\Gamma$, the inflaton particles decay into much
lighter relativistic particles reheating the Universe. After reheating,
the Universe becomes radiation dominated with the energy density scaling
as $\rg\propto a^{-4}$.

On the other hand, the evolution of the energy density of the vector
field, depends on its mass $\hat{m}$. As discussed in Sec.~\ref{sub:vCurv-fF-zero-mode},
if $\hat{m}\ll H_{*}$ the energy density scales as $\rw\propto a^{-4}$
until the vector field becomes heavy and starts oscillating. If $\hat{m}\gg H_{*}$,
however, the vector field has already started oscillating during inflation
and $\rw\propto a^{-3}$.

To avoid causing an excessive anisotropic expansion period the vector
field must be oscillating before it dominates the Universe and decays.
This requirement implies that \begin{equation}
\Gamma,\:\hat{m}>\gw,\: H_{\mathrm{dom}}\;,\end{equation}
 where $\gw$ is the decay rate of the vector field and $H_{\mathrm{dom}}$
is the value of the Hubble parameter when the vector field dominates
the Universe if it has not decayed already. Working as in Ref.~\cite{Dimopoulos2007},
we can estimate $H_{\mathrm{dom}}$ as \begin{equation}
H_{\mathrm{dom}}\sim\Omega_{\mathrm{end}}\Gamma^{1/2}\mathrm{min}\left\{ 1;\frac{\hat{m}}{H_{*}}\right\} ^{2/3}\mathrm{min}\left\{ 1;\frac{\hat{m}}{\Gamma}\right\} ^{-1/6}.\end{equation}
Similarly, if the vector field decays before it dominates, the density
parameter just before the decay is given by \begin{equation}
\Omega_{\mathrm{dec}}\sim\Omega_{\mathrm{end}}\left(\frac{\Gamma}{\gw}\right)^{1/2}\mathrm{min}\left\{ 1;\frac{\hat{m}}{H_{*}}\right\} ^{2/3}\mathrm{min}\left\{ 1;\frac{\hat{m}}{\Gamma}\right\} ^{-1/6}.\label{eq:Omega-dec}\end{equation}
 where $\Omega_{\mathrm{dec}}\equiv\left(\ow\right)_{\mathrm{dec}}$.
Combining the last two equations and using Eq.~\eqref{eq:zetaW}
we can express the inflationary Hubble scale as\begin{equation}
\frac{H_{*}}{\mpl}\sim\Omega_{\mathrm{dec}}^{1/2}\;\zeta_{W}\,\mathrm{min}\left\{ 1;\frac{\hat{m}}{H_{*}}\right\} ^{-1/3}\mathrm{min}\left\{ 1;\frac{\hat{m}}{\Gamma}\right\} ^{1/12}\left(\frac{\max\left\{ \Gamma_{W};H_{\mathrm{dom}}\right\} }{\Gamma}\right)^{1/4}.\label{eq:H-expr-gen}\end{equation}
The bound on the inflationary scale can be obtained by considering
that the decay rate of the vector field is $\gw\sim h^{2}\hat{m}$,
where $h$ is the coupling to the decay products. Then we can write
$\max\left\{ \gw;H_{\mathrm{dom}}\right\} \gtrsim h^{2}\hat{m}$.
Furthermore, we must consider the possibility of thermal evaporation
of the vector field condensate during the radiation dominated phase.
If this were to occur, all the memory of the superhorizon perturbation
spectrum would be erased. The bound on $h$, such that the condensate
does not evaporate before its decay, is given in Eq.~\eqref{eq:vCurv-Intro-bound-h}.

From Eq.~\eqref{eq:H-expr-gen} one can see that the parameter space
is maximized if the Universe undergoes prompt reheating after inflation,
i.e. if $\Gamma\rightarrow H_{*}$. To find the parameter space we
investigate two separate cases: when $\hat{m}\gg H_{*}$ and when
$\hat{m}\ll H_{*}$.

\subsubsection{The Statistically Isotropic Perturbation}

The statistically isotropic perturbation can be realized only in the
case when $\alpha=-4$. As mentioned before, if the mass of the vector
field at the end of inflation is larger than the Hubble parameter,
$\hat{m}>H_{*}$, then the field has started oscillating already during
inflation. In this case amplitudes of the longitudinal and transverse
perturbations are equal and therefore the curvature perturbation induced
by the vector field is statistically isotropic. We can assume, in
this case, that the vector field alone is responsible for the total
curvature perturbation in the Universe without the need to invoke
additional perturbations from other fields. Thus, we can set $\zg=0$
in Eq.~\eqref{eq:vCurv-fF-zeta-general} and write \begin{equation}
\zeta\sim\Omega_{\mathrm{dec}}\zw\;.\end{equation}
Using this and the lower bound on $h$ we find from Eq.~\eqref{eq:vCurv-Intro-bound-h}
the lower bound for the inflationary Hubble parameter \begin{equation}
\frac{H_{*}}{\mpl}\gtrsim\left(\frac{\zeta}{\sqrt{\Omega_{\mathrm{dec}}}}\right)^{4/5}\left(\frac{\hat{m}}{\mpl}\right)^{3/5},\label{Hbound0}\end{equation}
where we have taken into account that the parameter space is maximised
when the Universe undergoes prompt reheating, i.e. $\Gamma\rightarrow H_{*}$.
From this expression it is clear that the lowest bound is attained
when the vector field dominates the Universe before its decay, $\Omega_{\mathrm{dec}}\rightarrow1$,
and when the oscillations of the vector field commence at the very
end of inflation, i.e. $\hat{m}\rightarrow H_{*}$. With these values
we find the bounds \begin{equation}
H_{*}\gtrsim10^{9}\:\mathrm{GeV}\quad\Leftrightarrow\quad V_{*}^{1/4}\gtrsim10^{14}\:\mathrm{GeV}\,,\label{eq:constraint-H-heavy-vFd}\end{equation}
where $V_{*}^{1/4}$ denotes the inflationary energy scale and we
used that $\zeta\approx5\times10^{-5}$ from the observations of the
Cosmic Background Explorer.

In view of the above, we can obtain a lower bound for the decay rate
of the vector field. Indeed, using Eqs.~\eqref{eq:vCurv-Intro-bound-h}
and \eqref{eq:constraint-H-heavy-vFd} we find \begin{equation}
\gw\gtrsim\frac{\hat{m}^{3}}{\mpl^{2}}\gtrsim\frac{H_{*}^{3}}{\mpl^{2}}\gtrsim10^{-9}\:\mathrm{GeV}\,.\label{GAbound}\end{equation}

From the above we find that the temperature of the Universe after
the decay of the vector field is $T_{\mathrm{dec}}\sim\sqrt{\mpl\gw}\gtrsim10^{4}\,\mathrm{GeV}$,
which is comfortably higher than the temperature at BBN $T_{\mathrm{BBN}}\sim1\,\mathrm{MeV}$
(i.e. the decay occurs much earlier than BBN), and also higher than
the electroweak phase transition, i.e. the decay precedes possible
electroweak baryogenesis processes.

Since $\hat{m}>H_{*}$, Eq.~\eqref{eq:constraint-H-heavy-vFd} corresponds
to a lower bound on $\hat{m}$. An upper bound on $\hat{m}$ can be
obtained as follows. Because, $\hat{m}>H_{*}\gtrsim\Gamma$, Eq.~\eqref{eq:Omega-dec}
becomes \begin{equation}
\Omega_{\mathrm{dec}}\sim\Omega_{\mathrm{end}}\sqrt{\frac{\Gamma}{\gw}}\;.\label{Odec}\end{equation}
From Eq.~\eqref{eq:vCurv-Intro-bound-h} we have $\gw\gtrsim\hat{m}^{3}/\mpl^{2}$.
Combining this with the above we obtain \begin{equation}
\hat{m}^{3}\lesssim\left(\frac{\Omega_{\mathrm{end}}}{\Omega_{\mathrm{dec}}}\right)\Gamma\mpl^{2}.\label{mG}\end{equation}

Now, when $\alpha=-4$ we have $M\propto a^{3}$ during inflation.
Since the end of scaling occurs when inflation is terminated, for
$a<a_{\mathrm{end}}$ we can write \begin{equation}
\hat{m}=\left(\frac{a_{\mathrm{end}}}{a}\right)^{3}M\simeq\mathrm{e}^{3N_{\mathrm{osc}}}H_{*}\;,\label{mH}\end{equation}
where we considered that the field begins oscillating when $M\simeq H_{*}$
and $N_{\mathrm{osc}}$ is the number of remaining e-folds of inflation
when the oscillations begin. Inserting the above into Eq.~\eqref{mG}
we find \begin{equation}
N_{\mathrm{osc}}\lesssim N_{\mathrm{osc}}^{\mathrm{max}}\equiv\frac{2}{9}\left[\ln\left(\frac{\Omega_{\mathrm{end}}}{\Omega_{\mathrm{dec}}}\right)+\ln\sqrt{\frac{\Gamma}{H_{*}}}+\ln\left(\frac{\mpl}{H_{*}}\right)\right]<\frac{2}{9}\ln\left(\frac{\mpl}{\Omega_{\mathrm{dec}}H_{*}}\right)\,,\label{Nosc}\end{equation}
where in the last inequality we used that $\Omega_{\mathrm{end}}<1$
and $\Gamma\lesssim H_{*}$. Now, considering that $\hat{m}\gtrsim H_{*}$,
Eq.~(\ref{Hbound0}) gives \begin{equation}
\frac{\Omega_{\mathrm{dec}}H_{*}}{\mpl}\gtrsim\zeta^{2}.\label{Hbound1}\end{equation}
Hence, combining Eqs.~\eqref{Nosc} and \eqref{Hbound1} we obtain
\begin{equation}
N_{\mathrm{osc}}^{\mathrm{max}}<-\frac{4}{9}\ln\zeta=4.4\;.\label{Noscbound}\end{equation}
 Thus, in view of Eq.~\eqref{mH}, we obtain the bound $\hat{m}\lesssim\mathrm{e}^{3N_{\mathrm{osc}}^{\mathrm{max}}}H_{*}$,
which results in the following parameter space for $\hat{m}$: \begin{equation}
1\lesssim\hat{m}/H_{*}<10^{6},\end{equation}
where we used Eq.~(\ref{Noscbound}). The above range is reduced
if the decay of the curvaton occurs more efficiently than through
gravitational couplings, i.e. if $h>\hat{m}/\mpl$. Nevertheless,
we see that the parameter space in which the vector field undergoes
isotropic particle production and can alone account for the curvature
perturbation, is not small but may well be exponentially large. Indeed,
repeating the above calculation with $\gw\sim\hat{m}$ (i.e. $h\sim1$)
it is easy to find that \begin{equation}
N_{\mathrm{osc}}=\frac{2}{3}\left[\ln\left(\frac{\Omega_{\mathrm{end}}}{\Omega_{\mathrm{dec}}}\right)+\ln\sqrt{\frac{\Gamma}{H_{*}}}\right].\label{Nosc1}\end{equation}
Hence, using that $\Omega_{\mathrm{end}}<1$ and $\Gamma\lesssim H_{*}$
we obtain \begin{equation}
N_{\mathrm{osc}}^{\mathrm{max}}=-\frac{2}{3}\ln\Omega_{\mathrm{dec}}\lsim3.1\quad\Longleftrightarrow\quad1\lsim\hat{m}/H_{*}<10^{4},\end{equation}
where we used $\Omega_{\mathrm{dec}}\gsim10^{-2}$. This is because,
in the case considered, $\fnl$ is given by Eq.~\eqref{eq:vCurv-fF-fNL-zgg1},
so a smaller $\Omega_{\mathrm{dec}}$ would violate the current observational
bounds on the non-Gaussianity in the CMB temperature perturbations
(see the discussion in section~\ref{sub:fNL-observational-bounds}).

Still, it seems that, to obtain an exponentially large parameter space
for $\hat{m}$, we need $\rw$ not to be too much smaller that $V_{*}$
during inflation and also inflationary reheating to be efficient.
In the case of gravitational decay ($\gw\sim\hat{m}^{3}/\mpl^{2}$)
Eq.~(\ref{Nosc}) has a weak dependence on both $\Omega_{\mathrm{end}}$
and $\Gamma$: $\hat{m}\propto(\Omega_{\mathrm{end}}^{2}\Gamma)^{1/3}$,
which means that the allowed range of values for $\hat{m}$ remains
large even when $\Omega_{\mathrm{end}}$ and $\Gamma$ are substantially
reduced. This is not necessarily so when $\gw\sim h^{2}\hat{m}$,
with $h\gg\hat{m}/\mpl$. Indeed, in this case it can be easily shown
that $\hat{m}\propto h^{-2}\Omega_{\mathrm{end}}^{2}\Gamma$. Therefore,
if $\Gamma$ is very small it may eliminate the available range for
$\hat{m}$. Fortunately, the decay coupling $h$ can counteract this
effect without being too small.

\subsubsection{Statistically Anisotropic Perturbations}

If the vector field is not responsible for the total curvature perturbation
in the Universe, the parameter space is more relaxed. In this case,
the vector field may start oscillating after inflation and hence its
mass is $\hat{m}\ll H_{*}$. However, this means that the curvature
perturbation due to the vector field is strongly statistically anisotropic.
For this reason we can no longer set $\zg$ to zero in Eq.~\eqref{eq:vCurv-fF-zeta-general}
because the curvature perturbation present in the radiation dominated
Universe must be dominant. In other words, the parameter $\xi$ defined
in Eq.~\eqref{eq:vFd-ksi-def} needs to be very small, $\xi\ll1$.

In this case the total curvature perturbation is given in Eq.~\eqref{eq:vCurv-Intro-zeta-anisotropic}.
Inserting this into Eq.~\eqref{eq:H-expr-gen} and considering again
that the lowest decay rate of the vector field is through the gravitational
decay, $\mathrm{max}\left\{ \gw;H_{\mathrm{dom}}\right\} \geq\hat{m}^{3}/\mpl^{2}$
we find \begin{equation}
\frac{H_{*}}{\mpl}>\left(\frac{g\,\zeta^{2}}{\Omega_{\mathrm{dec}}}\right)^{3/4}\left(\frac{\hat{m}}{\mpl}\right)^{5/8}\left(\frac{\Gamma}{\mpl}\right)^{-3/8}\mathrm{min}\left\{ 1;\frac{\hat{m}}{\Gamma}\right\} ^{1/8}.\label{eq:Hbound-anisotropic}\end{equation}
The above suggests that the lower bound on $H_{*}$ is minimised for
prompt reheating with $\Gamma\rightarrow H_{*}$. Also, from observations
we know that the statistically anisotropic contribution to the curvature
perturbation must be subdominant. Thus, the vector field should not
dominate the Universe before its decay. Hence, using $\Gamma\rightarrow H_{*}$
and $\ow<1$ we obtain \begin{equation}
\frac{H_{*}}{\mpl}>\sqrt{g}\,\zeta\,\sqrt{\frac{\hat{m}}{\mpl}}\,.\end{equation}
From this expression it is clear that the parameter space for $H_{*}$
is maximised for the lowest mass value. The minimum mass of the vector
field can be estimated from the requirement that the field decays
before BBN. Because the lowest decay rate is the gravitational decay,
this condition reads $\hat{m}^{3}/\mpl^{2}\gsim T_{\mathrm{BBN}}^{2}/\mpl$,
with $T_{\mathrm{BBN}}\sim1\,\mathrm{MeV}$, which corresponds to
$\hat{m}\gsim10^{4}\;\mathrm{GeV}$. Using this, we find that the
parameter space for the vector curvaton model with the statistically
anisotropic curvature perturbations is \begin{equation}
H_{*}>g^{1/2}\;10^{7}\:\mathrm{GeV}\quad\Leftrightarrow\quad V_{*}^{1/4}>g^{1/4}\,10^{13}\:\mathrm{GeV}\,,\end{equation}
i.e. it is somewhat relaxed compared to the statistically isotropic
case (c.f. Eq.~(\ref{eq:constraint-H-heavy-vFd})) depending on the
magnitude of the statistical anisotropy in the spectrum, for which
$g\lsim0.3$ (see the discussion above Eq.~\eqref{eq:bound-on-g}).
This result is valid for both $\alpha=-1\pm3$ cases. From the above
it is evident that there is ample parameter space for the mass of
the vector field\begin{equation}
10\,\mathrm{TeV}\lsim\hat{m}\ll H_{*}.\end{equation}

\subsection{\boldmath Summary for the Massive $fF^{2}$ Model}

In section~\ref{sub:vCurvaton-fF2} we studied a particularly promising
vector curvaton model consisting of a massive Abelian vector field,
with a Maxwell type kinetic term and with varying kinetic function
$f$ and mass $m$ during inflation. The model is rather generic,
it does not suffer from instabilities such as ghosts and may be realized
in the context of theories beyond the standard model such as supergravity
and superstrings (see two tentative examples in Ref.~\cite{Dimopoulos_us(2009)_fF2}).

We have parametrised the time dependence of the kinetic function as
$f\propto a^{\alpha}$, where $a=a(t)$ is the scale factor. Our model
offers two distinct possibilities. If $\hat{m}<H_{*}$ (possible for
$\alpha=-1\pm3$) the vector field can only produce a subdominant
contribution to the curvature perturbation $\zeta$, but it can be
the source of statistical anisotropy in the spectrum and bispectrum.
In fact, non-Gaussianity in this case is predominantly anisotropic,
which means that, if a non-zero $\fnl$ is observed without angular
modulation, then our model is falsified in the $\hat{m}<H_{*}$ case.
The second possibility (possible for $\alpha=-4$ only) corresponds
to $\hat{m}\gsim H_{*}$. In this case the vector field can alone
generate the curvature perturbation $\zeta$ without any contribution
from other sources such as scalar fields. If $\hat{m}\gg H_{*}$,
particle production is isotropic and the model does not generate any
statistical anisotropy. The vector field begins oscillating a few
e-folds before the end of inflation but its density remains constant
until inflation ends. The parameter space for this case can be exponentially
large, i.e. $1\ll\hat{m}/H_{*}<10^{6}$. Significant non-Gaussianity
can be generated, provided the vector field decays before it dominates
the Universe, in which case $\fnl$ is found to be identical to the
scalar curvaton scenario. In other words, if $\hat{m}\gg H_{*}$,
our vector curvaton can reproduce the results of the scalar curvaton
paradigm. Finally, if $\hat{m}\sim H_{*}$ the vector field can alone
generate the curvature perturbation $\zeta$ but it can also generate
statistical anisotropy in the spectrum and bispectrum. In this case,
the anisotropy in $\fnl$ is subdominant and equal to the statistical
anisotropy in the spectrum, which is a characteristic signature of
this possibility. However, the allowed range for $\hat{m}$ values
in this case is very narrow, as shown in Eq.~\eqref{eq:vCurv-fF-m-range},
requiring accurate tuning of the initial conditions. We have also
found that inflation has to occur at energies of $V_{*}^{1/4}\gsim10^{14}\,\mathrm{GeV}$
in the (almost) isotropic and $V_{*}^{1/4}>g^{1/4}\,10^{13}\,\mathrm{GeV}$
in the anisotropic case.

\section{The End-of-Inflation Scenario\label{sec:vFd-SodYok-Scenario}}

\subsection{\boldmath Vector Field Perturbations and $\zeta$}

In this section we consider another model in which a vector field
influences the generation of the curvature perturbation. The model
avoids excessive large scale anisotropy in the Universe by a different
mechanism than the vector curvaton scenario described in section~\ref{sec:vFd-Vector-Curvaton}.
The idea is based on Ref.~\cite{Lyth(2005a)} which was summarized
in section~\ref{sub:End-of-Inflation-Scenario}, where it was shown
that in hybrid inflation models the generation of the curvature perturbation
can be realized due to the inhomogeneous end of inflation. Yokoyama
and Soda~\cite{Yokoyama_Soda(2008)} used this idea to generate the
anisotropic contribution to the total curvature perturbation. In their
model the anisotropy is generated at the end of inflation due to the
vector field coupling with the waterfall field. In other words the
scalar field $\sigma$ of section~\ref{sub:End-of-Inflation-Scenario}
is changed by the vector field $A_{\mu}$. In this section we calculate
the non-Gaussianity of the model in Ref.~\cite{Yokoyama_Soda(2008)}
using the formalism developed in section~\ref{sec:vFd-Perturbations-and-z}.

This scenario uses the conformal invariance breaking of the $U\left(1\right)$
vector field through the non-canonical kinetic function of the form
$f\left(t\right)F_{\mu\nu}F^{\mu\nu}$:\begin{equation}
S=\int\sqrt{-\mathcal{D}_{g}}\left(-\frac{1}{4}f\left(t\right)F_{\mu\nu}F^{\mu\nu}-\ldots\right)\d x4,\label{eq:SodYok-kinetic-term}\end{equation}
where $F_{\mu\nu}=\partial_{\mu}A_{\nu}-\partial_{\nu}A$. 

This action is only written for the conformal invariance breaking
term, the dots represent other terms which give inflation with practically
constant $H$, and generate $f\left(t\right)$ without having any
other effect on the evolution of the gauge field during inflation.
For the vector field to be gauge invariant any scalar field coupled
to $A_{\mu}$ must have zero expectation value (no spontaneous symmetry
breaking) with negligible quantum fluctuation around that value. 

This form of the conformal invariance breaking was considered in many
papers. Starting from Ref.~\cite{Ratra(1992)} such action was often
considered for the generation of the primordial magnetic fields (see
for example Refs.~\cite{Bamba_Yokoyama(2004),BambaSasaki2007,Martin_Yokoyama(2008),Seery(2008)})
and recently in Refs.~\cite{Dimopoulos2007,Dimopoulos_us(2009)_fF2,Dimopoulos_us(2009)_fF2_PRL}
it was considered for the generation of $\zeta$ by the vector field
(see section~\ref{sub:vCurvaton-fF2}). In these papers it was discovered
that a scale invariant perturbation spectrum of the physical, canonically
normalized vector field $W_{\mu}$ is obtained if $f\propto a^{2}$
(we showed in section~\ref{sub:vCurvaton-fF2} that this is the case
for $f\propto a^{-4}$ as well):\begin{equation}
\Pp=\left(\frac{H}{2\pi}\right)^{2}.\label{eq:SodYok-Pp}\end{equation}

In the end-of-inflation scenario of Soda and Yokoyama there are two
components of the curvature perturbation: one generated during inflation
and an anisotropic one, generated by a vector field at the end of
inflation:\begin{equation}
\zeta=\zeta_{\mathrm{inf}}+\zeta_{\mathrm{end}}.\end{equation}
 The first component is due to the perturbation of the light scalar
field, while the second one is due to the perturbation of the vector
field with the kinetic term in Eq.~\eqref{eq:SodYok-kinetic-term}.
Without parity violating terms the power spectra for left handed and
right handed polarizations are equal, while the longitudinal polarization
is absent for a massless vector field. In this situation we find that
parameters $p\left(k\right)$ and $q\left(k\right)$ defined in Eq.~\eqref{eq:vFd-p-q-def}
become\begin{equation}
p=-1\quad\mathrm{and}\quad q=0.\label{eq:SodYok-p-q}\end{equation}

$\zinf$ in this scenario is the statistically isotropic contribution
to the total curvature perturbation. In the slow roll inflation the
spectrum of the scalar field perturbation is $\mathcal{P}_{\phi}=\left(H/2\pi\right)^{2}$,
so that we get\begin{equation}
\mathcal{P}_{+}=\mathcal{P}_{\phi},\end{equation}
And the total isotropic part of the curvature perturbation from Eq.~\eqref{eq:vFd-z-Piso-def}
becomes\begin{equation}
\Pz{iso}=\mathcal{P}_{\phi}N_{\phi}^{2}\left(1+\xi\right),\label{eq:SodYok-Piso}\end{equation}
with $\xi$ given by\begin{equation}
\xi=\left(\frac{N_{W}}{N_{\phi}}\right)^{2}.\end{equation}
Using the expression for the anisotropy parameter $g$ in Eq.~\eqref{eq:vFd-z-g-def}
we find that in this scenario\begin{equation}
g=-\frac{\xi}{1+\xi}.\end{equation}

Taking into account Eq.~\eqref{eq:SodYok-p-q}, the vector $\mathcal{M}_{i}\left(\mathbf{k}\right)$
defined in Eq.(\ref{eq:vFd-M-def}) reduces to the simple form \begin{equation}
\boldsymbol{\mathcal{M}}\left(\mathbf{k}\right)=N_{W}\mathcal{P}_{\phi}\left[\hat{\mathbf{N}}_{W}-\hat{\mathbf{k}}\left(\hat{\mathbf{N}}_{W}\cdot\hat{\mathbf{k}}\right)\right].\label{eq:SodYok-Mi}\end{equation}

\subsection{Hybrid Inflation Model}

To calculate $f_{\mathrm{NL}}$ we consider a specific example of
the hybrid inflation with the potential\begin{equation}
V\left(\phi,\chi,A^{\mu}\right)=V_{0}+\frac{1}{2}m_{\phi}^{2}\phi^{2}-\frac{1}{2}m_{\chi}^{2}\chi^{2}+\frac{1}{4}\lambda\chi^{4}+\frac{1}{2}\lambda_{\phi}\phi^{2}\chi^{2}+\frac{1}{2}\lambda_{A}\chi^{2}A^{\mu}A_{\mu},\end{equation}
which contributes to terms in Eq.~\eqref{eq:SodYok-kinetic-term}
denoted by dots. Here $\phi$ is the inflaton and $\chi$ is the waterfall
field (compare this with the scalar field case in Eq.~\eqref{eq:end-of-infl-sFd-potential}).
The effective mass of the waterfall field for this potential is \begin{eqnarray}
m_{\mathrm{eff}}^{2} & = & -m_{\chi}^{2}+\lambda_{\phi}\phi^{2}-\frac{\lambda_{A}}{f}W_{i}W_{i},\label{eq:SodYok-meff-def}\end{eqnarray}
where we chose the Coulomb gauge with $W_{t}=0$ and $\partial_{i}W^{i}=0$
was chosen. Inflation ends when the inflaton reaches a critical value
$\phi_{c}$ where the effective mass of the waterfall field becomes
tachyonic. But one can see from Eq.\eqref{eq:SodYok-meff-def} that
the critical value is a function of the vector field $\phi_{c}=\phi_{c}\left(W\right)$.
With this in mind $N_{W}^{i}$ and $N_{W}^{ij}$ can be readily calculated:\begin{equation}
N_{W}^{i}=\frac{\partial N}{\partial\phi_{c}}\frac{\partial\phi_{c}}{\partial W_{i}}=N_{c}\frac{\lambda_{A}}{f\,\lambda_{\phi}}\frac{W_{i}}{\phi_{c}},\end{equation}
and\begin{equation}
N_{W}^{ij}=\frac{\partial N}{\partial\phi_{c}}\frac{\partial^{2}\phi_{c}}{\partial W_{i}\partial W_{j}}+\frac{\partial^{2}N}{\partial\phi_{c}^{2}}\frac{\partial\phi_{c}}{\partial W_{i}}\frac{\partial\phi_{c}}{\partial W_{j}}=\frac{N_{W}^{2}}{\phi_{c}N_{c}}\left(C^{2}\delta_{ij}-\hat{W}_{i}\hat{W}_{j}\right),\end{equation}
where we have defined \begin{equation}
N_{c}\equiv\frac{\partial N}{\partial\phi_{c}}\quad\mathrm{and}\quad C\equiv\sqrt{\frac{f\,\lambda_{\phi}}{\lambda_{A}}}\frac{\phi_{c}}{W},\label{eq:SodYok-definitions}\end{equation}
where $W\equiv\left|W_{i}\right|$ and $f$ are evaluated at the end
of inflation and we used the fact that $N_{cc}/N_{c}^{2}\sim N_{\phi\phi}/N_{\phi}^{2}\sim\mathcal{O}\left(\epsilon\right)$
under the slow roll approximation, where $\epsilon$ is the slow roll
parameter defined as $\epsilon\equiv\frac{1}{2}\mpl^{2}\left(V_{\phi}/V\right)^{2}$
(see Eq.~\eqref{eq:Infl-e-slow-roll-parameter}), with the prime
denoting derivatives with respect to the inflaton. As mentioned earlier
the total of perturbations consists of two components: perturbations
of the scalar and vector fields. This gives the following bispectrum
in the equilateral configuration\begin{eqnarray}
\mathcal{B}_{\zeta}^{\mathrm{equil}}\left(\mathbf{k}_{1},\mathbf{k}_{2},\mathbf{k}_{3}\right) & = & \mathcal{B}_{\phi}^{\mathrm{equil}}\left(\mathbf{k}_{1},\mathbf{k}_{2},\mathbf{k}_{3}\right)+\mathcal{B}_{W}^{\mathrm{equil}}\left(\mathbf{k}_{1},\mathbf{k}_{2},\mathbf{k}_{3}\right)=\nonumber \\
 & = & 3\mathcal{P}_{\phi}^{2}N_{\phi}^{2}N_{\phi\phi}+\left[\mathcal{M}_{i}\left(\mathbf{k}_{1}\right)N_{W}^{ij}\mathcal{M}_{j}\left(\mathbf{k}_{2}\right)+\mathrm{c.p.}\right]=\label{eq:SodYok-Bequil}\\
 & = & \mathcal{P}_{\phi}^{2}N_{\phi}^{4}\frac{\xi^{2}}{N_{c}\phi_{c}}3\left[\left(C^{2}-1\right)-\left(\frac{7}{8}C^{2}-1\right)W_{\bot}^{2}-\frac{3}{16}W_{\bot}^{4}\right].\nonumber \end{eqnarray}
The mixed term $\mathcal{B}_{\phi W}^{\mathrm{equil}}$ is absent
from Eq.(\ref{eq:SodYok-Bequil}) because in this model $N_{\phi W}^{i}=0$.
By using the expression for the isotropic power spectrum in Eq.~\eqref{eq:SodYok-Piso}
and the bispectrum in Eq.~\eqref{eq:SodYok-Bequil} from the definition
of $\fnle$ in Eq.~\eqref{eq:vFd-fNL-equil-def} we obtain \begin{equation}
\frac{6}{5}f_{\mathrm{NL}}^{\mathrm{equil}}=\eta g^{2}\left[\left(C^{2}-1\right)-\left(\frac{7}{8}C^{2}-1\right)W_{\bot}^{2}-\frac{3}{16}W_{\bot}^{4}\right],\label{eq:SodYok-fNL-end-of-infl-equil}\end{equation}
where the slow parameter $\eta$ is equal to $\eta\equiv\mpl^{2}V_{\phi\phi}/V=m_{\phi}^{2}\mpl^{2}/V_{0}$
and $\left.\mpl N_{c}=1/\sqrt{2\epsilon_{c}}\right.$, with $\epsilon_{c}$
being the $\epsilon$ parameter evaluated at the end of inflation.
Similarly, for the squeezed configuration we find\begin{equation}
\frac{6}{5}f_{\mathrm{NL}}^{\mathrm{local}}=\eta g^{2}\left[\left(C^{2}-1\right)-\left(C^{2}-1\right)W_{\bot}^{2}-\frac{1}{4}(\sin\varphi)^{2}W_{\bot}^{4}\right].\label{eq:SodYok-fNL-end-of-infl-local}\end{equation}
 In this equation $\varphi$ is the angle between the vectors $\mathbf{k}_{1}$
and $\mathbf{W}_{\bot}$ (see Figure~\ref{fig:vect-projections}).

We find that $f_{\mathrm{NL}}^{\mathrm{equil}}$ and $f_{\mathrm{NL}}^{\mathrm{local}}$
are functions of $W_{\bot}$, i.e. they are anisotropic and correlated
with the statistical anisotropy. Also the level of non-Gaussianity
is proportional to the anisotropy parameter squared, $f_{\mathrm{NL}}\propto g^{2}$,
as in the vector curvaton model. However, the angular modulation of
$\fnl$ in this scenario is different from the curvaton scenario.
From Eqs.~\eqref{eq:SodYok-fNL-end-of-infl-equil} and \eqref{eq:SodYok-fNL-end-of-infl-local}
we see the additional modulation term proportional to $W_{\perp}^{4}$.
This term is absent in the vector curvaton scenario.

As was mentioned earlier, in this model the vector field during inflation
is massless and, therefore, gauge invariant. The homogeneous value
of such vector field can be set to zero by an appropriate gauge choice.
However, as seen from Eqs.~\eqref{eq:SodYok-definitions} and \eqref{eq:SodYok-fNL-end-of-infl-equil},
\eqref{eq:SodYok-fNL-end-of-infl-local} calculated predictions do
depend on the homogeneous value of the vector field $W$. Therefore,
for this model as it stands, the interpretation of the results are
not clear. Although terms proportional to $W_{\perp}^{4}$ in $\fnl$
expressions do not depend on $C$ and consequently on the gauge choice.

\subsection{Summary of the End-of-Inflation Scenario}

In section~\ref{sec:vFd-SodYok-Scenario} we have considered a model
proposed in Ref.~\cite{Yokoyama_Soda(2008)}. In this model the energy
density of the vector field is subdominant throughout the history
of the Universe. However, it influences the generation of $\zeta$
by modulating the end of inflation through the coupling to the waterfall
field. The conformal invariance of the massless vector field is broken
by the time dependent kinetic function as in section~\ref{sub:vCurvaton-fF2}.
We consider a scale invariant perturbation spectrum of the vector
field with the kinetic function scaling as $f\propto a^{2}$. In this
model the vector field is gauge invariant, therefore its particle
production is anisotropic and the curvature perturbation generated
due to this field is statistically anisotropic. 

We have calculated the non-linearity parameter $\fnl$ in this model
and found that it is correlated with anisotropy in the power spectrum,
as in the vector curvaton scenario. In this model too $\fnl$ has
an angular modulation with the amplitude of the same order as the
isotropic part. In addition it has the modulation term, proportional
to $W_{\perp}^{4}$, which is absent in the vector curvaton model.

\pagebreak{}

\thispagestyle{empty}\onehalfspacing~

\pagebreak{}

\chapter{Summary and Conclusions}

The successes of the standard Hot Big Bang theory in explaining the
structure and evolution of the Universe since the very first second
until today is very impressive. The predictions for abundances of
the light elements are in a very good agreement with observations.
The origin and the process of formation of galaxies and galaxy clusters
are now well understood. However, to reproduce the observable Universe,
the initial conditions of the HBB model must be finely tuned. The
spatial curvature of the Universe must have been incredibly close
to zero near its birth for the Universe to have time to evolve to
the present state, and it must have started being extraordinary smooth
even in regions which were never in causal contact with each other.
In addition, in the framework of the standard HBB model, there are
no mechanisms to explain the origin of tiny primordial density perturbations
which are almost Gaussian, adiabatic and correlated on superhorizon
scales. Such perturbations are observed as temperature fluctuations
in the CMB sky and they seed the growth of large scale structure.

The fine tuning problems may be substantially alleviated by postulating
a period of accelerated expansion at the earliest stages of the evolution
of the Universe. This period is called inflation. In addition to solving
the flatness and horizon problems, the greatest achievement of the
inflationary paradigm is the explanation of the origin of the primordial
density perturbation which has the properties observed in the CMB
sky. According to this paradigm, the primordial density perturbation
originated as quantum fluctuations during the inflationary period.
In Chapter~\ref{cha:Scalars} we have demonstrated how the application
of quantum field theory on a curved space-time background may lead
to the amplification of quantum fluctuations and their conversion
into the classical field perturbation. This perturbation, consequently,
causes the perturbation in the curvature of space-time. Much later,
after inflation, when the wavelengths of the perturbation become smaller
than the horizon size, it seeds the formation of structure in the
Universe due to the process of gravitational instability.

To describe the formation and evolution of the cosmological perturbation
we have used a very important quantity: the curvature perturbation
$\zeta$. This quantity is constant throughout the history of the
Universe, except during those periods when the total pressure of the
Universe is not a unique function of the energy density. In other
words, when pressure is not adiabatic. To show how the classical field
perturbation, originating from quantum fluctuations, is related to
the curvature perturbation $\zeta$, we used the separate universes
approach. In this approach the evolution of the Universe on superhorizon
scales at each space point is treated as that of the separate, unperturbed
Universe with the locally defined expansion rate. The latter is determined
by the average energy density on the flat hypersurface at that point,
where the averaging is performed on a superhorizon scale of interest.

The statistical properties of $\zeta$ provide one of the main tools
in cosmology for observational tests of models of the very early Universe.
We have shown how these properties may be calculated using the $\delta N$
formalism. It was applied to calculate the power spectrum and the
bispectrum at tree level for three models, namely: the single field
inflation, the end-of-inflation and the curvaton scenarios. In the
treatment of these three models, we have assumed that $\zeta$ is
generated solely by quantum fluctuations of scalar fields. In Chapter~\ref{cha:Vectors}
we showed that quantum fluctuations of vector fields may contribute
or even generate the total curvature perturbation in the Universe
as well.

However, a massless, canonically normalized $U\left(1\right)$ vector
field cannot produce $\zeta$ because, being conformally invariant,
its quantum fluctuations are not amplified during inflation. And even
if they were amplified, such a field cannot dominate the Universe
without producing excessive large scale anisotropy, i.e. excessive
anisotropic expansion of the Universe, although in most scenarios
the vector field must dominate or nearly dominate the Universe to
generate $\zeta$. In section~\ref{sec:vFds-in-Cosmology} we discuss
possibilities of breaking the conformal invariance of vector fields
and introduce four mechanisms for the generation of $\zeta$ by vector
fields without producing an excessive large scale anisotropy.

In section~\ref{sec:vFd-Perturbations-and-z} we have extended the
$\delta N$ formalism to include perturbations of vector fields. In
contrast to the scalar field, which has one degree of freedom (DoF),
the massive vector field has three DoF. Therefore, in a theory with
a massive vector field we must consider quantum fluctuations for all
three of them. To calculate the evolution of each DoF they were decomposed
into the longitudinal and two circular polarization vectors. This
choice is advantageous because each polarization vector transforms
differently under the Lorentz group. Therefore, we can be sure that
they do not mix in the course of evolution. We found that in general
the amplification of quantum fluctuations is not the same for all
three DoF. In other words, the particle production of a vector field
is not in general isotropic. This results in different values of n-point
correlation functions for each polarization.

To quantify the anisotropy in the particle production we introduced
two parameters $p\left(k\right)$ and $q\left(k\right)$ in Eq.~\eqref{eq:vFd-p-q-def},
where $k$ is the wavevector. The $q$ parameter quantifies the difference
in the power spectra of two transverse polarization modes. It is non-zero
only in parity violating theories. The parameter $p$ quantifies the
difference in the longitudinal power spectrum and the average of the
transverse ones. If both parameters are equal to zero, the particle
production of the vector field is isotropic. However, if any of these
are non-zero, the particle production is anisotropic. The values of
$p$ and $q$ parameters are determined by the mechanism which brakes
the conformal invariance. 

If the vector field with anisotropic particle production generates
or affects the curvature perturbation, the latter is statistically
anisotropic, i.e. statistical properties of $\zeta$ are not invariant
under rotations. The power spectrum of such perturbation will have
an angular modulation. To the lowest order we can express it as \cite{Ackerman_etal(2007)}\begin{equation}
\Pz{}=\Pz{iso}\left[1+g\left(\hat{\mathbf{n}}\cdot\hat{\mathbf{k}}\right)^{2}\right],\end{equation}
where $\Pz{iso}$ is the isotropic part of the spectrum, $\hat{\mathbf{n}}$
is the unit vector along the preferred direction and $g$ parametrizes
the amount of modulation. In the vector field models $\hat{\mathbf{n}}$
is in the direction of the homogeneous vector field. $\Pz{iso}$ in
these models may be solely due to the vector field, if $g$ satisfies
the observational bounds, or it may be dominated by some other, statistically
isotropic source of $\zeta$. The present observational bound on the
anisotropy in the spectrum of $\zeta$ is $g\lesssim0.3$ (see the
discussion above Eq.~\eqref{eq:bound-on-g}). The value of $g$ is
determined by the mechanism which generates the curvature perturbation
and by the value of $p$. In this thesis we consider two such mechanisms:
the vector curvaton and the end-of-inflation scenarios.

The vector curvaton scenario, first proposed in Ref.~\cite{Dimopoulos2006},
uses the fact that a heavy vector field oscillates rapidly with the
frequency much larger than the Hubble parameter. The time averaged
pressure of such field is zero and the energy density decreases with
the scale factor as $a^{-3}$. Thus, the heavy vector field acts as
pressureless, isotropic matter and can dominate the Universe without
producing excessive large scale anisotropy. In accord with the curvaton
scenario, the vector field dominates (or nearly dominates) the Universe
after reheating, when the latter is radiation dominated. The curvaton
imprints its perturbation spectrum and decays before the BBN. The
perturbation spectrum of the vector curvaton field is acquired during
inflation, when the field is light and its energy density is negligible
compared to the inflaton one. During this period the values of parameters
$p$ and $q$ are determined, depending on the mechanism of conformal
invariance breaking.

In section~\ref{sub:vCurv-generic-fNL} the general predictions for
the non-linearity parameter $\fnl$ are derived in the vector curvaton
scenario with $p\ne0$ and $q\ne0$. First, we find that $\fnl$ has
an angular modulation, similarly to the power spectrum. The amplitude
of the modulation is parametrized by $\mathcal{G}$ given by\begin{equation}
\fnl=\fnli\left(1+\mathcal{G}\: W_{\perp}^{2}\right),\end{equation}
where $W_{\perp}$ is the projection of the unit vector of the preferred
direction onto the plane of vectors $\mathbf{k}_{1}$, $\mathbf{k}_{2}$
and $\mathbf{k}_{3}$ which were used to calculate the bispectrum.
The preferred direction is determined by the direction of the homogeneous
vector field. Therefore, we find that both, the power spectrum and
$\fnl$ have the same direction of angular modulation. Another important
prediction of the vector curvaton scenario is that the magnitude of
$\fnl$ is correlated with the anisotropy in the power spectrum, i.e.
$\fnli\propto g^{2}$. 

We calculated $\fnl$ in the squeezed and equilateral configurations
(Eqs.~\eqref{eq:vCurv-fNL-equil-iso-gen}, \eqref{eq:vCurv-fNL-curG-equil}
and \eqref{eq:vCurv-fNL-local-iso-gen}, \eqref{eq:vCurv-fNL-curG-local})
and found that only the equilateral configuration is sensitive to
the parity violating terms in the Lagrangian of the theory. If the
theory is parity conserving, isotropic parts of $\fnl$ are equal
in both configurations. Therefore, the detection of different values
of $\fnli$ in the squeezed and equilateral configurations would indicate
parity violation. However, the amplitude of the angular modulation
$\mathcal{G}$ is not equal in the squeezed and equilateral configurations
for both parity violating and conserving theories. In addition, the
anisotropic part of $\fnl$ dominates over the isotropic part if $p>1$.
Although presently there are no observational constraints on the values
of $\mathcal{G}$ in the squeezed and equilateral configurations,
the detection of them would allow a unique determination of $p$ and
$q$, and therefore, would constraint very tightly the possible conformal
invariance breaking mechanisms for the vector field during inflation.

If, on the other hand, the particle production is isotropic, i.e.
$p=0$ and $q=0$, the predictions of the vector curvaton scenario
do not differ from the standard scalar curvaton case. However, this
offers a possibility to generate the total curvature perturbation
in the Universe solely by the vector field, without directly invoking
scalar fields at all. But even if the particle production is anisotropic
with $\left|p\right|\lesssim0.3$ and any value of $q$, the vector
field can still generate the total $\zeta$ with the amount of statistical
anisotropy satisfy observational bounds.

To find the values of $p$ and $q$, we consider two mechanisms of
breaking the conformal invariance. In the first one a massive Abelian
vector field is non-minimally coupled to gravity through the Ricci
scalar, see Eq.~\eqref{eq:vCurv-RA-Lagran-gen}. We calculate the
perturbation power spectra for all three polarizations and find that
the scale invariance is achieved if the non-minimal coupling constant
is equal to $1/6$ and the bare mass of the vector field is much smaller
than the Hubble parameter. Because the given Lagrangian is parity
conserving, the parity violation parameter is $q=0$. The other anisotropy
parameter is $p=1$ in this model. As was discussed after Eq.~\eqref{eq:vFd-z-g-def},
because $p>0.3$ such a vector field cannot produce the total curvature
perturbation in the Universe, without violating observational bounds
on statistical anisotropy. Therefore, the dominant contribution to
$\zeta$ must come from some other, statistically isotropic source.
In the context of the vector curvaton scenario this means that the
vector field must decay before dominating, while the dominant contribution
to $\zeta$ must be present in the radiation dominated background
before the curvaton decay.

We find that, in non-minimally coupled vector curvaton model, the
isotropic part of $\fnl$ is equal to $2g^{2}/\ow$, where $\ow<1$
is the density parameter of the vector field just before its decay.
The amplitudes of anisotropic parts are $1$ and $9/8$ times the
isotropic part in the squeezed and equilateral configurations respectively.
After taking into account all cosmologically relevant bounds we find
that the parameter space for this model is\begin{equation}
H\gtrsim g\:10^{10}\:\mathrm{GeV}\;\Longleftrightarrow\; V^{1/4}\gtrsim g^{1/2}\:10^{14}\:\mathrm{GeV},\end{equation}
where $H$ is the inflationary Hubble parameter and $V^{1/4}$ is
the energy scale of the inflation. From this result it is clear that
the parameter space is large enough for a successful realization of
this scenario in particle physics models.

Another model considered in this thesis is of the vector curvaton
with time dependent kinetic function and mass, see Eq.~\eqref{eq:vCurv-fF-Lagrangian}.
As in the previous model we calculate the superhorizon perturbation
spectra for all three polarizations and find that they are scale invariant
if the mass varies with the scale factor as $m\propto a$ and is smaller
than the Hubble parameter when cosmological scales exit the horizon,
while the kinetic function scales as $f\propto a^{-1\pm3}$.

We assume that degrees of freedom, which modulate the time dependence
of the kinetic function and mass, are stabilized at the end of inflation.
Therefore, the vector field mass becomes constant at that moment,
i.e. $m=\mathrm{constant}\equiv\hat{m}$. Since any constant value
in front of the kinetic function may be absorbed into the definition
of the vector field, we may set $f=1$ at the end of inflation, and
the field becomes canonically normalized. As we saw, the scale invariant
perturbation spectra may be achieved if the kinetic function $f$
is increasing as well as decreasing. If the vector field is a gauge
field, then $f$ is the gauge kinetic coupling. In this case it is
inversely proportional to the gauge coupling $e$ as $f\propto e^{-2}$.
Therefore, an increasing $f\propto a^{2}$ (small during inflation)
would correspond to the strongly coupled regime, while $f\propto a^{-4}$
would correspond to the weak coupling. Therefore, only the second
case may be realized in the particle physics models.

First we calculate the anisotropy in the particle production for $f\propto a^{-4}$
and find that it depends on the mass of the vector field at the end
of inflation $\hat{m}$. Since the Lagrangian of this model has no
parity violating terms, $q=0$. But the value of $p$ depends on $\hat{m}$.
If the vector field is light at the end of inflation $p\ne0$ and
is given by $p=\left(3H/\hat{m}\right)^{2}-1$. If, on the other hand,
the vector field is heavy, $p=0$. Therefore, the light vector field
generates the statistically anisotropic curvature perturbation, while
the heavy field generates the statistically isotropic one.

The isotropic part of $\fnl$ in the $p\gg1$ case is equal to $\fnli=\left(2g^{2}/\ow\right)\cdot\left(3H/\hat{m}\right)^{4}$.
The amplitude of the $\fnl$ angular modulation in this regime is
$\left(3H/\hat{m}\right)^{2}$ and $\frac{1}{8}\left(3H/\hat{m}\right)^{4}$
times larger than the isotropic part in the squeezed and equilateral
configurations respectively.

If the vector field is heavy at the end of inflation, $p=0$ and the
generated curvature perturbation is statistically isotropic. Such
a vector field may generate the total curvature perturbation in the
Universe without the need of scalar field contribution. In this regime
the standard curvaton scenario predictions for the non-Gaussianity
are valid, i.e. if the curvaton decays before domination $\fnl\approx3/\left(2\ow\right)$.
In the opposite case, when it decays being dominant, the generated
$\zeta$ is Gaussian.

For this model, when the vector curvaton is light at the end of inflation,
the allowed range of inflationary Hubble parameter and energy scale
is \begin{equation}
H>g^{1/2}\:10^{7}\:\mathrm{GeV}\quad\Longleftrightarrow\quad V^{1/4}>g^{1/4}\:10^{13}\:\mathrm{GeV},\end{equation}
while the allowed region for the vector field mass at the end of inflation
is \begin{equation}
10\:\mathrm{TeV}\lesssim\hat{m}\lesssim H.\end{equation}
For the heavy field, and consequently statistically isotropic curvature
perturbation, the analogous bounds are\begin{equation}
H>10^{9}\:\mathrm{GeV}\quad\Longleftrightarrow\quad V^{1/4}>10^{14}\:\mathrm{GeV}\end{equation}
and\begin{equation}
1\lesssim\hat{m}/H\lesssim10^{6},\end{equation}
were we considered that the vector field produces the total curvature
perturbation. Although for the statistically isotropic case the parameter
space is somewhat reduced, in both cases it is large enough for a
successful implementation in realistic particle physics models.

So far we have discussed only the case $f\propto a^{-4}$. In the
case of increasing kinetic function with $f\propto a^{2}$, the same
results apply, but the vector field has to be light and can produce
only statistically anisotropic $\zeta$.

In the final section~\ref{sec:vFd-SodYok-Scenario} of this thesis
we calculate the non-Gaussianity in the end-of-inflation scenario
introduced in Ref.~\cite{Yokoyama_Soda(2008)}. The conformal invariance
of the vector field in this model is broken by the time varying kinetic
function, similarly as in the vector curvaton case discussed above.
However, in this model the vector field is massless; therefore, it
has only two degrees of freedom and particle production is necessarily
anisotropic. In the end-of-inflation scenario the vector field is
always subdominant. However, it influences the generation of $\zeta$
through a coupling to the waterfall field of hybrid inflation. In
this way the end of inflation is spatially modulated by the vector
field, i.e. the hypersurface of the synchronous end of inflation does
not coincide with the uniform density hypersurface. We calculated
the non-Gaussianity for this model and found that the amplitude of
angular modulation of $\fnl$ is larger than the value of the isotropic
part, as in the curvaton scenario. However, in contrast to the curvaton
scenario, in this model $\fnl$ has an additional modulation, proportional
to $W_{\perp}^{4}$, where the latter is the projection of the preferred
direction onto the plane of three $\mathbf{k}$ vectors, used to calculate
the bispectrum. 

In summary, we have shown that a vector field can influence or even
generate the total curvature perturbation in the Universe. If the
particle production of the vector field is isotropic, the generated
curvature perturbation by such field is statistically isotropic. Then
the vector field may generate the total $\zeta$ in the Universe without
the direct involvement of scalar fields. In this case observational
predictions for the curvature perturbation are the same as for models
with scalar fields. If, on the other hand, the particle production
of the vector field is anisotropic, the generated contribution to
$\zeta$ by such field is statistically anisotropic. In this case
observational signatures, very distinct from the scalar field case,
will be present: anisotropic power spectrum and $\fnl$, where the
magnitude and the preferred direction of the latter is correlated
with the anisotropy in the spectrum.

Until recently CMB analyses were performed assuming statistical isotropy
of the curvature perturbation. Our results suggest a new observable:
statistical anisotropy. Therefore, it is desirable to reanalyze CMB
maps without imposing rotational invariance \emph{a priory}. Although
current measurements might not be sensitive enough to constraint the
statistical anisotropy in the primordial curvature perturbation (see
Refs.~\cite{Groeneboom_etal(2009)anisotropy2,Rudjord_etal(2009)anisotropic_fNL}),
with an advent of the Planck data the situation will improve considerably.
For example, according to Ref.~\cite{Pullen_Kamionkowski(2007)}
the lowest detectable value of $g$ from WMAP data is $\left|g\right|\gtrsim0.1$.
With an expected performance of the Planck satellite this bound will
be reduced to $\left|g\right|\gtrsim0.02$. Planck measurements will
be much more sensitive to non-Gaussianity as well. Current WMAP bound
is $\left|\fnl\right|\lesssim100$, in case of no detection with Planck
data it will be reduced to $\left|\fnl\right|\lesssim5$, very close
to the cosmic variance limit \cite{Komatsu_Spergel(2001)fNL_bounds}.
Presently there are no observational constraints on the angular modulation
of $\fnl$. With such increase in sensitivity of measurements in a
very near future one expects that anisotropy in the spectrum and bispectrum
will be discovered or constrained very tightly. In case of the discovery,
with the magnitude and anisotropy of $\fnl$ proposed above, it will
be a smoking gun for a vector field contribution to the primordial
curvature perturbation.

Another important advancement in this direction will be the confirmation
or falsification of the presence of the {}``Axis of Evil'', which
suggests that low multipoles of the CMB are aligned along one direction
\cite{Land_Magueijo(2005)_AxisOfEvil}. Presently the statistical
significance of the discovery of the {}``Axis of Evil'' is still
debatable. However, its confirmation will have profound implications:
this would prove the existence of the preferred direction in the Universe.
Such direction cannot be accounted for by scalar fields, but for vector
fields it is natural.

It is necessary for vector field models to be confronted with observations,
in addition, the treatment presented in this thesis should be extended
in several directions. We have investigated only two mechanisms of
breaking the conformal invariance of massless Abelian vector fields:
non-minimal coupling to gravity and the time varying kinetic function.
In the literature on primordial magnetic fields there are many more
mechanisms proposed to brake this invariance. It would be desirable
to explore which of them may give scale invariant perturbation spectra
from quantum fluctuations of vector fields.

Even more so, one would also like to understand the generation of
perturbations from vacuum fluctuations in the anisotropically inflating
Universe. In the scalar field dominated Universe it is natural to
assume isotropic expansion, provided inflation lasted long enough
before cosmological scales exit the horizon, so that according to
the no-hair theorem, initial anisotropy was inflated away. In the
presence of light vector fields, the backreaction on the expansion
of the Universe might not be negligible, generating the large scale
anisotropy. We neglected such backreaction in vector curvaton models
because the vector field energy density is negligible during inflation.
However, the anisotropic expansion can be easily accommodated within
these models or it might be obligatory in others.

We considered two scenarios for the generation of the curvature perturbation:
vector curvaton and end-of-inflation. However, the developed formalism
may be easily extended to include other scenarios that have already
been explored for the contribution of the scalar field perturbation.

Three toy models were presented in this thesis to generate the curvature
perturbation by vector fields. Ultimately any model of the early Universe
must be firmly rooted in realistic particle physics theories. In the
context of inflationary model building `particle physics theories'
can mean two things. It may be the realization of the inflationary
expansion of the Universe in some fundamental theory, currently the
best developed of which is string theory. For the cosmological aspects
of string theory one may see Refs.~\cite{Kallosh(2007)Inflation&Strings,Baumann_etal(2009)Inflation&Strings}
and references therein. Or particle theory may mean an effective field
theory, which accurately describes the Nature at the energies when
cosmological scales exit the horizon. In this direction an extensive
effort exists in explaining the inflationary epoch in the context
of supersymmetry and supergravity (for reviews see Refs.~\cite{Lyth_Riotto(1999),Alabidi_Lyth(2005)review}). 

From this point of view a particularly attractive is the vector curvaton
model with time varying kinetic function $f\left(t\right)$ and mass
$m\left(t\right)$, presented in section~\ref{sub:vCurvaton-fF2}.
These functions, $f$ and $m$, cannot have an explicit time dependence
but must be modulated by some dynamical DoF during inflation. It might
be an inflaton itself, or some other field. A vector field with such
Lagrangian is very natural in string theory, where parameters such
as masses and kinetic functions are modulated by scalar fields called
moduli. Moduli are not fundamental scalar fields, they parametrize
the size and shape of the manifold on which extra dimensions are compactified.
But from our four dimensional perspective they act as scalar fields.
In Ref.~\cite{Dimopoulos_us(2009)_fF2} it was shown that the modulus
field with an exponential potential (which is reasonable for a modulus
field) can play a role of a single DoF driving inflation as well as
modulating the time dependence of the vector field kinetic function
and mass.

From the effective field theory point of view, the time varying kinetic
function and mass is very general in supergravity theories. In this
case $f$ is the gauge kinetic function, which is a holomorphic function
of the scalar fields of the theory. In supergravity the potential
of these scalar fields receive a correction from the K\"{a}hler potential
such that their mass become $m_{\varphi}^{2}\sim H^{2}$ \cite{Dine_etal(1995)SUSYbraking,Dine_etal(1996)baryogenesis,Lyth_Moroi(2004)masses}.
In the inflationary model building this is known as the $\eta$ problem.
Therefore, scalar fields fast-roll during inflation and one expects
a considerable evolution of the gauge kinetic function, which is modulated
by these fields. Indeed, in Ref.~\cite{Dimopoulos_us(2009)_fF2}
it was shown that an expectation of $\dot{f}/f\sim H$ is quite generic.
The time dependence of the mass in these theories may be modulated
by the same or additional DoF through the Higgs mechanism. In the
same work it was shown that the required scaling of the gauge field
mass, i.e. $m\propto a$, can be achieved if the mass of the Higgs
field is $m_{H}\sim H$. This, again, is very reasonable due to corrections
from the K\"{a}hler potential. However, to be a gauge field, the
gauge coupling constant of the vector field must be small. As was
discussed above, this means that only the kinetic function with $f\propto a^{-4}$
scaling is applicable in this case. Fortunately, this is a case which
have the richest phenomenology.

In the context of implementing vector curvaton scenario in the particle
physics theories it is important to note that models considered in
this thesis involve only Abelian vector fields. However, most of gauge
bosons in simple extensions of the Standard Model (SM) are non-Abelian.
Therefore, an investigation of the particle production and the generation
of the curvature perturbation by non-Abelian vector fields is desirable
(a related work can be found in Refs.~\cite{Bartolo_etal(2009)_Bispectrum,Bartolo_etal(2009)Trispectrum}).

The investigation of the very early Universe is exciting for two reasons.
First, it offers a possibility to understand the origin and history
of the observable structure in the Universe. Secondly, it serves as
the giant laboratory to constrain theories of the fundamental physics.
From the second point of view the research in cosmology is complimentary
to the research in particle physics which can be tested by large experiments,
such as LHC. At the time of writing LHC just started operating and
everyone is looking forward with excitement for new discoveries. First
of all, the detection of the Higgs boson is expected. This would prove
the existence of fundamental scalar fields in Nature. If it is not
discovered, the particle physics models without a fundamental Higgs
field will become favorable, such as technicolor. But for inflationary
model building until very recently only scalar fields were considered
for the generation of the curvature perturbation. If such a field
is not discovered, alternative models will become more attractive.
However, currently the only alternatives being explored in the literature
are vector fields.

Another exciting possibility is the discovery of signatures of physics
beyond SM. This will have a profound significance for particle physics
as well as early Universe theories. If these signatures will be compatible
with the supersymmetry, it will be a strong assurance that investigation
of supersymmetric or supergravity models of inflation is the fruitful
direction.

From the astronomy side a large contribution towards the particle
physics theories will be provided by the observations of recently
launched Planck satellite. The most relevant questions for this thesis
which Planck is expected to answer are: does the primordial curvature
perturbation have a detectable level of non-Gaussianity and statistical
anisotropy? If non-Gaussianity and anisotropy is detected and if it
is of the form suggested in this thesis, it will prove the non-negligible
contribution of vector fields to the primordial curvature perturbation.
This will provide a new observable allowing to probe the gauge field
content of the effective field theory which governs the physics at
energies when cosmological scales exit the horizon.

\appendix

\chapter{Calculation of $\boldsymbol{W_{\bot}}$ in Equilateral Configuration\label{cha:Appendix-A-calculation} }

\begin{figure}
\begin{centering}
\includegraphics[width=7cm]{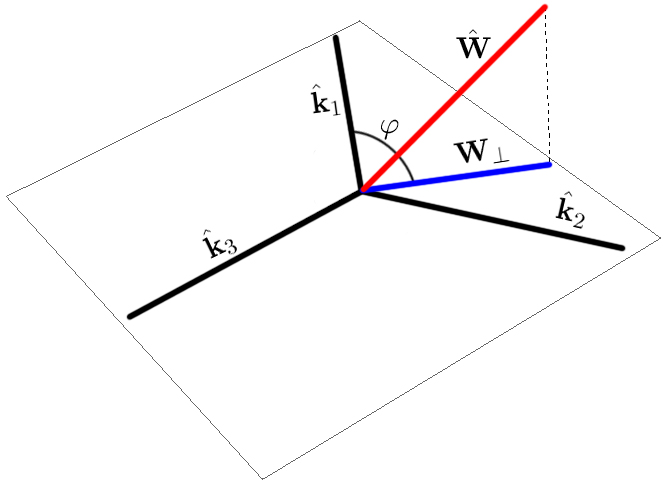}
\par\end{centering}

\caption{\label{fig:vect-projections}$W_{\bot}$ is the projection of the
unit vector $\hat{\mathbf{W}}$ into the plane of three vectors $\hat{\mathbf{k}}_{1}$,
$\hat{\mathbf{k}}_{2}$ and $\hat{\mathbf{k}}_{3}$. $\varphi$ is
the angle between $W_{\bot}$ and $\hat{\mathbf{k}}_{1}$. In this
figure the equilateral configuration of $\hat{\mathbf{k}}_{a}$ is
shown.}

\end{figure}

First note that in the equilateral configuration the unit vectors
$\hat{\mathbf{k}}_{a}$ satisfy $\hat{\mathbf{k}}_{1}+\hat{\mathbf{k}}_{2}=-\hat{\mathbf{k}}_{3}$,
where $a=1,2,3$. If we define scalar products of the unit 3-vector,
$\hat{\mathbf{A}}$, with each $\hat{\mathbf{k}}_{a}$ as $W_{a}\equiv\hat{\mathbf{W}}\cdot\hat{\mathbf{k}}_{a}$,
then in the equilateral configuration $W_{1}+W_{2}=-W_{3}$ and\begin{equation}
\begin{array}{l}
W_{1}^{2}+W_{2}^{2}+W_{3}^{2}=2\left(W_{1}^{2}+W_{1}W_{2}+W_{2}^{2}\right);\\
W_{1}W_{2}+W_{2}W_{3}+W_{3}W_{1}=-\left(W_{1}^{2}+W_{1}W_{2}+W_{2}^{2}\right);\\
W_{1}^{2}W_{2}^{2}+W_{2}^{2}W_{3}^{2}+W_{3}^{2}W_{1}^{2}=\left(W_{1}^{2}+W_{1}W_{2}+W_{2}^{2}\right)^{2}.\end{array}\label{eq:As}\end{equation}
 Let us define a vector $\mathbf{W}_{\bot}$ which is the projection
of $\hat{\mathbf{W}}$ to the plane containing vectors $\hat{\mathbf{k}}_{1}$,
$\hat{\mathbf{k}}_{2}$ and $\hat{\mathbf{k}}_{3}$ (see Figure~\ref{fig:vect-projections}).%
 Then the scalar product of these vectors and $\hat{\mathbf{W}}$
is the same as the product with $\mathbf{W}_{\bot}$: \begin{equation}
\hat{\mathbf{W}}\cdot\hat{\mathbf{k}}_{a}=\mathbf{W}_{\bot}\cdot\hat{\mathbf{k}}_{a}.\end{equation}

Without loss of generality we can assume that the angle between $\mathbf{W}_{\bot}$
and $\hat{\mathbf{k}}_{1}$ is $\varphi$: \begin{equation}
W_{1}\equiv\hat{\mathbf{W}}\cdot\hat{\mathbf{k}}_{1}=\mathbf{W}_{\bot}\cdot\hat{\mathbf{k}}_{1}=W_{\bot}\cos\varphi,\label{eq:A1}\end{equation}
 where $W_{\bot}=\left|\mathbf{W}_{\bot}\right|$. In equilateral
configuration the angle between vectors $\hat{\mathbf{k}}_{1}$ and
$\hat{\mathbf{k}}_{2}$ is $2\pi/3$, and $W_{2}$ becomes\begin{equation}
W_{2}\equiv\mathbf{W}_{\bot}\cdot\hat{\mathbf{k}}_{2}=W_{\bot}\cos\left(\varphi+\frac{2\pi}{3}\right)=-W_{\bot}\left(\frac{1}{2}\cos\varphi+\frac{\sqrt{3}}{2}\sin\varphi\right).\label{eq:A2}\end{equation}
 From the last two equations we get\begin{equation}
W_{1}^{2}+W_{1}W_{2}+W_{2}^{2}=\frac{3}{4}W_{\bot}^{2}.\label{eq:Apr_value}\end{equation}
 Putting this result back into Eq.~\eqref{eq:As} we find\begin{equation}
\begin{array}{l}
W_{1}^{2}+W_{2}^{2}+W_{3}^{2}=\frac{3}{2}W_{\bot}^{2};\\
W_{1}W_{2}+W_{2}W_{3}+W_{3}W_{1}=-\frac{3}{4}W_{\bot}^{2};\\
W_{1}^{2}W_{2}^{2}+W_{2}^{2}W_{3}^{2}+W_{3}^{2}W_{1}^{2}=\frac{9}{16}W_{\bot}^{4}.\end{array}\label{eq:A_values}\end{equation}

\chapter{Scale Invariant Perturbation Spectrum of the Vector Field with Time
Varying Kinetic Function\label{cha:AppendixB-Scale-Inv}}

In section~\ref{sub:vCurv-fF-Power-Sp} it was stated that the vector
field with the time varying kinetic and mass terms in Eq.~\eqref{eq:vCurv-fF-Lagrangian}
acquires a scale invariant spectrum if the kinetic function scales
as $f\propto a^{\alpha}$, where $\alpha=-1\pm3$, and the mass scales
as $m\propto a^{\beta}$, where $\beta=1$. Here we will prove this
result following Ref.~\cite{Dimopoulos_us(2009)_fF2}, where it was
derived by Dr. K. Dimopoulos.

The equation of motion of the transverse modes is calculated in Eq.~\eqref{eq:vCurv-fF-EoM-wp}.
For convenience let us rewrite it here using the conformal time\begin{equation}
w''_{+}+2\frac{a'}{a}w'_{+}+\left[-\frac{1}{4}(\alpha+4)(\alpha-2)\left(aH\right)^{2}+\left(aM\right)^{2}+k^{2}\right]w_{+}=0,\label{eq:AppendixB-EoM-wp}\end{equation}
where primes denote derivatives with respect to the conformal time
$\tau$. This equation is simpler that the one of the longitudinal
mode. Thus we will find the value of $\alpha$ first, and then consider
the equation for the longitudinal mode to determine $\beta$.

The value of $\alpha$ can be readily deduced by noting that Eq.~\eqref{eq:AppendixB-EoM-wp}
reduces to the equation of motion of a scalar field in Eq.~\eqref{eq:FRW-massive-sFd-eq}
if%
\footnote{In Fourier space Eq.~\eqref{eq:FRW-massive-sFd-eq} becomes $\phi_{k}''+2\frac{a'}{a}\phi_{k}'+\left[\left(am\right)^{2}+k^{2}\right]\phi_{k}=0$.%
} \begin{equation}
\alpha=-1\pm3.\label{eq:AppendixB-alpha}\end{equation}
Then in subsection~\ref{sub:Fd-Perturbation-Inflation} it was calculated
that the scalar field acquires a scale invariant perturbation spectrum
if the field is effectively massless and is initially in the Bunch-Davies
vacuum state. Therefore, by analogy we conclude that transverse modes
of the vector field acquire the scale invariant perturbation spectrum
if it is effectively massless, has Bunch-Davies initial conditions
and the kinetic function scales as $f\propto a^{2}$ or $f\propto a^{-4}$.

However, the perturbation spectrum of the longitudinal mode depends
not only on the scaling of the kinetic function but on the scaling
of the mass as well, i.e. on the value of $\beta$. To determine this
parameter let us rewrite Eq.~\eqref{eq:vCurv-fF-EoM-wl} as\begin{equation}
w_{||}''-\frac{4-\alpha+2\beta}{\tau}w_{||}'+\left[-\frac{1}{2}\left(\alpha-2\right)\left(2-\alpha+2\beta\right)\tau^{-2}+k^{2}\right]w_{||}=0,\label{eq:AppendixB-EoM-wll}\end{equation}
where we have also taken into account that the field has to be light
for the perturbation spectrum of transverse modes to be scale invariant
and we used the substitution $\left.\tau=-\left(aH\right)^{-1}\right.$
valid in the de Sitter space-time. This equation can be solved using
the vacuum initial conditions. For the longitudinal mode they are\begin{equation}
\lim_{k\tau\rightarrow-\infty}=\gamma\frac{a^{-1}}{\sqrt{2k}}\mathrm{e}^{-ik\tau},\label{eq:AppendixB-vacuum-cond}\end{equation}
where $\gamma$ is the Lorentz boost factor defined in Eq.~\eqref{eq:vCurv-fF-Lorentz-boost}.
For the light vector field it is equal to \begin{equation}
\gamma=\frac{k/a}{M},\end{equation}
where $M$ is the mass of the physical vector field and is defined
in Eq.~\eqref{eq:vCurv-fF-M-def}. Solving Eq.~\eqref{eq:AppendixB-EoM-wll}
with initial conditions in Eq.~\eqref{eq:AppendixB-vacuum-cond}
we find\begin{equation}
w_{||}=\frac{k}{aM}\frac{\sqrt{-\tau\pi}}{2a}\frac{\mathrm{e}^{-i\frac{\pi}{2}\left(\nu-\frac{3}{2}\right)}}{\sin\left(\pi\nu\right)}\left[J_{\nu}\left(-k\tau\right)-\mathrm{e}^{i\pi\nu}J_{-\nu}\left(-k\tau\right)\right],\end{equation}
where $J_{\nu}$ denotes Bessel function of the first kind, and\begin{equation}
\nu=\frac{1}{2}\sqrt{2(\alpha-2)(2-\alpha+2\beta)+(5-\alpha+2\beta)^{2}}.\label{eq:AppendixB-niu}\end{equation}
At late times, when the mode exits the horizon, the dominant term
of the above solution approaches\begin{equation}
\lim_{k\tau\rightarrow-0}w_{||}=-\frac{1}{\Gamma\left(1-\nu\right)}\frac{\sqrt{-\tau\pi}}{a}\frac{\mathrm{e}^{i\frac{\pi}{2}\left(\nu+\frac{3}{2}\right)}}{\sin\left(\pi\nu\right)}\left(\frac{H}{M}\right)\left(\frac{k}{2aH}\right)^{1-\nu}.\end{equation}
With this solution we find that the power spectrum is given by\begin{equation}
\Pl=\frac{k^{3}}{2\pi^{2}}\left|\lim_{k\tau\rightarrow-0}w_{||}\right|^{2}=\frac{16\pi}{\sin^{2}\left(\pi\nu\right)\left[\Gamma\left(1-\nu\right)\right]^{2}}\left(\frac{H}{M}\right)^{2}\left(\frac{H}{2\pi}\right)^{2}\left(\frac{k}{2aH}\right)^{5-2\nu}.\end{equation}
The above expression becomes scale invariant if $\nu=5/2$, and $\Pl$
becomes\begin{equation}
\Pl=9\left(\frac{H}{M}\right)^{2}\left(\frac{H}{2\pi}\right)^{2}.\end{equation}

Using Eqs.~\eqref{eq:AppendixB-niu} and \eqref{eq:AppendixB-alpha}
we find that $\nu=5/2$ is achieved if\begin{equation}
\beta=-\frac{1}{2}\left(3\pm5\right).\end{equation}
However, the value $\beta=-4$ must be disregarded. This can be seen
using the definition of the mass $M$ in Eq.~\eqref{eq:vCurv-fF-M-def}\begin{equation}
\frac{M}{k/a}\propto a^{-3-\alpha/2}.\end{equation}
The above expression is a decreasing function of $a$ with any value
of $\alpha$ in Eq.~\eqref{eq:AppendixB-alpha}. Thus, with $\beta=-4$
the vector field is massive at early times, which is contradictory
to the requirement for scale invariance of the perturbation spectrum
of transverse modes. Therefore, only the value $\beta=1$ is allowed.\pagebreak{}

\thispagestyle{empty}\onehalfspacing~

\pagebreak{}

\bibliographystyle{D:/Programos/TeX_editors/BibTeX_files/phcpc}
\clearpage\addcontentsline{toc}{chapter}{\bibname}\bibliography{D:/PhD/References/myCosmoRef}

\end{document}